University Of Trento

Doctoral Thesis

# A torsion pendulum ground test of the LISA Pathfinder Free-fall mode

*Author:*
Giuliana Russano

*Supervisor:*
William Joseph Weber
Rita Dolesi

*A thesis submitted in fulfilment of the requirements for the degree of Doctor of Philosophy*

*in the*

LISA Pathfinder Trento group
Department of Physics

May 14, 2015

*"If your experiment needs statistics, you ought to have done a better experiment"*

Ernest Rutherford

*To right the unrightable wrong*
*To love pure and chaste from afar*
*To try when your arms are too weary*
*To reach the unreachable star*

from MAN OF LA MANCHA (1972)

UNIVERSITY OF TRENTO

# *Abstract*

Faculty Name
Department of Phisics

Doctor of Philosophy

**Thesis Title**

by Giuliana Russano

eLISA is the main project for a space interferometer observatory to search for gravitational waves at very low frequencies, from roughly 0.1 $mHz$ to 1 $Hz$. A set of freely falling test masses placed at $10^6$ $km$ apart are the mirrors of the interferometer, joined by a laser link. The gravitational radiation will be detected by measuring the time-varying changes of optical path-length between masses. The mirrors of the interferometer are cubic proof bodies kept inside orbiting spacecrafts. In order to reach the observatory sensitivity requirement they have to be orbiting in a free fall condition, following a geodesic trajectory within an acceleration noise below 3 $fm/s^2/\sqrt{Hz}$.

Each spacecraft has the aim to protect proof masses from external disturbances and also houses the GRS or gravitational reference sensor, that includes a capacitive position sensor, that is an housing around the test masses that senses the relative position with the spacecraft. The sensor itself is a potentially dominant source of force noise that contaminates the free-fall and whose measurement is of fundamental importance in preparation for eLISA.

Given the technological complexity of the eLISA mission, the need of an accurate testing of the free fall and of the knowledge of the noise in a space environment is considered a mandatory phase for the project. For this reason, the LISA Pathfinder mission was developed.

The LISA Technology Package (LTP) is the technological demonstrator for the eLISA mission and will fly aboard the satellite LISA Pathfinder. An arm of eLISA is squeezed into one spacecraft with only two masses in free fall, whose distance is measured by an interferometer. Two position sensors, rigidly mounted on board the spacecraft, surround the free falling proof masses without mechanical contact to them and allow the measurement of relative displacement between the three bodies. The aim of LISA Pathfinder mission is to test the feasibility of the free-fall and its purity, by measuring the residual acceleration between the two test masses at a level of $30 fm/s^2\sqrt{Hz}$ at 1 $mHz$. The differential acceleration noise level that LISA Pathfinder will detect would represent an upper limit for the eLISA acceleration noise budget.

In such a sense the Pathfinder can be considered as a differential accelerometer with the main goal of qualifying the acceleration noise in a space environment. The satellite dynamical scheme, is based on the "drag-free" control system, that is a well known way to reduce disturbances on freely falling test masses. The reference position sensor monitors the position and attitude of the test mass relative to the spacecraft, and this information is used by the control system which commands the spacecraft to follow the orbit of the test mass. Because the satellite can't follow the two masses at the same time, the second mass must be forced to follow either the other one or the spacecraft, by applying small electrostatic forces. The capacitance position sensor can also apply electrostatic forces on the second mass to prevent it from accelerating away because of the existence of the differential steady gravitational pull of the spacecraft on the two test masses. The actuation is needed to compensate for the difference in the gravitational field produced by the satellite in the positions of the two masses.
The fluctuations in the amplitude of the voltages applied to actuate the mass are likely to be the dominant source of differential force noise acting on the test masses. By reducing this source of noise it would be possible to reach a sensitivity well below the noise budget task of the mission.
To do that a particular actuation control scheme was developed, named free-fall mode (or drift mode). It consists of controlling one of the two masses periodically for a short time, producing very short impulses to kick it along the sensitive degree of freedom, instead of a continuous control force. So the mass is allowed to move freely, accelerated around the center of the GRS housing, the rest of the time. The displacement data during this phase, free of electrostatic actuation, is then used to estimate the power spectral density of the remaining noise sources affecting the free-fall motion.
A free-fall mode parallel testing has been successfully implemented on torsion pendulum facility at the University of Trento and the results of the measurement campaign are ready to be shown and discussed. The torsion pendulums are equipped with an LPF-like test mass enclosed in a Gravitational Reference Sensor prototype and have been used successfully, during the recent years, to measure the small forces relevant to the free-fall purity in eLISA and LPF. The pendulum torque sensitivity is around $1fNm/\sqrt{Hz}$ at $1mHz$, corresponding to an equivalent acceleration of $50fm/s^2\sqrt{Hz}$, and thus near the LPF specification. The pendulum can thus allow a quantitatively significant test of the free-fall mode. External torques on the suspended test mass provide the ability to mimic the LPF gravity gradients that must be compensated in orbit, making the torsion pendulum an useful test bed for the LPF actuation and the free-fall mode.
The implementation on ground of the free fall mode is very similar to that in-flight. To simulate a large DC acceleration, the pendulum can be rotated by an angle $\Delta\Phi$ with respect to the inertial sensor, such that a torque is required to keep it centered. This is analogous to the differential actuation force, needed to compensate the self-gravity difference with LISA Pathfinder. The actuation forces can be applied continuously or periodically like in the free-fall mode. The differential forces applied by a pair of electrodes to produce this torque can be made similar to the force levels required in flight, but can be varied at will on ground by choosing the pendulum rotation, something that is not possible in orbit. The on-ground experiment will also allow more flexibility to explore different control strategies, by varying flight and impulse time or control points, and different dynamic configurations, thanks the possibility to have a variable stiffness.
The preliminary part of the work consisted in the characterization of the pendulum background torque noise level in absence of any applied force, when the pendulum is not rotated and no forces are required to keep it centered. The measured angular displacement is then converted into torque by means of the pendulum dynamical model. Then the pendulum is rotated by a large angle to simulate the DC acceleration and a torque is applied continuously to keep the test mass centered. The measured angular displacement is again converted into torque and the contribution from the noisy electrostatic actuation produces an excess in noise power relative to the first configuration. Finally, the free fall control scheme is employed to control the position of pendulum with the impulse-control scheme described above. Quasi-parabolic flights are alternated to kick phases where the actuation is used to invert the pendulum motion and allow the next flight.

Final task is the implementation of the data analysis algorithms to estimate the spectrum of the acceleration of the controlled test mass, and the estimation of system parameters such as the force gradients, stiffness terms, etc.

It is a very challenging goal, because the data are affected by the presence of kicks, that produce a big signal in torque and that must be removed from data producing gap in the data. These are source of aliasing due to high frequency component that can affect the measurement frequency of interest.

To analyze the on-ground data experiment, two main techniques have been developed and implemented. The first technique consists of a sinusoidal fit to the angular time series during each flight to calculate the pendulum equilibrium point, converted afterwards in torque to estimate the power spectral density. The second technique is to convert the angle of the pendulum instantaneously into torque by double differentiation of data, followed by a Blackman-Harris low pass filter and finally in rejecting the data in which the actuation impulses are present. In the end we estimate the power spectrum.

The "free-fall" mode has been tested at the level of $2fNm/\sqrt{Hz}$, at the frequencies of $1mHz$, corresponding to an acceleration of about $100fm/s^2/\sqrt{Hz}$. This level of noise in torque is to be compared with that measured by keeping the pendulum centered with a force of constant DC actuation, which is at a lower level to $1fNm/\sqrt{Hz}$ at $1mHz$.

The discrepancy observed is still under investigation, to determine if its origin is intrinsic to the technique of analysis with pulses and then with gaps in the data.

Moreover, due to high dynamical range of the pendulum angular displacement during each flight, some non linearity in the model of torque was found. The characterization of the pendulum dynamic and sensing non-linearities is one of the main parts of this work.

The ground tests have nevertheless allowed to develop and test quantitatively this innovative method of noise reduction due to actuation noise, contributing to its development for the LISA Pathfinder mission.

Therefore, in Chapter 1 we give an overview about the gravitational wave detection principles, the eLISA mission and in particular to the description of the technological demonstrator mission LISA Pathfinder. Moreover, we focused on the the main characteristics of the Gravitational Reference Sensor for eLISA and LISA Pathfinder, on its design and noise performances and on its electrostatic model.
In the Chapter 2 we will show in detail the free-fall actuation mode concept as will be implemented on board of LTP, paying particularly attention to the acceleration sensitivity requirement for LISA Pathfinder, that show the importance of implementing a free-fall scheme able to reduce the electrostatic actuation noise and the aim of the related ground testing.
The 3rd Chapter presents the one test mass torsion pendulum, as GRS and free fall mode testing facility, with its main features and operation principle and its relevant noise sources of disturbances. We will discuss about the limit and performance of the facility in torque noise floor measurement and in the actuation authority of the used FM FEE electronic, characteristic that will be a basic element of the implementation of free-fall mode on torsion pendulum.
In Chapter 4, the free fall mode on torsion pendulum is presented, with the basic measurement concept. The equations of pendulum dynamic during free fall are then developed, with also the description of controller loop and of the home made FEE circuit.
Chapter 5 address the problem of data analysis technique developed for the laboratory free fall mode, focusing on the two analysis techniques mentioned above, sine fit and instantaneous torque conversion. Preliminary results of data analysis techniques were showed, applied on simulated data coming from a simple harmonic oscillator simulator of pendulum dynamics during free fall mode.
Finally, in Chapter 6 we present results from the real pendulum free fall experiment. Non linearity in the pendulum dynamics and other problems of implementation of the free fall mode arose during the testing campaign, and we address these problems and their solutions in the same chapter. Conclusive phase show torque noise spectra and discussion about their interpretation and results. In Chapter 7 we will summarize the most interesting and innovative aspect of the work presented throughout this thesis: the success in the free fall mode implemented on torsion pendulum facility, but also its limits and features to be improved.

# Contents

# List of Figures

# List of Tables

# Abbreviations

| | |
|---|---|
| **BH** | **B**lackman **H**arris |
| **CSD** | **C**ross **S**pectral **D**ensity |
| **FEE** | **F**ront **E**nd **E**lectronics |
| **GRS** | **G**ravitational **R**eference **S**ensor |
| **IFO** | **I**nterferometerical **O**ptic |
| **IS** | **I**nertial **S**ensor |
| **LPF** | **L**ISA **P**ath**F**inder |
| **LTP** | **L**ISA **T**echnology **P**ackage |
| **PSD** | **P**ower **S**pectral **D**ensity |
| **TM** | **T**est **M**ass |
| **SC** | **S**paceCraft |

*Dedicated to Agata, my little twinkling star.*

# Chapter 1

# eLISA and LISA Pathfinder: gravitational wave detection

Gravitational wave detection is one of the most attractive targets of modern international astrophysical research. The scientific history of gravitational waves starts in the first two decades of the twentieth century, when the German physicist Albert Einstein published his General Relativity Theory, introducing a completely new way of thinking about the relationship between matter, space and time. According to his theory, mass acts on the space-time, dictating how it curves. Compact concentrations of matter and energy modify the intimate structure of spacetime, warping it and changing the distance between points, as compared with a reference ruler, such as the wavelength of a light beam.

During 1916, a few years after the first formulation of the General Relativity theory, Einstein deduced that the information about the variation of the curvature had to propagate through the space at the speed of light by means of waves. He discovered the existence of waveform solutions of the field equations, in which a ripple of space-time propagates through the empty space, as an independent entity, with speed equal to that of light. These are the gravitational waves that carry information on how the change in time of the distributions of matter and energy affect the curvature of space-time.

Gravitational waves interact very weakly with matter and can go through anything without losing intensity significantly. This makes them a powerful tool for the investigation of distant regions, in extreme conditions, but also makes them very difficult to detect. Gravitational waves are, therefore, a fundamental prediction of General Relativity that still has not found a direct experimental proof. Studying this new form of energy will convey rich new information about the behavior, structure, and history of the physical universe, and about physics itself.

In this chapter we will present the basics of the gravitational waves detection and the main principles of what will be the first space interferometer capable of their detection at very low frequency, *eLISA*. The measurement will be achieved by detecting the relative acceleration of free-flying test masses in a constellation of three spacecraft separated by 1 million km. The level of free-fall required for eLISA sets a limit on the parasitic acceleration on the test mass of $3fm/s^2\sqrt{Hz}$ at 0.1 mHz. Due to hard and challenging technology aspects of the mission, a technological demonstrator was designed and will fly at the end of 2015, *LISA Pathfinder*. It will be able to demonstrate a level of free-fall within one order of magnitude of the LISA performance in acceleration noise and frequency. We will describe its main features and will focus on the *Gravitational Reference Sensor*, which is used for nm-level control of the surrounding spacecraft but, more importantly, must define the environment in which TM free-fall near the $fm/s^2$ level is possible.

## 1.1 Gravitational waves and their detection

The idea on which gravitational waves detection is based is the measurement of extremely small changes in distances. It arises from the principle that gravitational waves produce tiny fluctuations in the distance between the masses in free fall, isolated from all other forces except gravity. Such fluctuations can be measured by using the technique of laser interferometry.

The first steps in the detection of these gravitational signals have been made with several experiments on ground as the first resonant bars in the 60s and later with laser interferometers, Michelson type, as LIGO (Laser Interferometer Gravitywave Observatory) and VIRGO.

The principle of operation of a ground interferometer considers the coherent light of a laser, send it along two orthogonal paths to distant free-falling test bodies that act as mirrors, and then recombine the two beams to form an interference pattern. If we assume that the mirrors of the interferometer are very close to a free falling conditions, the effect of gravitational waves is a lengthening and shortening of the arms of the instrument and therefore a change in the optical path of the light. In this way, the interference pattern will be changed. An advantage of this scheme is that it is non-resonant, that is, the natural frequencies of the mirrors suspended are much lower than those of gravitational waves. Consequently, rather than respond to only one of the frequencies of the incident wave, the mirrors simply follow the trend of the wave, whatever its form. In this way it is not limited to only record the passage of the wave, but it also get information about its features. The final oscillations is measured by a photon detector placed in the output from the interferometer and is a simple function of the phase difference of the two light beams divided by the beamsplitter, which propagate in the two arms and which are recombined at the output.

In this scenario, in the 90s the idea of using laser interferometry in space, on a larger scale and in a much more quiet environment than the ground, was born. In a space configuration it is used a laser interferometry between free falling masses at very long distances, measuring the variations in light travel time along the arms due to the tidal deformation of space-time produced by gravitational waves.

eLISA (Laser Interferometer Space Antenna) will be the first large-scale space mission to survey the Universe with gravitational waves [1]. It can be thought of as a modified setup of a Michelson interferometer in space, with an arm length of 1 million km between three spacecraft orbiting the sun and exchanging laser light beams. This will allow the observation of most of the interesting sources of gravitational waves in the $mHz$ frequency range, emitted by coalescing binary black holes and ultra-compact galactic binaries. A greater length of the arms can amplify the effect of gravitational waves, thereby eLISA can measure their signature, that is a fractional squeezing of space-time perpendicular to the direction of propagation, with an amplitude $h = \Delta L/L$ of the order of $10^{-20}$, measuring displacements of the order of fractions of picometers.

## 1.2 eLISA mission overview

eLISA basic measurement idea is to place proof masses in space in true geodesic motion, 1 million km distance apart and to measure interferometrically the distance variations induced by a passing gravitational wave.

Its design considers four proof masses, $46\,mm$ AuPt cube, which serve as mirror for the interferometer, shielded by three spacecraft from the external disturbances in a triangular configuration as in figure 1.1, which form a single Michelson interferometer configuration. The spacecraft follow independent heliocentric orbits without any station-keeping and form a nearly equilateral triangle in a plane that is inclined by 60° to the ecliptic, as visible in figure 1.3.

Each spacecraft contains identical units in which a Gravitational Reference Sensor (GRS) hosts a free-falling test mass that acts both as the end point of an interferometric sensor, and as a geodesic reference test particle [1]. The GRS is a capacitive sensor, made by an cubic electrode housing that hosts 18 electrodes, as shown in figure 1.5, in order to measure the relative position of the proof mass with respect to the spacecraft at $nm/\sqrt{Hz}$ level. It can also provide nN-level electrostatic force actuation on all non-interferometric degrees of freedom.

Each payload includes an optical bench, a telescope for receiving and transmitting light and a laser source, as in figure 1.2. Each telescope of the mother satellite points at one of the distant spacecraft at the other two corners of the triangle.

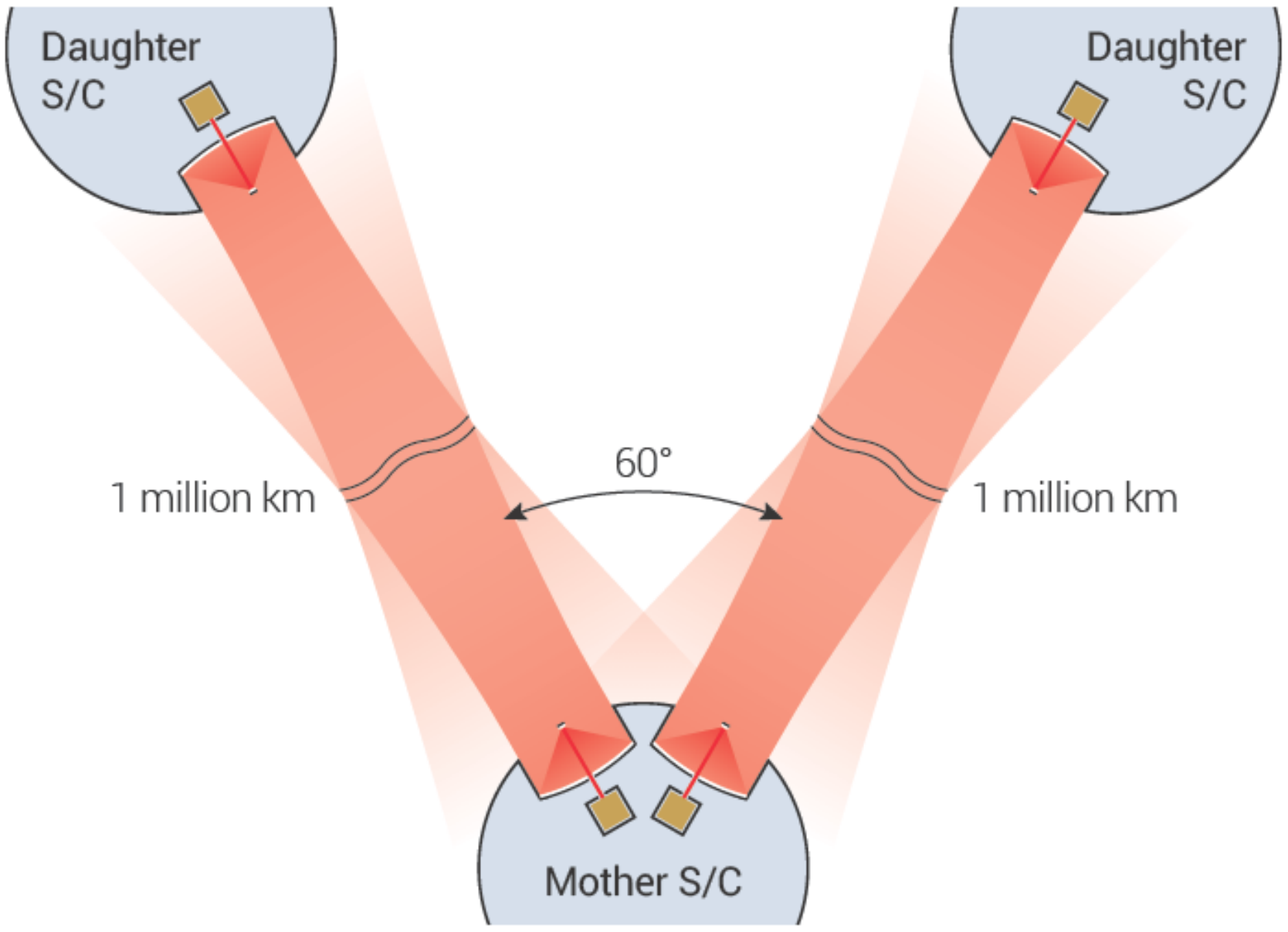

FIGURE 1.1: eLISA configuration. One mother and two daughter spacecraft exchanging laser light form a two-arm Michelson interferometer. [1]

While the proof masses are freely falling in space, the hosting spacecraft must keep as stationary as possible with respect to them and avoid any interference with their geodesic motion. This purpose is obtained measuring and correcting the relative position between spacecraft and proof mass with a drag-free control loop, by which the spacecrafts are actively controlled to remain centered on the test masses along the interferometric axes, without applying forces on the test masses along these axes. The GRS provides the control signals for the drag-free control loop and special microthrusters force the spacecraft to follow the displacements of the freely falling proof mass along that degree of freedom.

### 1.2.1 eLISA sensitivity

In figure 1.4 it is showed the eLISA strain sensitivity in term of noise power spectral density of the instrument. The gravitational wave sensitivity will be limited at low frequency by residual stray acceleration noise $g_n$ in the orbits of the free falling test mass, at a level of

$$S_{g_n}^{1/2} \leq 3\frac{fm}{s^2\sqrt{Hz}}. \tag{1.1}$$

at frequencies from 0.1 to roughly 10 mHz. The acceleration noise is typically divided into two categories. The first one includes a contribution from random position independent stray forces $f_{str}$, such as unshielded non-gravitational external forces, disturbances generated by sources on board the spacecraft, thermal noise and displacement sensor back-action. The second category comprises forces that originate from the spring-like coupling or "stiffness", of the test-mass with respect to the noisy motion of the spacecraft, with spring constant $m\omega_p^2$, and arising from position dependent forces due to the position sensor noise and the finite gain of the drag-free loop. The closed loop residual acceleration, with respect to an inertial frame, such as that provided by the optical wavefront, can be thus expressed as [8]:

$$\ddot{x} = g + \omega_p^2 \left( x_n + \frac{F_{str}}{M\omega_{DF}^2} \right) \tag{1.2}$$

where $M\omega_{DF}^2$ is the drag-free force to displacement gain, and is valid in the limit of high drag-free gain, such that $\omega_{DF}^2 \gg \omega_p^2$. The term $x_n + \frac{F_{str}}{M\omega_{DF}^2}$ is the residual jitter in the spacecraft position around the TM. Much of this term can be subtracted, in the limit that the spacecraft jitter is dominated, at the $nm/\sqrt{Hz}$ level, by the stray force on the satellite $F_{str}$ and not the interferometric position sensor, which is more precise.
At higher frequencies position noise, essentially laser shot noise, is dominant, while above about $5\,mHz$, arm length measurement noise dominates.
Finally, both the satellite and the position sensor themselves could produce force disturbances on the free falling test masses. In order to keep them along the stabilized geodesic orbit, the gravitational reference sensor should perform the position measurement with sufficiently high precision but minimizing the residual force disturbances on the test mass. We will describe its main features in the next section.

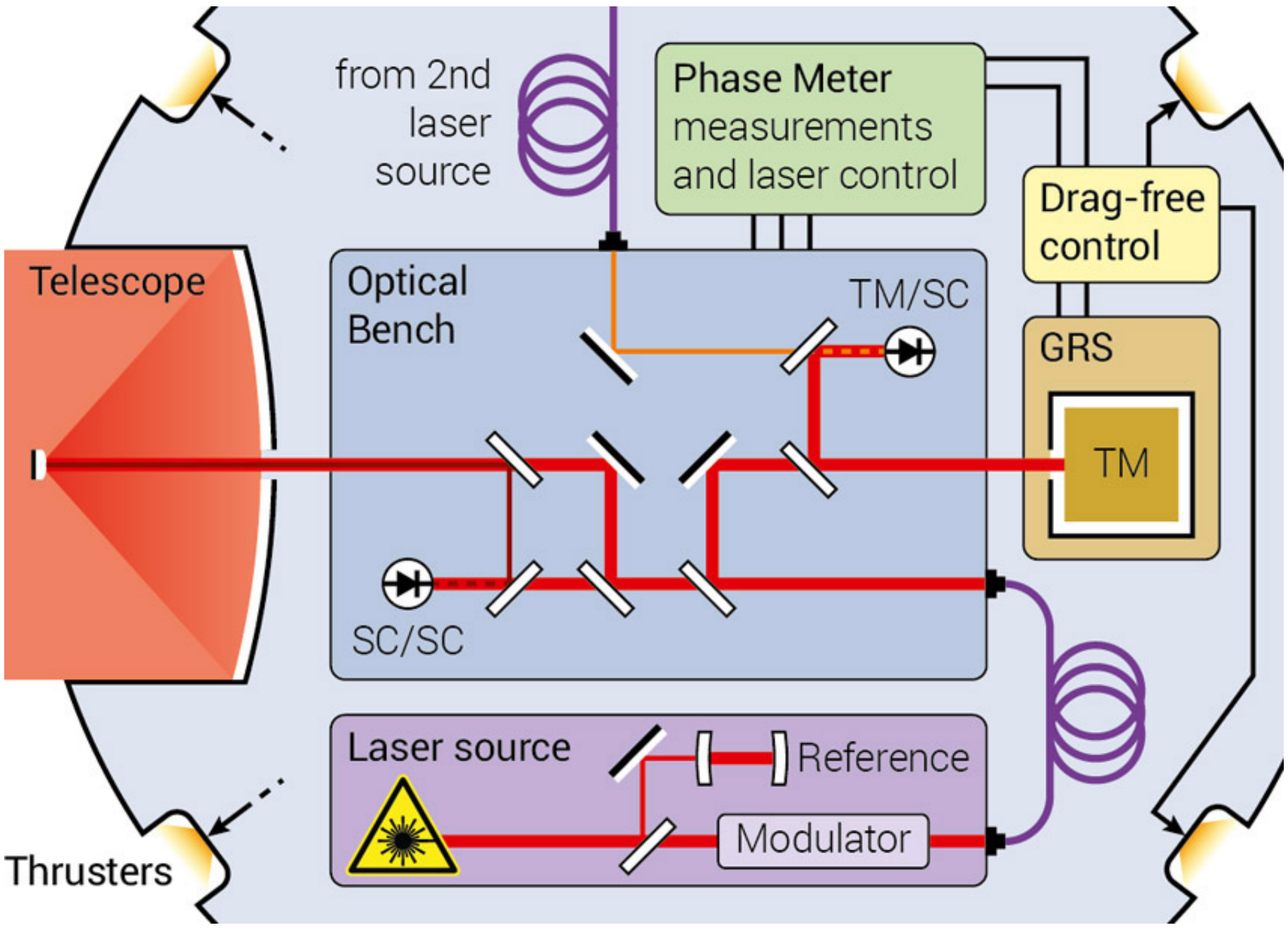

FIGURE 1.2: eLISA payload. Each payload unit contains a 20 cm telescope, the test mass enclosed inside the Gravitational Reference Sensor (GRS) and an optical bench hosting the interferometers. [1]

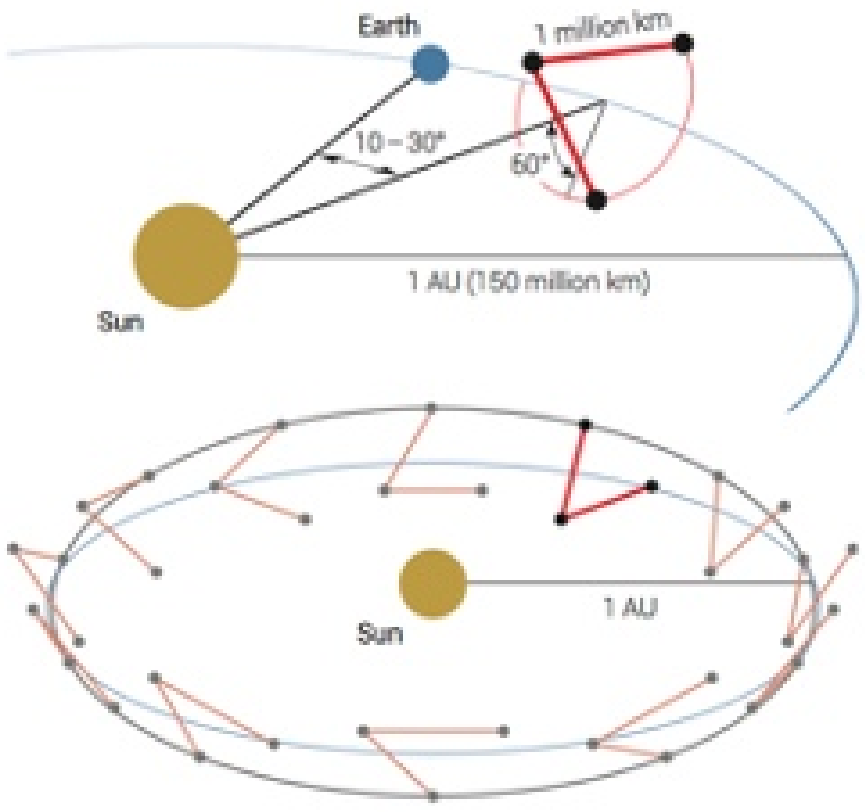

FIGURE 1.3: eLISA set of three orbits in a near-equilateral triangular formation [1].

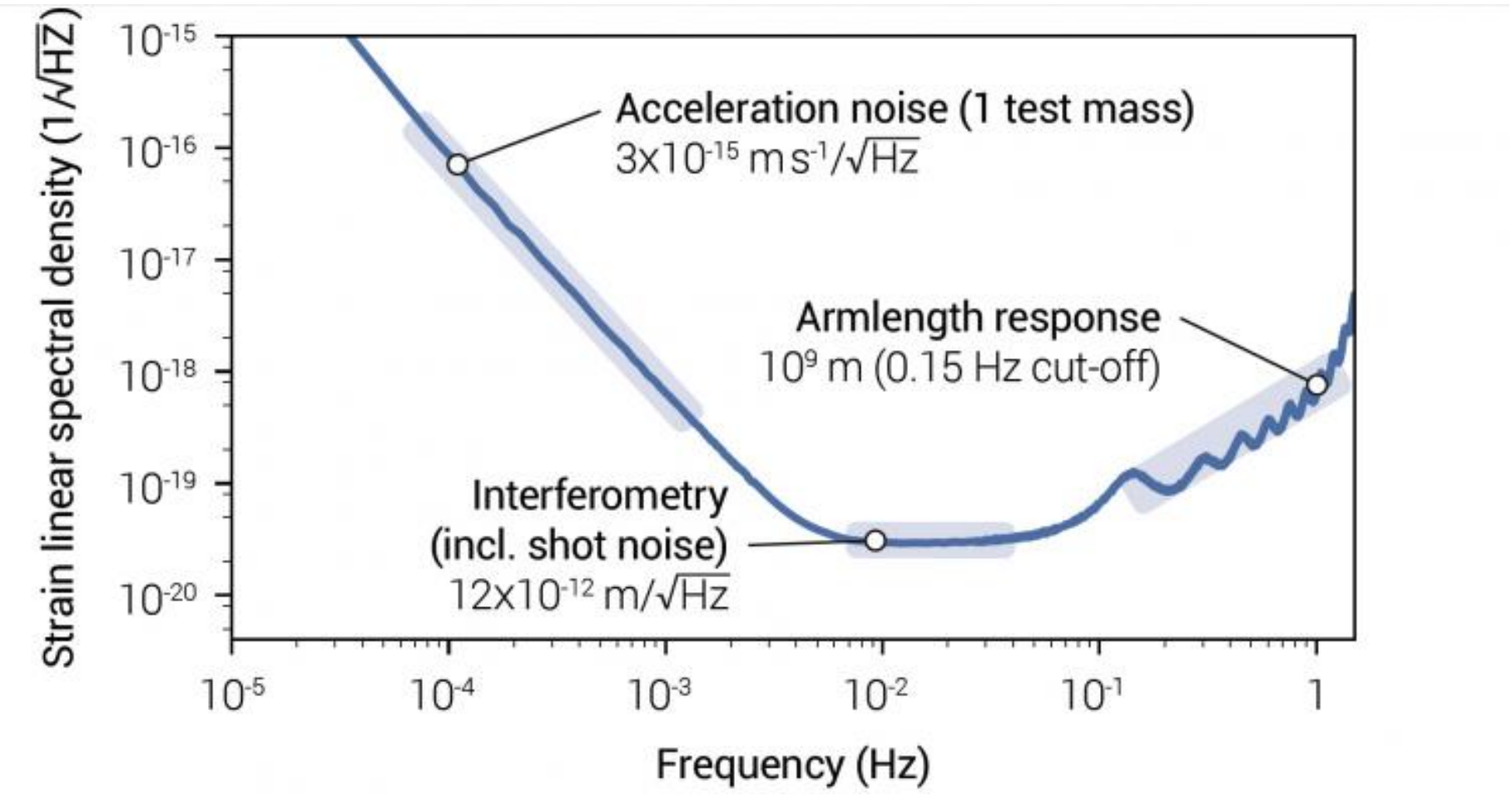

FIGURE 1.4: eLISA sensitivity in terms of strain resolution. The noise spectrum is plotted as a linear spectral density. [1]

## 1.3 The Gravitational Reference Sensor

The Gravitational Reference Sensor is a capacitive readout developed to measure proof mass position, whose performance must meet the requirement of LISA, in terms of position noise, residual couplings and force noise. Its geometrical configuration is showed in figure 1.5. A set of 18 electrodes are hosted by a cubic electrode housing and measure the mass position in six degrees of freedom and orientation relative to the spacecraft [9]. Six pairs of them are used as sensing electrodes (green electrodes in figure 1.5), also used to apply electrostatic forces to the proof masses by modulating the amplitude of an audio frequency carriers (30 - 270 Hz) applied to the electrodes. The actuation is necessary on degrees of freedom other then the sensitive in LISA configuration, while the sensor drive the drag-free loop only along the interferometric axis, sensitive to the gravitational waves.
The readout circuit, or GRS front-end electronics (FEE), has the scheme of a capacitive resonant bridge (see figure 1.6). The test mass motion modulates the difference between the two capacitances couples $C_1$ and $C_2$ facing the test mass, inducing a difference between the currents flowing through the inductance arm of the circuit that can be read out by the low noise pre-amplifiers as a current flowing through the low losses transformer [10].

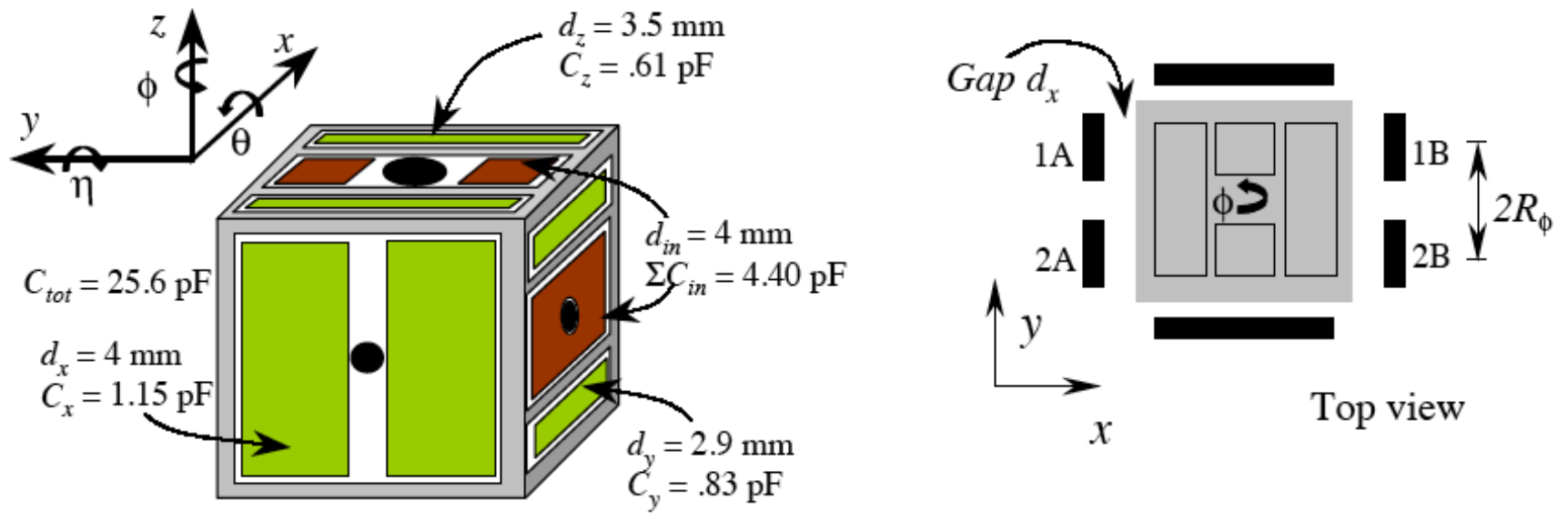

FIGURE 1.5: GRS scheme. Electrode geometry and dimensions refer to the current design adopted for the LTP flight test. Sensing electrodes are green, injection electrodes are red, guard ring surfaces are gray [2].

The capacitance bridge current is provided by capacitively polarizing the TM with an oscillating voltage at resonance frequency $\omega_0 = 2\pi \cdot 100\,kHz$, with roughly $0.6\,V$ amplitude, by applying voltages on the six injection electrodes, 2+2 on the z faces plus 1+1 on the y faces (red electrodes in figure 1.5). The $100\,kHz$ injection bias is also the reference for the phase sensitive detector at the output of differential pre-amplifiers. The signals are extracted by the phase sensitive detectors and demodulated to provide only the component at $\omega_0$, rejecting the electrical noise at different frequencies. This allows the measurement of the effective test mass displacement even in an electrostatically noisy environment and in presence of actuation voltages. The demodulated signals are then A/D converted and processed by the on board computer. Finally, the sum of the signal in the two channels provides the translational displacement of the test mass with respect to the center of the electrode housing, while the difference provides the test mass rotation.

The GRS, has an almost symmetric electrodes configuration, and works with $4.0\,mm$ gaps for the x interferometric axis, sensitive to the gravitational signal, and $3.5\,mm$ and $2.9\,mm$ respectively for the y and z axes, as also evidenced in figure 1.5.

Because the Gravitational Reference Sensor is the closest object to the test masses, it is expected to be a major source of stray forces. It is, thus, important to model all of different sources of stiffness and noise, in force and position, produced by this sensor, as will be done by the technology demonstration mission LISA Pathfinder [11].

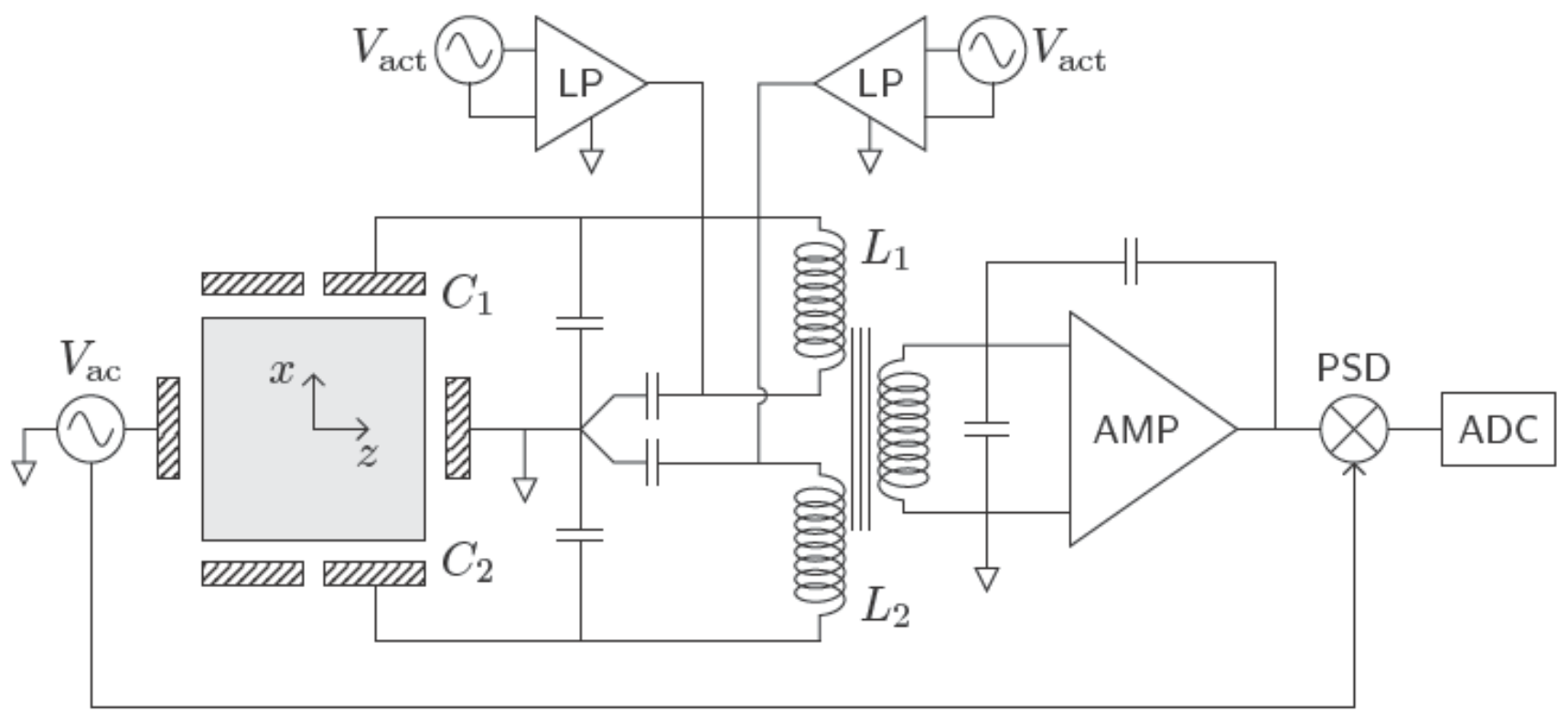

FIGURE 1.6: Scheme of the capacitive resonant bridge readout and actuation circuitry.

## 1.4 LISA Pathfinder, technology demonstrator

Given the technological complexity of the systems required by eLISA, was considered necessary a demonstration mission of the feasibility of the drag-free control for eLISA test masses, in order to put also an overall upper limit on all sources of random force noise. This is the LISA Pathfinder mission (LPF), and will fly at the end of 2015.
LISA Pathfinder has been designed as one single spacecraft (SC) that enclose two test masses (TM) that are not in mechanical contact with it and so are nominally in free fall, providing a downscaled version of one of the eLISA arms [12].
The two cubic proof masses, made of gold-platinum, have 46 $mm$ edge and weight 1.98 $kg$. They are also separated by a nominal distance of $\sim 38\,cm$, as visible in figure 1.7.
The relative displacement of TMs and SC along the sensitive axis, is measured by means of laser interferometers at pm-level of accuracy. More precisely, one interferometer measures the displacement $x_1$ of one of the TMs relative to SC, and the second measures the relative displacements $x_{12}$ between the TMs and serves as the main probe of the drag-free performance.
Each proof mass, with its own electrode housing, is enclosed in a vacuum chamber vented to space, reaching a pressure below $1^{-5}\,Pa$. The laser interferometer light passes through the vacuum chamber wall through an optical window.
LTP will be also equipped with the most crucial aspects of the eLISA technology. One of these is a set of precision $\mu N$ thrusters needed for the drag-free compensation of external disturbances: the spacecraft will follow one of the two test masses in its geodesic motion along x, whereas the second test mass will follow the other according to the interferometer signal, by means of the electrostatic suspension, as we will largely discuss in the next chapter.
Others important apparatus are: the Caging Mechanism, which holds the test masses in place during the launch phase and releases the test mass for the scientific measurement phase [13]; the apparatus called Charge Management System, which by means of UV light and suitable driving voltages controls and remove the net charge accumulated on the test mass due to the exposure to the cosmic ray radiation.

LTP aims to demonstrate the quality of free fall within one order of magnitude from LISA such that the residual acceleration noise of the test masses is proven to be below

$$S_{\Delta g}^{1/2} \leq 3 \cdot 10^{-14} \left[1 + \left(\frac{f}{3\,mHz}\right)^2\right] \frac{m}{s^2\sqrt{Hz}}. \tag{1.3}$$

In order to meet stray force and stiffness requirements of the mission, the position sensor displacement sensitivity must satisfy a measurement noise of $1.8\,nm/\sqrt{Hz}$ at $1\,mHz$ in the three displacement d.o.f, and $200\,nrad/\sqrt{Hz}$ for angular d.o.f in the measurement bandwidth.

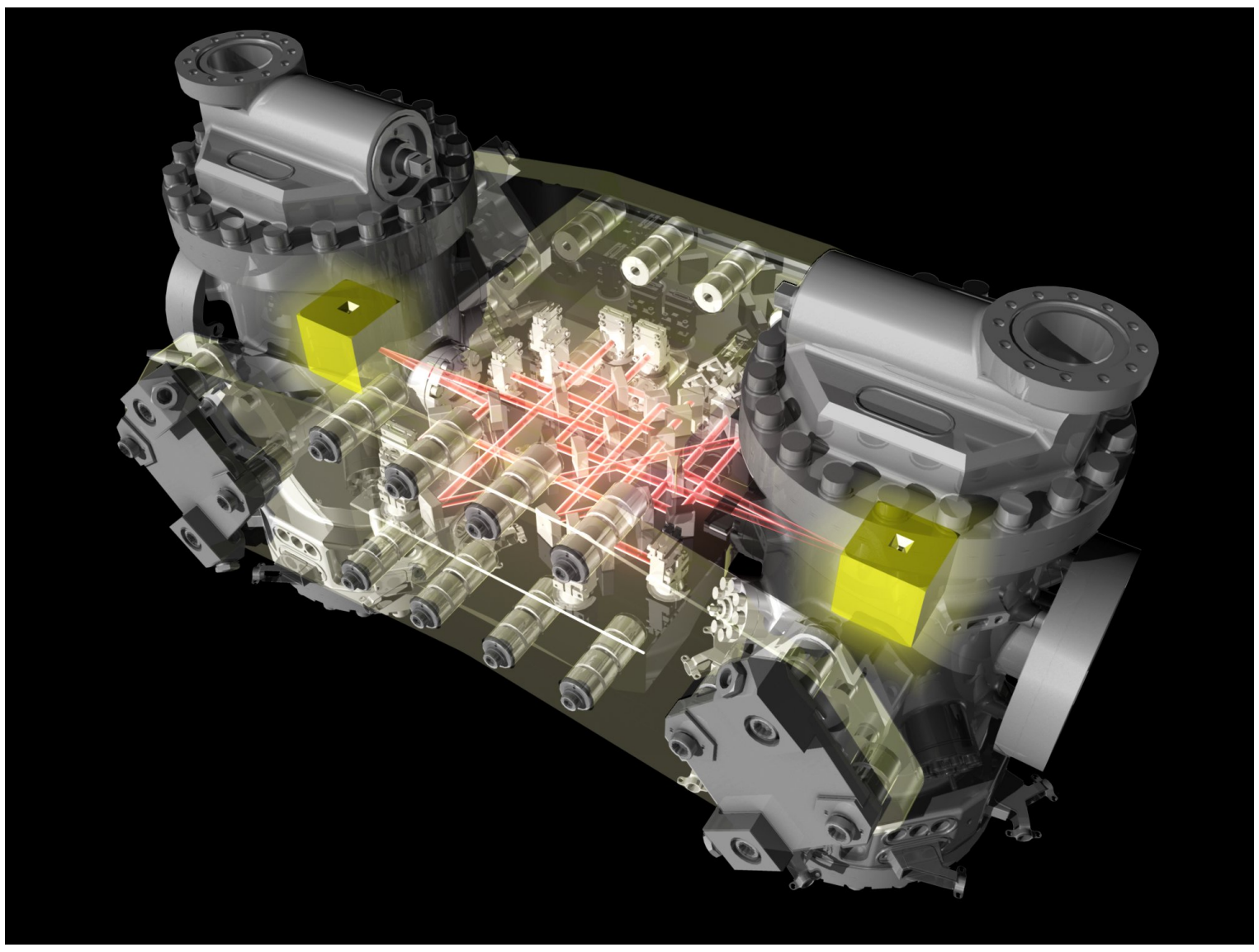

FIGURE 1.7: LISA Technology Package

# Chapter 2

# The LPF differential accelerometry measurement and the Free-Fall mode

The LISA Technology package is, in its essence, a differential accelerometer, and will be capable of measuring small acceleration acting on a mass in free fall, at the *femto-g* level. It will measure the noise in the relative stray test mass acceleration and its requirement is to place an overall differential acceleration noise upper limit of $30\,fm/s^2\sqrt{Hz}$ at $1\,mHz$.

LPF configuration requires an actuation force along the sensitive degrees of freedom, which is needed to compensate the static spacecraft self-gravity imbalance between the two test masses, introducing a source of noise in the final budget that is an important limiting factor at the low frequencies of interest for the mission. This is not a source of noise for eLISA configuration, in which actuation is performed only along d.o.f. other then the sensitive ones.

In general, if a large DC force acts on the free falling mass, which arises from the spacecraft self-gravity, it is possible to implement a new technique to eliminate, at least most of the time, the applied actuation force and the associated force noise: the free-fall mode. Actuation forces are limited to short impulses, with the possibility of measuring the actuation-free acceleration noise in the several hundred second free-flights between these impulses. So the mass is letting freely to move, with its time-averaged position at the center of the GRS, the rest of time. The electrostatic actuation is than turned off between the kicks and the test mass displacement during the free phase can be used for a spectral estimation of the remaining noise sources affecting the free-fall motion.

In this chapter we will explain the basics of LISA Pathfinder acceleration measurement and how the acceleration noise budget is dominated by the electrostatic actuation noise.

Finally we will explain the main scheme of the free fall control mode.
The free fall mode is a very powerful method to reduce noise from electrostatic suspension systems.

## 2.1 Differential accelerometry with combined drag-free and electrostatic suspension control

LISA Pathfinder can be considered as a differential accelerometer aiming to demonstrate the possibility to perform a relative acceleration measurement between two test mass in the same spacecraft, and with the main goal of measuring the acceleration noise in a space environment.
Most space applications requiring high resolution measurement of acceleration between spacecraft and other test bodies are based on electrostatic accelerometers [14]. The baseline of such instruments is a metallic mass that serves as a geodesic reference particle, that we usually call test mass, enclosed in a structure hosting electrodes, named electrode housing, which sense mass position and actuate it with electrostatic forces. The mass is thus electrostatically suspended, forced to follow the spacecraft, in all axes. The applied control force can then be used to calculate the stray forces acting on the satellite and to thus reconstruct in analysis its geodesic orbit.
Scientific mission, like eLISA, work with acceleration sensors used in drag-free closed loop control system in order to compensate the thrust due to all non-gravitational disturbances acting on the satellite. In this way, the geodesic reference mass is not forced, thus allowing it to truly follow a geodesic orbit and removing force noise which inevitably accompanies any test mass actuation.
In order to protect the purity of the test mass free fall, the satellite shields test mass and high precision thrusters are employed to center itself about the test mass, controlling it on the basis of the feedback signal coming from the test mass position sensor. This is the basis mechanism of the technique of drag-free control loop, conceived of in the 1960's, and that is employed for the acceleration disturbance reduction [15].
LISA Pathfinder and the geodesy mission GOCE [16] are single spacecraft differential accelerometers which combine the drag-free and accelerometer techniques. One test mass serves as the reference for the drag-free control of the satellite, while a second test mass must be forced to follow the first (or the satellite) to compensate any residual differential acceleration between the two TM. GOCE is the most sensitive differential accelerometer to date, with its differential acceleration measurements of order $10^{-10}\, m/s^2\sqrt{Hz}$. GOCE measures the difference in acceleration between TM displaced along three axes, thus measuring the 9 component tensor $dg_i/dq_j$. LPF measures one of these components, $dg_x/dx$, at a lower precision of $30\, fm/s^2\,\sqrt{Hz}$ at $1\, mHz$. This is possible by reducing each source of stray forces arising in the interaction between the sensor and masses but also because LPF will be located at the first Lagrangian point (L1), not in low earth orbit like GOCE. This reduce drastically the gravity difference felt by the two TMs from order $micron/s^2$ for GOCE to $nm/s^2$ for LPF.
Both systems are only partially drag-free, and the need to apply forces to the second mass to compensate the residual differential acceleration $\Delta g$ is quantitatively an important limitation. We will see that in the Pathfinder configuration, the actuation noise is a main noise source along the sensitive axis, that must be reduced.

### 2.1.1 LPF as a differential accelerometer

Because of the two masses are both hosted in the same satellite, it is impossible for the spacecraft to follow contemporary two of them. So, one of the proof masses is considered the reference (say $TM1$), it defines the inertial frame, it is left free to move in a geodetic motion, and the spacecraft follows it along a single axis, such as the sensitive $x$ measurement axis in LPF. The spacecraft is thrusted to follow the $TM1$ in the *drag-free* control loop, as we can see in the Figure 2.1 on the left. The optical path in figure, $o_1$, is the interferometer output which measures relative displacement of $TM1$ with respect to $SC$, $o_1 = x_1 - x_{SC}$, and used for drag-free control. The second mass, (say $TM2$), must be forced to follow either the other one or the spacecraft, by applying small electrostatic forces $F_{ES}$, by means of the capacitance actuation, possibly within a closed loop control scheme at low frequency, like appears in the Figure 2.1 on the right. This control loop is usually called the *electrostatic suspension*. The interferometric displacement $o12$, is the differential displacement of the two $TMs$, $o_{12} = x_2 - x_1$. Both $o1$ and $o12$ are required to achieve $9\,pm/\sqrt{Hz}$ or better readout noise, with $o12$ critical to the sensitivity of the main differential acceleration LPF measurement.

Different causes can deviate the masses from their purely geodesic motion, and they can be external or internal. The external ones can be compensated by acting on the satellite relative position by using thruster. The internal forces are related to the gravitational coupling both among the masses, that the masses and satellite. The local gravitational field between $TM1$ and $TM2$ brings the masses to each other, while, the $TMs$ - spacecraft field produce differential displacement between the involved bodies, and also a translation that can again be compensated by thruster moving. To counteract the undesired differential shift between the test particles due to the local gravity field, the only way is act on one of the test mass itself. By actuating on the $TM2$ in the suspension loop, would mean impress a constant acceleration but also a fluctuating force, due to actuation force voltage fluctuations, that produce undesired accelerations that must be reduced, as we will show below.

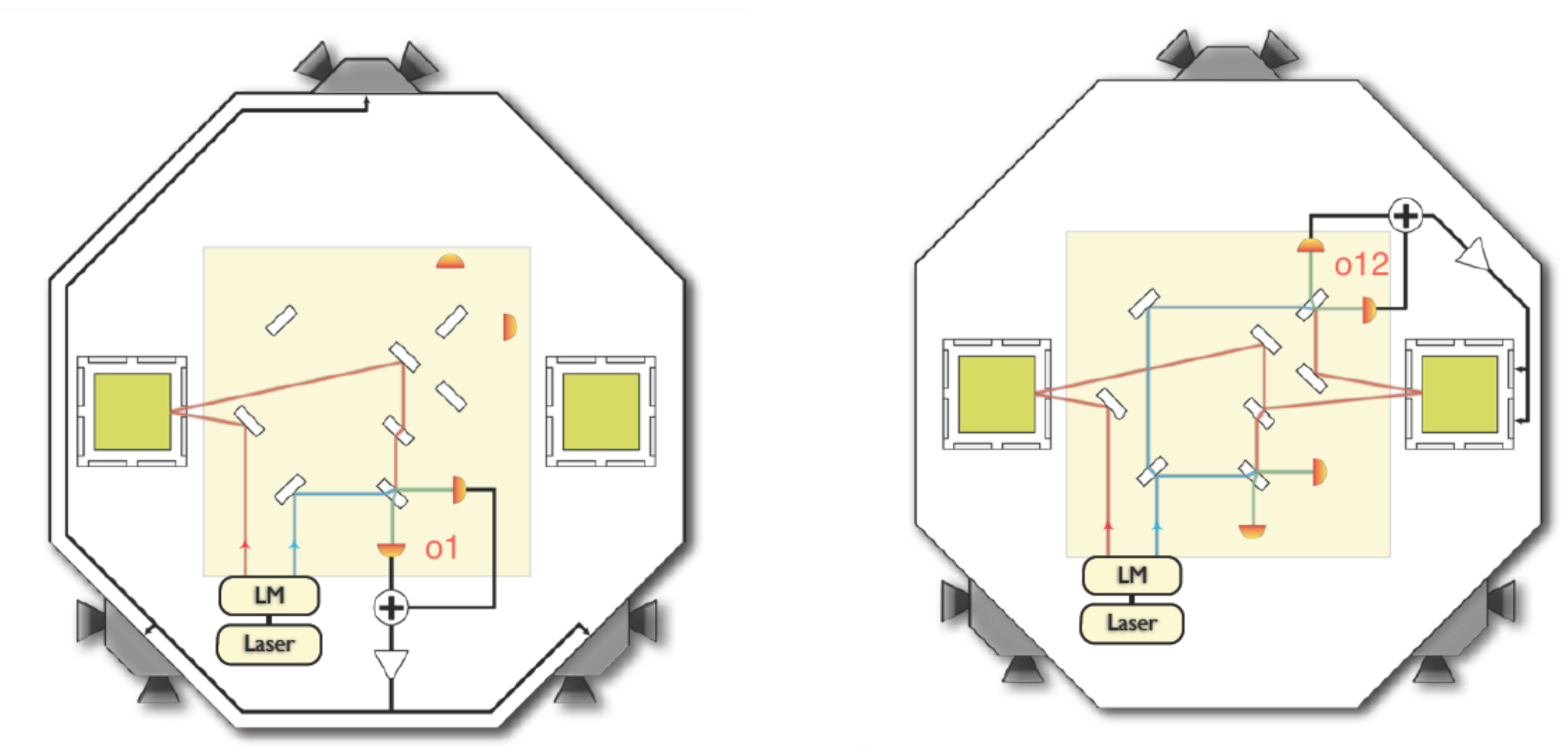

FIGURE 2.1: On the left: Drag-free on $TM1$ The interferometer signal path $o_1$ is showed. On the right: Low frequency control on $TM2$. The interferometer signal $o_{12}$ is outlined. [3]

The measured differential displacement of the two $TMs$, provide the acceleration of $TM2$ relative to $TM1$, or rather the difference in the stray forces per unit mass acting on the two $TMs$

$$\Delta g \equiv \frac{(F_2 - F_1)}{m}. \tag{2.1}$$

The observable $\Delta g$, is the differential acceleration that would be present between the two $TMs$ in the absence of any applied forces or any elastic coupling to the satellite, and with both TM centered such that the stiffness term in equation 1.2 is zero, and is equivalent to a difference in the local gravitational field.
If we apply the Newton's equations to the system of the satellite and the two $TMs$,

$$m\ddot{x}_1 = F_1 - m\omega_{1p}^2(x_1 - x_{SC}) \tag{2.2}$$

$$m\ddot{x}_2 = F_2 - m\omega_{2p}^2(x_2 - x_{SC}) + F_{ES} \tag{2.3}$$

where $F_{ES}$ is the applied electrostatic forces on $TM2$, $\omega_{1p}^2$ and $\omega_{2p}^2$ are the effective resonant angular frequencies associated with the elastic coupling, and therefore also with any steady force gradients, between each $TM$ and the spacecraft, is it possible to subtract the two equation si that the acceleration $\Delta\ddot{x} = \ddot{x}_2 - \ddot{x}_1$ is:

$$\Delta\ddot{x} = \Delta g - \Delta\omega_p^2(x_1 - x_{SC}) - \omega_{2p}^2(x_2 - x_1) + \frac{F_{ES}}{m}. \tag{2.4}$$

We can thus construct an observable quantity to estimate $\Delta g$, if we call the relative displacement interferometer readout $o_{12} = x_2 - x_1 + n_{12}$ and the $TM1$ displacement relative to the optical bench $o_1 = x_1 - x_{SC} + n_1$, where we have take into account also the noise components in the signals detected, which are respectively $n_{12}$ and $n_1$, so that the equation of motion can be written as:

$$\Delta\hat{g} \equiv \ddot{o}_{12} - \frac{F_{ES}}{m} + (\omega_{2p}^2 - \omega_{1p}^2)o_1 + \omega_{2p}^2 o_{12} \quad (2.5)$$

$$= \Delta g + \ddot{n}_{12} + (\omega_{2p}^2 - \omega_{1p}^2)n_1 + \omega_{2p}^2 n_{12}. \quad (2.6)$$

This equation represents a time domain estimator of the differential stray force per unit mass $\Delta g$ and can be considered as the sum of the real differential stray acceleration $\Delta g \equiv (F_2 - F_1)/m$ plus a noise component due to the displacement noise in the interferometry signals [3]. We are interested in the measurement of the real acceleration $\Delta g$ which is the observable $\Delta\hat{g}$ contaminated by a readout noise term.
The presence of the applied control force, $\frac{F_{ES}}{m}$, contributes to the differential force noise $\Delta g$, because along with the desired, commanded actuation force $F_{ES}$ there is a stochastic, fluctuating force that comes from noise in the electrostatic actuator.
The electrostatic control force $F_{ES}$ can be expressed via its frequency-dependent controller gain, $\omega_{ES}^2$

$$F_{ES} = -m\omega_{ES}^2 o_{12}. \quad (2.7)$$

Electrostatic actuation is, thus, source of acceleration noise that originates in the need to compensate the DC acceleration imbalance, that we will call $\Delta g_{DC}$, as we will see, the free fall mode is a way to reduce this source of noise.

## 2.2 Acceleration noise budget

The LISA Pathfinder goal is that the noise in the differential acceleration measurement $\Delta\hat{g}$ would be less than

$$S_{\Delta\hat{g}}^{1/2} \leq 3 \cdot 10^{-14} \frac{m}{s^2\sqrt{Hz}} \quad (2.8)$$

at $1mHz$. This is the requirement sets for the gravitational free fall purity, within one order of magnitude from the requirement for $eLISA$. This number play the role of a constraint on the spectral noise density of residual stray forces and interferometry noise.

In figure 2.2 the total differential acceleration requirements and a budget for all the noise component is shown [4]. These are different contributions of the individual subsystems and noise sources. An extensive ground testing and analysis of the flight hardware systems and known environmental noise has been used to form a current best estimate of the differential acceleration noise at 1 mHz that is 3 times below the required one, as is evidenced in figure 2.3 [3].

This figure shows the current best estimate for the differential acceleration noise in LPF, based on ground test campaigns of flight hardware and system modeling. This current estimate is obtained considering the configuration known as "SCI mode 1", in which the spacecraft follows $TM1$, as read by $o_1$, along the direction of the interferometer arm, and the distance between the proof masses, as read by $o_{12}$, is used as the error signal to actuate electrostatically the second proof mass along the same axis [3].
It is important to pay attention to the term related to the $X$ actuation noise budget. At low frequency it is clear that the noise due to the actuation adds up as an additional noise source.
This is the force noise due to the fluctuations in the amplitude of the applied electrostatic actuation voltage of the electrostatic suspension loop. If the amplitude of the audio frequency voltages used to apply the electrostatic actuation force fluctuates, then this feedback force fluctuates, adding to the noise in $\Delta g$ that is expected to limit the sensitivity at the lowest frequencies. This actuation noise can be related to the instability of actuation voltage references or in the amplifier gain.
Any instability in the applied electrostatic actuation forces $F_{ES}$, produces an acceleration noise related to the actuation voltage stability noise $S^{\frac{1}{2}}_{\delta V/V}$

$$S^{1/2}_{\Delta g} \approx 2\lambda \Delta g_{DC} S^{\frac{1}{2}}_{\delta V/V}, \tag{2.9}$$

considering that the applied force is $F \propto V^2$ so that $\delta F/F \propto 2\delta V/V$, as we will se in the next section in detail. The proportionality factor $\lambda$ here is equal to 1 if the actuation amplitudes of actuation amplifier fluctuate in correlated fashion. If not, must be also considered that the same electrodes are used to control rotation around the $z$ axis orthogonal to $x$ (figure 1.5), and the presence of this torque actuation $N_\Phi$, on both $TMs$, can produce force noise in $x$.
The DC force generated by the electric field fluctuates because itself fluctuates, so that the noise contributes directly to the measured differential acceleration $\Delta g$. We will explain more precisely how this happen, with electrostatic consideration about the actuation in the next section. Here we report only that the overall actuation amplitude stability, including the DC reference stability, amplifier gain stability, and other contributions, is required to be stable within

$$S^{\frac{1}{2}}_{\delta V/V} < 2 \cdot 10^{-6}/\sqrt{Hz}, \tag{2.10}$$

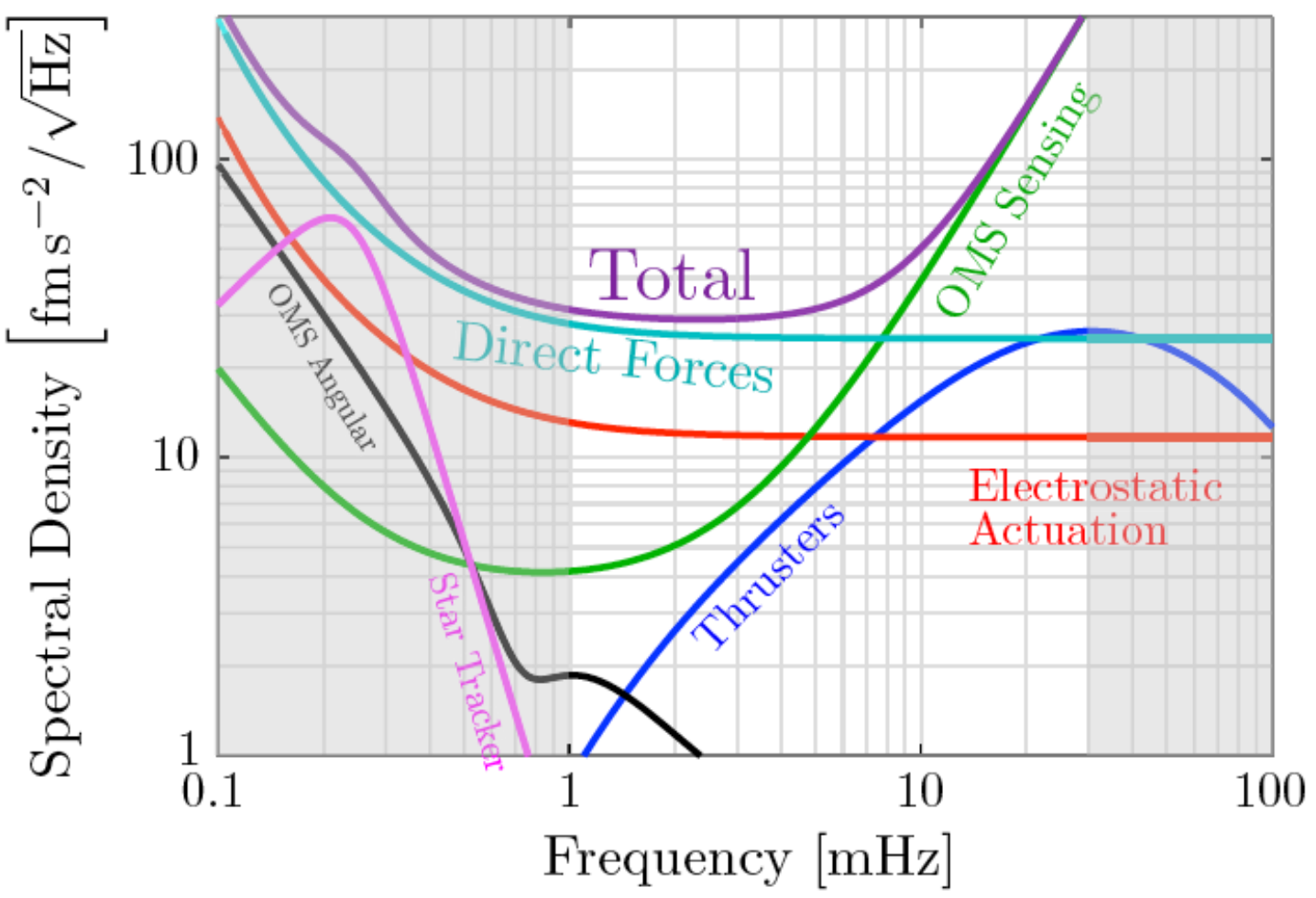

FIGURE 2.2: LISA Pathfinder total differential acceleration requirements and a budget for different contributions of the individual subsystems and noise sources. The gray areas are out of the measurement band for the LPF but are of interest for eLISA. [4]

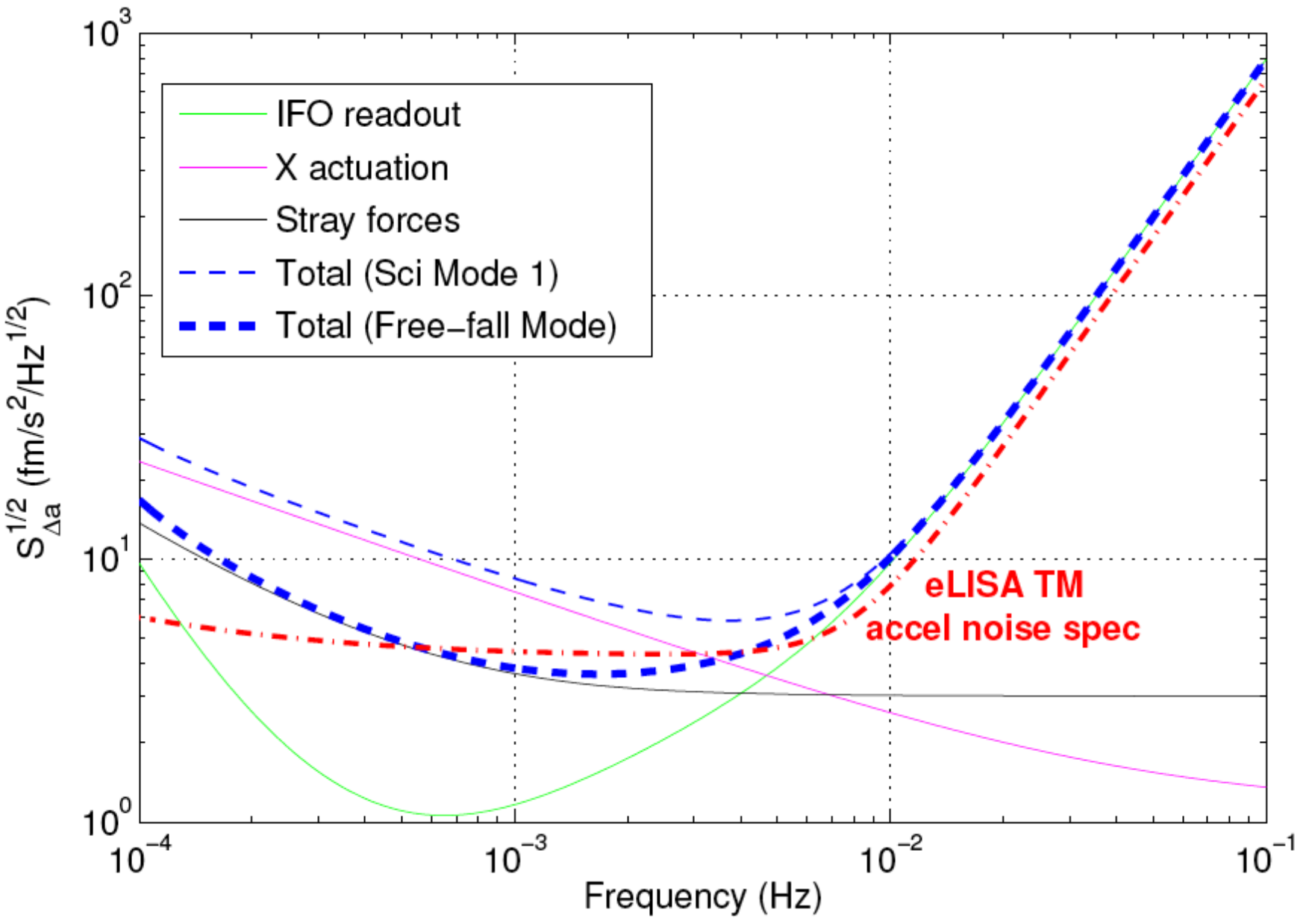

FIGURE 2.3: Current best estimate for the LPF differential acceleration noise measurement. Blue curves show the total noise with and without the actuation force. [3]

The estimated differential acceleration noise level of $7.5\, fm/s^2\sqrt{Hz}$ at $1\, mHz$ visible in figure 2.3 considers the allotted gravitational balancing tolerances along $x$ equal to $\Delta g_{DC} \approx 0.65 nm/s^2$ and the estimated gravitational torque around $z$ axis less then $2\, nrad/s^2$. Following these specifications, this noise source would give an acceleration noise contribution of roughly $3\, fm/s^2\sqrt{Hz}$. Without the DC acceleration contribution and considering only the presence of the $\Phi$ actuation, this acceleration noise contribution falls to $1\,fm/s^2\sqrt{Hz}$, which is an important consideration for motivating the actuation noise reduction with the free-fall mode, that we will address shortly.
The actual measured stability with the inertial sensor Front End Electronics (FEE) gives $3-7ppm/\sqrt{(Hz)}$ at $1\,mHz$ [17] [18] and the observed actuation fluctuations appear to be uncorrelated between different electrode channels.

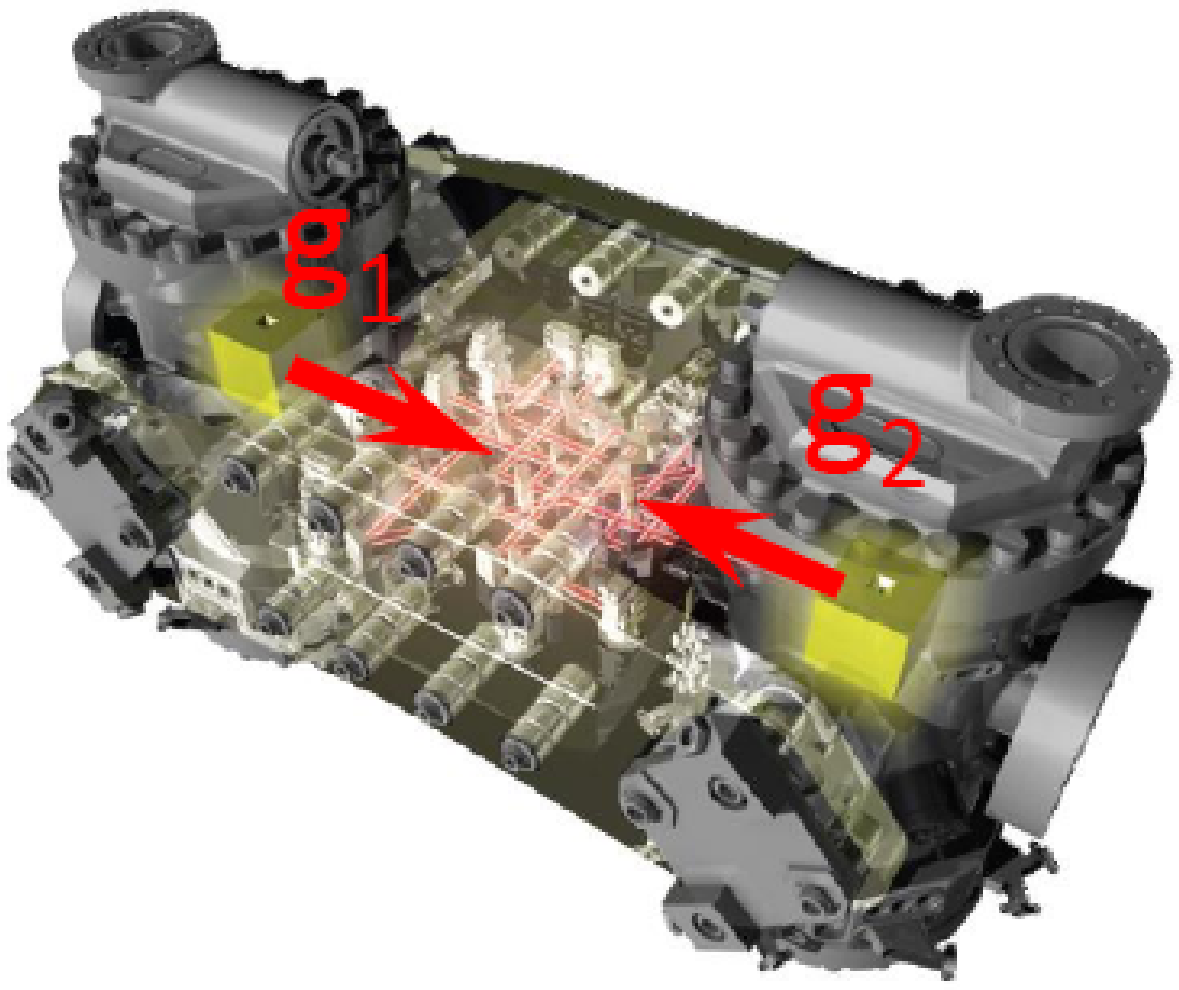

FIGURE 2.4: Test mass - spacecraft self-gravity.

Static compensation of the spacecraft gravity imbalance is mandatory to meet the $0.65\,nm/s^2$ residual DC acceleration requirement and can be done in part by using compensation block masses to minimizes the residual gravitational imbalance and stiffness [19]. The expected residual gravitational accelerations $\Delta g_{DC} \approx 0.65nm/s^2$ represents the uncertainty on the mass compensation after have used the compensation block masses.
The other part of the noise reduction is carried out by the free fall actuation mode, a way to control the $TM2$ aimed at remove completely the actuation noise limiting factor on sensitive axis and allowing to get closer the performances expected for $eLISA$.

## 2.3 The GRS actuation noise

We can explain the noise due to the actuation, giving the basics of the electrostatic and actuation models of the capacitive sensor.
To allow the production of actuation forces, a voltage difference are established between TM and various electrodes to which we apply actuation voltages, that, as we said in section 1.3, altogether form a system of conductors. In figure 1.6 is shown the *GRS* actuation circuitry scheme, that allows application of audio frequency AC and DC voltages on each sensing electrodes. For our purpose, we consider only the sensitive axis $x$.

Actuation forces arise from TM-electrode voltage differences. So, if we call $V_i$ and $V_{TM}$ respectively the electrodes and test mass potentials, it is possible to write the general formula of the DC force applied (as we will discuss later in equation 3.13), from standard energetic considerations of the electrostatic field of conductors [1]:

$$F_i = \frac{1}{2}\sum_i \frac{\partial C_i}{\partial x}(V_i - V_{TM})^2. \tag{2.11}$$

The test mass potential depends on the charge value of the mass $Q_{TM}$ and on the electrodes capacitance $C_i$, as well as by the potential of each surface $V_j$ in this way:

$$V_{TM} = \frac{q}{C_{tot}} + \frac{1}{C_{tot}}\sum_j C_j V_j. \tag{2.12}$$

The sums are on all the electrodes facing the test mass and the total capacitance $C_{tot}$ includes both the housing and the electrodes capacitances (as discussed in the electrostatic model described in section 3.2.1).

When the voltages are fixed by the electronic circuitry, it is possible to expand the capacitance derivative, respect to a reference position, and rewrite the electrostatic force between the mass and any electrode as:

$$F_{xi} = \frac{1}{2}\left[\left(\frac{\partial C_i}{\partial x}\right)_0 + \left(\frac{\partial^2 C_i}{\partial x^2}\right)_0 x\right](V_i - V_{TM})^2. \tag{2.13}$$

The first term, that depends on the first capacitance derivative, gives the applied force, while the second term determines the dependence of the applied force from the test mass position and depends on the second order capacitance derivative. The second terms, essentially, expresses the electrostatic spring-like coupling between the test mass and all the surfaces of the electrode housing and is named the electrostatic stiffness associated to actuation, $\omega^2_{p,act} \propto -\partial F/\partial x$. This act as a negative spring, it will also make the test mass unstable towards the sensor surfaces. So the electrostatic actuation will add both force noise and, as the applied forces depend on position, a significant contribution to the negative parasitic stiffness.

It is possible to apply AC or DC voltages to electrodes, with a constant stiffness actuation model [20], to apply any force in some range $[-F_{MAX}, +F_{MAX}]$ while holding the force gradient constant.

Each electrode can be used to actuate two d.o.f.s, i.e. $x$ and $\phi$ for the x-facing electrodes. The actuation voltage can thus be separated into a contribution for x-actuation and one for $\phi$-actuation. If we apply $\pm V_{x1}$ and $\pm V_{x2}$ respectively on the right and the left side of the electrode housing (see figure 2.5), we will have $(\sum_j C_j)V_{TM} = C_{tot}\sum_i V_i = 0$. So no changes are produced on $V_{TM}$ caused by the actuation voltage.
If the $TM$ is centered and not rotated and if we want to apply only force keeping the TM average voltage at zero, the mean force along $x$ is given by:

$$F_x = \frac{1}{2}\sum_i \frac{\partial C_i}{\partial x}\left\langle (V_i - V_{TM})^2 \right\rangle \tag{2.14}$$

$$\approx \frac{1}{2}\sum_i \frac{\partial C_i}{\partial x}\left\langle V_i^2 \right\rangle \tag{2.15}$$

[1] Note that the sum is made on all electrodes and the partial derivative $\frac{\partial C_i}{\partial x} \equiv \frac{\partial C_{i,TM}}{\partial x} + \frac{\partial C_{i,H}}{\partial x}$ sum of contribution of capacitance of electrodes x w.r.t. test mass and housing. The second term $\frac{\partial C_{i,H}}{\partial x}$ is typically less then the first of roughly 10% (see section 3.2.1)

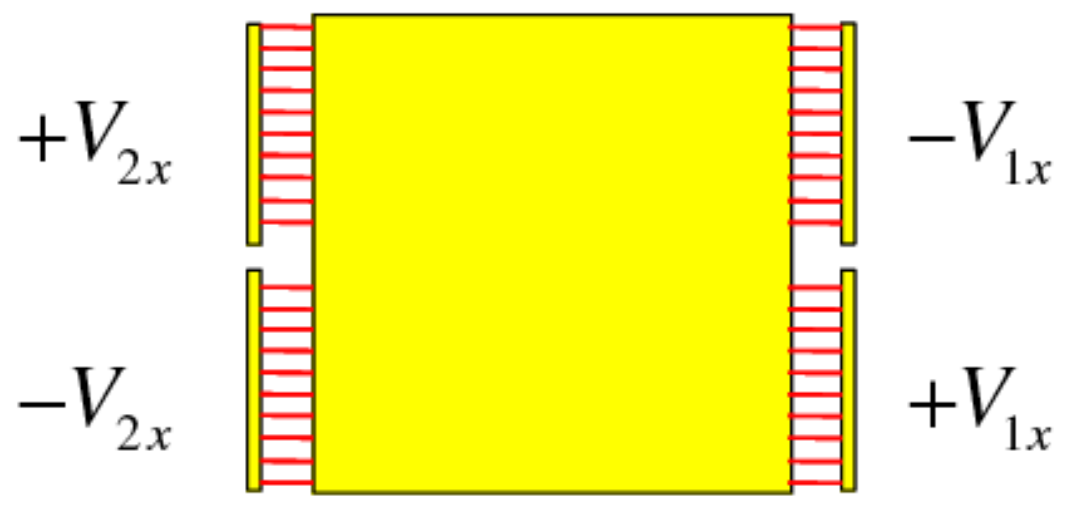

FIGURE 2.5: Scheme of actuation voltages applied on $x$ electrodes.

If $V_{x1} = V_{x2}$, the force is zero. If $V_{x1} > V_{x2}$ (or vice versa), from the equation 2.11 it would be:

$$F_x = \left|\frac{\partial C_x}{\partial x}\right| \left(V_{x1}^2 - V_{x2}^2\right) \tag{2.16}$$

As consequence the stiffness will be:

$$\omega_{p,act}^2 = -\frac{1}{m}\frac{\partial F}{\partial x} = -\frac{1}{m}\left|\frac{\partial^2 C_x}{\partial x^2}\right| \left(V_{x1}^2 + V_{x2}^2\right). \tag{2.17}$$

We can thus hold the stiffness constant by keeping constant $V_{x1}^2 + V_{x2}^2 \equiv V_{MAX}^2$. This gives a range of forces $[-F_{MAX}, +F_{MAX}]$ at constant stiffness

$$\omega_{p,act}^2 = -\frac{F_{MAX}}{m}\frac{\left|\frac{\partial^2 C_x}{\partial x^2}\right|}{\left|\frac{\partial C_x}{\partial x}\right|} \approx -\frac{2\Delta g_{MAX}}{d} \tag{2.18}$$

where the expression of the first and second capacitance derivatives has been replaced as they are from the parallel plate model $\partial C_x/\partial x = \pm C_x/d$ and $\partial^2 C_x/\partial x^2 = 2C_x/d^2$ (see section 3.2.1 and reference [20]), showing up the dependence on the gap between the electrodes and the test mass $d$.
The actuation scheme can be considered at constant stiffness because the equation 2.17 defines a family of circles in the $V_{1x}-V_{2x}$ plane, with radius of constant stiffness $\sqrt{|\omega_{p,act}^2|}$ in the space $V_{x,i}$.

Accordingly to what it has been said about the electrostatic compensation of the gravity imbalance, the term $\frac{F_{MAX}}{m}$ is the actuation used to counteract the gravity-dominated DC force imbalance $\Delta g_{DC}$ and the relative stability of the quantity $< V_{x,i}^2 >$ is the relative stability of the applied force.
An "in band" fluctuation of the actuation drive voltage amplitude $\delta V_x$ is related to a force fluctuation because $F_x \propto V_{x,i}^2$, so that

$$\frac{\delta F_x}{F_x} = 2\frac{\delta V_x}{V_x} \tag{2.19}$$

which translates finally into acceleration noise, according to equation 2.9:

$$S_{\Delta g}^{1/2} = 2\lambda \Delta g_{DC} S_{\delta V/V_{act}}^{\frac{1}{2}} \tag{2.20}$$

and $S_{\delta V/V_{act}}^{\frac{1}{2}}$ is the noise spectral density of the actuation voltage fluctuation. Both the stiffness and the actuation force noise along the $x$ direction, are thus proportional to the residual imbalanced $\Delta g_{DC}$.
It is worth to note here that the factor $\lambda$ can be considered equal to 1 for amplifiers actuation noise correlated. Otherwise in the acceleration noise $S_{\Delta g}^{1/2}$, depends not just from $\Delta g_{DC}$, but also from the maximum allowable actuation $\Delta g_{MAX} = \frac{F_{MAX}}{m}$,and the applied torques and torque authority on the two TM.
As already said in the acceleration noise budget section, the current design specification estimate a noise density $S_{\delta V/V}^{\frac{1}{2}} \approx 3 - 7\, ppm/\sqrt{Hz}$ to have a gravitational balancing to $\Delta g < 0.65\, nm/s^2$.

## 2.4 Actuation noise and the free-fall mode

As we said, the capacitive actuation noise is likely to be the limit of our experiment sensitivity at the lower end of the bandwidth. But the electrostatic actuation is necessary to compensate of the static field experienced by the $TM2$.
To reduce this source of noise, the free-fall mode idea is to actuate the test mass only 1% of the time with 100 times the DC force, so that the average suspension force would be the same. It is possible to use the actuation forces over a short period of time by switching on and off the suspension force with a constant frequency and a low duty cycle of 1%. In this way the control on the mass is performed periodically with a series of discrete kicks by using large suspension forces, rather than to use a continuous applied force.
Between two kicks, the TM is thus free of applied force and thus also free of the actuation noise from x actuation. It will follow an approximately parabolic trajectory respect to the other mass, and this displacement can be used to estimate the differential acceleration as desired. From the equation of motion 2.6 it is possible to derive the equation during a phase of drift between two kicks, when both $TMs$ are free to move and no actuation is applied

$$\ddot{o}_{12} = \Delta g + \Delta\omega_p^2 o_1 + \omega_{2p}^2 o_{12} + \Delta n \tag{2.21}$$

where all noise component were grouped in the term $\Delta n$. The term which comprises $F_{ES}$ in the observable $\Delta\hat{g}$ is zero and the also actuation noise associated with $F_{ES}$ is absent from $\Delta g$. The stiffness term multiplying $o_{12}$ is now much larger, as the displacement in $o_{12}$ is now large, of order $5\mu m$ as shown in figure 2.7.

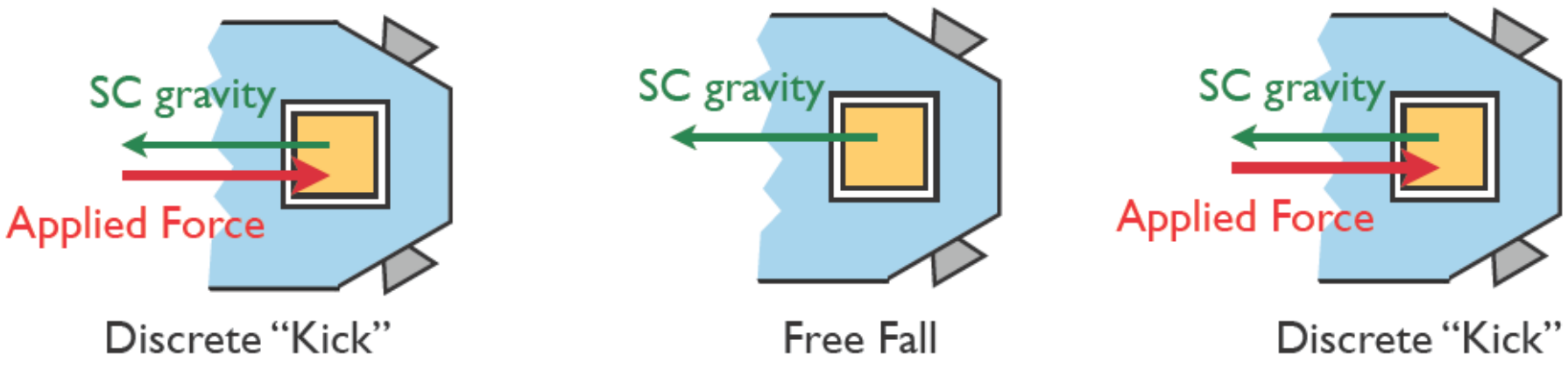

FIGURE 2.6: Free-fall mode scheme [5].

The negative stiffness $\omega_p^2$ makes the dynamics unstable with a time of $500\,s$ but the constant force $F$ brings the TM out of the interferometer sensing range in even less time ($400\,s$) and it would go out of range if we assume a certain $\Delta g$ [21]. As a conclusion, the $TM2$ can have a free motion, only during short range of time, where no kicks are applied. The free fall mode, therefore, consists of a series of repeated quasi-parabolic flights with a duration fixed by the controller designed for the LTP, of $350\,s$ along $x$, periodically alternated with very short kicks $\sim 1\,s$, as it is possible to see in the left part of Figure 2.7, that shows a segment of simulated data [6]. The cycle frequency is $2.85\,mHz$.

Degrees of freedom other than $x$ are controlled through capacitive actuation of drag-free controlled in this scheme.

An observer tracks the motion of the $TM2$ during the free flight and estimates the impulse required to maintain the mass position. The amplitude of the subsequent kick is adjusted to deliver this impulse and the process is repeated [21].

The right part of Figure 2.7 shows a series of discrete kicks separated by nearly constant acceleration during the free flight segments.

An important assumption is made about the disturbance force noise in $\Delta g$ and the readout noise contribution. It is considered stationary from one kick to another, so that the drift phases can be used to estimate the noise spectrum at frequency below the kick frequency. This assumption is possible if we consider that the actuation use little power to have a short and smooth transitory behavior.

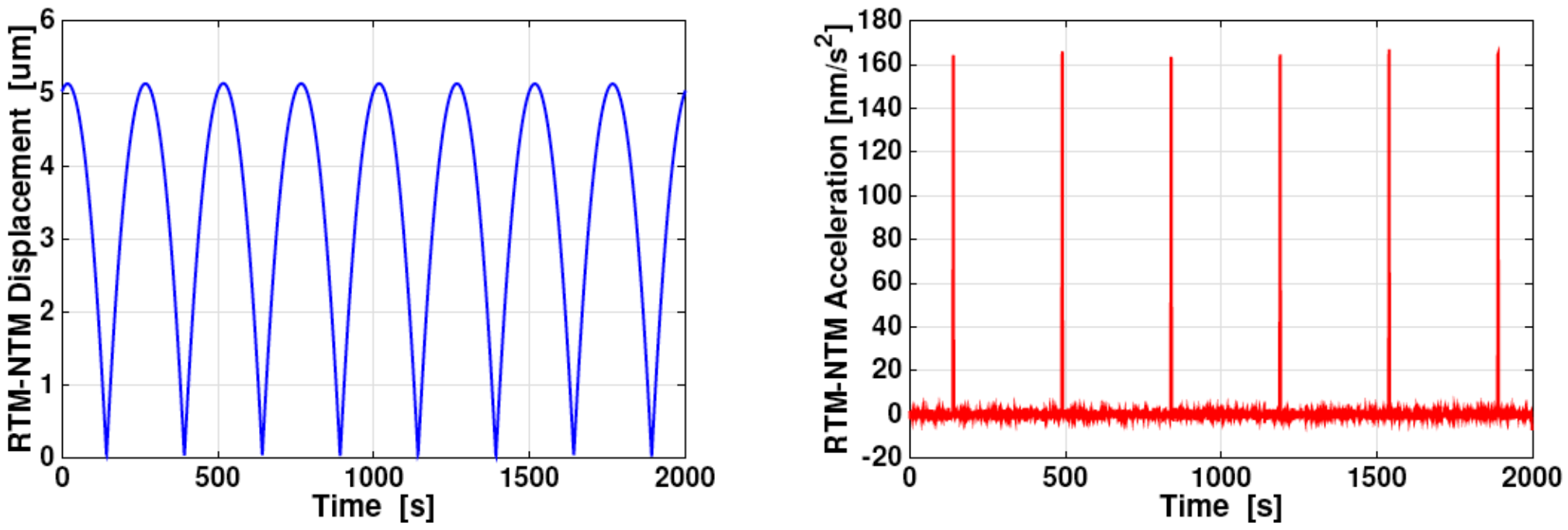

FIGURE 2.7: Free-fall mode time series from simulations. Here, RTM indicates the reference test mass $TM1$ and NTM the non-reference test mass $TM2$. On the left: time series of differential displacement of the two $TMs$. On the right: relative acceleration time series. [6]

# Chapter 3

# Testing free fall on ground: torsion pendulum facility

The eLISA and LTP level of performance cannot be verified on ground, due to the presence of the large Earth gravity, furthermore, their force noise isolation and reduction requirements, at low frequencies, are well below any previous free-fall experiment. So each known source of noise should be tested and characterized and needs experimental investigation. This is the reason why a ground based test bench for small force measurements was developed and realized.
An upper limit to all parasitic forces that act on the proof-mass surfaces (electrostatics and electromagnetics, thermal and pressure effects, etc) has been established during the last years with pendulum facilities, configured for low frequency (mHz and sub-mHz region) and high sensitivity measurements of torque introduced by the sensor on the suspended test mass. We describe here the most relevant disturbances and discuss the characterization of their torque noise contribution to the overall pendulum torque noise. Moreover, the possibility of test the free fall mode with the appropriate accuracy and sensitivity on ground, is provided by means of torsion pendulum test bench, as we will explain in this chapter.

## 3.1 Single TM facility

Torsion pendulum and torsion balances have been widely used in weak forces measurements and the main reasons are that they can measure forces in a plane always perpendicular to the gravity field, and they are weakly coupled to the seismic noise. So a pendulum is the best instrument to measure the weak forces exerted by the eLISA capacitive sensor prototypes on a test mass that is suspended by a thin torsion fiber, so that it is nearly free along one rotational degree of freedom.
As said before, the required sensitivity in the measurement of small forces on ground is the $fN$ level, at frequency between 0.1 and 10 $mHz$, a very challenging task considering that is about $10^{16}$ orders of magnitude less than gravity on Earth. So it is necessary to decouple the measurement process from the local gravitational field, and this is simply done by suspending the weight of the inertial member with a fiber, hangs the mass parallel to local gravitational field. With a correct choice of fiber dimensions and material, a very low torsional spring constant and high quality factor can be achieved, that result in a lower torque thermal noise and thus in an increased sensitivity. It is clear that the only sensitive degree of freedom is the rotational one around the fiber axis.
The equation of motion of the pendulum along the rotational d.o.f. $\phi$ is written as:

$$I\ddot{\Phi}(t) + \gamma\dot{\Phi}(t) + \Gamma\Phi(t) = N(t) \tag{3.1}$$

where $I$ is the pendulum inertial moment, $\gamma$ is the dissipative term and we are assuming a structural damping, with thus $\gamma(\omega) = \frac{\Gamma\delta}{\omega}$, where $\Gamma$ is the fiber torsional elastic constant, and $N$ is the applied torque. This equation gives an equilibrium angle $\phi_{EQ} = \frac{N_{DC}}{\Gamma}$ that is included in $N(t)$ as the fiber torque $\Gamma\Phi_{EQ}$.
In the frequency domain the equation can be rewritten as:

$$-I\omega^2\Phi(\omega) + \Gamma(1 + i\delta)\Phi(\omega) = N(\omega) \tag{3.2}$$

where the dissipative term has been defined as $\delta = 1/Q$, that is the loss angle, with $Q$ quality factor of the oscillator. The moment of inertia can be written in term of $\omega_0$, the natural resonant angular frequency, $I = \Gamma/\omega_0^2$, considering to be in the approximation of small $\delta$ (high quality factor).
The equation 3.2 can be rewritten as:

$$\Phi(\omega) = \frac{N(\omega)}{\Gamma(1 + i\delta) - I\omega^2} = \frac{N(\omega)}{\Gamma\left(1 - (\frac{\omega}{\omega_0})^2 + \frac{i}{Q}\right)} \tag{3.3}$$

Introducing the pendulum transfer function $H(\omega)$, it will be:

$$\Phi(\omega) = H(\omega)N(\omega) \quad with \quad H(\omega) = \frac{1}{\Gamma\left(1 - (\frac{\omega}{\omega_0})^2 + \frac{i}{Q}\right)} \tag{3.4}$$

Each force, or rather torque, acting on the pendulum at frequency $\omega$, can thus be converted in angular displacement at the angular frequency by means of the pendulum transfer function and then detected as deflection of the pendulum angular rotation. Measuring the angular motion, it is thus possible, through the knowledge of the pendulum parameters, to estimate the external torque exciting the pendulum or just applying the equation 3.1.
It is clear that the pendulum angular measurement is affected also by the noise, $\Phi_{meas}(\omega) = \Phi(\omega) + \Phi_n(\omega) = H(\omega)N(\omega) + \Phi_n(\omega)$, so that the sensitivity of the pendulum, in the torque measurement, it will be affected by the angular readout noise and by the mechanical thermal noise as predicted by the fluctuation-dissipation theorem [22]. The maximum torque sensitivity of the pendulum is reached when it is limited by:

$$S_{N_{th}}(\omega) = 4k_B T \frac{\Gamma}{\omega Q} \tag{3.5}$$

where $k_B$ is the Boltzmann constant, and $T$ is the temperature.
In order to improve the sensitivity by keeping the thermal noise low and the quality factor $Q$ high, the pendulum is operated in high vacuum and the test mass is suspended from a fiber with intrinsically low mechanical losses, made by fused silica. This material choice was well demonstrated an improvement of force sensitivity over what can be achieved with thermal noise-limited tungsten over a wide range of frequencies, during the past years [24]. It also possible to lower the torsional stiffness $\Gamma$, that depends on the fiber radius $r$ via the relation:

$$\Gamma = \frac{\pi r^4}{2L}\left(F + \frac{mg}{\pi r^2}\right) \tag{3.6}$$

where $L$ is the length of the fiber, $F$ in this case refers to the elastic modulus of the material employed, $m$ is the pendulum mass and $g$ the local gravitational field. All these parameters are chosen to minimize the torsional constant in order to be sensitive to very weak forces [25].
Finally, the noise floor can be expressed as:

$$S_{N_{meas}} = S_{N_{th}}(\omega) + \frac{S_{\Phi_n}(\omega)}{|H(\omega)|^2} \tag{3.7}$$

The intrinsic noise limit of the pendulum is dominated at low frequency by the pendulum thermal noise, that masks external contributions up to its value. The high frequency is instead dominated by the readout sensitivity, that goes as the $\omega^{-2}$ factor of the transfer function 3.4, unable to distinguish pendulum movement below a certain value. The sensitivity in angle and torque is shown in figure 3.1.
Moreover, the final estimate of torque is not done in the frequency domain, but directly in the time domain from the equation 3.1, by converting the angular time series into an instantaneous applied torque $N(t)$. This because, by dividing the power spectral density of the pendulum angular position $S_{\Phi_n}$ by the square modulus of the pendulum transfer function as in equation 3.7, can produce problem around the most sensitive frequency we are interested in, the resonance. In fact, the transfer function has a narrow peak at resonance frequency $\omega = \omega_0$, as visible in figure 3.1, as does $S_{\Phi_n}$. Dividing the two quantity easily lead to unwanted artifacts around $\omega_0$.
The measured torque $N(\omega)$ can be converted into a force $F(\omega)$ by means of a suitable conversion arm-length $R_\Phi$, depending on the nature of the noise source involved [26]:

$$F(\omega) = \frac{N(\omega)}{R_\Phi}. \tag{3.8}$$

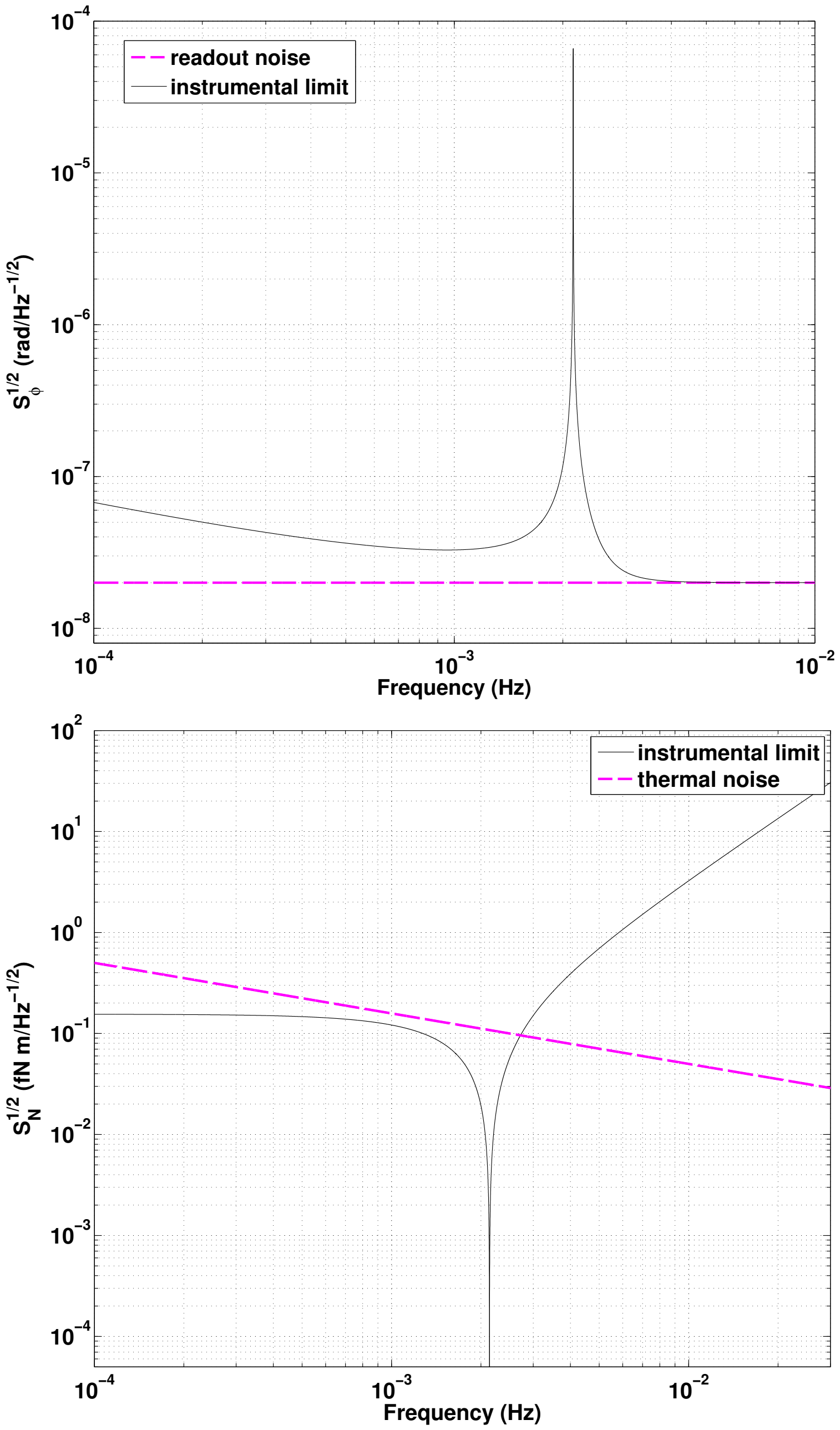

Figure 3.1: Pendulum sensitivity curves. On the left: angular sensitivity, obtained considering an angular white noise level $S_{\Phi_n} = 20\, nrad/\sqrt{Hz}$ at $10\, mHz$. On the right: torque sensitivity obtained considering a withe torque level of $S_N^{1/2} = 1.4\, fNm/\sqrt{Hz}$. Thermal limit is obtained considering moment of inertia $I = 4.31^{-5} kg\, m^2$, quality factor $Q = 1 \cdot 10^6$ and pendulum period $T_0 = 468\, s$.

For our purposes, we are interested in the noise source coming from electrostatic interaction. Any electrostatic interaction between sensing circuit noise and the sensor excitation produces back action forces and torque noise proportional to the net acting force noise. For these sources of noise the right armlenght is $R_\Phi = 10.75\,mm$, that is the nominal half separation between adjacent electrodes.
Similarly, it is possible to convert to an equivalent noise in acceleration caused by surface forces for a eLISA / LPF proof mass of $\sim 2\,kg$. This means that a measured torque of $\sim 1 fN\,m/\sqrt{Hz}$ at mHz frequency, corresponds to an acceleration of $\sim 50\,fm/s^2\sqrt{Hz}$, a factor 2 above LISA Pathfinder specification. The torsion pendulum facility can thus allow a quantitatively significant test of the purity of the free fall, because the measured pendulum angular noise in the eLISA and LISA Pathfinder measurement band, establishes an upper limit on the contribution of noisy surface forces, as we will explain in the next sections, where we conclude also at which level it is able to test quantitatively the feasibility of the free-fall mode.

## 3.2 Experimental apparatus

The torsion pendulum test bench is based on a lightweight ($80g$) and hollow eLISA test mass, of $46\,mm$ side and gold coated aluminum, suspended by a thin torsion fiber of $1\,m$ length, made by fused silica of $\approx 38\,\mu m$ of diameter [27]. The test mass has no electrical contact with the surrounding environment, and is electrically insulated from the rest of the inertial member by means of a quartz ring, in order to provide a very high electrical resistance and also reduce the stray capacitance of the test mass to ground.
We will describe schematically the various part and devices of torsion pendulum facility.

- *GRS* The gold coated aluminum mass hangs inside a prototype of the Gravitational Reference Sensor (GRS), with a gap between $TM$ and electrodes of $4\,mm$, as already showed in figure 1.5. This prototype is an engineering model very similar to the final gravitational reference sensor design for the LTP experiment, employing Sapphire electrodes. For typical injection bias of $5\,V$ at $100\,kHz$, the noise floor of the sensor is roughly 100 $nrad/\sqrt{Hz}$ at frequencies of 10 $mHz$. The aspect of the homemade GRS FEE that is most important to our measurements, the actuation circuitry, is described below in Section 3.2.2.

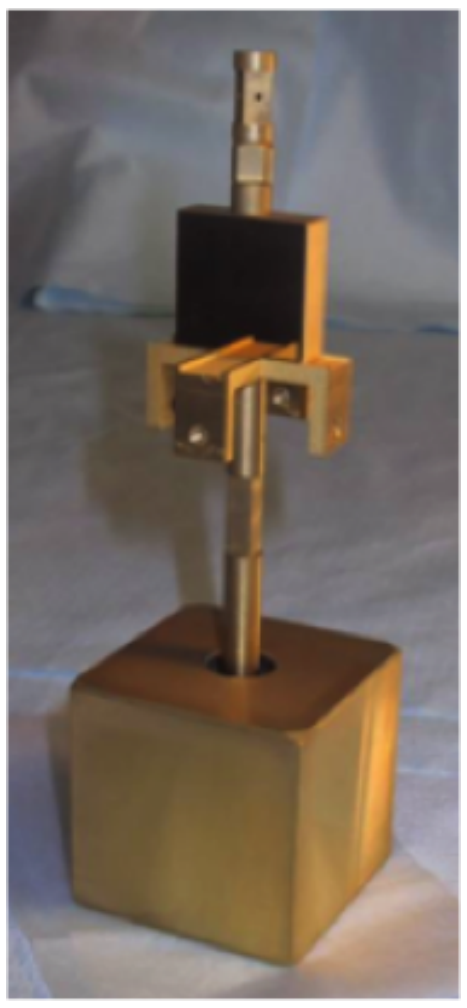

FIGURE 3.2: The real test mass suspended in the torsion pendulum facility.

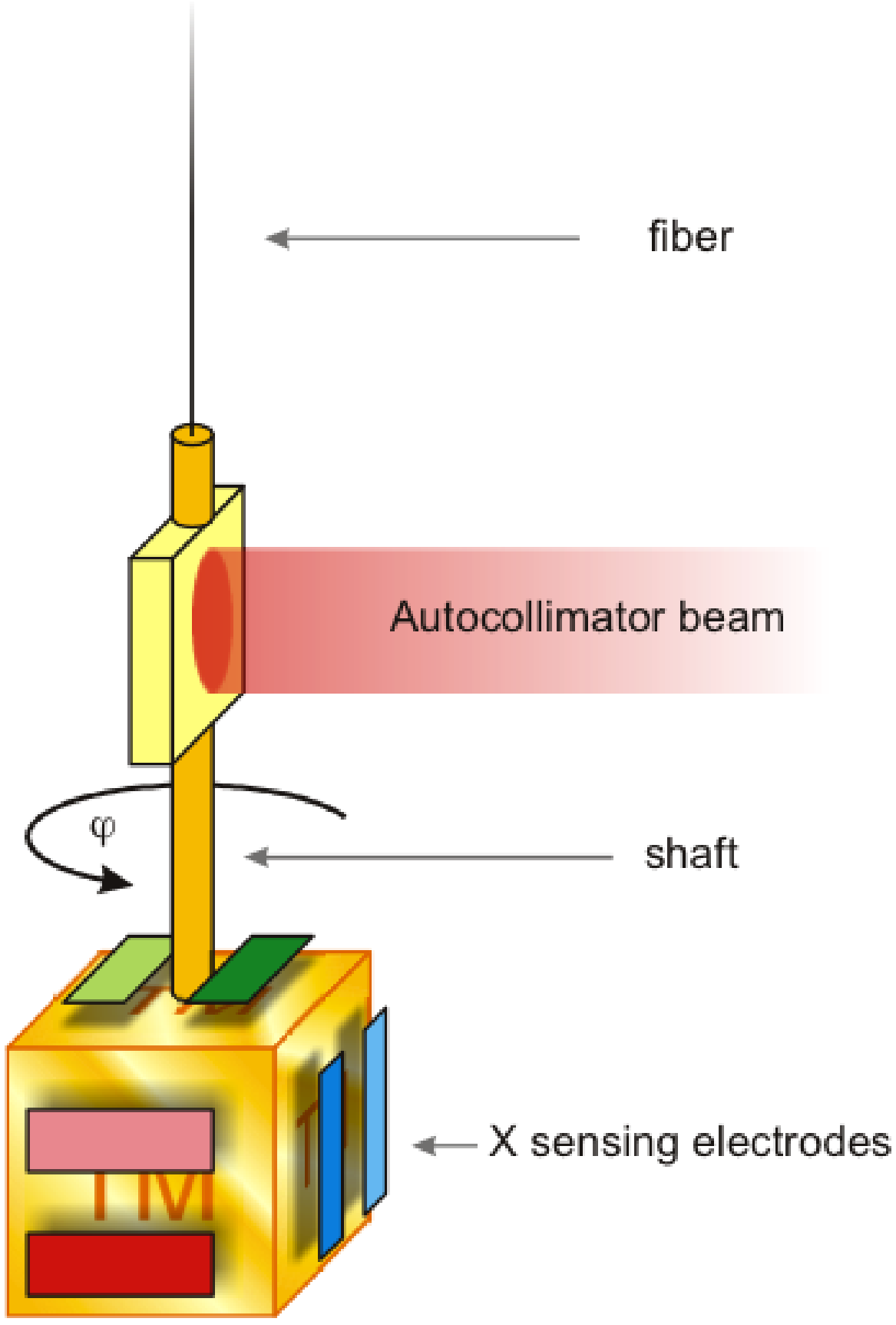

FIGURE 3.3: Scheme of the one test mass facility.

- *Autocollimator* The supporting shaft is also made of gold coated aluminum. It

carries a gold coated mirror, used for the independent optical readout of the pendulum angular position. It is based on a commercial two axis autocollimator *Möller-Wedel ELCOMAT vario*, that is used as a reference sensor, monitoring the angular d.o.f. of twist $\Phi$ and swing $\eta$, as well as the GRS. A LED light source reflects on the mirror and is collected by a CCD camera. It has a bit resolution of $50\,nrad$ LSB and a full scale range of about $\pm 5\,mrad$. It was measured that its resolution in the angular measurement with a fixed mirror is to the best of $100\,nrad/\sqrt{Hz}$ at $1\,mHz$ [24]. The optical readout is also used for calibration of the capacitive readout in $\Phi$ and $\eta$ angular d.o.f..
The intrinsic sampling frequency of the optical readout is $50\,Hz$, typically averaged to give $10\,Hz$ sampling. It is connected via a serial communication port with the data acquisition computer.

- *Pendulum parameters* The inertial member has a measured moment of inertia $I = 4.3 \cdot 10^{-5} kg\,m^2$ [28]. The pendulum typical oscillation period is measured to be about $460\,s$, in a condition where no electrical field is applied to the electrodes. Considering that the torsional stiffness is related to the pendulum period by the formula $\Gamma = I(2\pi/T_0)^2$, its value is near to $8.04 \cdot 10^{-9} N\,m/rad$. Purposely applied or parasitic electrical fields can decrease this number by introducing a corresponding electrostatic stiffness $\Gamma_{ES}$, as it is produced mainly by the injection sensing bias and possible applied actuation or DC bias voltages. In typical operating conditions, with a 100 kHz injection bias of $6\,V_{RMS}$ amplitude, the pendulum period raise up to $468\,s$. With this choice of material and dimensions of fiber, the mechanical losses and the energy dissipation are very low, and the energy decay time is of the order of $10^8 s$ corresponding to a quality factor of the oscillator $Q$ around one million. In the next chapters we will discuss about the total amount of the stiffness $\Gamma$ as sum of all the electrostatic contribution to the rotational stiffness provided by different electrostatic sources.
  With values and parameters reported before for fiber radius and quality factor, the thermal noise of such a pendulum, from the equation 3.5, is around $0.14\,fN\,m/\sqrt{Hz}$ at $1\,mHz$, at room temperature (as also shown in figure 3.1).

- *Vacuum and thermal parts* All the part described [25] are enclosed in a cylindrical vacuum chamber of about $50\,litres$ capacity and $350\,mm$ of radius, where the residual pressure is kept to about $1-2 \cdot 10^{-5}\,Pa$ by means of a $50\,l/s$ turbo molecular pump.
  The torsion fiber is instead enclosed in an $80\,cm$ long vacuum tube, mounted on the top of the vacuum vessel (see figure 3.4). To avoid any dangerous effect due to environmental vibration, the whole system is kept on a laboratory floor platform separated from the rest of the floor. This gives isolation from seismic noise produced by human activity in the surrounding of the experiment. All the connected systems, like primary pump and temperature control system, sit on a separated platform to obtain an additive vibrational isolation.

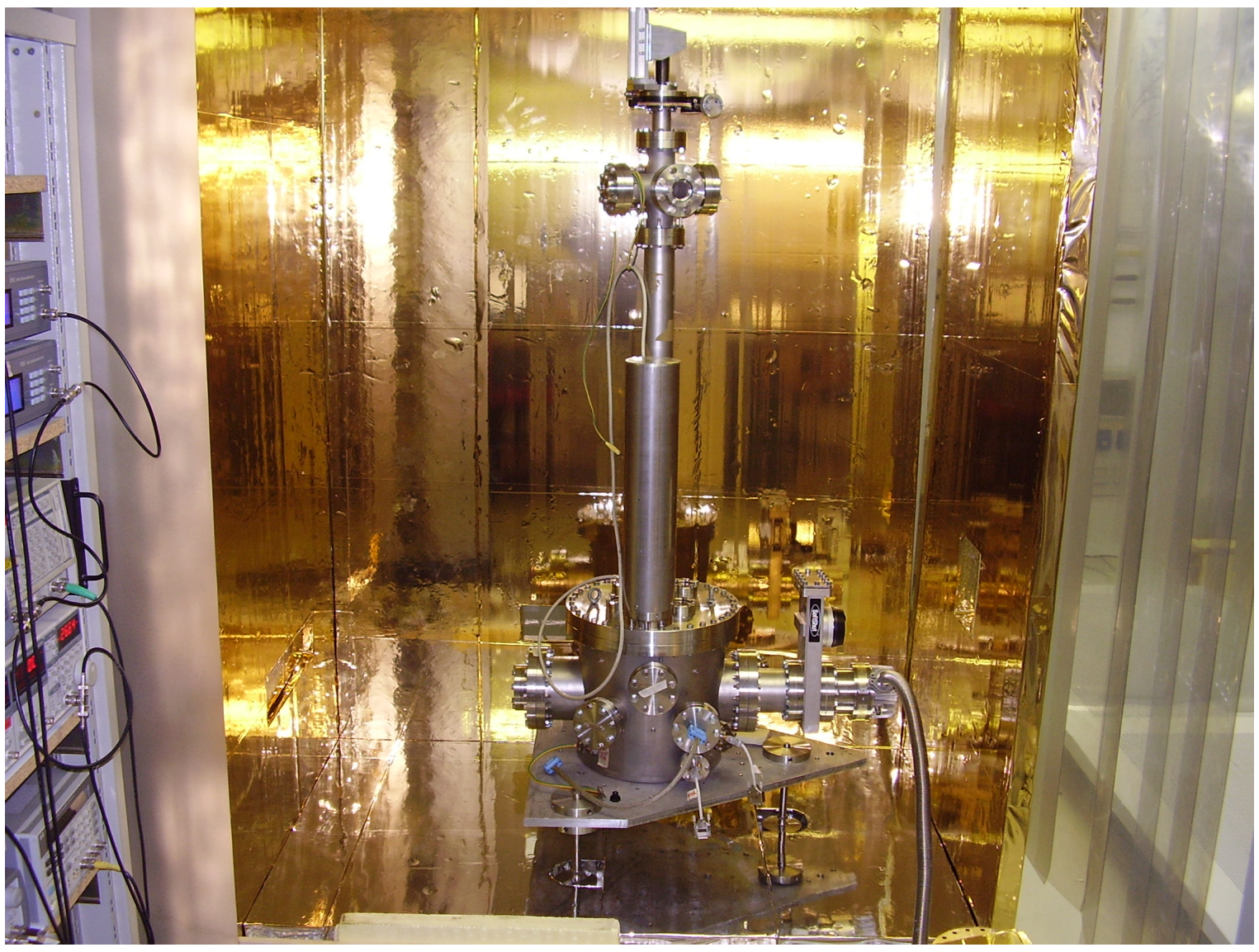

FIGURE 3.4: Torsion pendulum vacuum chamber and the thermal box that surround it.

The entire apparatus is also enclosed inside a thermally isolated box made by styrofoam panels (the gold panels in figure 3.4). Moreover, to stabilize the temperature inside the thermal box, water circulate in a radiator through a coil in thermal contact with the metallic plate supporting the experiment. Temperature gradients are also reduced forcing air circulation within the thermal box with a set of fans. Thermal stability more than one order of magnitude better than daily laboratory temperature fluctuations is achieved. The temperature of the experiment is measured by several $PtAu100$ thermometers read by a digital multimeter obtaining a measurement resolution about $2\,mK/\sqrt{Hz}$.

- *Micromanipulator* The torsion pendulum has also a two d.o.f manual micromanipulator, mounted on top of the vacuum tube, for vertical and rotational alignment. The micro-positioning system allows centering of the sensor about the suspended test mass with a trial and error manual resolution of order $100\,\mu rad$ in $\Phi$ rotation, according to the capacitive sensor signal itself. This operation has been done before starting the free fall mode testing campaign, and at the end, as we will explain in the next chapter.

- *Data acquisition system* All experimental data are continuously recorded by a dedicated quasi real time data acquisition system, developed in *Labview*. Sensor data are sampled at $10\,Hz$ and measured by a PCI-ADC NI-3032 counter with several $ms$ of jitter determined by interaction with the data acquisition computer windows.

### 3.2.1 GRS electrostatic model

The GRS electrostatic model comprises the set of equations that describes mathematically the inertial sensor electrostatic forces and torques. Here we presents a general approach to the electrostatic model used in the whole thesis work.
A double approach has been used to describe the model, an analytic one that follows the basics of a parallel plate electrostatic model of the sensor capacitance surfaces, and a finite element (FE) model analysis provided by Astrium Germany, whose results are used to verify the approximations of the current electrostatic modeling applied by performance analyses, simulations, and on-board algorithms [7].
According to simple model of infinite parallel plate capacitors, conducting surfaces facing the test mass have a capacitance

$$C = \frac{\varepsilon_0 A}{d}, \tag{3.9}$$

where $A$ is the overlapping area between parallel conductors, $\varepsilon_0$ is the vacuum dielectric constant and $d$ the distance from the mass. In this case each border effects are neglected, and parallel translation of plates don't affect capacitance.
Capacitance derivatives, used to calculate both the sensor sensitivity and actuation forces, follows directly from the infinite plates approximation. This approach, considers forces and torques exerted on the TM computed taking into account only the gradient of the electrode to test mass capacitance ($C_{ELi,TM}$) w.r.t. the force direction.
An improved model has been recently introduced considering all the set of test mass and sensor surfaces, comprising electrodes and guard-rings surfaces, in a schematic configuration of sensor conductors and capacitances like in figure 3.5 [7]. This scheme form a system of conductors, for which fundamental laws are valid and from standard energetic considerations of the electrostatic field of conductors, the general form of the force arising in the system is defined as [7]

$$F_q = \frac{1}{2}\sum_{i=1}^{N}\sum_{j=i+1}^{N}\frac{\partial C_{i,j}}{\partial q}(V_i - V_j)^2, \tag{3.10}$$

where q is a generalized coordinate of all possible d.o.f., $V_i$ and $V_j$ are conductor potentials, varying in a system of N conductors, and $C_{i,j}$ is the corresponding capacitance between them. In the system of N conductors considered here, the relevant housing surfaces are those inside the GRS facing the TM and individual electrodes, known also as guard ring surfaces. This equation can thus be developed considering all the contributions of capacitance gradients between each electrode and its surrounding housing ($C_{EL_i,H}$), between the test mass and its surrounding housing ($C_{TM,H}$) and also the in-between electrode capacitance gradient ($C_{EL_i,EL_j}$):

$$F_q = \frac{1}{2}\sum_{i=1}^{18}\frac{\partial C_{EL_i,TM}}{\partial q}(V_i - V_{TM})^2 + \frac{1}{2}\frac{\partial C_{TM,H}}{\partial q}V_{TM}^2 \quad (3.11)$$

$$+ \frac{1}{2}\sum_{i=1}^{18}\frac{\partial C_{EL_i,H}}{\partial q}V_i^2 \quad (3.12)$$

$$+ \frac{1}{2}\sum_{i=1}^{18}\sum_{j=i+1}^{18}\frac{\partial C_{EL_i,EL_j}}{\partial q}(V_i - V_j)^2. \quad (3.13)$$

The second and the third lines of the equations are new terms that were not in the original electrostatic models used by LPF, which only considered the TM - electrode and TM - housing capacitances. Moreover, we don't use in the final computation the last line because the sum of EL-EL capacitance derivatives is negligible. The test mass potential $V_{TM}$ used in above equation is defined by

$$V_{TM} = \frac{\sum_{k=1}^{18}\frac{\partial C_{EL_k,H}}{\partial q}V_k + Q_{TM}}{C_{TOT}} \quad (3.14)$$

where $Q_{TM}$ is the test mass charge and $C_{TOT} = \sum_{k=1}^{18} C_{EL_k,TM} + C_{TM,H}$. By using this new approach where all the surfaces are taking into account prevents an error on force and torque estimation typically of order 10% [7] respect to the simple parallel plate model.

FE analysis model, starting with a detailed geometrical model of the sensor housing and test mass, from multiple analyses runs applying a 3D field simulation software, computed all forces and torques in the six d.o.f., as well as all capacitances in-between the TM, the 18 electrodes, and the housing.

According to this complete electrostatic model, we can define effective capacitance derivative of $x$ electrodes as sum of derivative w.r.t. $\Phi$ of electrodes respect to housing and respect to test mass, as:

$$\frac{\partial C_x}{\partial q} = \frac{\partial C_{x,EL-TM}}{\partial q} + \frac{\partial C_{x,EL-H}}{\partial q}, \quad (3.15)$$

and this will be used in all equations in the next chapters.

We will express the equation in 3.13 for force and torque during the present thesis work, where necessary for the used degrees of freedom to explain different electrostatics effects acting on our system.

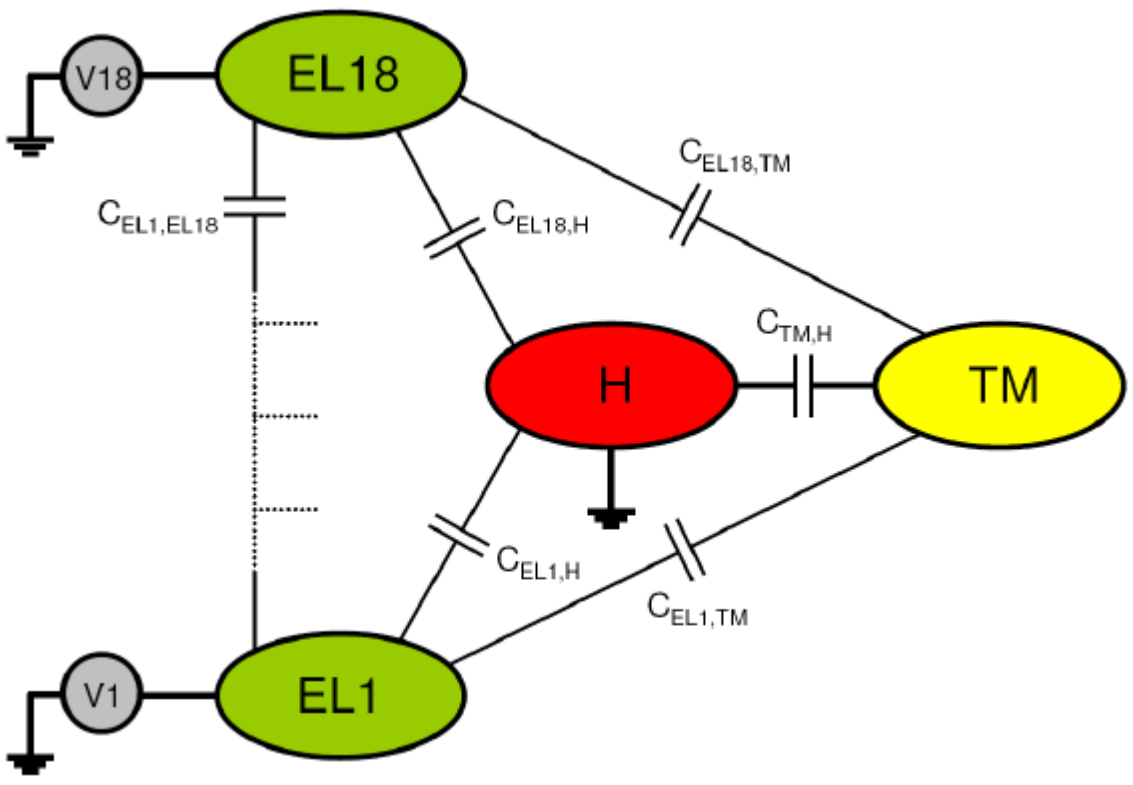

FIGURE 3.5: GRS conductors and capacitances scheme [7]. H indicates electrode Housing surfaces, while TM is the test mass surface and $EL\#$ stands for the number of electrodes.

### 3.2.2 GRS front-end electronics

General mode of functioning of the GRS readout circuit has been described in section 1.3. Here, particular attention will be paid to the description of the front-end electronic devices used as actuation circuitry of the capacitive position sensor on torsion pendulum facility, to give the idea of the authority and of the sensitivity with which the free fall experiment can be implemented.

Actuation voltages are obtained applying audio frequency and DC voltages $V_{act}$ directly to the $x$ electrodes across the transformer inputs and it is driven by a computer 16 bit PCI-DAC, a National Instruments 6703, with dynamical range $\pm 10\,V$ and resolution $312.5\mu V$. A switch driven by a 205 $Hz$ external clock alternatively transmits the applied DAC voltage with +1 and -1 gain, by switching between the output of unity gain follower and inverting amplifiers. DC and audio signals are then summed and low-pass filtered, with a cutoff around 1 $kHz$ to avoid interference with the 100 $kHz$ readout before being applied to the sensor [29].

Finally, considering that the maximum voltage that can be applied by DAC to the each electrode is $V_{DC} = \pm 10V$, which in the usual torque actuation configuration is employed on a diagonal pair of $x$ electrodes biased with opposite phase audio voltages, the maximum torque obtained is $|N_{max}| = \left| \frac{\partial C_x}{\partial \Phi} V^2_{RMS,MAX} \right| \approx 200pN\,m$.

The squarewave amplitude commanded is attenuated by a RMS attenuation factor $f_{att}$, measured to be 0.85, so that $< V^2_{RMS,MAX} >= f^2_{att}\, V^2_{DC}$. This maximum authority, in presence of a torsional spring constant $\Gamma = 8\,nN\,m/rad$, allow control of the test mass rotational degree of freedom over a range of $\approx 2.5\,mrad$.

Torque authority for torsion pendulum facility is reduced respect the LTP actuation FEE authority in flight, where it is possible to apply roughly 11 $V$ in science mode and 135 $V$ AC, applied half the time, in wide range [18].

The given authority sets a constraint about the pendulum free fall mode implementation as we will largely discuss in chapter 4, about torque authority and actuator characteristics.
Audio voltages are also used for PID control of the pendulum torsional mode, such as for rapid damping or to modify the torsion pendulum equilibrium position, while DC biases are applied for electrostatic characterization of the sensor electrodes and to measure the test mass charge.

## 3.3 Upper limit on measured torque

In this section we present the torsion pendulum sensitivity in terms of power spectral density of the external torque noise acting on the test mass during a measurement in quiet conditions.
In figure 3.6 a typical angular deflection time series is showed, for both autocollimator and capacitive sensor. The $\approx 480\,s$ free mode oscillation is visible. Torsion pendulum oscillation amplitude is less then $1\,\mu rad$ in a typical measurement. The angular time series are then converted in torque by direct application of equation of motion and processed to calculate power and cross spectral densities.
The statistical tool used to describe the sensitivity is the power spectral density (PSD) [30]. Figure 3.7 shows the typical angular noise $S_{\Phi}^{1/2}$ of the torsion pendulum as measured by the sensor and by the optical readout, compared with the intrinsic thermal limit. The angular noise follows the behavior described in the equation 3.7, in which thermal noise is factor 6 lower at low frequency below the resonance, and the almost flat readout noise dominates the frequency band above $10\,mHz$. The angular noise limit is around $30\,nrad/\sqrt{Hz}$ at $10\,mHz$.
Peaks visible at multiples of resonance frequency have been traced to the autocollimator non-linearity and will be addressed later in chapter 6. In figure 3.8 the angular noise is converted into an instantaneous applied torque $N(t)$ through the torsion pendulum equation of motion 3.1, in time domain, that can be rewritten:

$$N(t) = I\ddot{\Phi}(t) + \beta\dot{\Phi}(t) + \Gamma\Phi(t). \tag{3.16}$$

This is to point out that the quantity to estimate at time $t_i$ is $N(t)$, obtained approximating the first and second derivatives of $\Phi$ in $t_i$ by means of a parabolic fit to the 5 adjoining points at times $t_{i-2} \dots t_{i+2}$. The obtained values can be then substituted in equation 3.16 to calculate $N(t)$. This method get rids of both transients and free oscillation amplitude, giving an instantaneous estimate of $N(t)$ independent from the initial conditions. Otherwise, the torque noise is thus not estimated via the angular noise and the frequency domain transfer function as in equation 3.4, but rather from the PSD of the torque time series calculated as in equation 3.1. In either case, accurate estimation relies on knowledge of pendulum parameters $I$, $T_0$ and $Q$. This technique (eqn 3.16) allows us to replace, if needed, the elastic force $\Gamma\Phi$ with a more complicated position dependent restoring force (with non linear dependences form angular position $\Phi^2$, $\Phi^3$ terms). The presence of possible noise sources, originating in the G.R.S. itself or from coupling to environmental disturbances, can prevent the reaching of the torque sensitivity limit and resolve the fused silica pendulum thermal noise. It is important to identify and characterize these disturbances and eventually, if they are both significant and well-characterized, subtract them from the torque time series, as we will see in the next section.

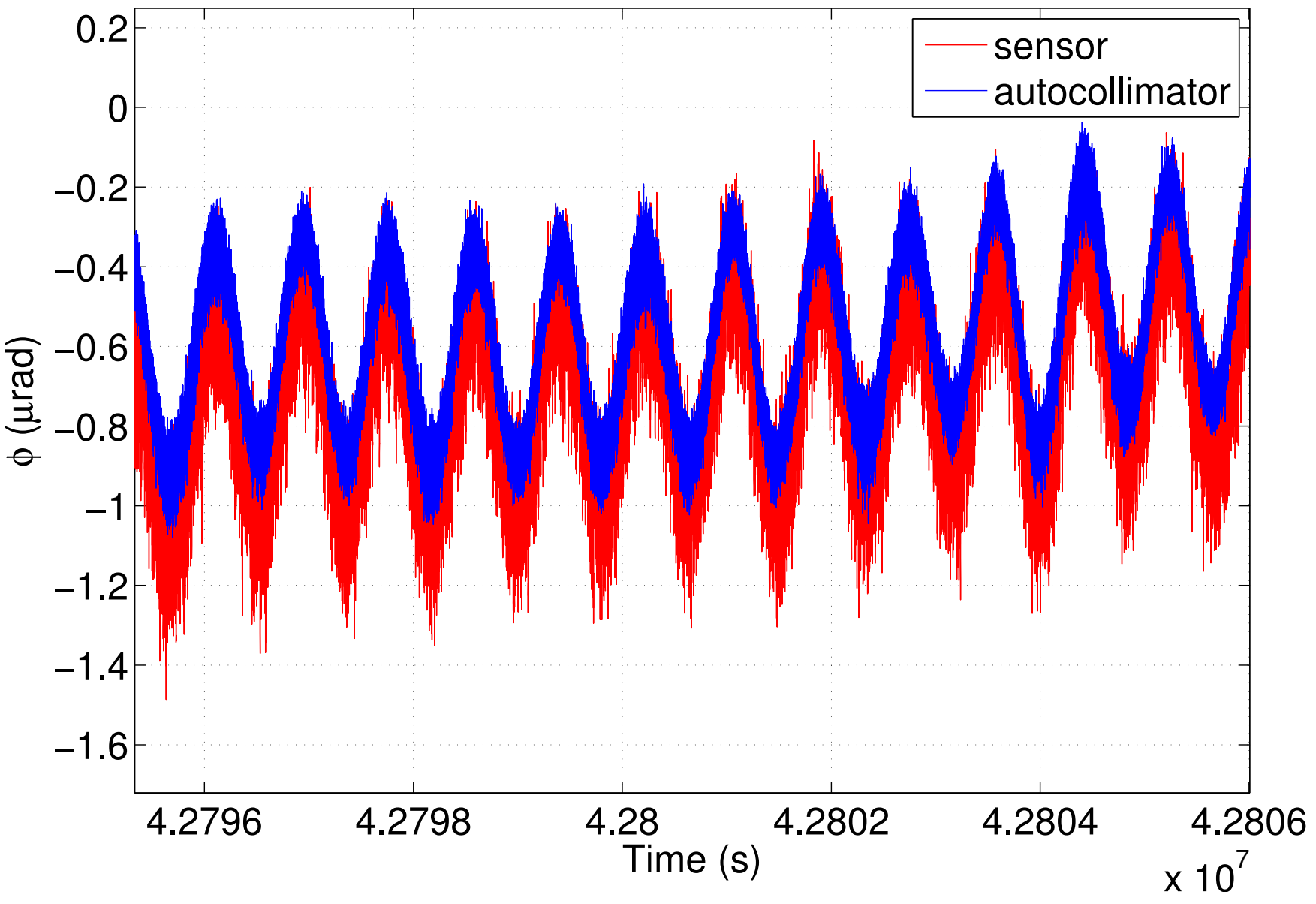

FIGURE 3.6: Typical pendulum angular deflection during a 10000 s of a torque noise measurement run. Both the readout are able to resolve the torsion pendulum free oscillation.

### 3.3.1 Relevant environmental noise sources

The thermal limit is a lower limit on the pendulum sensitivity and on the resolution with which we can characterize stray torques at low frequency. We measure noise above this limit and it is possible to investigate at least several environmental noise sources that have been observed in the past to produce significant torque noise and can couples to the thermal limit [23]. We identify three disturbance categories: *trans-twist noise*, or the coupling to the laboratory floor tilt, *magnetic field fluctuations* and *temperature fluctuations*.

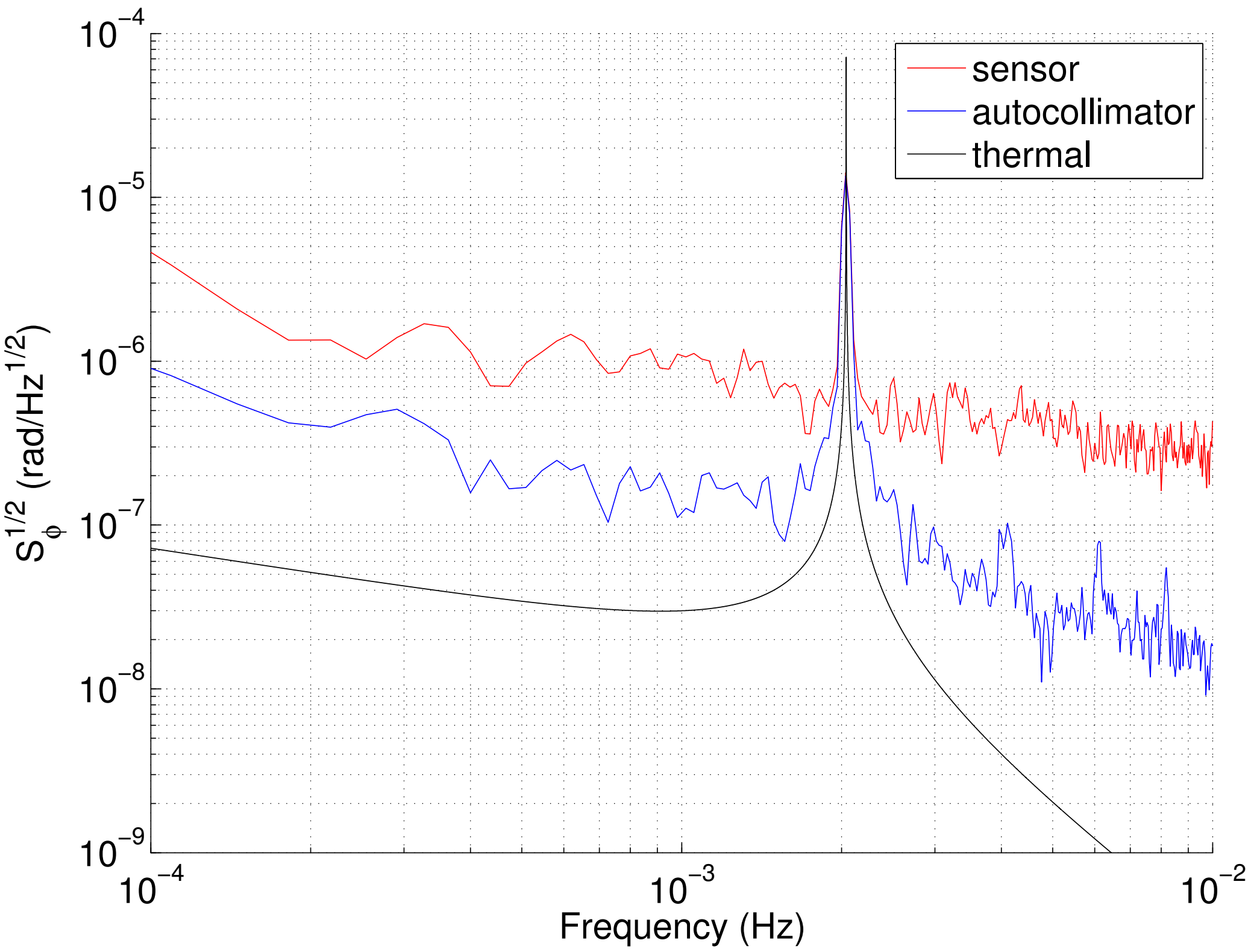

FIGURE 3.7: Angular noise spectrum of one weekend noise data. The red curve is from sensor readout, the blue one is from autocollimator and the black curve is the thermal limit.

In cases of linear coupling of some disturbance into the measured pendulum torque, each disturbance $A$ can be considered in terms of its contribution to the pendulum torque as a $\Delta N = \frac{\partial N}{\partial A}\Delta A$. Their fluctuation can have also a contribution on noise $S_{N_A}^{1/2} = |\frac{\partial N}{\partial A}|S_A^{1/2}$. By estimating the coefficients relative to the disturbance, then the corrected torque $N_{corr}$ is given by the difference between the measured torque $N_{meas}$ and the sum on all the known disturbances $\frac{\partial N}{\partial A_i}\Delta A_i$.

$$N_{corr} = N_{meas} - \sum_i \frac{\partial N}{\partial A_i}\Delta A_i. \tag{3.17}$$

The coupling factors can be characterized by experiments in which the external source is modulated at an high enough level to induce a well resolved signal in the pendulum twist. Otherwise it can be possible to calculate the coupling factors by minimizing the residual torque noise obtained with the disturbance subtraction. This minimization is performed using the least square fit method in order to evaluate the coefficients with which to subtract off the effects of different noise fields. This allows to minimize the noise spectrum in a certain frequency range, checking for any correlation (with a cross correlation analysis) between the measured torque and the known disturbances sources.

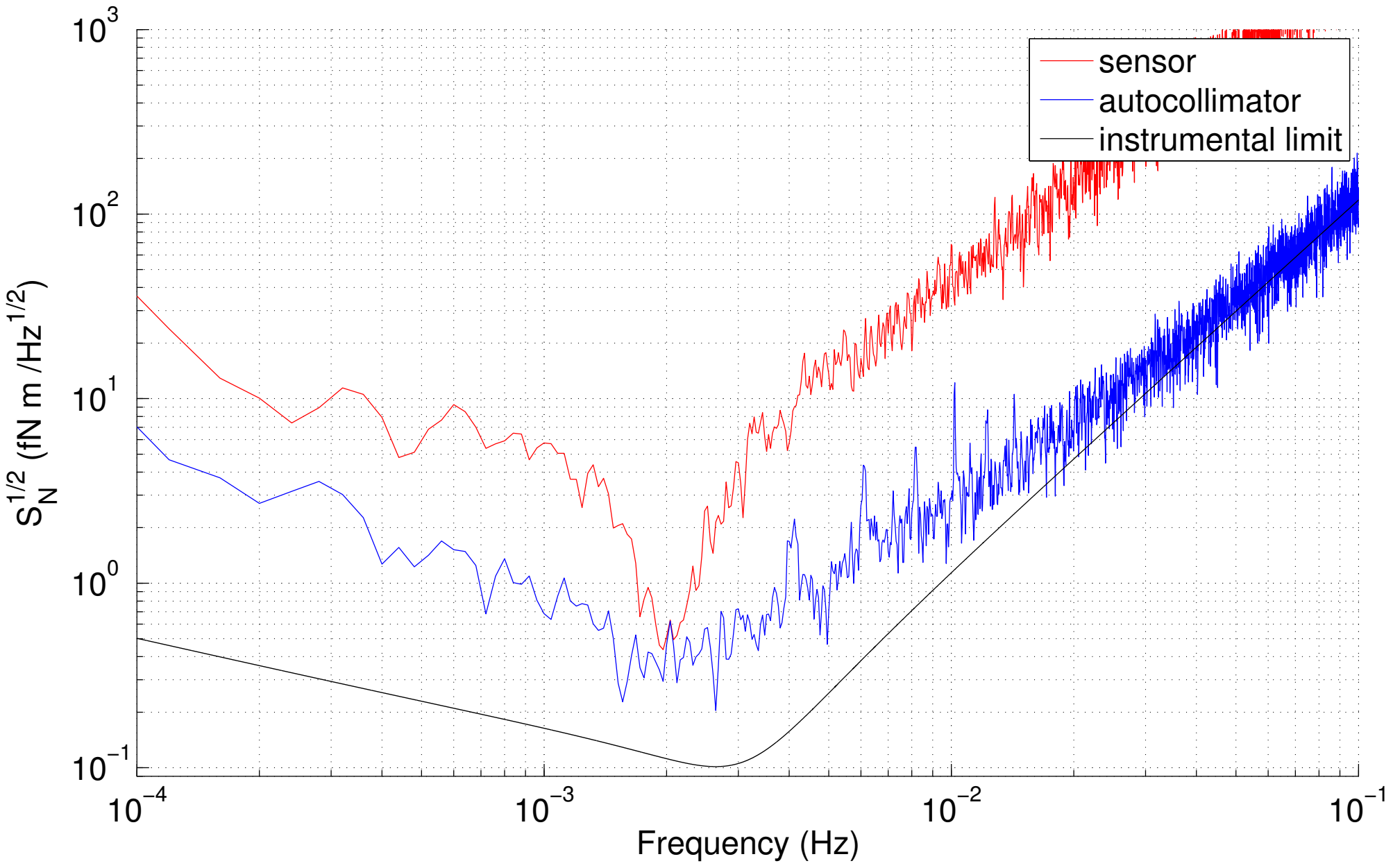

FIGURE 3.8: Torque noise spectrum of one weekend noise data. The red curve is from sensor readout, the blue one is from autocollimator and the black curve is the thermal limit.

**Tilt-twist** The tilt-twist effect originates in a stray interaction between the suspended pendulum and the position sensor itself, and was observed affecting pendulum noise during the past years [31], with downgrade version of the facility. Any tilt motion of laboratory floor, any structure relaxation or even laboratory temperature fluctuation, can produce a torque on the pendulum. The same effect is produced either by translating the sensor housing with respect to the suspended test mass and viceversa by holding the sensor housing and tilting the apparatus. This is proportional, to first order, to the relative translation displacement of the TM with respect to apparatus:

$$\Delta N_{tt} \approx \frac{\partial N}{\partial \eta}\Delta\eta + \frac{\partial N}{\partial Y}\Delta Y \tag{3.18}$$

where $\Delta\eta$ and $\Delta Y$ are the tilt angle and the translational d.o.f. along the $Y$ direction, measured by the sensor.

The coupling to floor tilt was measured by purposely tilting the apparatus by moving a weight on the floor platform, where the pendulum sits. For each step, a bilinear fit is performed in $\eta$ and $Y$ to estimate the contribution of $\Delta\eta$ and $\Delta Y$. Then a least square fitting is performed to extract the coupling coefficients $\frac{\partial N}{\partial \eta}$ and $\frac{\partial N}{\partial Y}$ and finally the corrected torque is obtained subtracting the relative contributions as $N_{corr} = N_{meas} - N_{tt}$.

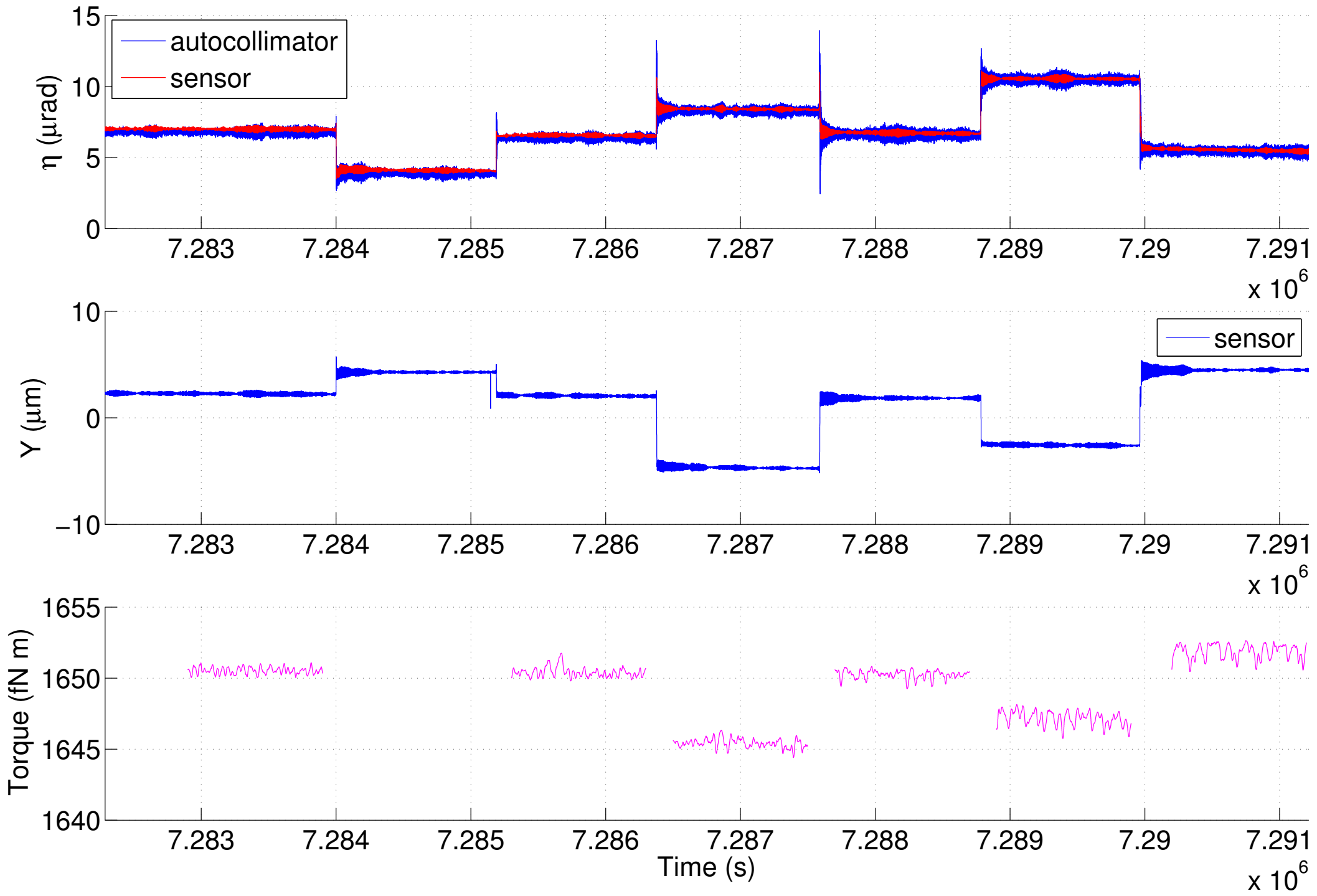

FIGURE 3.9: $\eta$, Y and torque time series during a tilt twist measurement.

We typically measure coupling coefficients $\frac{\partial N}{\partial \eta}$ and $\frac{\partial N}{\partial Y}$ with $0.02\, nNm/m$ precision, and the measured values are typically $\frac{\partial N}{\partial \eta} \approx 0.1\, nN\, m/rad$, and $\frac{\partial N}{\partial Y} \approx 0.6\, nN\, m/rad$, with variations of 70% observed over the 1.5 years of measurements.

We do not know if the measured coupling comes from some residual electrostatic or other interaction inside the GRS, or from some coupling at the fiber suspension point, but this noise source is no longer a limit to the noise of the pendulum, being very close to the thermal limit at frequencies below $1\, mHz$, as visible in figure 3.10. It will become relevant if we hope to one day reach the thermal noise limit.

**Temperature** Environmental temperature fluctuations can induce a torque on the pendulum as they may cause structural distortion of the apparatus, changes in the fiber equilibrium angle and, in general, mechanical deformations. They can also produce electronics drifts like variation in the readout electronics gain, or induce a tilt motion of the whole experimental apparatus, introducing a fake defection signal in the readout. Laboratory floor motion or thermal relaxations of the vacuum vessel, can act as the trans-twist coupling effect, producing a similar torque noise. This is proportional, again

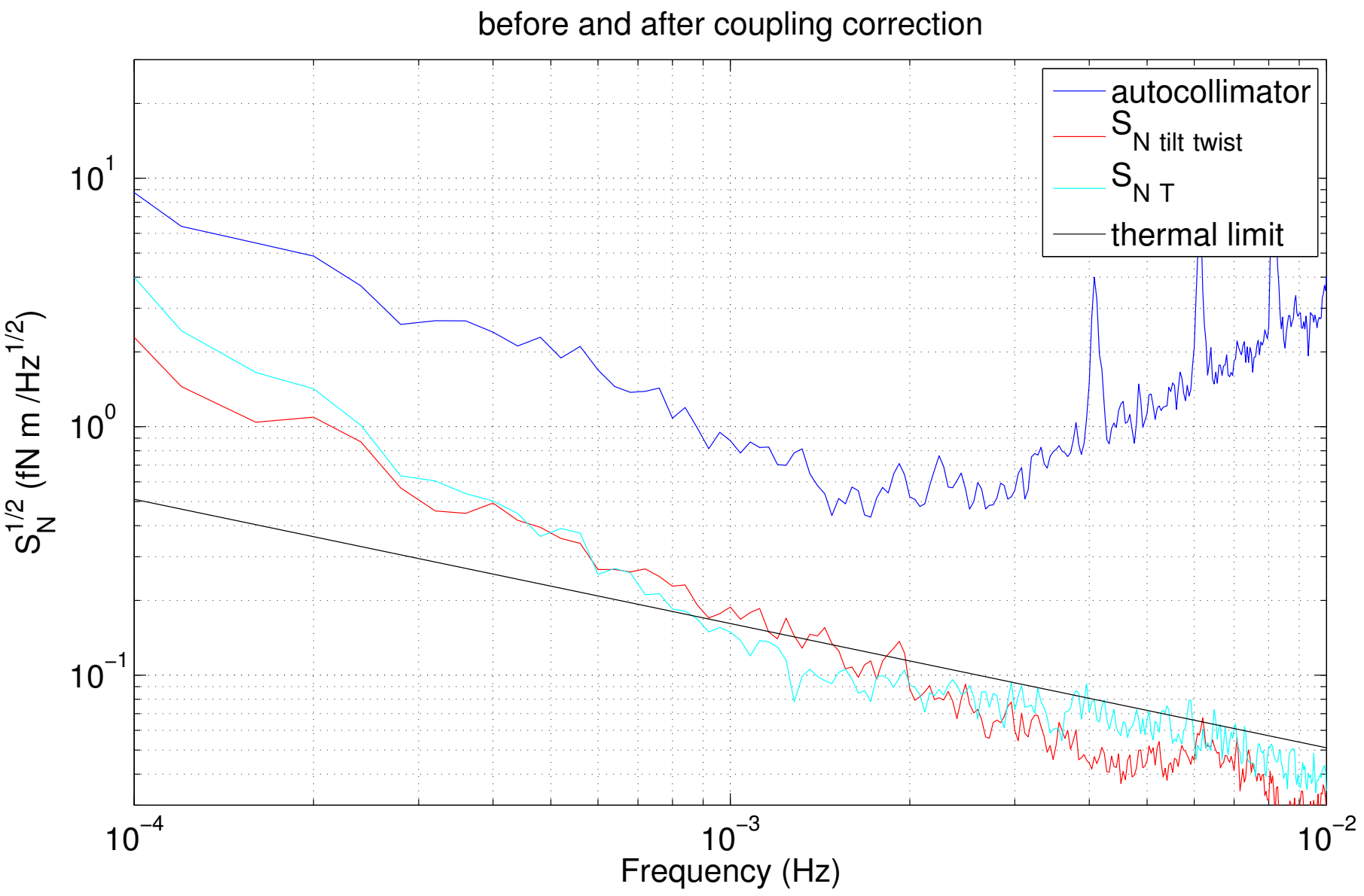

FIGURE 3.10: Torque noise spectrum of one weekend noise data. The red curve is the contribution of the tilt-twist coupling to torque. The cyan curve is the contribution of the temperature coupling to torque.

with the same linear behavior, to

$$\Delta N_T \approx \frac{\partial N}{\partial T_{platform}} T_{platform} + \frac{\partial N}{\partial T_{fiber\ tube}} T_{fiber\ tube}. \tag{3.19}$$

It is possible to calculate the coupling factors by minimizing the residual torque noise obtained with the disturbance subtraction, starting from the temperature measured time series. In this case we have minimized the noise due to the effect of the temperature variation of platform in coincidence to fiber tube thermal variation, in the frequency range $0.4 - 1\, mHz$. This because the fiber temperature variation can change the value of the torsional spring constant and the temperature fluctuations of the vacuum tube supporting the torsion fiber was observed to induce variations in the angular equilibrium point.

The coupling factors estimated are $\frac{\partial N}{\partial T_{platform}} = -4 \pm 12\, fN\, m/K$ and $\frac{\partial N}{\partial T_{fiber\ tube}} = 130 \pm 70\, fN\, m/K$. As such, the temperature coupling would appear to be barely resolvable. Additionally, as seen in figure 3.10, coupling at this level produces an effect that is well below the measured noise floor.

It is worth to note that the minimization is performed on data already corrected for tilt-twist coupling in order to found a connection between different weekend measurements. Other torque effects caused by temperature fluctuations among the others thermometers used in the facility were evaluated, but any correlation was founded, so their coupling factors can be considered negligible.

**Magnetic field** The local magnetic field $\vec{B}$ can exert a torque on the pendulum by direct interaction with its residual magnetic moment $\vec{m}_\perp$ that has two components in the horizontal plane $xy$, so that $\vec{N}_B = \vec{m}_\perp \times \vec{B}$. The magnetic field fluctuations will induce a torque noise of order $S_{N_B}^{1/2} \approx \sqrt{m_x^2 S_{B_y} + m_y^2 S_{B_x}}$. In order to avoid any coupling with this effect, all the parts of the pendulum are made by non-magnetic materials and a $\mu$-metal shield surrounds the apparatus inside the vacuum chamber. Anyway, it is possible to measure the coupling factors in the $x$ and $y$ directions, with a magnetometer placed outside the $\mu$-metal shield, in order to evaluate any stray magnetic contribution, by modulating a sinusoidal external magnetic field at $3\,mHz$ frequency, by means of a round coil of 45 $cm$ radius and 40 turns, alternatively placed orthogonally to the preferred directions, and observing the coherent deflection of the pendulum rotation angle in phase with the oscillating magnetic field.
The measured magnetic noise along the two directions is $S_B^{1/2} \approx 60\,nT/\sqrt{Hz}$ at $1\,mHz$ and the measured torque is proportional to the applied magnetic field as:

$$N_B \approx \frac{\partial N}{\partial B_x} B_y - \frac{\partial N}{\partial B_y} B_x \tag{3.20}$$

The measured coupling factors are of the order $\frac{\partial N}{\partial B_x} \approx 0.22 \pm 0.01\,n\,Am^2$ and $\frac{\partial N}{\partial B_y} \approx 0.135 \pm 0.002\,nA\,m^2$. So, the stray torque contribution is $S_{N_B}^{1/2} = \sqrt{m_x^2 S_{B_y} + m_y^2 S_{B_x}} \approx 0.016\,fN\,m/\sqrt{Hz}$ at $1\,mHz$, a factor ten below the thermal limit, so it is not a limiting factor for the pendulum's torque sensitivity performance.

We note in figure 3.10 that the trans-twist and temperature coupling effects cannot explain most of the excess pendulum noise. As such, we will not perform any subtraction for such effects from our torque time series. It is still partly related to the pendulum systematic effects. Any unexplained excess, even if possibly due to other intervening mechanisms irrelevant to eLISA and LTP, can in principle come from the presence of the GRS, and should thus be considered as an upper limit to the disturbances originating inside it. Some of the noise can be related to the noise in the readout, and we will show in the next section how we can separate this noise source from true force noise.

## 3.4 Readout noise and cross spectral density analysis

The cross spectral density (CSD) between torque measured by the two readout, the optical and the capacitive one, is used as a tool to reduce the noise [23].
The real pendulum motion should be in principle detected by all the sensors if they are correctly calibrated. The angular position measured from each sensor can be thought as a combination of the real angular position of the pendulum $\Phi(t)$ plus a noise $\Phi_{n_i}(t)$ that is combination of the read out noise and a possible contribution due to the motion of sensor respect to the TM, $\Phi_i(t) = \Phi(t) + \Phi_{n_i}(t)$. The angular part of noise should have a part of totally correlated noise among the sensors, and another that depends on the used sensor. The same idea can be applied when we convert the angular displacement in torque, procedure that involve only linear operations. It is thus possible to write for the two readout:

$$N_{ac}(t) = N(t) + N_{ac}^{noise}(t) \tag{3.21}$$

$$N_{sensor}(t) = N(t) + N_{sensor}^{noise}(t) \tag{3.22}$$

We can define two combination of the measured $N_{ac}$ and $N_{sensor}$:

$$N^{+}(t) = \frac{N_{ac}(t) + N_{sensor}(t)}{2} = N(t) + \frac{N_{ac}^{noise}(t) + N_{sensor}^{noise}(t)}{2} \tag{3.23}$$

$$N^{-}(t) = \frac{N_{ac}(t) - N_{sensor}(t)}{2} = \frac{N_{ac}^{noise}(t) - N_{sensor}^{noise}(t)}{2} \tag{3.24}$$

If we consider $N_{ac}^{noise}(t)$ and $N_{sensor}^{noise}(t)$ ideally stationary and zero mean random process, we can calculate the power spectral densities:

$$S_{N^{+}}(\omega) = S_N(\omega) + \frac{S_{N_{ac}^{noise}}(\omega) + S_{N_{sensor}^{noise}}(\omega)}{4} \tag{3.25}$$

$$S_{N^{-}}(\omega) = \frac{S_{N_{ac}^{noise}}(\omega) + S_{N_{sensor}^{noise}}(\omega)}{4} \tag{3.26}$$

$S_N(\omega)$ is the power spectral densities of the true torque acting on the TM, and can be estimated considering the definition of $N^{+}$ and $N^{-}$, by subtracting the last two equations

$$S_N(\omega) = S_{N^{+}}(\omega) - S_{N^{-}}(\omega) = Re(S_{N_{ac},N_{sensor}}(\omega)). \tag{3.27}$$

The power spectral density $S_N(\omega)$ of the correlated part of the signal computed is thus equal to the real part of the cross spectral density between the torque time series obtained by the two different sensors output. It is worth to note that this determination of the PSD of pendulum torque noise allow to use the readout of two partially independent sensors to get rid of the uncorrelated noise read by each of them.

The CSD is computed using the *Welch periodogram method* [30], given that we process time series of finite length. This method is an unbiased estimator of power and cross spectral density, from finite length discrete real time series and its expectation value should be the external torque power spectral density, defined real and positive. The uncertainty on the estimated CSD obtained from a single stretch of data is given by:

$$\delta S_{N_{ac},N_{sensor}}(\omega) = \sqrt{S_{N_{ac}N_{ac}}(\omega) S_{N_{sensor}N_{sensor}}(\omega)} \tag{3.28}$$

where $S_{N_{ac}N_{ac}}(\omega)$ and $S_{N_{sensor}N_{sensor}}(\omega)$ are the PSD of single processes. In general, the uncertainty of the CSD depends on the readout noise contribute from each sensor. Consequently, the cross correlation technique for readout noise rejection gives optimal results when the readout noise of the two detectors is comparable. When one of the two contributes much larger readout noise than the other, the cross correlation still provides a good estimation of the external torque power spectral density, but with a larger uncertainty than the PSD of the output of the less noisy detector, due to the noise contribution of the noisiest sensor. The ability of extracting from noisy measurements the external torque acting on the torsion pendulum is limited by the noisiest of the readouts. It is possible to use some strategies to reduce the uncertainty of the spectral estimator. First of all it is possible to divide the data string from which the power spectral density is performed into $n$ segments of the same length. Then, the best estimate of the cross spectral density is then computed as the *average* of the CSD obtained from each segment, obtaining a reduction of the uncertainty by a factor $\sqrt{n}$. However, dividing the time series into shorter segments for averaging, therefore limits the minimum frequency and the frequency resolution of the obtained cross spectral density because they are related to the inverse of the time duration of the data segment. Because we are interested to measure the performance of our instruments in the low frequency region, by averaging to reduce the uncertainty on the estimation of the spectral densities we need to perform measurements longest possible. For our frequency region of interest, that extends below 0.1 mHz, we need data stretches of at least 10000 s. Usually we choose 25000 s long spectral windows for discarding the first three points of the spectrum known to be biased with a Blackman Harris windowing function.

The application of a normalized window function like the Blackmann Harris 3rd order function, together with the operation of data detrending, is another operations that ensure the data stretch smoothly approaches zero at its ends, avoiding artifacts in the estimated power spectral density introduced by the unavoidable truncation of the data string to a finite length.

To obtain maximum information from the time series, it is then possible to use the head and tail of each data segment more than once in the estimation of the cross spectral density. This is done dividing the time series in *overlapping* segments. The amount of overlapping chosen for Blackman Harris windows, to maximize the information and avoid the introduction of some correlation between the PSD computed from adjacent windows, is 0.66% for our system data.

Another important operation is to perform a *frequency binning*, in order to produce a smoother estimate of the spectrum. The frequency range is divided into logarithmically spaced bins and, the spectrum points belonging to the same bin, are averaged. For pendulum data is performed a singularly frequency binning, in order to obtain 10 frequency bin per decades. Binned spectra are then averaged into groups of ten consecutive data stretches and uncertainty in each frequency bin computed from the standard deviation of the samples.
All the operations performed on pendulum data allow to reduce the problematic of the computation of spectral estimates uncertainties that is peculiar of low frequency experiments.

## 3.5 Excess torque noise

As a conclusion of torsion pendulum facility presentation, we show the external torque noise best estimate measured in the recent past. This is the remaining unknown source of noise attributed to the test mass interaction with the GRS, when all other identified disturbances have been subtracted, and is considered as an upper limit to the excess torque noise we can measure with torsion pendulum facility. The excess noise is computed subtracting the thermal noise background from the measured external torque.
The measurements are realized mainly during the weekend, to remove the environmental noise produced by the presence of human activity during the work day and reach high sensitivity performances. In order to have an high statistics to reduce spectral estimation uncertainty, as explained in the previous section, we need to average over long time measurements repeatedly in the time. If the torsion pendulum torque noise can be considered stationary in the period of time covered by the measurements, it is possible consider non consecutive measurements to form a single estimate of the external torque power spectral density.
Figure 3.11 shows data obtained for seven long weekend of noise runs performed in the summer of 2013. The cross correlation technique has been used to reduce the spectrum uncertainties at low frequency and it is the plotted as green points in the figure. It is clear that the torsion pendulum facility is able to reach a sensitivity around $0.8\, fN\, m/\sqrt{Hz}$ at $1\, mHz$. It is useful to convert the noise level from torque to acceleration, to directly compare with the eLISA and LISA Pathfinder requirements, where the relevant experimental quantity is a translation acceleration noise. It is done, as explained in section 3.1, and results, compared with missions requirements are shown in figure 3.12. As conclusion, the torsion pendulum facility is able to rule out a large class of TM surface disturbances at level of $30\, fm/s^2\sqrt{Hz}$ at $1\, mHz$, within a factor 1.5 of LISA Pathfinder goal.

Finally, it is possible also to show the noise stationarity during time for our facility. This is a necessary requirement to statistically compare sets of data acquired over a long range of time. Figures 3.13 shows the difference between torque power spectral densities from two groups of measurements taken from the whole set of August and September 2013 shown in figure 3.11 and 3.12. We subtract two group of three weekends of noise runs in order to show the reproducibility of background noise level over time. The difference is consistent with zero within the estimated uncertainty in a small part of the frequency region of interest, and is compared with the LPF requirement, suitably converted into an equivalent torque noise on a single mass. The possible error due to non-stationarity is around $0.2\, fN^2m^2/Hz$, that converted in to an effective differential acceleration noise, is roughly the level of LISA Pathfinder goal. However, we are able to detect extra acceleration noise acting on our instrument within LISA Pathfinder specifications in the frequency region of interest near $1\, mHz$, either from actuation noise or by the free fall experiment.

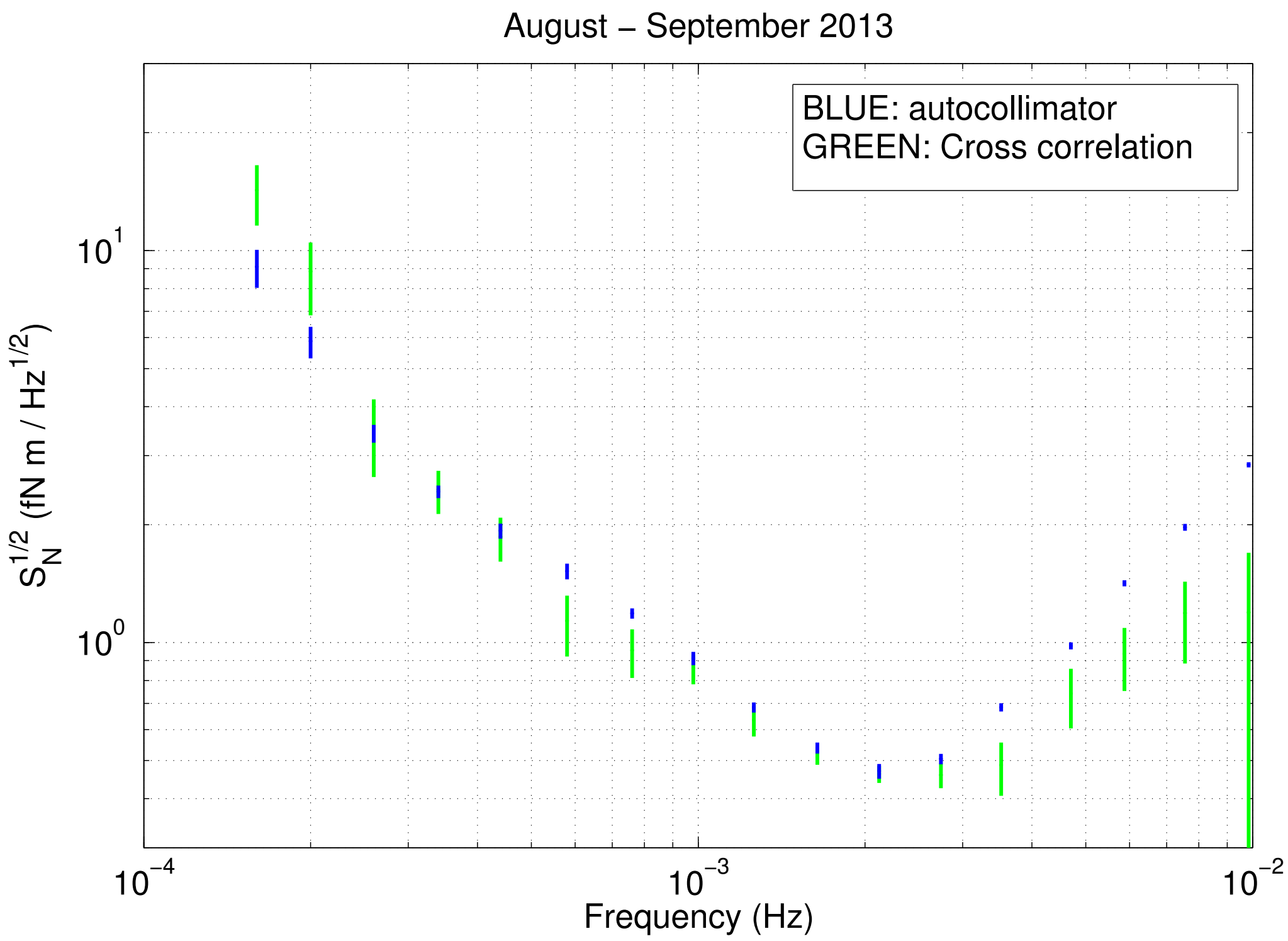

FIGURE 3.11: Estimated toque noise excess spectral density for a set of noise run performed during August and September 2013. Blue points are autocollimator data. Green points represent cross correlation between the sensor and the autocollimator.

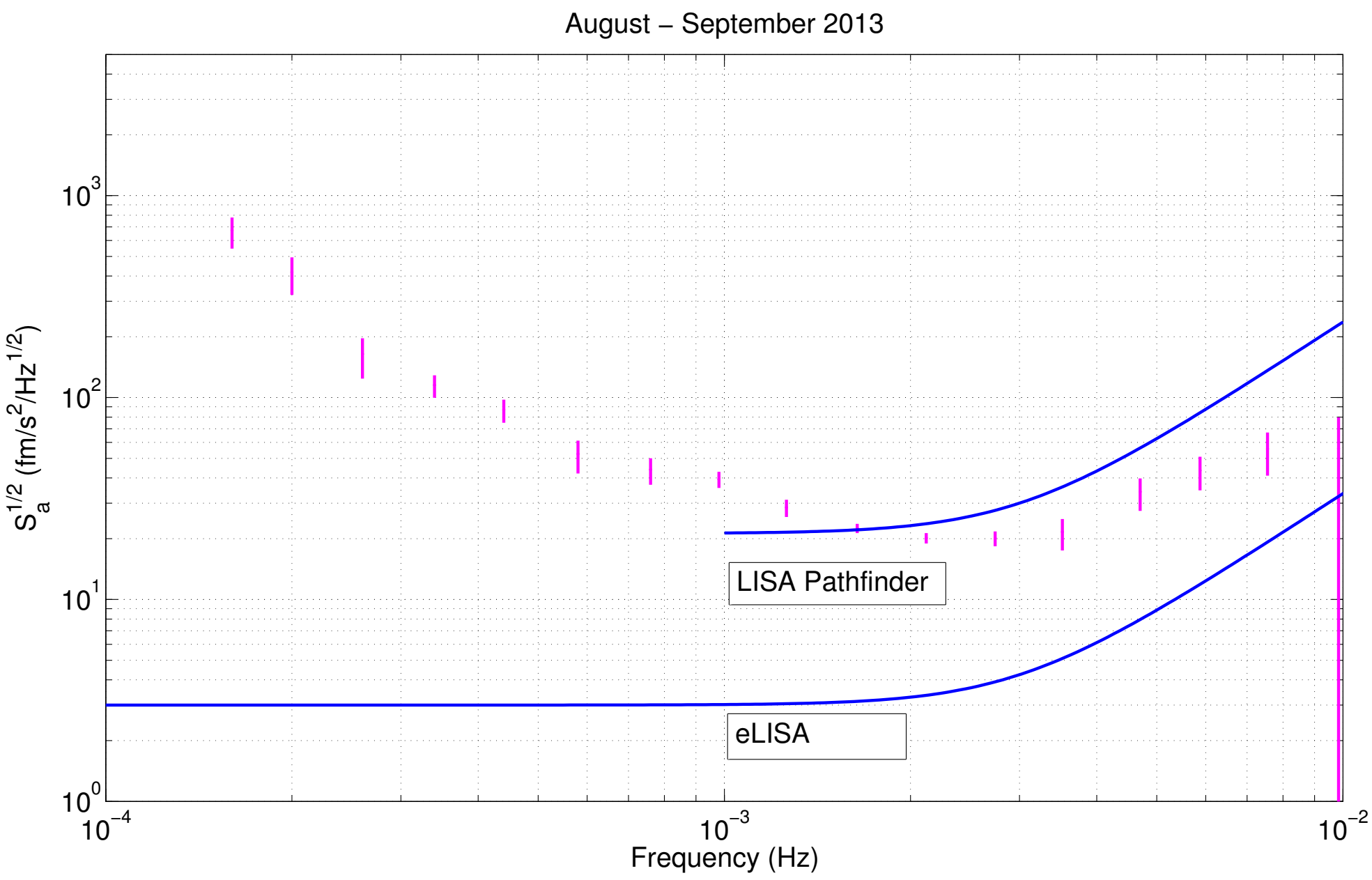

FIGURE 3.12: Estimated acceleration upper limits obtained from torsion pendulum measurements as cross correlation between sensor and autocollimator, opportunely converted from torque noise measurements. This is compared with eLISA and LISA Pathfinder specifications.

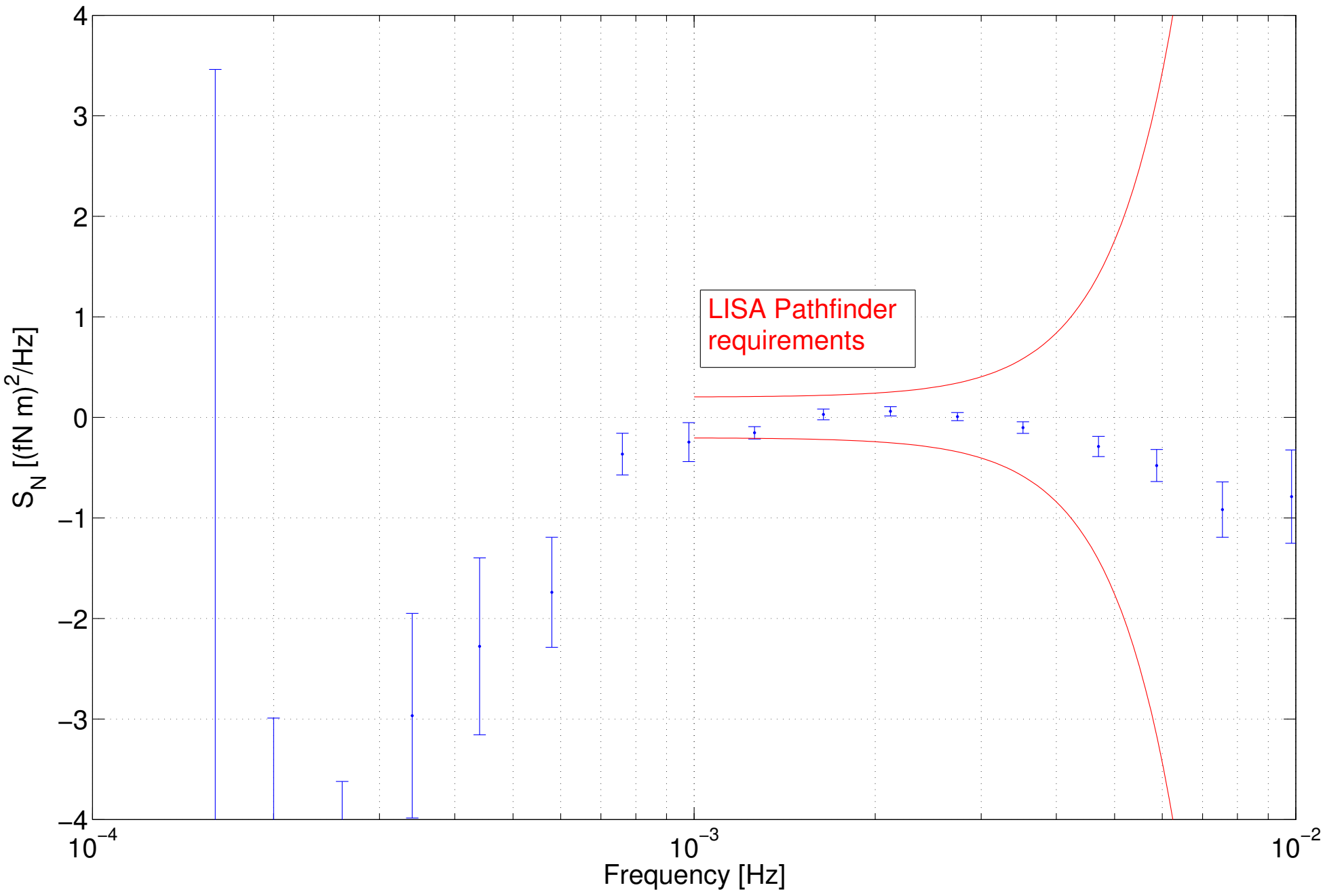

FIGURE 3.13: Difference between the external torque noise power spectral densities estimated from two groups of the same set of data compared to the LPF requirements converted into an equivalent torque noise on a single mass.

# Chapter 4

# Free fall mode on torsion pendulum

As shown in chapter 2, actuation noise is potentially the dominating noise source in the LPF differential acceleration measurement and the free-fall mode represents a possible way to eliminate this noise component. However, this has not been demonstrated experimentally. Small force testing of the relevant GRS configuration, presented in chapter 3, has all been performed with continuous torsion pendulum measurements in which the TM is essentially still at the GRS center without actuation.
Moreover, many aspects like much larger dynamic range associated with the free-fall mode or data analysis change introduced by gaps in the usable data, require experimental verification before application in flight. Our torsion pendulum provides a similar system for such a test, with a quantitatively interesting possible level of sensitivity.
The measured torque noise sensitivity of $0.8\, fNm/\sqrt{Hz}$ at $1\, mHz$ corresponds to an effective LPF TM acceleration noise level $40\, fm/s^2\sqrt{Hz}$. Moreover, the torque authority of the actuation circuit used for testing on ground is around $200\, pNm$ and corresponds to an equivalent differential applied force in DC of roughly $20\, pNm$, corresponding to $2\, nN$ forces applied by each of 2 electrodes, as we will explain later in this chapter. These characteristics, make the pendulum a good instrument to test the feasibility of the free fall mode.
The in-flight free fall mode test is designed to measure the intrinsic test mass acceleration noise in the absence of applied electrostatic forces. This is a technique of disturbance reduction and on ground it is possible to implement a similar but simplified experiment with the objective to establish a facility for performing a parallel testing. The on-ground experiment aims to demonstrate, with sufficient sensitivity, the feasibility of the free-fall mode and the level of force noise measurement that can it allows. Additionally, if the sensitivity is sufficient, we aim to measure the actuation force noise and to verify the possibility of actually achieving a lower noise level with the free-fall mode.
This is done thanks to the possibility to effectively simulate the LPF differential DC gravitational acceleration with a large external torque on our pendulum. By rotating pendulum respect to the gravitational reference sensor, it is necessary to apply external torques on the suspended mass to hold it centered, and this means to mimic the LPF gravity gradients that must be compensated in orbit. Moreover, the on-ground experiment has the advantage to tune the effective DC gravitational imbalance, to allow more flexibility to explore different control strategies, by varying flight and impulse time or control points, and different dynamic configurations made possible by having a variable stiffness.

## 4.1 Measurement concept

In flight a differential gravity imbalance must be compensated to hold the two masses at a near constant distance. For this reason the second test mass must be actuated to avoid being accelerated away, applying a force to compensate the differential acceleration $\Delta g_{DC}$. On ground a similar DC torque $N_{DC}$ can be introduced on torsion pendulum if it is rotated with respect to the inertial sensor housing, in order to simulate a large DC acceleration that require a force to be compensated, so that the test mass can be held centered. In this way the condition on torsion pendulum are very close to the in flight configuration and allow to implement the free fall control mode.
An ideal experiment to test our free-fall mode sensitivity would then consist of three measurements: measure the noise floor of torsion pendulum in nominal conditions; measure the external torque noise in presence of a continuous actuation force in order to measure the contribution from the actuation fluctuation that affect the noise estimation; finally, measure the noise floor during the free fall mode to compare with the expected one with no forces applied.

- *Background force noise measurement.* This is the measurement of the torque noise floor of torsion pendulum as presented in section 3.5 and showed in figure 3.11. On suspended mass acts a real external torque $N(t)$ so that:

$$I\ddot{\Phi} = -\Gamma\Phi - \gamma\dot{\Phi} + N(t) \tag{4.1}$$

  The signal detected from the pendulum angular deflection coming from the readout, is sum of the real displacement, plus a term of noise

$$\Phi_m = \Phi + \Phi_n. \tag{4.2}$$

We can thus define an observer for the external torque N acting on the pendulum

$$\hat{N}_1 \equiv I\ddot{\Phi}_m + \Gamma\Phi_m + \gamma\dot{\Phi}_m \tag{4.3}$$

and includes a real torque $N_1(t)$ as in equation 4.1, plus a term of readout noise components

$$\hat{N}_1 = N_1(t) + (\ddot{\Phi}_n + \Gamma\Phi_n + \gamma\dot{\Phi}_n). \tag{4.4}$$

The measured noise is thus $S_{\hat{N}_1}$. This noise measurement is obviously performed without any applied electrostatic actuation, when the pendulum suspension point, and thus the test mass, are centered with respect to the gravitational reference sensor electrode housing.

- *Torque noise with applied DC actuation force.* The idea is to measure again the torque noise, but in a different condition, where the pendulum suspension point is rotated and the TM is held in the GRS center with a constant applied torque. The torsion pendulum is, in this configuration, rotated with respect to the GRS electrode housing by an angle $\Delta\Phi$ to simulate a large DC acceleration acting on mass. It is thus necessary to apply a DC torque $N_{DC} = -\Gamma\Delta\Phi_{EQ}$ to roughly hold the test mass centered respect the sensor housing. We remember that $\Gamma$ is the pendulum elastic torsion constant (see section 3.1). This DC force is analogous to the bias $\Delta g_{DC}$ on the non-reference $TM2$ suspension in LPF, needed in orbit to compensate the self-gravity difference. Unlike Pathfinder, the bias for the torsion pendulum can be tuned by adjusting $\Delta\Phi$, effectively allowing us to set the $N_{DC}$ to a chosen level.

  With this constant DC force applied, the pendulum equation of motion became in principle

$$I\ddot{\Phi} = -\Gamma_2\Phi - \gamma\dot{\Phi} + N(t) + N_{ACT}(t), \tag{4.5}$$

The stiffness $\Gamma_2$ is not the same in equation 4.1. It has an additive contribution due to the actuation, as we will explain in section 4.2.3. The torque measured by the readout is then

$$\hat{N}_2 \equiv I\ddot{\Phi}_m + \Gamma_2\Phi_m + \gamma\dot{\Phi}_m - \hat{N}_{ACT} \tag{4.6}$$

where $\hat{N}_{ACT}$ is the constant commanded torque applied to keep the mass at center. Measured torque include again the real torque $N_2(t)$ plus a noise component so that the measured total torque is

$$\hat{N}_2 = N_2(t) + (N_{ACT}(t) - \hat{N}_{ACT}) + (\ddot{\Phi}_n + \Gamma_2\Phi_n + \gamma\dot{\Phi}_n). \tag{4.7}$$

The true actuation torque $N_{ACT}(t)$ has two component

$$N_{ACT}(t) = \alpha \hat{N}_{ACT} + \delta N_{ACT}(t). \tag{4.8}$$

This is the commanded torque $\hat{N}_{ACT}$ multiplied by an actuator calibration factor $\alpha$, plus the noise of the actuator of the GRS front-end electronics, $\delta N(t)$. By substituting in equation 4.7

$$\hat{N}_2 = N_2(t) + (\delta N_{ACT}(t) + (\alpha - 1)\hat{N}_{ACT}) + (\ddot{\Phi}_n + \Gamma\Phi_n + \gamma\dot{\Phi}_n). \tag{4.9}$$

The measured torque noise in this phases is $S_{\hat{N}_2} = S_{N_2} + S_{N_{ACT}}$, which include the contribution from the actuation fluctuations,

$$S^{1/2}_{N_{ACT}} = 2N_{DC}S^{1/2}_{\delta V/V}, \tag{4.10}$$

analogous to the "in band" fluctuation of the actuation drive voltage amplitude that translates into acceleration noise in flight, in the limit that the two actuators have correlated fluctuations, as we saw in equation 2.20, $S^{1/2}_{\Delta g} = 2\Delta g_{DC}S^{\frac{1}{2}}_{\delta V/V_{act}}$. The contribution on torque noise coming from the noisy electrostatic actuation produces an excess in the noise power compared to the first configuration, that we want to account for with the explained second measurement.

- *Free fall torque noise with applied impulse.* Using the same pendulum rotation angle as in the previous experiment, the free fall control scheme is employed to control the position of the TM. Torque impulses are applied periodically with a frequency $f_{exp} = 1/T_{exp}$. The length of flights is defined as $T_{fly}$, while the impulse is long $T_{imp}$, so that the experiment time is $T_{exp} = T_{fly} + T_{imp}$ and $T_{fly}$ can be connected to the impulse duty cycle $\chi \equiv \frac{T_{imp}}{T_{exp}}$.
  In order to hold the test mass centered on average in between two impulses, the applied torque must be equal, to $N_{DC} = -\Gamma\Delta\Phi_{EQ}$ of the previous experiment. So, the amplitude of each impulse is $N_{imp} \approx -\Gamma\Phi_{EQ}/\chi$.
  We can write again the measured torque from the readout

$$\hat{N}_3 \equiv I\ddot{\Phi}_m + \Gamma\Phi_m + \gamma\dot{\Phi}_m \tag{4.11}$$

$$= N_3(t) + (\ddot{\Phi}_n + \Gamma\Phi_n + \gamma\dot{\Phi}_n), \tag{4.12}$$

where $N_3(t)$ is the true torque in between two impulses, during which, the actuation $N_{ACT}(t)$ is no longer present and the test mass is actuated only during the impulse phases. Data that are free from continuous actuation, are analyzed to estimate the free fall torque noise, $S_{\hat{N}_3}$ and then compared with both the background $S_{\hat{N}_1}$ and continuous actuation cases $S_{\hat{N}_2}$, with the aim to recover the actuation free torque noise $S_{\hat{N}_1}$ in the impulses experiment and resolve the excess actuation noise in $S_{\hat{N}_2}$.

## 4.2 Pendulum dynamics during free fall

During the free fall mode experiment, the free motion of torsion pendulum follows the dynamic of the free oscillation of a simple harmonic oscillator, truncated by the applied impulses. The pendulum dynamics in between two impulses, is a free oscillation around its equilibrium point $\Phi_{EQ}$, as showed in figure 4.1.

It is possible to calculate the angular dynamics of what we could define as an ideal flight, in which the TM traces a trajectory around the GRS center (and thus with a time average angle of zero) repeatedly in every flight. We start from the basic pendulum equation of motion, simplified neglecting the contribution of the dissipation term

$$I\ddot{\Phi}(t) = -\Gamma\Phi(t) + N(t). \tag{4.13}$$

If N(t) is constant in time $N = N_{DC}$, this has a solution

$$\Phi(t) = A\cos\omega_0 t + B\sin\omega_0 t + \Phi_{EQ}, \tag{4.14}$$

where $\Phi_{EQ} = N_{DC}/\Gamma$. We set $t = 0$ at center of flight so that it lasts from $-T_{fly}/2$ until $+T_{fly}/2$ and introduce conditions for position and velocity $\Phi(0) = \Phi_0$ and $\dot{\Phi}(0) = 0$. In this way $A = \Phi_0 + \Phi_{EQ}$ and $B = 0$, so that

$$\Phi = (\Phi_0 - \Phi_{EQ})\cos\omega_0 t + \Phi_{EQ}. \tag{4.15}$$

Moreover, during the flights, the time average of $\Phi(t)$ between $-\frac{T_{fly}}{2}$ and $\frac{T_{fly}}{2}$ is chosen to be zero, so that:

$$< \Phi(t) >= \frac{1}{T_{fly}}\int_{-\frac{T_{fly}}{2}}^{\frac{T_{fly}}{2}} \Phi(t)dt = 0. \tag{4.16}$$

This means to choose time average pendulum angle being equal to some setpoint value $\Phi_{set}$, that in this case is zero. Finally, with a little algebra, this ideal free motion is given by

$$\Phi(t) = \Phi_{EQ}\left(1 - \frac{\omega_0\frac{T_{fly}}{2}}{\sin\omega_0\frac{T_{fly}}{2}}\cos\omega_0 t\right). \tag{4.17}$$

This solution is related to the pendulum period $T_0$ and to the pendulum stiffness $\Gamma$, via the angular frequency $\omega_0 = 2\pi/T_0$ that is equal to $\frac{1}{2\pi}\sqrt{\frac{\Gamma}{I}}$. The solution is also symmetric about $t = 0$ and by design has a null average angle. This solution diverges in amplitude at $T_{fly} = T_0$, while in the limit of $T_{fly} = T_0/2$, will be a free fall with $\Phi(t) = \Phi_{EQ}$ for $t = T_{fly}/2$. It is thus important to choose correctly the flight time duration related to the pendulum period as we will see below. In the limit $\frac{1}{T_{fly}} \gg \frac{1}{T_0}$, this would reduce to constant angular acceleration, $\frac{1}{2}(\frac{\Gamma\Phi_{EQ}}{I})t^2$, and thus to a parabolic motion characteristic of constant acceleration.

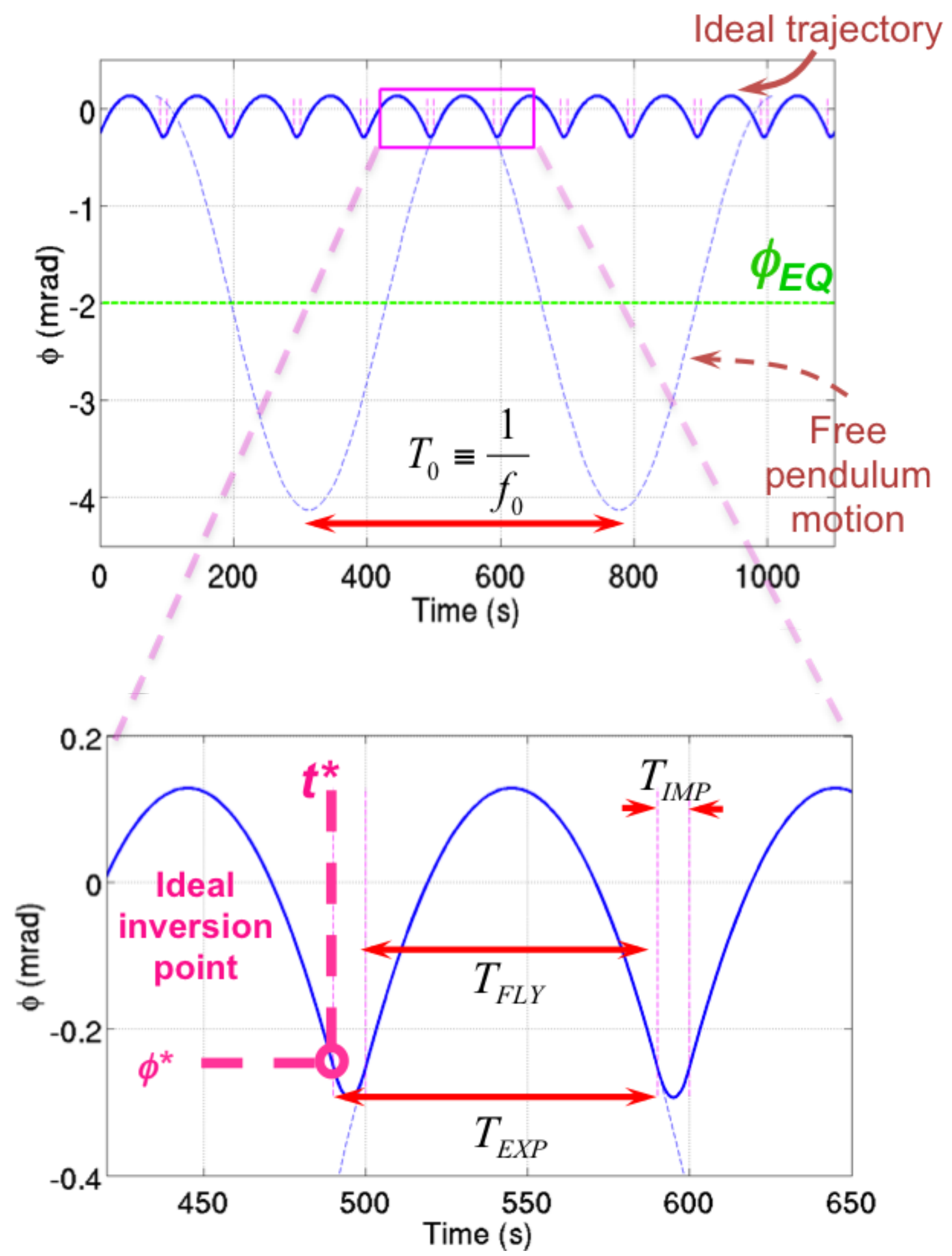

FIGURE 4.1: Pendulum trajectory during the free fall mode.

Deriving the equation 4.17 it is possible to obtain the angular velocity of suspended mass

$$\dot{\Phi}(t) = \Phi_{EQ}\frac{\omega^2\frac{T_{fly}}{2}}{\sin\omega\frac{T_{fly}}{2}}\sin\omega t. \tag{4.18}$$

There is an ideal impulse torque value such that the pend motion is periodically forced by the impulses applied, to come back to a single initial position with a chosen velocity, to allow a periodic flight. Ideally, the value of impulse $N_{imp}$ can be calculated starting from the equation of motion during an impulse phase:

$$I\ddot{\Phi}(t) = -\Gamma(\Phi(t) - \Phi_{EQ}) - \Gamma_{ACT}\Phi(t) + N_{imp} \tag{4.19}$$

whose solution is

$$\Phi(t) = C\cos\omega_0 t + D\sin\omega_0 t + \frac{N_{imp} + \Gamma\Phi_{EQ}}{\Gamma + \Gamma_{ACT}} \tag{4.20}$$

with $\omega_0 = \sqrt{\frac{\Gamma+\Gamma_{ACT}}{I}}$. $\Gamma_{ACT}$ is the induced stiffness of the actuation due to the electrostatic, that sum to the fiber stiffness. Considering, this time, that $t = 0$ corresponds to the beginning of the impulse phase, and considering that $\dot{\Phi}^*$ and $\Phi^*$ are velocity and position at that time and come from the end of the previous free fall phase, the impulse to apply can be derived imposing $\dot{\Phi}(0) = \dot{\Phi}^*$, $\Phi(0) = \Phi^*$ and $\dot{\Phi}(\frac{T_{imp}}{2}) = 0$.
With some calculation to determine $C$ and $D$ parameters, the impulse to apply became:

$$N_{imp} = (\Gamma + \Gamma_{ACT})\left(\Phi^* - \frac{\dot{\Phi}^*}{\omega_0\tan(\omega\frac{T_{imp}}{2})}\right) - \Gamma\Phi_{EQ}. \tag{4.21}$$

Using the ideal velocity $\dot{\Phi}^*$ and position $\Phi^*$ as solutions of the free fall phases equation 4.17, it is possible to derive the definitive impulse necessary to drive the mass at the right position at the beginning of the next flight. In principle, the ideal flight can be repeated over and over by starting with the correct initial conditions and repeatedly applying the ideal $N_{imp}$ as in figure 4.4. However, any imperfection in the initial conditions, or any torque noise, will drive the pendulum away from this ideal motion, as in figure 4.3. To avoid instability, a control scheme has been implemented, made by an observer that estimates the pendulum position and velocity before each impulse. Then a controller estimates the impulse intensity needed to reach the initial point for the next cycle, by using the pendulum dynamic constants as well as flight and impulse times. Controller mechanism will be described in the next section.

### 4.2.1 Controller

The purpose of the controller is precisely to calculate the right impulse to apply to bring the mass to the ideal position to start the next flight. Given initial conditions different from the ideal, the controller should, after a certain number of kicks, stabilize the position of the mass and converge to a single repeated applied torque impulse corresponding ideally to that calculated in equation 4.21. This allows to make flights on average around the set-point desired. Controller scheme is showed in figure 4.2. An observer estimates pendulum position and velocity at the instant of the beginning of the impulse, i.e. the last 100 points before the end of the free fall phase (corresponding to 10 s), using a linear LSQ fit to a model of quadratic time dependence of the pendulum position. Then controller calculates the impulse intensity that will be opportunely converted in voltages then applied by the GRS FEE actuation circuit to a diagonal couple of $x$ electrodes (as in scheme 4.5 and as we will explain in detail later).

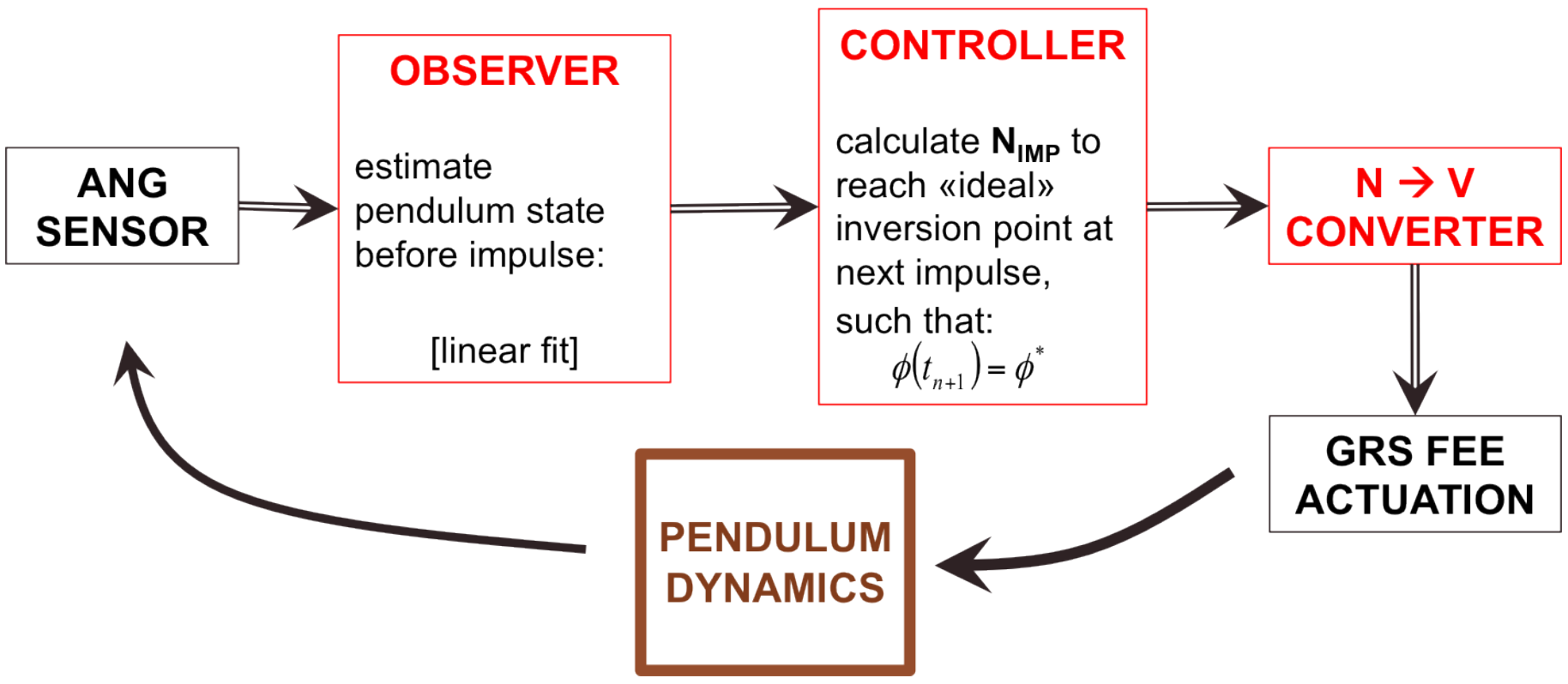

FIGURE 4.2: Scheme od controller working flow.

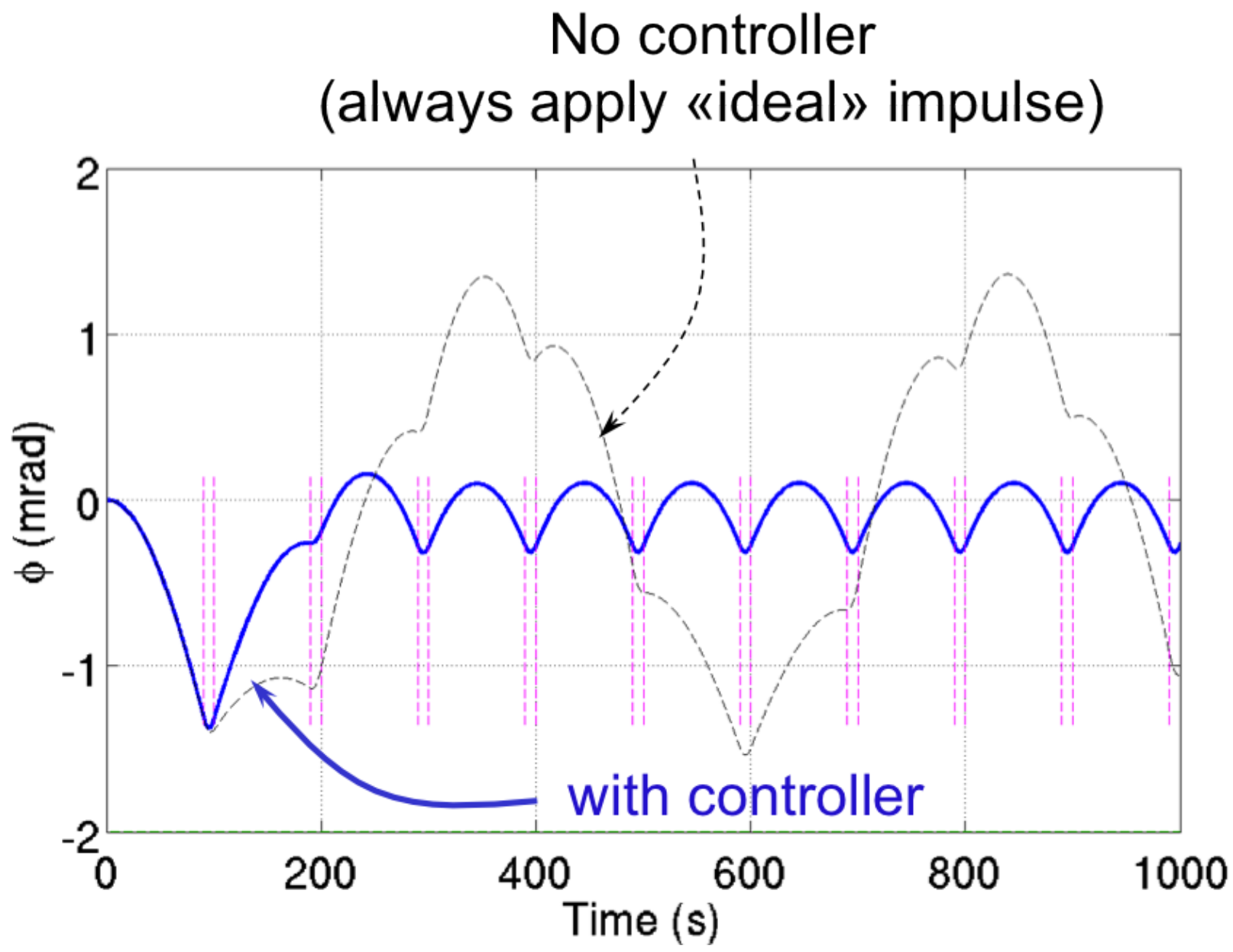

FIGURE 4.3: Simulated free fall angular displacement with control scheme in an open and closed loop. The action of controller allow to stabilize the mass position after the initial phase.

### 4.2.2 Labview controller

Observer and controller are implemented in laboratory using Labview control routine, in order to evolve pendulum motion in time, starting from initial condition that are position $\Phi^*$ and velocity $\dot{\Phi}^*$ of the pendulum at the end of free phase, following the equation described in section 4.2. In figure 4.3 a time series during the free fall mode obtained from simulated data, is showed. It is clear from figure 4.3, that the use of controller in a closed loop is mandatory, in simulation as in the real case, to reach the convergence in time of the pendulum motion towards a stable free fall displacement. In the ideal case, the achievement of stability depends only on the initial conditions and there is only one value of $\Phi_0$ and $\dot{\Phi}_0$ such that $\Phi(0) = \Phi(T_{fly})$ and $\dot{\Phi}(0) = -\dot{\Phi}(T_{fly})$. From equations 4.17 and 4.18, ideal position and velocity values are:

$$\Phi(0) = \Phi(T_{fly}) = \Phi_{EQ}\left(1 - \frac{\omega_0 \frac{T_{fly}}{2}}{\sin \omega_0 \frac{T_{fly}}{2}}\right) \tag{4.22}$$

$$\dot{\Phi}(0) = -\dot{\Phi}(T_{fly}) = -\Phi_{EQ}\omega_0^2 \frac{T_{fly}}{2}. \tag{4.23}$$

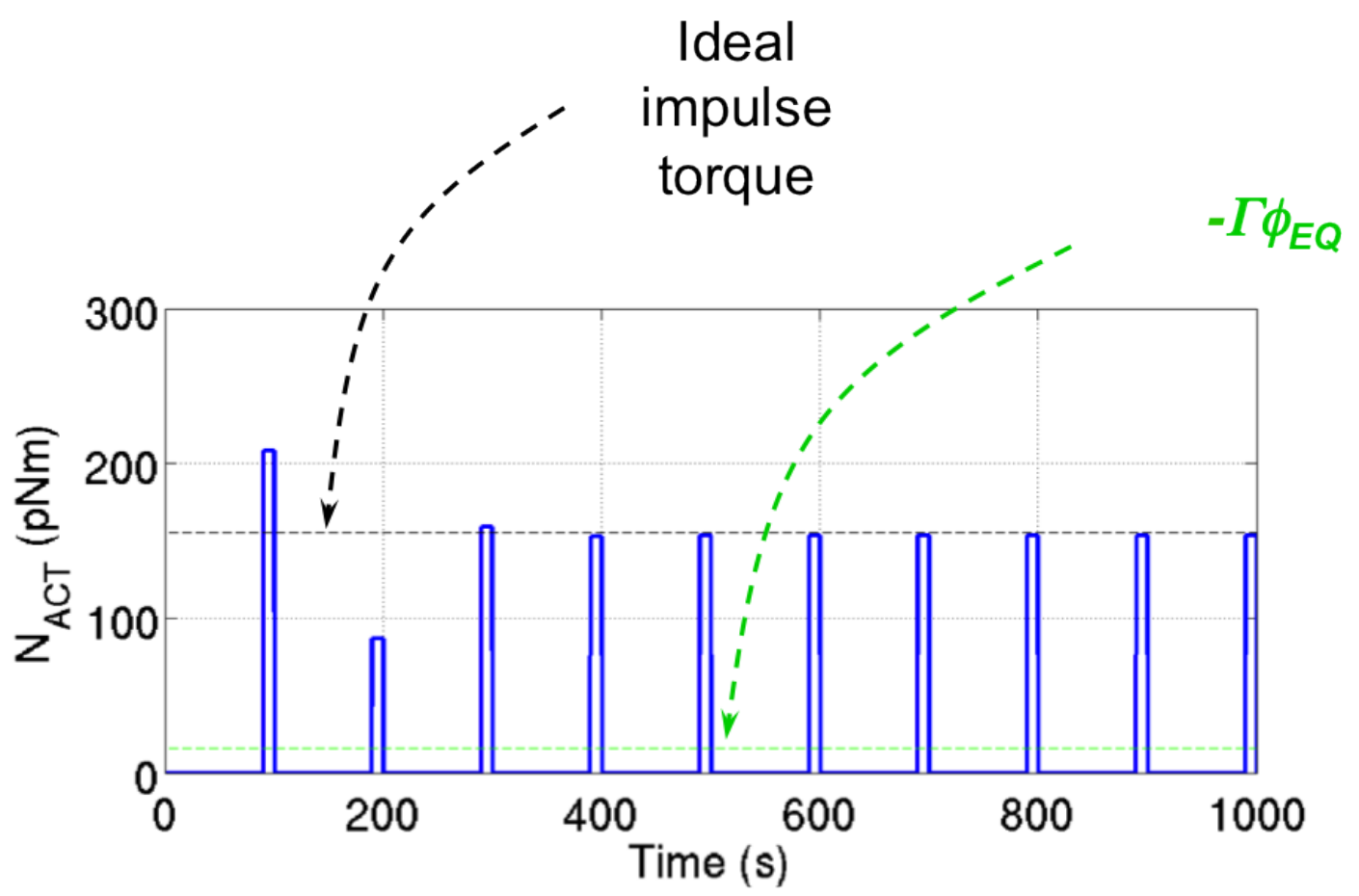

FIGURE 4.4: Controller impulse amplitude applied on a simulated harmonic oscillator. After a certain time, the impulse value stabilize to the ideal case.

Independently from the boundary conditions, the controller should ensure that the mass reach the ideal position expressed by equation 4.22. Otherwise the motion is observed to oscillate with much larger amplitude than needed.

There is only a limitation to the possible torque that the actuator can produce that correspond at $V_{MAX} = \pm 10\,V$, maximum voltages available. If the initial conditions define an impulse greater than this value, the actuator can provide only the maximum value of the design, considering that the maximum authority is $N_{MAX} \approx 200pNm$, as we will explain in the next section.

Moreover, from equations 4.22 and 4.23 it is clear that the main parameters that can affect the convergence towards a stable motion are the equilibrium angle $\Phi_{EQ}$ and the duty cycle $\chi$, or $T_{fly} = (1-\chi)T_{exp}$ that must be chosen so that $N_{imp} \leq N_{MAX}$.

### 4.2.3 GRS FEE actuator

As explained in section 3.2.2, the actuation circuit can apply voltages to the $x$ electrodes, in order to produce a constant DC actuation and then a constant torque $N_{DC}$ on mass as in the proposed second measurement, or to produce large impulses $N_{imp}$ to kick the pendulum during the free fall mode. The circuit produces a low-pass filtered square wave, applied with opposite phase to diagonal pairs of x electrodes to nominally maintain the TM at zero induced potential. For a positive torque, $\pm V_{1\Phi}$ are applied to $EL1$ and $EL3$, as can be seen in figure 4.5, with $\pm V_{2\Phi}$ applied to $EL2$ and $EL4$ for negative torques. In general, the torque can be expressed:

$$N_{ACT} = \left| \frac{\partial C}{\partial \Phi} \right| (< V_{1\phi}^2 > - < V_{2\phi}^2 >) \tag{4.24}$$

where $< V_{i,\phi}^2 >$ are time average square voltages applied as in figure 4.5. This is slightly lower than the squarewave amplitude commanded by DAC because of the presence of low pass filter that attenuate of an RMS factor measured to be $f_{att} = 0.85$, so that $< V_{RMS,MAX}^2 >= f_{att}^2 V_{MAX}^2$. The allowed maximum torque applicable, for maximum allowed DAC voltages $V_{MAX} = 10\,V$ is thus $N_{MAX} = \left| \frac{\partial C}{\partial \Phi} \right| < V_{RMS,MAX}^2 > \approx 200\,pNm$, assuming that the capacitance derivative with respect to $\Phi$ is equal to $\left| \frac{\partial C}{\partial \Phi} \right| = 2.84 \pm 0.06\,pF/rad$, as measured during the testing campaign.
The corresponding rotational stiffness can be written as

$$\Gamma_{ACT} = - \left| \frac{\partial^2 C}{\partial \Phi^2} \right| (< V_{1\phi}^2 > + < V_{2\phi}^2 >). \tag{4.25}$$

The maximum value of torque $N_{MAX}$ allows also to give an estimate of the actuation stiffness

$$|\Gamma_{ACT}| = \frac{\frac{\partial^2 C}{\partial \Phi^2}}{\frac{\partial C}{\partial \Phi}} N_{MAX} \approx 2.81\,nNm/rad. \tag{4.26}$$

considering the measured value of $\frac{\partial^2 C}{\partial \Phi^2} = 28.14 \pm 0.03\,pNm/rad^2$.
The actuation scheme allows a range of torques $N_{ACT}$ while holding the stiffness $\Gamma_{ACT}$ constant [20], by choosing $V_{1\phi}^2$ and $V_{2\phi}^2$ such that $< V_{1\phi}^2 > + < V_{2\phi}^2 > \equiv V_{Auth}^2 = const.$ The resulting available torque range is thus $[-N_{Auth}, +N_{Auth}]$ where

$$N_{Auth} = \left| \frac{\partial C}{\partial \Phi} \right| V_{Auth}^2. \tag{4.27}$$

In case of the free fall experiment, this set the maximum allowable $N_{imp}$, can't overcome $\chi N_{MAX}$. Considering a duty cycle of 10%, it is not possible to get over $20\,pNm$ of total DC torque, corresponding to $2\,nN$ forces applied by each of 2 electrodes. The maximum authority allows to set also the maximum angle to which torsion pendulum can be rotated, considering the constraint about the fiber stiffness $\Gamma = I\sqrt{2\pi/T_0} = 8\,pNm/rad$ for a period of $T_0 = 460\,s$. With $N_{MAX} \approx 200\,pNm$, the corresponding maximum angle of rotation is about $\chi N_{MAX}/\Gamma \approx 2.5\,mrad$. The bias for the torsion pendulum can be tuned by adjusting $\Delta\Phi$ up to the level provided by the geometry of the sensor, effectively allowing us to set the $N_{DC}$ to a desired level.

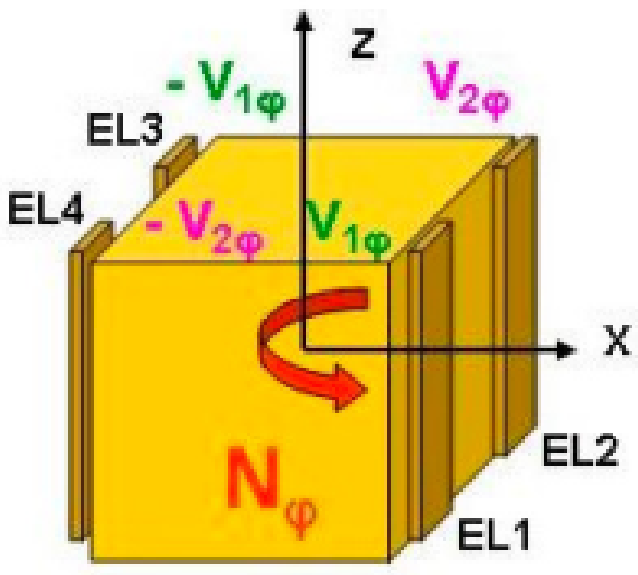

FIGURE 4.5: Scheme of x electrodes able to produce a torque $N_\Phi$ on mass. The actuation voltages $V_{i,\Phi}$ are applied on diagonal pairs of electrodes. Choosing $V_{2\Phi}^2 = 0$ will produce a positive torque on pendulum.

In the current free fall mode implementation, furthermore, the value of maximum authority $N_{Auth}$ is reduced by choosing $180\,pNm$. In this case, the induced stiffness due to actuation is, from 4.26, around $2\,nNm/rad$ that corresponds to commanded RMS voltage of $V_{Auth,commanded} = \sqrt{\Gamma/\frac{\partial^2 C}{\partial\Phi^2}} \approx 8.5\,V$.

### 4.2.4 Simulated data with a harmonic oscillator simulator

Pendulum dynamic during free fall experiment can be simulated implementing a simple harmonic oscillator simulator. It is a state space representation of the pendulum system equations used to compare the ideal expected motion during free fall with the real dynamic during the implemented experiment, prepared in MATLAB by our group.

It can reproduce the time series evolution during free fall experiment starting from free experimental parameters like total duration of one cycle $T_{exp}$, duty cycle $\chi$ (related to the impulse time duration as $T_{imp} = \chi T_{exp}$), pendulum angle of equilibrium, $\Phi(0)$ and $\dot{\Phi}(0)$ which are initial values of angle and angular velocity. All the pendulum intrinsic dynamic parameters like moment of inertia and fiber stiffness can be fixed in the model. An example of simulated free fall time series is shown in figure 4.6 on top. The panel below shows the corresponding torque time series, calculated from the equation of motion 3.1, and the impulse phases are visible.

Many options can be implemented to simulate real case of interest. A suitable level of white noise can be added as time series of noise in torque or readout. Though a simplification of the true pendulum noise sources, these allow an idea of dynamics and analysis issues with noise levels similar to those of the pendulum. Moreover, it is possible to restrict the value of $N_{imp}$ to the maximum allowed by the torsion pendulum actuation board.

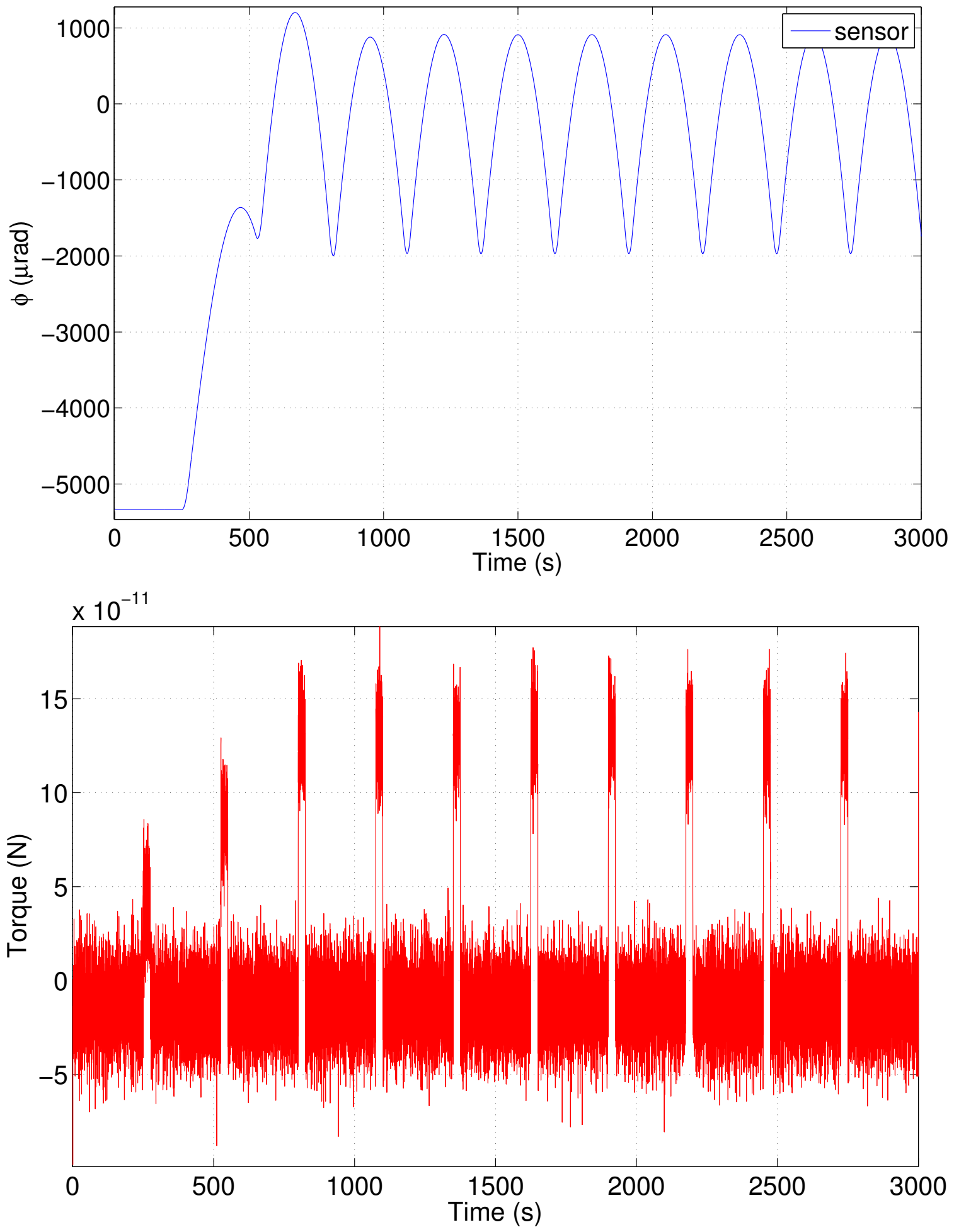

FIGURE 4.6: Simulated time series of angular motion during free fall experiment and the corresponding value of $I\ddot{\Phi}$.

The pendulum simulator will be used as an important test of the dynamics and control in different conditions, as well as the limits of data analysis techniques developed for the free fall measurements.

## 4.3 Experimental parameters and implementation

As explained until now, the on-ground experiment performed with torsion pendulum has the advantage to allow more flexibility to explore different control strategies, by varying flight and impulse time or control points, and different dynamic configurations made possible by having a variable stiffness.

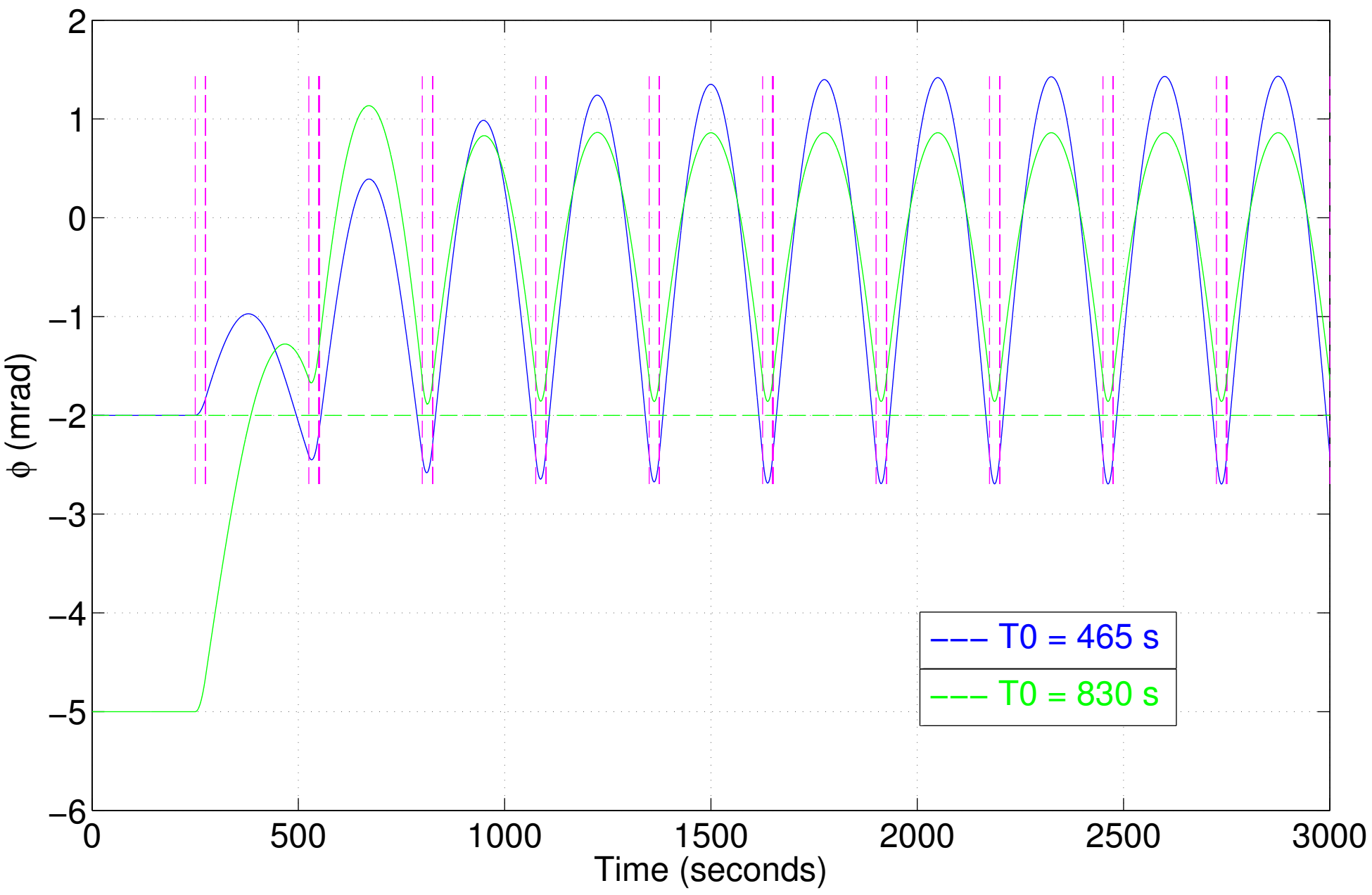

FIGURE 4.7: Time series of two simulated free fall experiments. Blue curve is obtained using $T_0 = 465\,s$ and $T_{fly} = 250\,s$. Green curve corresponds to $T_0 = 830\,s$ and $T_{fly} = 250\,s$.

The current configuration used to perform the free fall test provides to rotate the torsion pendulum of an angle $\Delta\Phi \approx -1.8\,mrad$ with respect to the electrode housing, to simulate a large DC acceleration. Such a rotation angle requires a DC torque, in nominal operating conditions, of roughly $14.4\,pNm$ to keep the test mass centered (considering that $\Gamma_{fib} \approx 8\,nNm/rad$), analogous to a differential force of roughly $14.4\,nN$.
It is possible to soften electrostatically the pendulum by applying DC constant voltages to lengthen the pendulum period from roughly $465\,s$, without applied fields, to as much as $T_0 \approx 830\,s$, to allow flight times comparable to those foreseen for LPF, that are of $350\,s$. As seen from pendulum dynamics during free fall, amplitude of motion diverges with $T_{fly}$ approaching $T_0$ as in equation 4.17, and this is the reason why we lengthen the period from 465 s to roughly 830 s, in order to avoid a diverging pendulum amplitude by performing free fall mode with large flight times. As is visible in figure 4.7, simulating a free fall run by using a $T_{fly} = 250\,s$ and a period of $T_0 = 465\,s$ (blue curve), means to have an higher dynamic range compare to the case with period $T_0 = 830\,s$ (green curve). We choose $T_{fly} = 250\,s$ and hence a $T_{imp} = 25\,s$ considering a duty cycle of 10%.

Pendulum softening is done by applying DC bias of $9.5\,V$ on electrodes placed on $Y$ faces of the sensor (see figure 1.5). This means to add an electrostatic stiffness on pendulum with the effect of reduce the system strength. From electrostatic considerations, it is

$$N_{\Phi,y} = \frac{1}{2}\sum_i \frac{\partial C_i}{\partial \Phi} V^2_{DC,y}. \tag{4.28}$$

With a suitable capacitance derivative expansion around a zero angle, stopping at the first order, the corresponding stiffness will be

$$\Gamma_Y = -\frac{\partial N_{\Phi,y}}{\partial \Phi} = -V_{DC,y}^2 \frac{\partial^2 C_y}{\partial \Phi^2} \tag{4.29}$$

where the subscript $C_y$ indicates that the capacitance are referred to those of the electrodes $y$ with respect to test mass or housing. $\frac{\partial C_y}{\partial \Phi}$ for the Y electrodes is zero by symmetry so that the net torque on center is zero.

With electrostatic softening of the pendulum, the total stiffness as sum of fiber and Y DC bias contribution, became $\Gamma \approx 2.47\, nNm/rad$, while the required torque to center mass is around $N_{DC} \approx 14.6\, pNm$, so that the equilibrium point became $\Phi_{EQ} \approx 5.9\, mrad$.

With this values, it is possible to put an estimate on the allowable resolution in the measurement of actuation fluctuations of $S_{\delta V/V}^{1/2}$ from the equation 4.10. With a torque noise level of $S_N^{1/2} \approx 1\, fNm/\sqrt{Hz}$ at $1\, mHz$, and a DC allowable torque level of $N_{DC} \approx 14.6\, pNm$, relative fluctuations in the applied actuation bias would become a dominant noise source at a level of $S_{\delta V/V}^{1/2} \approx 3 \cdot 10^{-5}/\sqrt{Hz}$ corresponding to a relative voltage fluctuations at the $30\, ppm/\sqrt{Hz}$ level.

Real time data series from pendulum free fall experiment are shown in figure 4.8, where we observer flight times of 90 and $250\, s$, using pendulum periods of 482 and $830\, s$, respectively, the latter with three different setpoints. Most of the measurement presented in this thesis are performed with the choice $T_{fly} = 250\, s$ and $T_{imp} = 25\, s$, and the cycle frequency is $\omega_k = 1/T_{exp} = 3.6\, mHz$. It is worth to note here how large the dynamic range spanned from pendulum is in this case, with a $\Delta\Phi \approx 3\, mrad$. This picture also shows the stability of used controller and the convergence to desired setpoint.

We can note also that the net torsional spring for pendulum systems is relatively large and positive, $\omega_0^2 = (2\pi/T_0)^2 \approx 1.6 \cdot 10^{-4}\, s^{-2}$, compared to the small, negative stiffness relevant to flight conditions of $\approx -10^{-6}\, s^{-2}$.

On ground actuation voltage needed to apply to keep the test mass centered $N_{DC}$ can vary by the rotation of the pendulum as desired, while the $\Delta g_{DC}$ in LPF cannot be changed. Also the stiffness can always be altered with fields. Moreover, changing the setpoint changes the effective $N_{DC}$, as we can move towards or away from the equilibrium angle. This variability of measurement parameters will help us to validate measurement and analysis techniques in different conditions.

To conclude, it can be noted that applied actuation forces on torsion pendulum are similar to the flight ones, using the same x electrodes, albeit in a different configuration, paired diagonally for torque rather than force. In flight, the involved electrodes are along one side of test mass, while torsion pendulum actuation is performed with a diagonal pair of x electrodes as in figure 4.5.

Finally, we can also note that we obtain similar DC force levels $\approx 1nN$ by using an impulse duty cycle that is 10 times larger then in-flight.

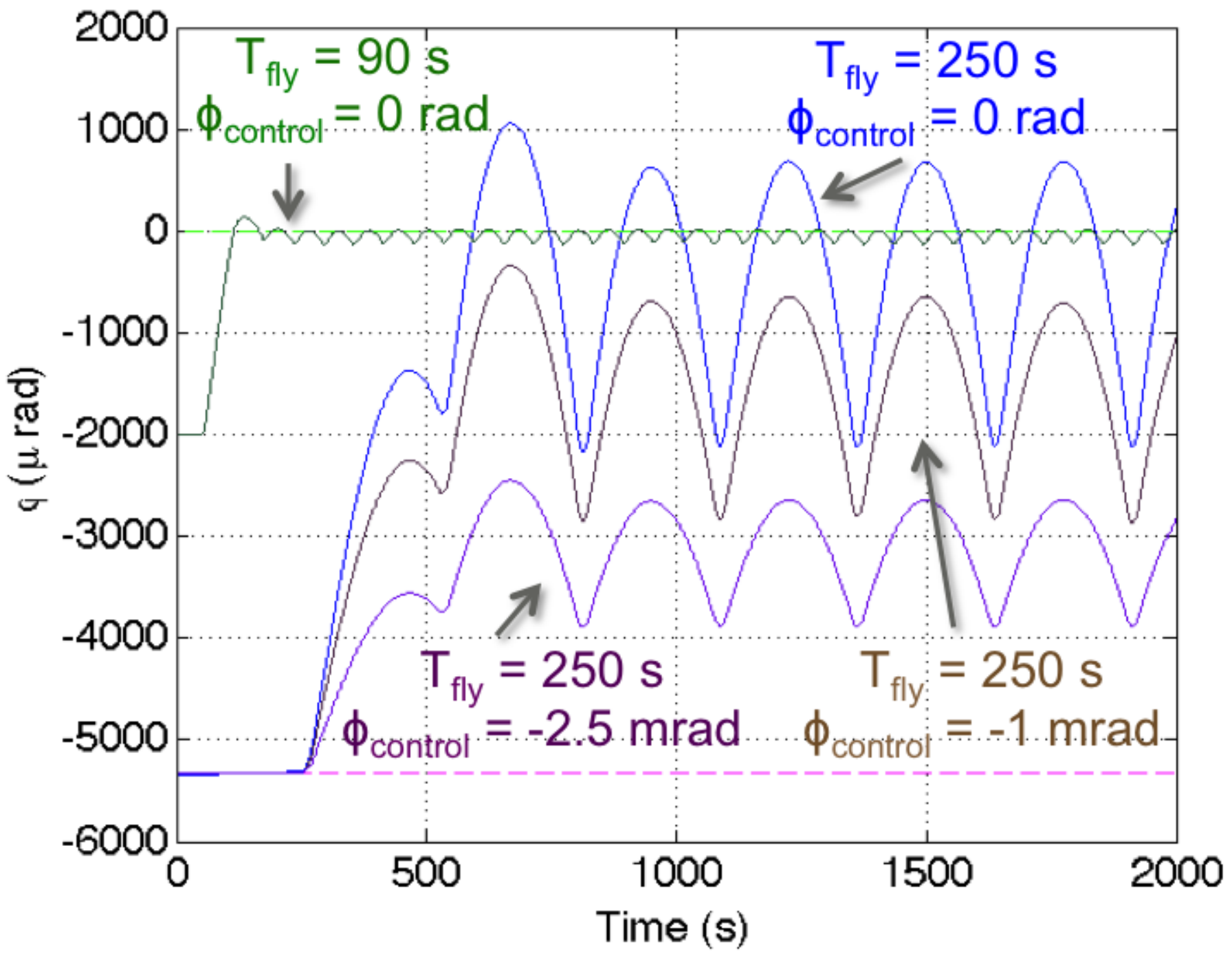

FIGURE 4.8: Free fall angular trajectories during the free fall mode operated at different setpoints and with different time flight durations. The action of controller allow to stabilize the mass position after the initial phase.

Differences between pendulum free fall implementation and in flight experiment, are summarized in table 4.1.

| | LTP | Torsion pendulum |
|---|---|---|
| Equation of motion | $\Delta\ddot{x}_{12} = \Delta g - \Delta\omega^2 (x_1 - x_{SC}) - \omega_{2p}^2 \Delta x_{12}$ | $I\ddot{\Phi} = -I\omega_0^2\Phi - \gamma\dot{\Phi} + N(t)$ |
| Stiffness | $\omega_0^2 \approx -10^{-6}\, s^{-2}$ | $\omega_0^2 = (2\pi/T_0)^2 \approx 1.6 \cdot 10^{-4}\, s^{-2}$ |
| DC force | $f = m\Delta g_{DC} \approx nN$ | $N_{DC} = \Gamma\Phi_{EQ} = I\omega_0^2\Phi_{EQ} \approx 14.4\, pNm \Longrightarrow 1.4\, nN$ |
| Experimental frequency | $2.86\, mHz$ | $3.6\, mHz$ |
| $T_{fly}$ | 350 s | 250 s |
| $T_{imp}$ | 1.5 s | 25 s |
| Employed $x$ electrodes | parallel couple on one side TM | diagonal couple |

TABLE 4.1: Differences in free fall implementation parameters between LISA Technology Package and torsion pendulum facility.

# Chapter 5

# Data analysis techniques

The free fall experiment provides a measurement that includes data segments that are free of actuation and thus also free of the associated force noise. The corresponding acceleration data in flight, and torque data on ground, are then used to estimate the experiment disturbance spectrum. Different analysis algorithms have been developed to estimate the power spectral density of the signal detected in presence of displacement data with gaps coinciding with the applied impulses that appear periodically and which need to be removed from the analyzed data. Here we present two analysis techniques that we have implemented to calculate the torque time series for our laboratory experiment. One involves a time-domain fitting of the angular displacement, with one data point per flight. The other, closely aligned with a technique that is being implemented for LPF [21], uses the "instantaneous" conversion into torque, followed by Blackman-Harris low-pass filtering and decimation. We explain here both techniques, which are then applied to data in chapter 6.

## 5.1 Free-fall data analysis concepts

Free fall measurements data have some peculiarity: they contain large motion in term of angular displacement but also big kicks, and then big torque peaks (as shown in figure 4.6), which need to be removed introducing gaps in the data.
The main idea is to use the torque measured during the free phases, when the free fall mode is applied, to retrieve noise spectrum estimates at frequencies below the kick-frequency $\omega_k$, particularly at frequencies around and below $1\,mHz$. This idea is based on the assumption that the disturbance force noise is stationary over time scales much longer than the flight - impulse cycle time $T_{exp}$.
If we use the whole torque time series, including gaps, for a direct spectral estimation,

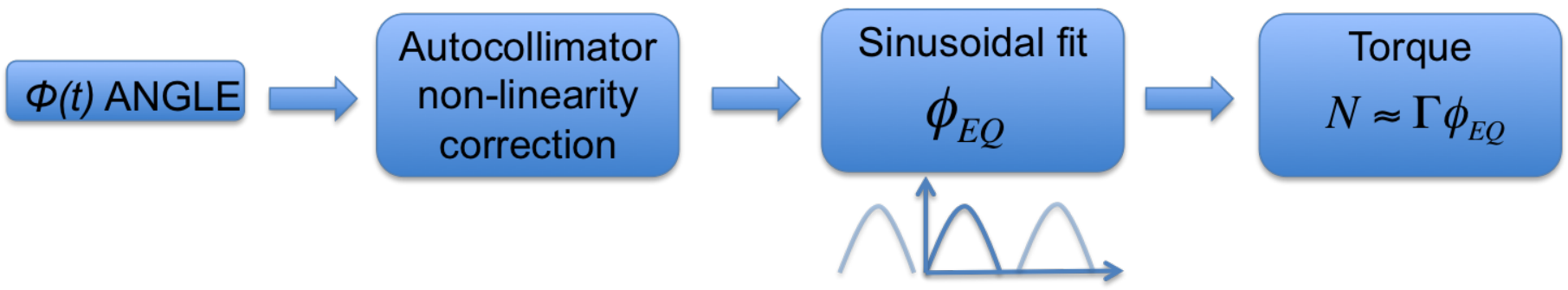

FIGURE 5.1: Sinusoidal fit technique flow chart.

we will observe some problem in the final evaluation. First of all the frequency band over $1\,mHz$ is polluted with peaks produced by data gaps at multiple of $f_{exp} = 1/T_{exp} = 3.6\,mHz$, in case of $T_{exp} = 275\,s$ which means flights of $250\,s$ alternated with impulses of $25\,s$. Moreover, a possible aliasing problem at low frequency can be caused from high frequency noise component and the presence of missing data in the gaps can also be related to an excess final noise, as we will addess later.

Two main techniques were developed to extract the true low frequency torque time series and fluctuations from free fall measurement on torsion pendulum, both with the aim of allow the final power spectral density estimation. These employ, respectively, a sinusoidal fit to each flight and an instantaneous conversion in torque, followed by a low pass filter, technique.

The proposed data analysis is performed on real data coming from the optical readout autocollimator, and not on gravitational reference sensor data, the latter being more noisy, as already seen in section 3.3 for noise floor estimation, and this has an high impact using the analysis technique developed. Moreover, the cross spectral density estimation (section 3.4) is not implemented as yet for free fall experiment data, because of high sensor noise and its non-linearity, which has not yet been investigated.

**Sinusoidal fit technique** The sinusoidal fit techniques consists of performing a linear least squares fit to each free fall phases, when no actuation is applied. The schematic flow chart of the technique is shown in figure 5.1. Angular data are preprocessed before fitting, performing, first of all, a correction on angular data, due to the non-linearity of the autocollimator response, as we will explain in section 6.2.1. At the beginning and end of each flight, $T_{cut} = 2\,s$ of data are cut to eliminate transient effect between free and impulse phases. The used model is:

$$\Phi(t) = \Phi_{EQ} + A\cos\omega_0 t + B\sin\omega_0 t. \tag{5.1}$$

This is a solution to equation 4.1 in the event that $N(t)$ is constant for the duration of the flight, with $\Phi_{EQ} = N/\Gamma$. This is a simple way to recover pendulum equilibrium angle for each flight, obtaining one data point per experiment cycle, as we see in figure 5.2. It must be noted that this solution is exact only if there is a constant value of the external torque, otherwise it must be considered a change in equilibrium angle that doesn't follow the 5.1. This works only for a truly harmonic oscillator, with $dN/d\Phi = -\Gamma$ for all $\Phi$ angles, as it assumes the solution to the equation of motion 5.1.
The corresponding torque value for each equilibrium point is calculated by multiplying by the stiffness

$$\hat{N}_{sin} = \Gamma\,\Phi_{EQ}, \tag{5.2}$$

as directly comes from pendulum equation of motion 3.1 for constant torque, neglecting the dissipation term. In this technique there is one point per flight, resulting in a time series that is uniformly sampled at $1/T_{exp}$. The stiffness is the only important parameter for torque resolution, and its determination depends mainly on the pendulum period $T_0$ knowledge.

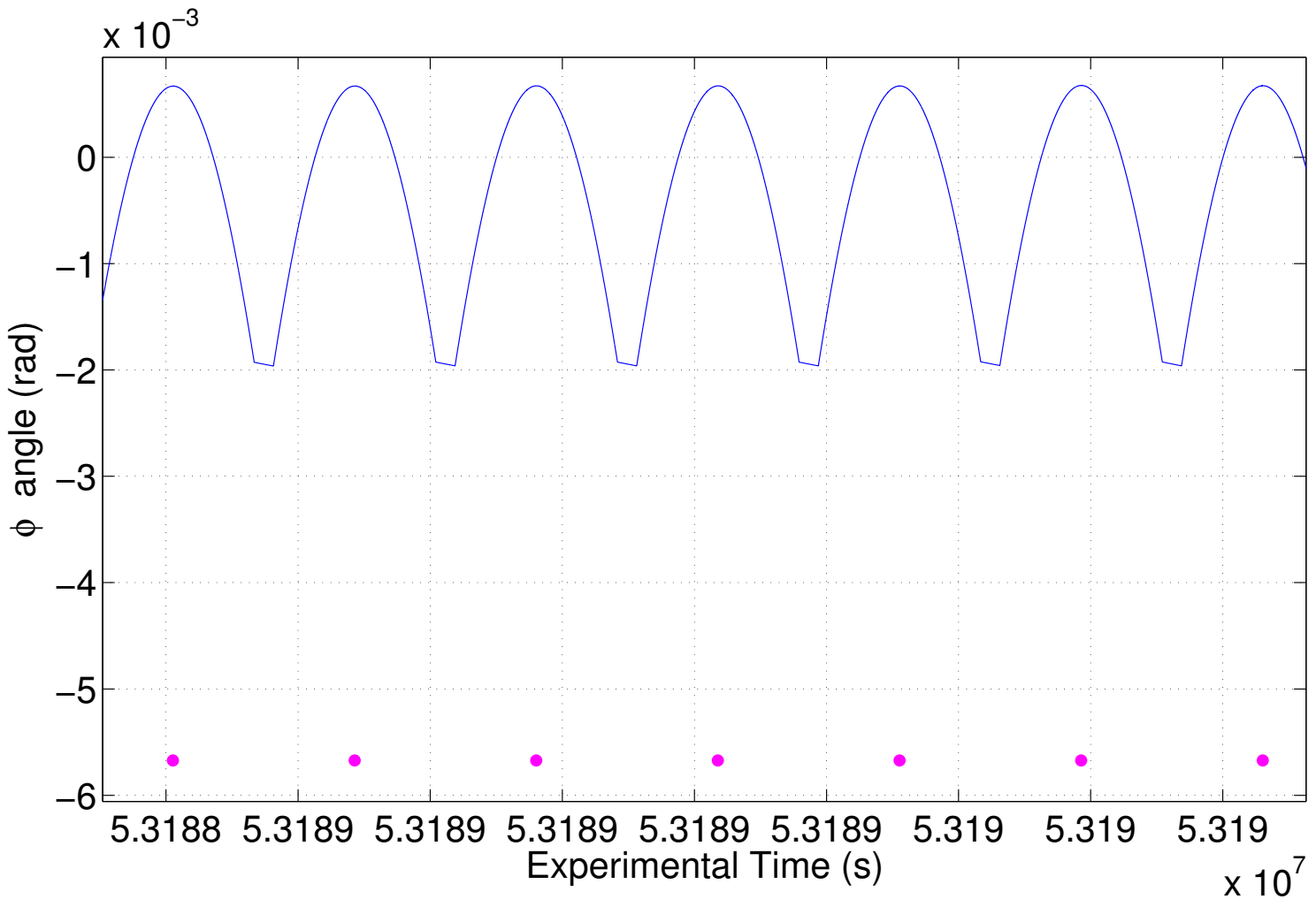

FIGURE 5.2: Pendulum equilibrium angle (magenta points) estimated for each flight (blue line) by performing a sinusoidal fit.

**Instantaneous conversion in torque** This analysis technique consists of an instantaneous conversion of angular data into torque by double differentiation of data, followed by a Blackman-Harris low pass filter and finally in rejecting the data in which the actuation impulses are present. We briefly name the techniques Blackman-Harris (BH) low pass filter later in the text.
The flow chart of this technique is shown in figure 5.3. First of all, pendulum angular data time series are linearly interpolated at the sampling frequency (nominally 10 $Hz$) in order to obtain equally spaced data, because of the possible presence of missing samples. Then they are corrected for the autocollimator non-linearity (that we address later in section 6.2.1) and converted in torque by means of pendulum equation of motion 3.16, where $N(t)$ is obtained approximating the first and second derivatives of $\Phi(t)$ in $t_i$ by means of a parabolic fit to the 5 adjoining points at times $t_{i-2} \dots t_{i+2}$. This fit can be implemented as a 5 point FIR filter for constant sampling time, hence requiring the interpolation of samples onto a uniform time grid. Time torque conversion need pendulum dynamics parameters as period $T_0$, moment of inertia $I$ and quality factor $Q$, following the equation:

$$\hat{N}_{BH} = I\ddot{\Phi} + \Gamma\Phi_m + \frac{\Gamma}{\frac{2\pi}{T_0}Q}\dot{\Phi} \tag{5.3}$$

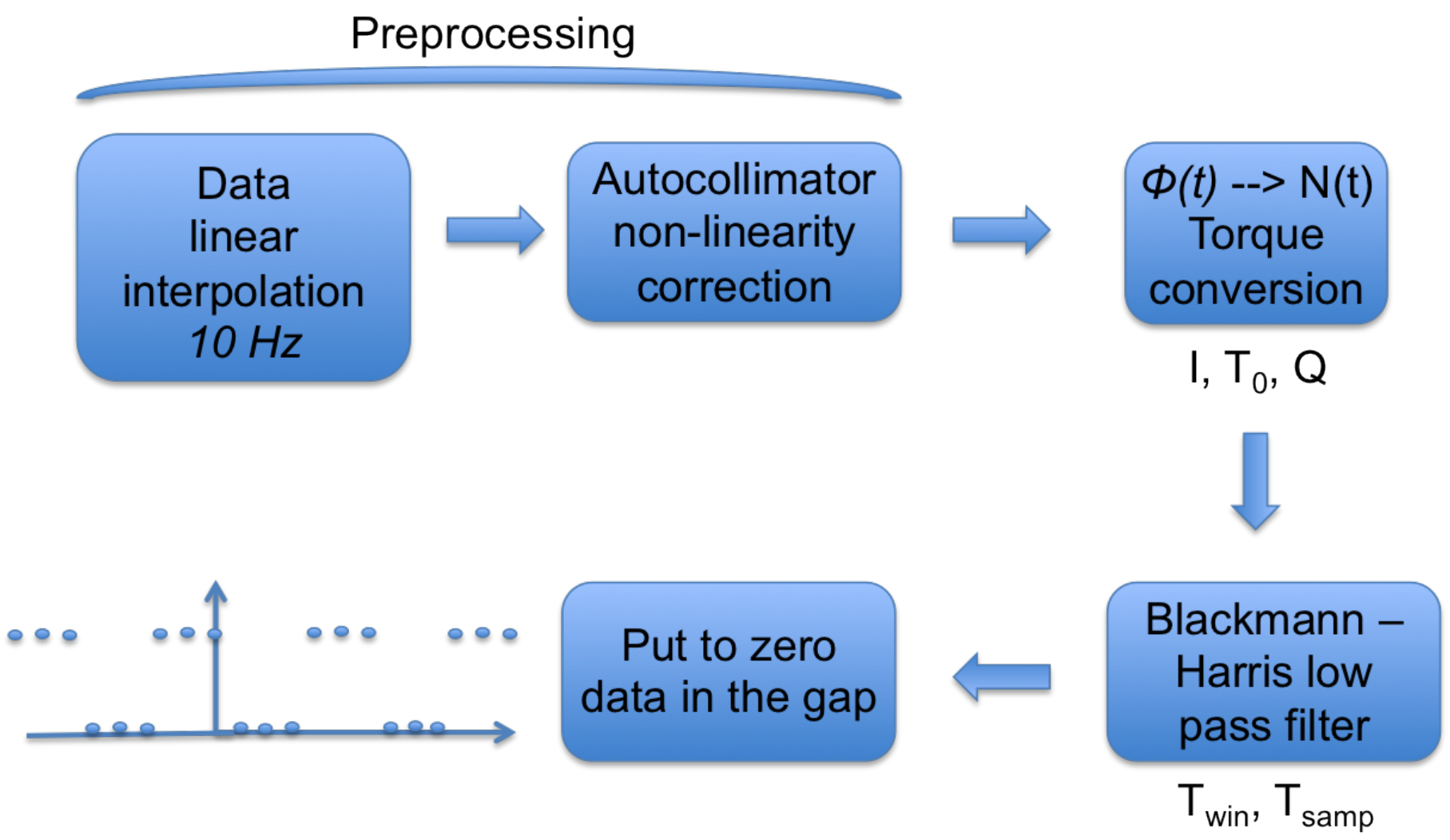

FIGURE 5.3: Instantaneous conversion in torque and Blackman-Harris low pass filtering technique flow chart.

The obtained torque time series is then filtered by multiplication by a normalized Blackman-Harris (BH) window of length $T_{win}$

$$w(t) = \frac{1}{a_0\, T_{win}} \sum_{j=0}^{3} a_j \cos 2\pi j \frac{t}{T_{win}}{}^{1}. \tag{5.4}$$

This is a low pass filtering operation on torque data with a chosen sampling frequency $T_{samp}$. Torsion pendulum data are sampled at 10 $Hz$, so it is important to choose window sampling time in order not to produce aliasing problems, also with the frequency of the experiment $1/T_{exp}$.

An integer number of total samples $n_{tot}$ per experimental times $T_{exp}$ is chosen to define $T_{samp}$ like

$$T_{samp}\, n_{tot} = T_{fly} + T_{imp}. \tag{5.5}$$

The length of the BH window is related to the number of samples $n_{keep}$ that can be kept per experimental time so that

$$T_{win} = T_{fly} - 2T_{cut} - (n_{keep} - 1)T_{samp} \tag{5.6}$$

[1] $a_0 = [.35875; -0.48829; 0.14128; -0.01168]$

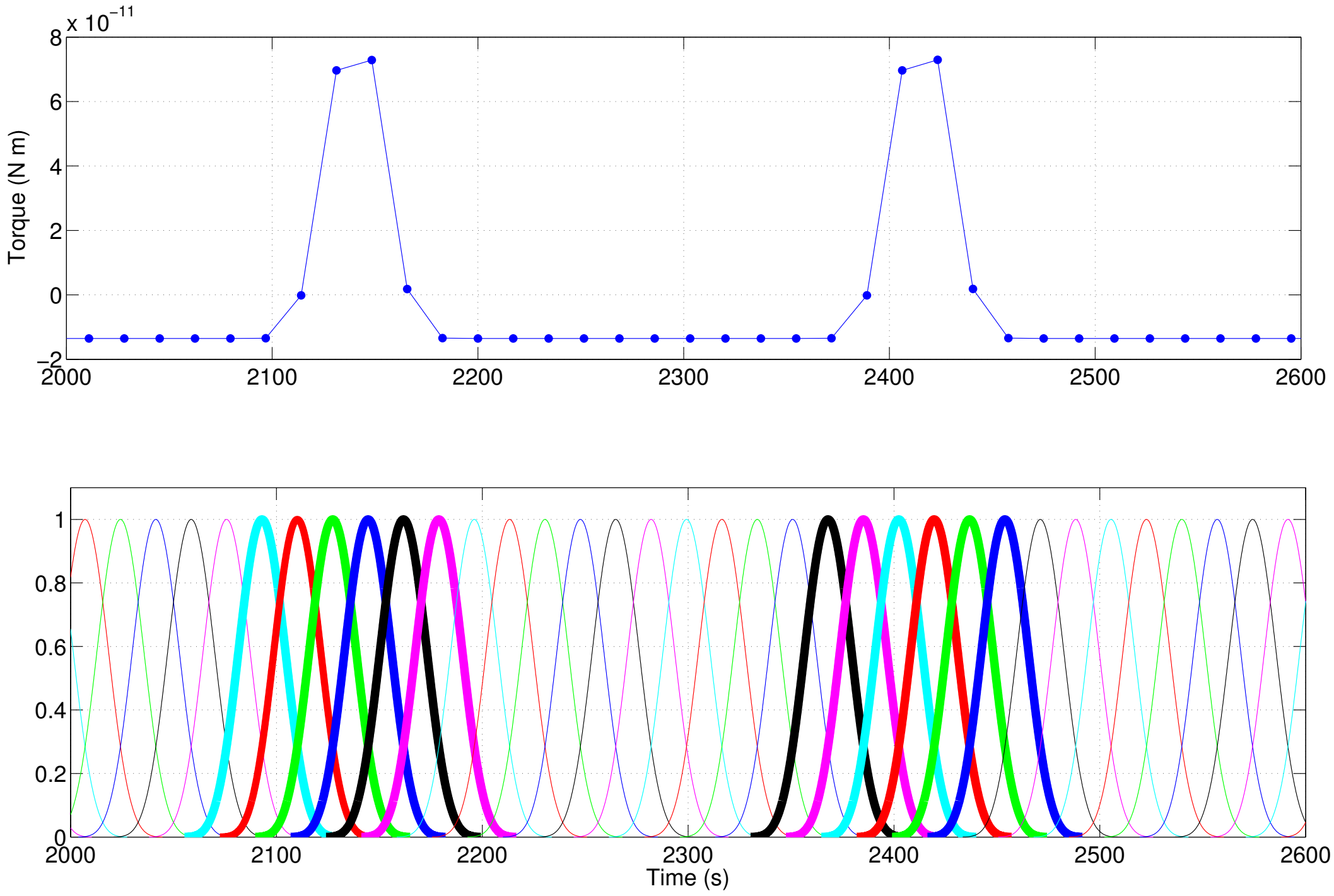

FIGURE 5.4: On top, pendulum angular deflection converted in torque time series, during free fall mode. Peaks corresponding to impulse phase are visible. On bottom, Blackman-Harris windows used to filter data. Bold line correspond to data in the impulse phases.

where $T_{cut}$ is the number of seconds cut at the beginning and end of each free phase of one flight. In this way we can define a period $T_a \equiv T_{fly} - 2T_{cut}$, corresponding to the length of each flight really analyzed.

The idea is to perform the analysis only for $n_{tot}$ and $n_{keep}$ that give an oversampling factor $T_{win}/T_{samp}$ in a range that we fix to be between 4 and 6.

If we opt for keeping $n_{keep} = 10$ samples over a total of $n_{tot} = 16$, in case of $T_{fly} = 250\,s$ and $T_{imp} = 25\,s$, will be $T_{win} = 91.3125\,s$ and $T_{samp} = 17.1875\,s$, with an oversampling factor of $T_{win}/T_{samp} = 5.3127$. An example of used windows is in figure 5.4.

Torque data points corresponding to windows that have some overlap with the time of the impulse are contaminated by the applied torque. We will set these points to zero as in figure 5.5. This is again a filter operation that can cause aliasing problem.

The BH filter, which transfer function we plot in figure 5.6, is chosen for its attenuation of high frequency lobes, to avoid down-converting high frequency acceleration noise into our mHz frequency band.

Finally, torque time series of data simulated with the simple harmonic oscillator simulator presented in section 4.2.4, and analyzed with both the techniques described, are shown in figure 5.7. Low pass filtering has been performed fixing $T_{win} = 91.3125\,s$ and $T_{samp} = 17.1875\,s$.

We can note that for instantaneous conversion into torque technique, the sampling rate $(1/T_{samp} = 1/17.1875\,s = 58\,mHz)$ is higher then the sinusoidal fit technique $(1/275\,s = 3.63\,mHz)$, but with the constraint that the data that we analyze are not longer uniformly sampled because of gaps. In fact, on a total of 16 data points, only 10 of these are kept, while the others are put to zero as visible in figure 5.5.

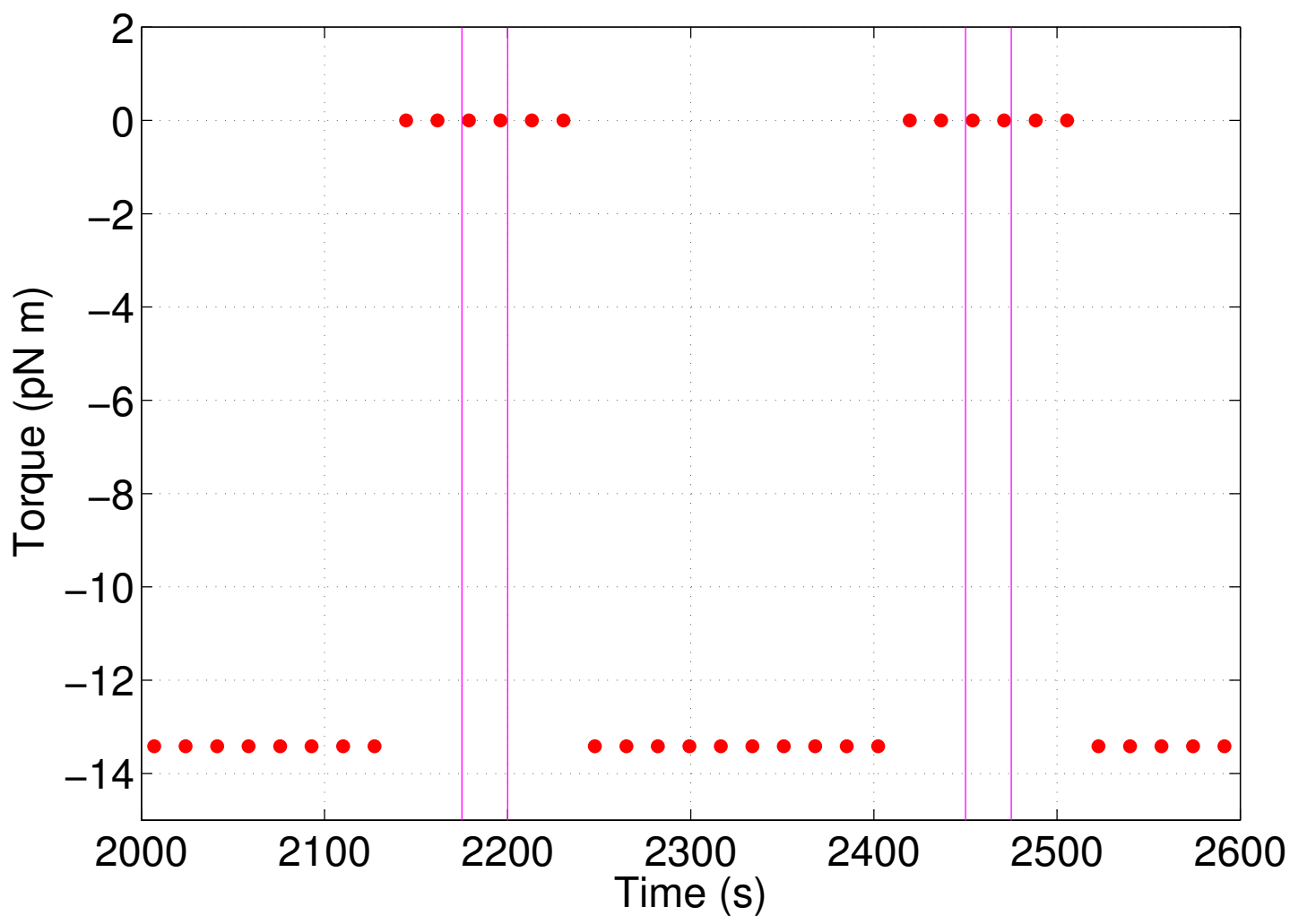

FIGURE 5.5: Torque time series after low pass filter and set to zero data in the gaps. Pink lines denote beginning and end of each impulse.

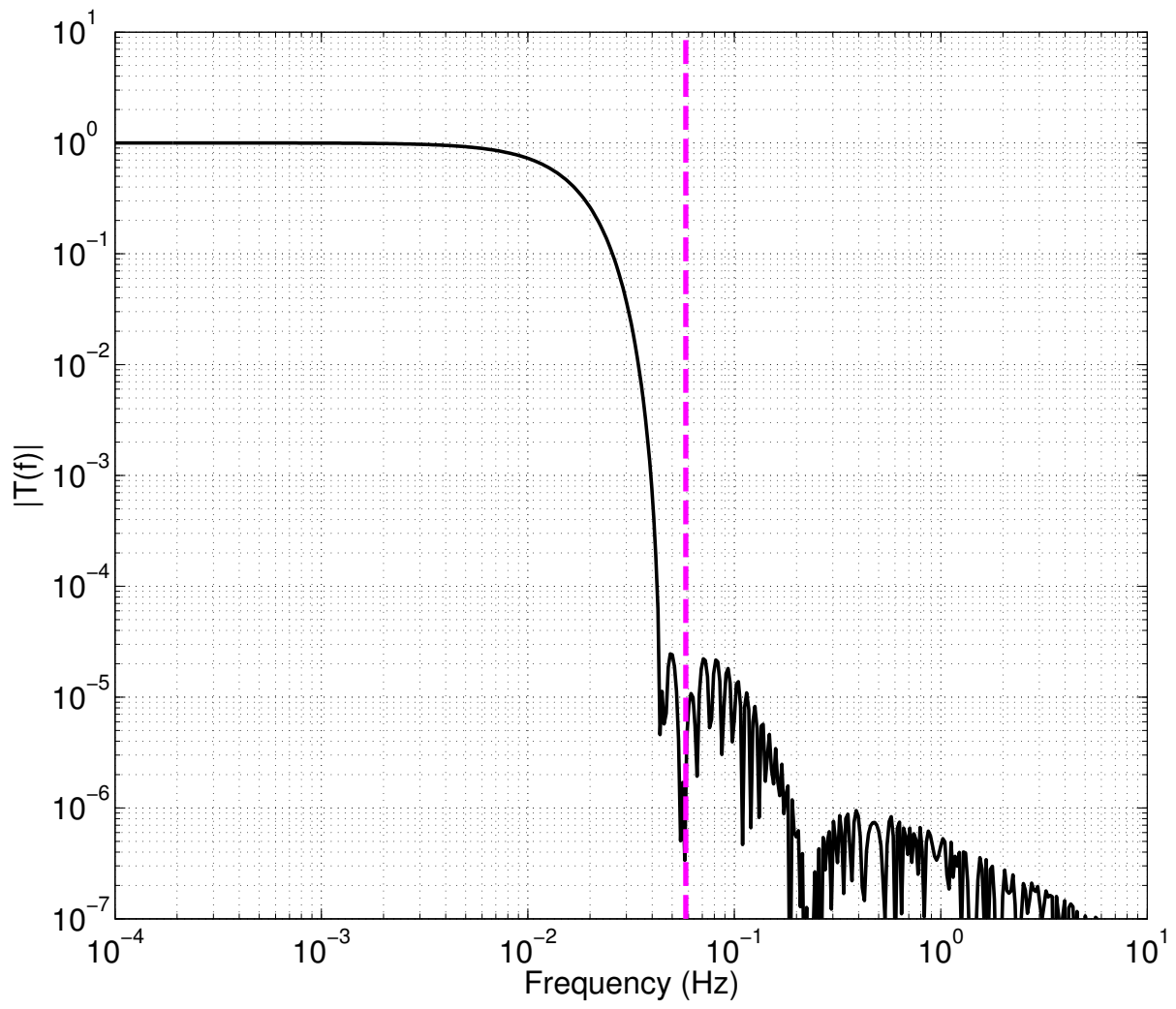

FIGURE 5.6: Blackman-Harris transfer function for $T_{win} = 91.3125\,s$. Instantaneous torque conversion technique sampling frequency $58\,mHz$ is outlined from magenta line.

Moreover, respect to the sinusoidal fit for which the equation of motion is solved for $N_{elastic} = \Gamma\Phi$, for the BH filtering technique, it is no longer obligatory to use the exact simple harmonic oscillator equation of motion, but it is possible to implement a more exotic "quasi elastic" coupling $\Gamma\Phi$ in the equation 5.3 like $N_{elastic} = -\Gamma\Phi + a\Phi^2 + b\Phi^3$, with an effective $\Gamma$ that is dependent on $\Phi$.

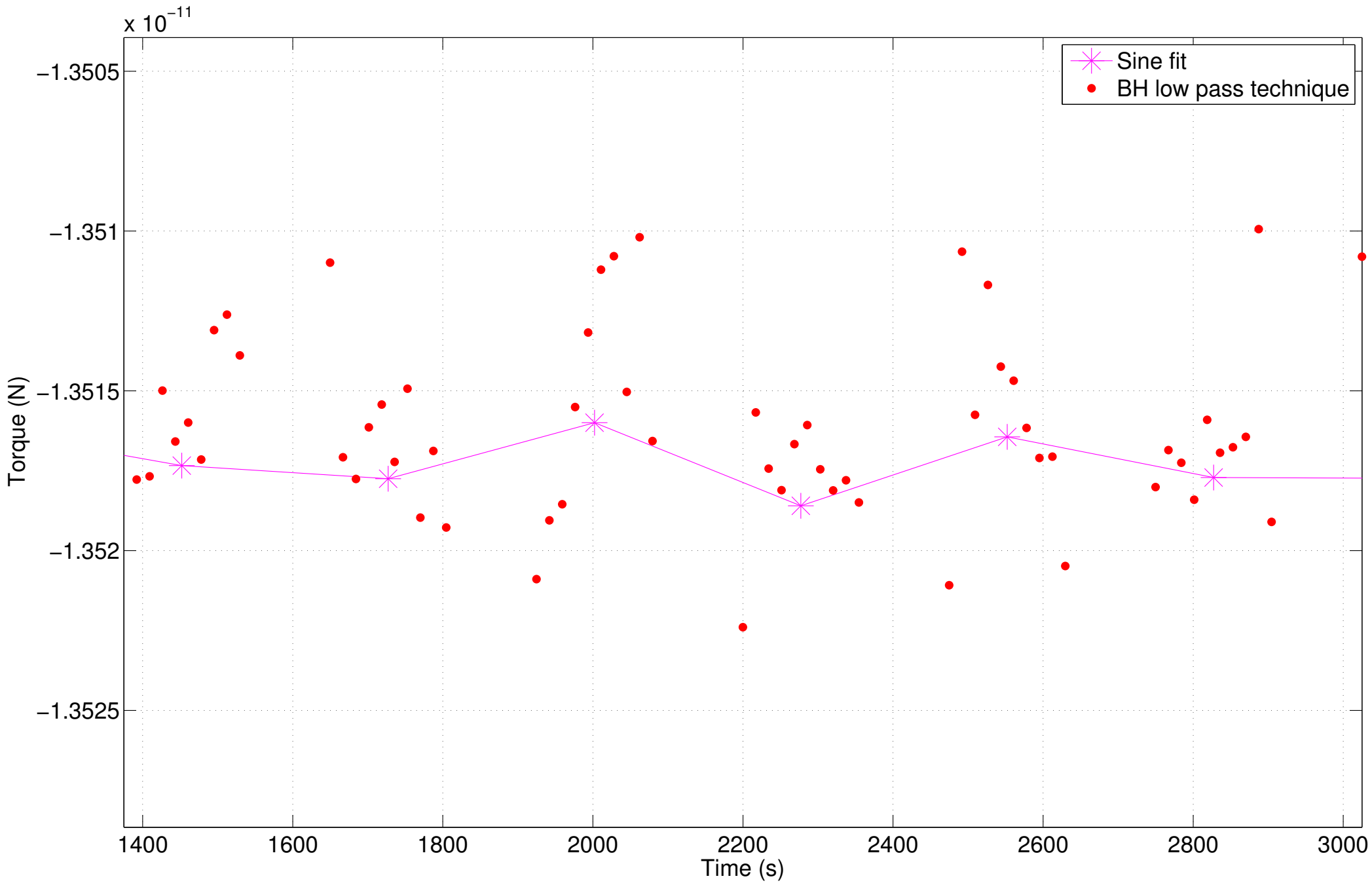

FIGURE 5.7: Torque time series obtained applying the two techniques of free fall data analysis described in the text. Analyzed data are produced with the harmonic oscillator simulator.

### 5.1.1 Estimate noise spectrum of simulated data

Data analysis techniques described above were employed on simulated data in order to understand how they work and their effect on known data.

As we said in section 4.2.4, it is possible to insert, in simulated free fall data, a desired level of white noise, in both torque and angle.

A free fall run has been simulated by considering a pendulum dynamic with a period of $823\,s$ and thus a relative positive stiffness of around $2.9\,nNm/rad$. Moreover, free fall parameters employed are: $T_{fly} = 250\,s$, $T_{imp} = 25\,s$, set-point angle equal to zero, equilibrium point of $-5.33\,mrad$ and measurement duration of $100000\,s$, similar to the conditions of a real free-fall experiment. The white noise level of torque injected in the simulation is of $1.4\,fNm/\sqrt{Hz}$, similar to that measured with the pendulum at mHz frequencies, while the angular noise injected is of the order of $20\,nrad/\sqrt{Hz}$, which corresponds roughly to the measured for the autocollimator noise at high frequencies like $10\,mHz$. Finally we analyze the simulated free fall run with both analysis techniques developed in order to estimate torque time series and calculate the noise spectra.

Power spectral densities of the recovered torque time series are performed as explained in section 3.4, with the Welch periodogram method. In order to reduce the uncertainty on the estimation of the spectral densities we need to perform measurements longest possible, so that it is possible to divide the data series in shorter stretches, that are multiplied by a Blackman Harris spectral windows and then averaged. Stretches are also overlapped by 66%. Blackman-Harris windows length is chosen to be 27500 $s$ in order to have 100 flights of 250 $s$ length, on each stretches.
Some processing is performed for both techniques to obtain the final spectrum, as will explain below.

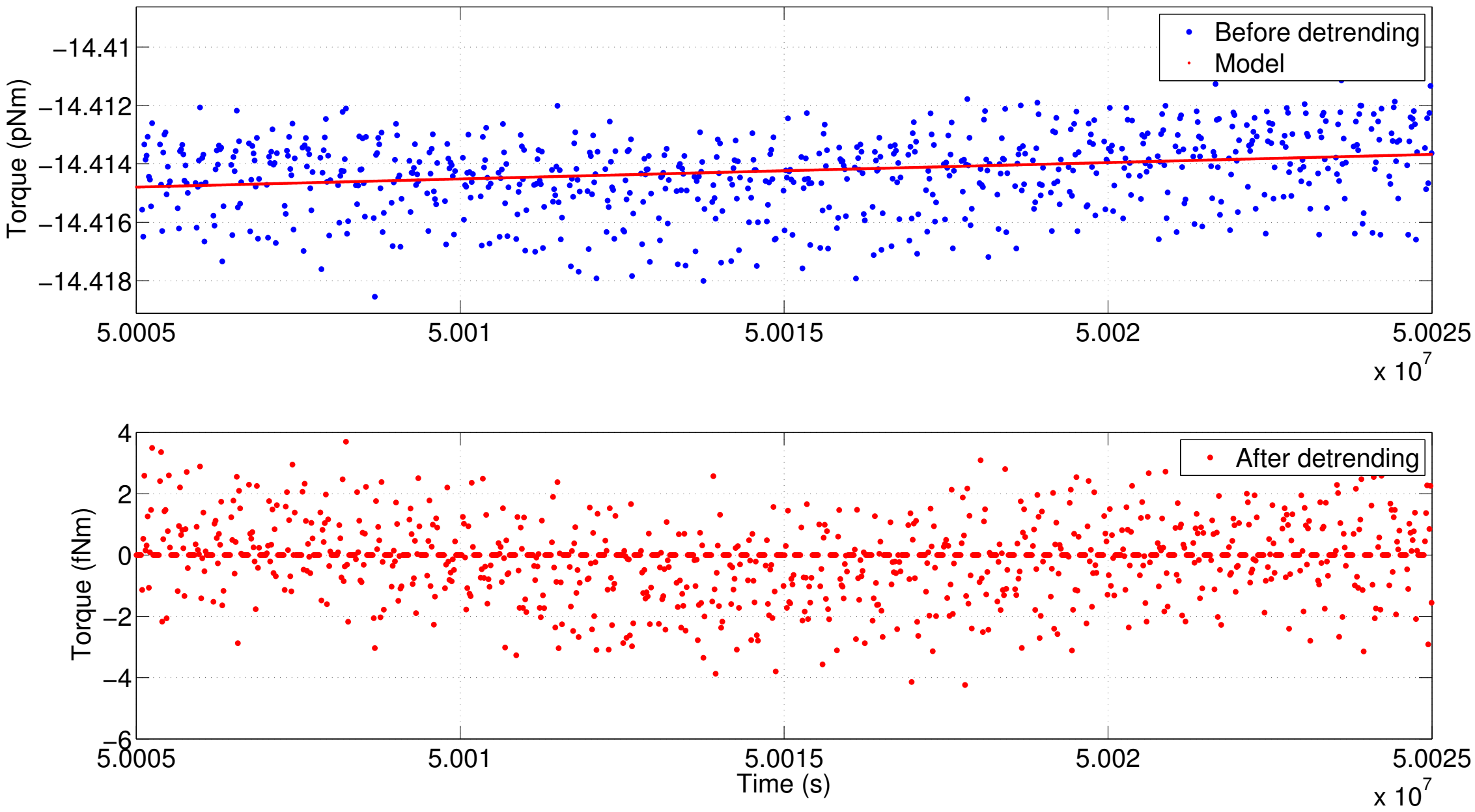

FIGURE 5.8: Torque time series of data from a real free fall run, before and after data detrending over the spectrum window length of 27500 s.

**Instantaneous conversion in torque spectra** Data analyzed with the BH low pass filtering technique are detrended before performing the spectrum. This means to perform a linear fit to all good data points over the spectrum window length, and setting to zero all points from windows that overlap with the impulse, and then subtract from the data the linear trend. This intends to subtract the DC component of the square waves obtained after putting zero data in the gaps as showed in figure 5.8. This reduces the height of peaks at high frequency in the final spectrum as visible in figure 5.9, in case of real data.
Moreover, the power spectrum must be multiplied by a scaling factor equal $n_{tot}/n_{keep}$, that is 1.6 in case of $n_{tot} = 16$ and $n_{keep} = 10$, that provides a normalization necessary because of the presence of zeros in the data.

Finally, spectrum is also normalized for the BH window transfer function, shown in figure 5.6, used to low pass filter data, after the linear trend subtraction.

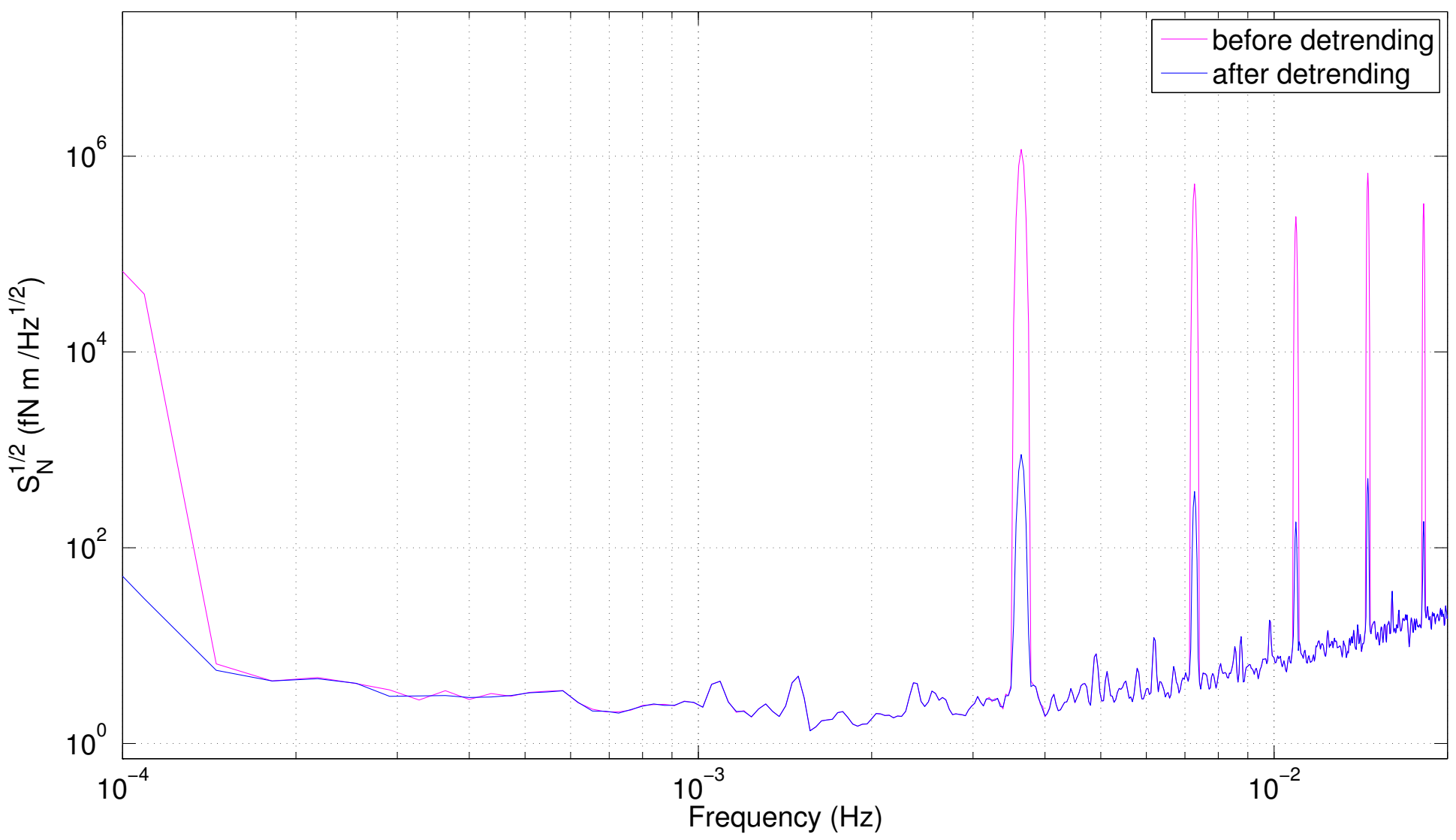

FIGURE 5.9: Torque noise spectra for a free fall real run before and after data detrending. Spectra are averaged over a window length of 27500 s and overlap of 66%.

**Sinusoidal fit spectra** In case of sinusoidal fit techniques, the spectrum is obtained after performing a normalization dividing for a transfer function calculated as we will explain in a while.

It must be considered how the the sinusoidal fit analysis technique works. First of all the pendulum filters an external torque $N(\omega)$ with its transfer function $H(\omega)$ of equation 3.4, so that $N(\omega)$ became $\phi(\omega)$. Then the sinusoidal fit performs a linear least square fit to the model in equation 5.1

$$\Phi(t) = \Phi_{EQ} + A\cos\omega_0 t + B\sin\omega_0 t \tag{5.7}$$

such that $\Phi(\omega)$ goes in $\Phi_{EQ}(\omega)$.

This process can be shown to be convolution of the time series in $\Phi$ with a linear filter function, that provides a decimation to one point per flight. This is an exact solution of the equation of motion with $\Phi_{EQ} = \frac{N}{\Gamma}$, if we neglect the dissipation term and if the torque $N(t)$ is constant over the time range analyzed.

Finally, the angle is reconverted back into torque by multiplying by $\Gamma$, so that $\hat{N}_{sin} = \Gamma\,\Phi_{EQ}$ as in equation 5.2.

The linear fit to the three parameter model in equation 5.7 can be considered equivalent to a linear filter. If we approximate the data as continuous and consider data analyzed over the period $T_a \equiv T_{fly} - 2T_{cut}$ as defined in equation 5.6, then the output value of

$\phi_{EQ}(t)$ is given by

$$\phi_{EQ}(t) = \frac{1}{T_a}\int_{t-T_a/2}^{t+T_a/2} \left(C_{11} + C_{22}\cos\omega_0 t'\right)\phi\left(t'\right)dt' \tag{5.8}$$

$C_{11}$ and $C_{12}$ are the relevant elements of the covariance matrix resulting from fit, and have values

$$C_{11} = -\frac{\frac{1}{2}\left(1+\operatorname{sinc}\omega_0 T_a\right)}{\left(\operatorname{sinc}\frac{\omega_0 T_a}{2}\right)^2 - \frac{1}{2}\left(1+\operatorname{sinc}\omega_0 T_a\right)} \tag{5.9}$$

$$C_{12} = \frac{\operatorname{sinc}\frac{\omega_0 T_a}{2}}{\left(\operatorname{sinc}\frac{\omega_0 T_a}{2}\right)^2 - \frac{1}{2}\left(1+\operatorname{sinc}\omega_0 T_a\right)}. \tag{5.10}$$

The approximation of continuous data is justified if we consider to have many points per cycle, because in the $T_a = 246\,s$ period of data analyzed ($T_{fly} = 250$ s and we choose $T_{cut} = 2$ s), there are roughly 2500 points.
We can then Fourier transform the equation 5.8 to give the frequency response of the sinusoidal fit filter function:

$$\begin{aligned}\tilde{\phi}_{EQ}(\omega) &= H_{sine}(\omega)\,\tilde{\phi}(\omega) \\ &= \tilde{\phi}(\omega)\left[C_{11}\operatorname{sinc}\frac{\omega T_a}{2} + \frac{C_{12}}{2}\left(\operatorname{sinc}\frac{(\omega+\omega_0)T_a}{2} + \operatorname{sinc}\frac{(\omega-\omega_0)T_a}{2}\right)\right]\end{aligned} \tag{5.11}$$

Substituting the calculation of the coefficients $C_{11}$ and $C_{12}$ in equation 5.10, we can see that the transfer function $H_{sine}$ goes to 1 in the limit of zero frequency. Multiplying the combined pendulum and sinusoidal fit transfer functions by $\Gamma$ to convert into torque, we have the net transfer function between external torque and our sinusoidal fit torque observable, which is thus

$$H(\omega) = \Gamma H_{pend}(\omega)\,H_{sine}(\omega) \tag{5.12}$$

and is plotted in figure 5.10. It can be noted that the transfer function is continuous and we sample this, once per flight, and thus with a frequency $f_{samp} = 1/T_{exp} \approx 3.64$ mHz, with $T_{exp} = T_{fly} + T_{imp} = 275$ s. The finite input response filter allows us to make this sampling in such a way that there is no overlap of the impulse time with the analyzed pieces of data.

#### 5.1.1.1 Comparing noise spectra

Finally, in figure 5.11 we compare noise spectra results obtained applying both our techniques to the simulated free fall run, with respect to the level of theoretical torque noise injected, that has a level of $1.4\, fNm/\sqrt{Hz}$. Both analysis techniques are able to recover the white noise level a factor 1.4 above the noise injected, at frequency lower then $2\, mHz$. We perform different simulations in order to explain if this effect is due to aliasing effects.

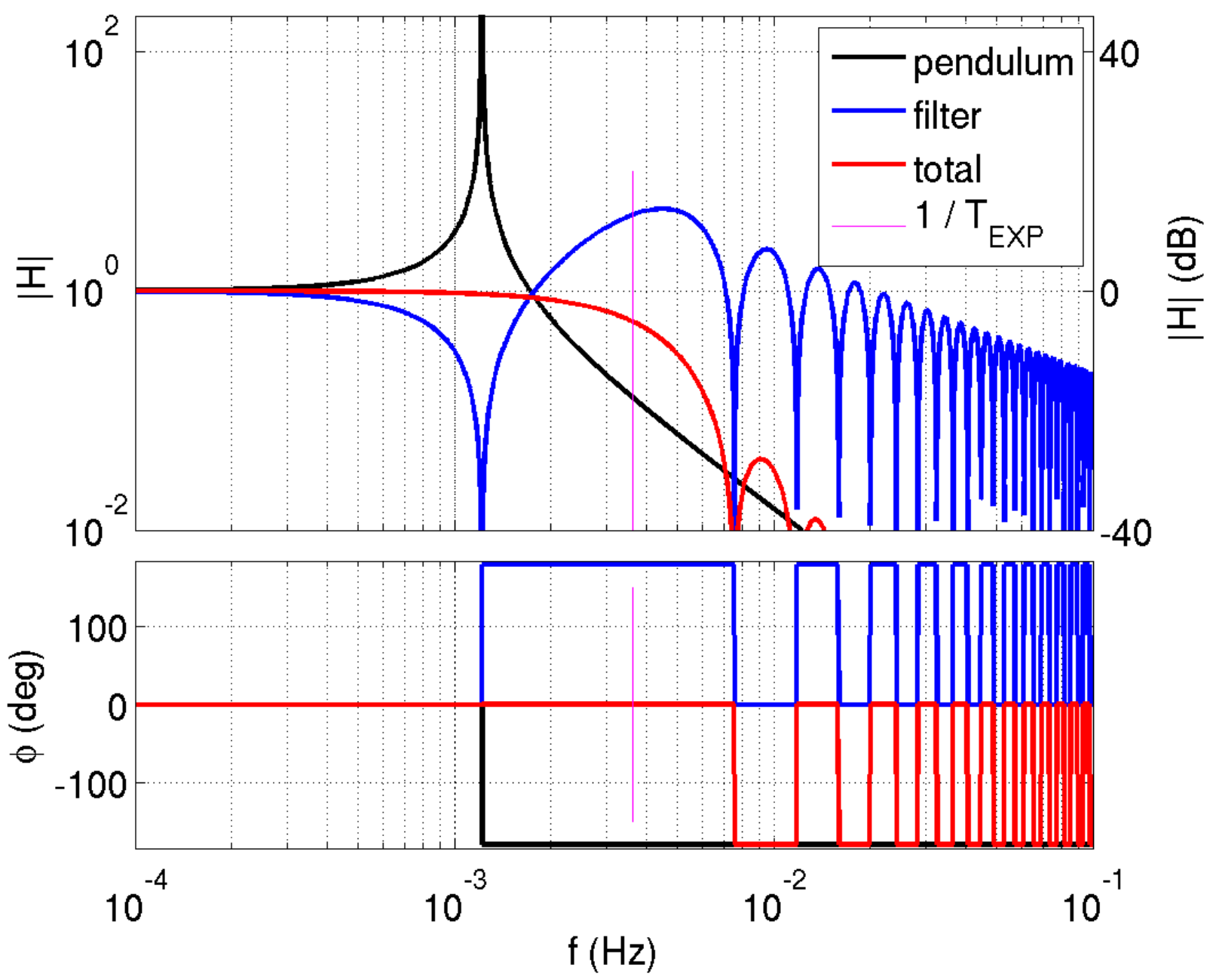

FIGURE 5.10: Transfer function between external torque and measured output torque using our torsion pendulum and the sine-fit analysis method. The pendulum transfer function has been multiplied by Γ in order to be unitless, as is natural because the final analysis step multiplies by Γ. Also shown is the frequency at which we sample the filter, which is $\frac{1}{T_{exp}}$.

Problems arise if we inject an higher sensing noise level, which should affect only the high frequency portion of the torque noise spectrum, into the simulator data. We compare torque noise spectra of different free fall simulations made by using white angular noise level of 20, 50 and 80 $nrad/\sqrt{Hz}$, analyzed with BH filtering technique, and results are shown in figure 5.12. There is a noise component, at low frequency, that scales with level of high frequency readout noise. This is due to the presence of an aliasing effect that is still to study and explain.

Another important comparison can be done by performing the same analysis also on simulated pendulum data in the absence of impulses, with the DC torque and the actuation set to zero. The free-fall analysis thus inserts artificial gaps in the otherwise continuous data. This allows to investigate whether the aliasing effect is due to the impulse control scheme or rather to the presence of gaps in the data and analysis scheme. From figure 5.13, but also 5.15 and 5.16, it is clear that the gapped-data analysis techniques obtain the same noise whether we have data with impulses or the impulse free data.

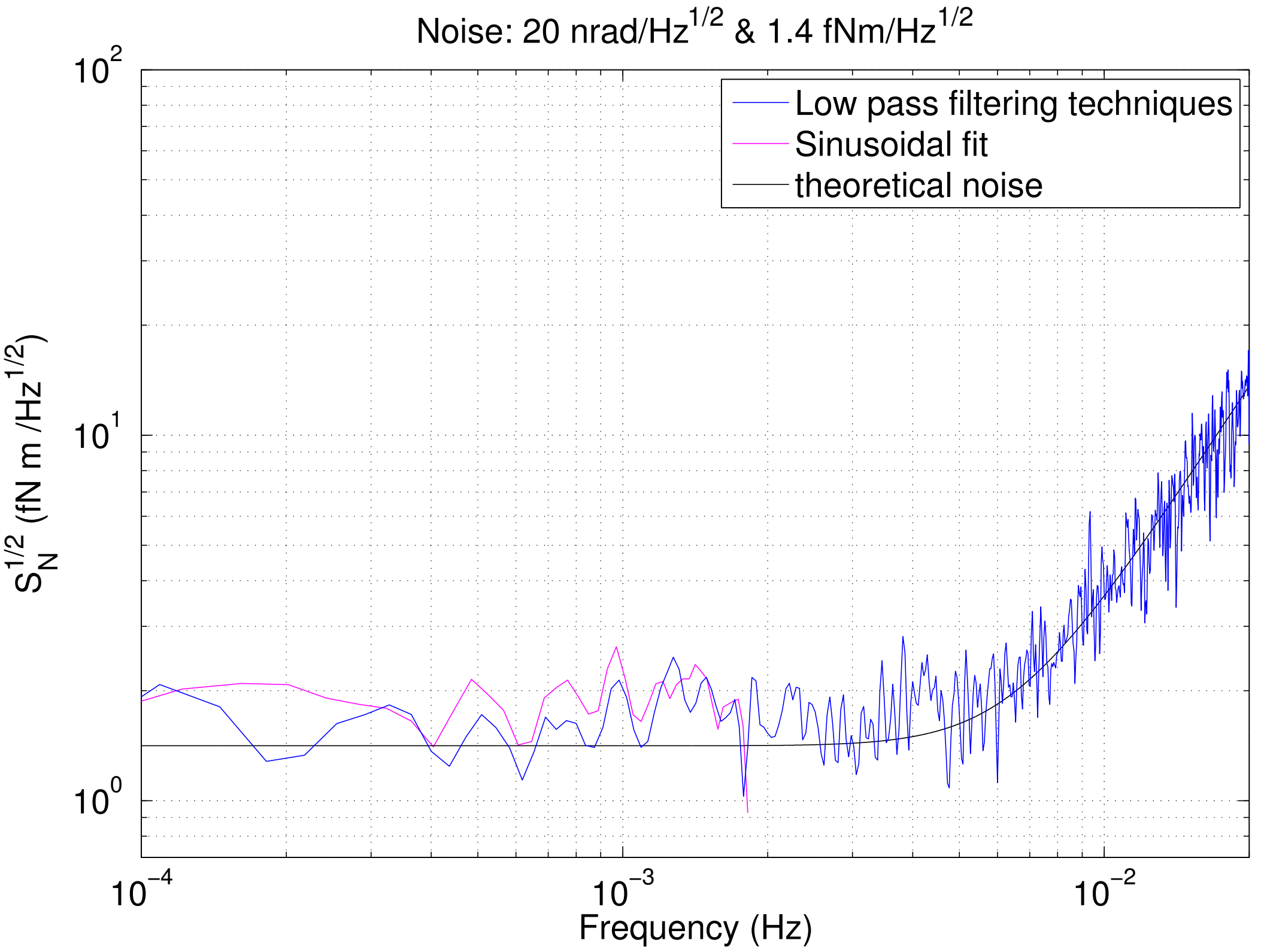

FIGURE 5.11: Torque noise spectrum obtained analyzing simulated free fall data with both techniques described in the text. Analyzed data are produced with the harmonic oscillator simulator in which white noise levels are injected: $1.4\, fNm/\sqrt{Hz}$ for torque and $20\, nrad/\sqrt{Hz}$ for angular noise.

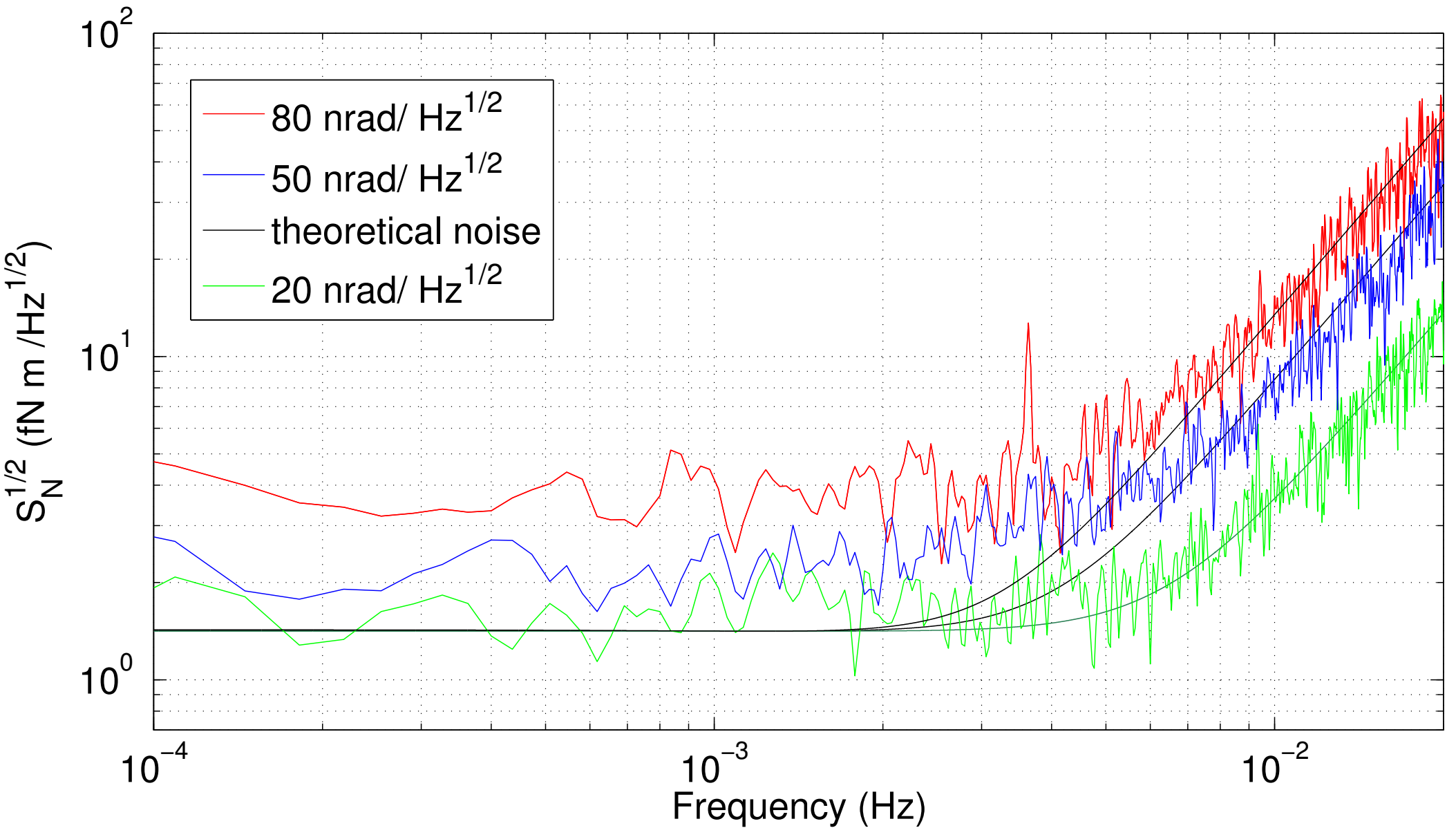

FIGURE 5.12: Torque noise spectrum obtained analyzing free fall simulated data with the low pass filtering technique. Analyzed data are produced with the harmonic oscillator simulator using white noise levels similar to torsion pendulum experiment, are injected: torque noise of $1.4\, fNm/\sqrt{Hz}$ and 20, 50 and $80\, nrad/\sqrt{Hz}$ for readout noise.

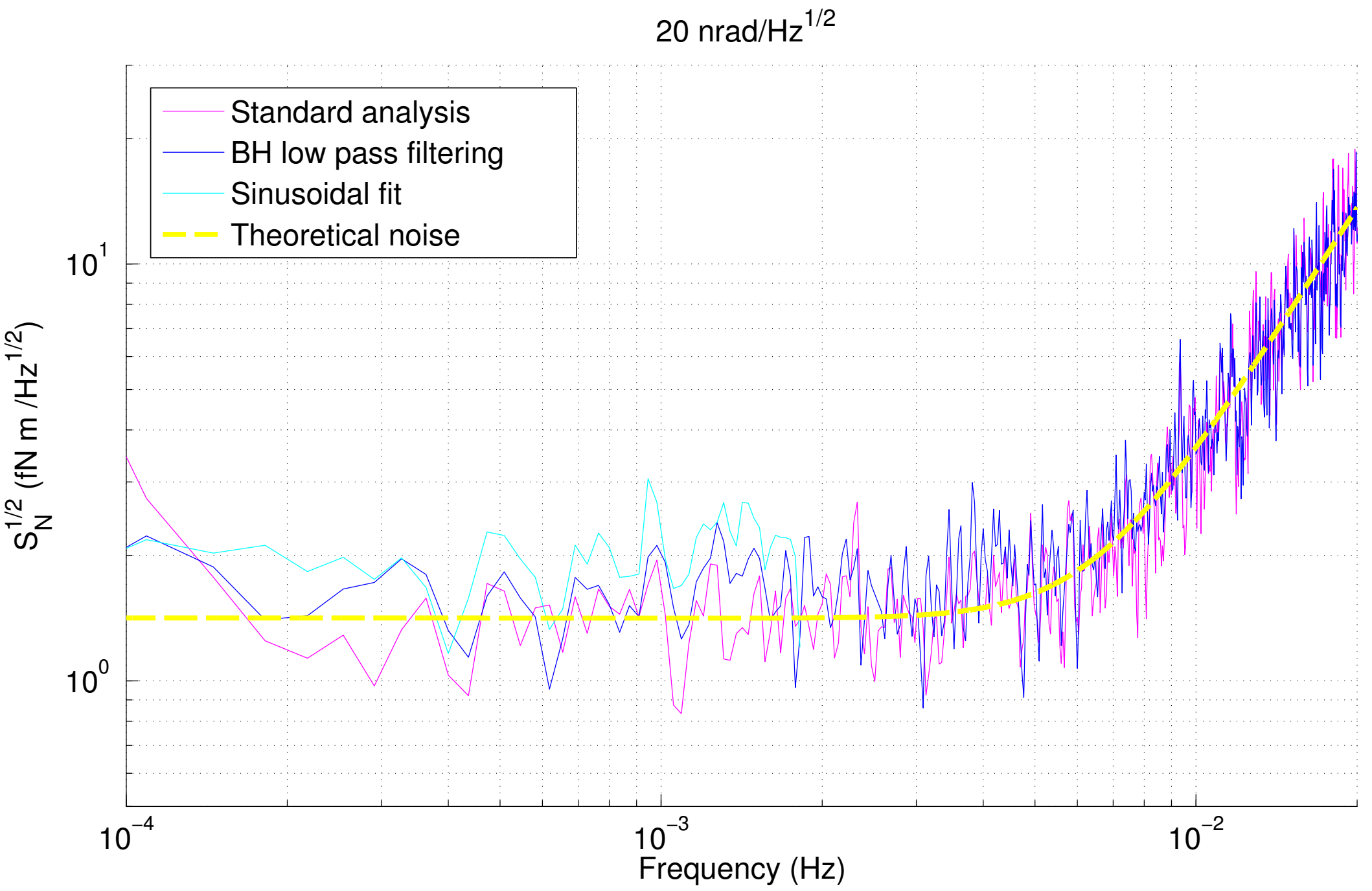

FIGURE 5.13: Torque noise spectra obtained analyzing a simulated noise run with period $T_0 = 819.69\,s$, without a DC torque and without applied impulse torques, with free-fall data analysis techniques described and with the standard analysis. Torque noise level injected in the data is $1.4\,fNm/\sqrt{Hz}$. Angular noise is $20\,nrad/\sqrt{Hz}$.

The aliasing effect visible in figure 5.12 is still present if we analyze with free fall analysis techniques no-impulse data with an high sensor noise level injected, $80\,nrad/\sqrt{Hz}$, as visible in figure 5.14. It is thus clear that the standard analysis (without gaps) extracts the expected noise level, and the gapped analysis techniques do not.
The wrong noise estimation is then not related to the high dynamical range of free fall data with respect to background noise measurements, but to the presence of the gaps.

We also perform the same analysis on simulated data by injecting white noise levels similar to those expected for the LISA Technology Package system. The simplified simulation performed again with the simple harmonic oscillator simulator, requires to map the LISA Pathfinder free-fall experiment in order to enter it in the torsion pendulum free fall simulator. The $\Phi$ angle will be substituted by

$$x_0 = \frac{\Delta g_{DC}}{(\frac{2\pi}{T_0})^2} \tag{5.13}$$

where we have chosen $\Delta g_{DC} \approx -1nm/s^2$, because simulation requires a dynamic with a very low and positive stiffness, on the contrary to the negative stiffness expected for LPF. The sign of the stiffness does not have a large impact on the analysis, but is chosen to make the simulation easier and to keep the stiffness term of the same order of magnitude as in flight. Period of oscillation is chosen to be $T_0 = 4400\,s$. Free fall experiment parameters are $T_{exp} = 275\,s$ and $T_{imp} = 2.75\,s$, with a duty cycle of 1%.

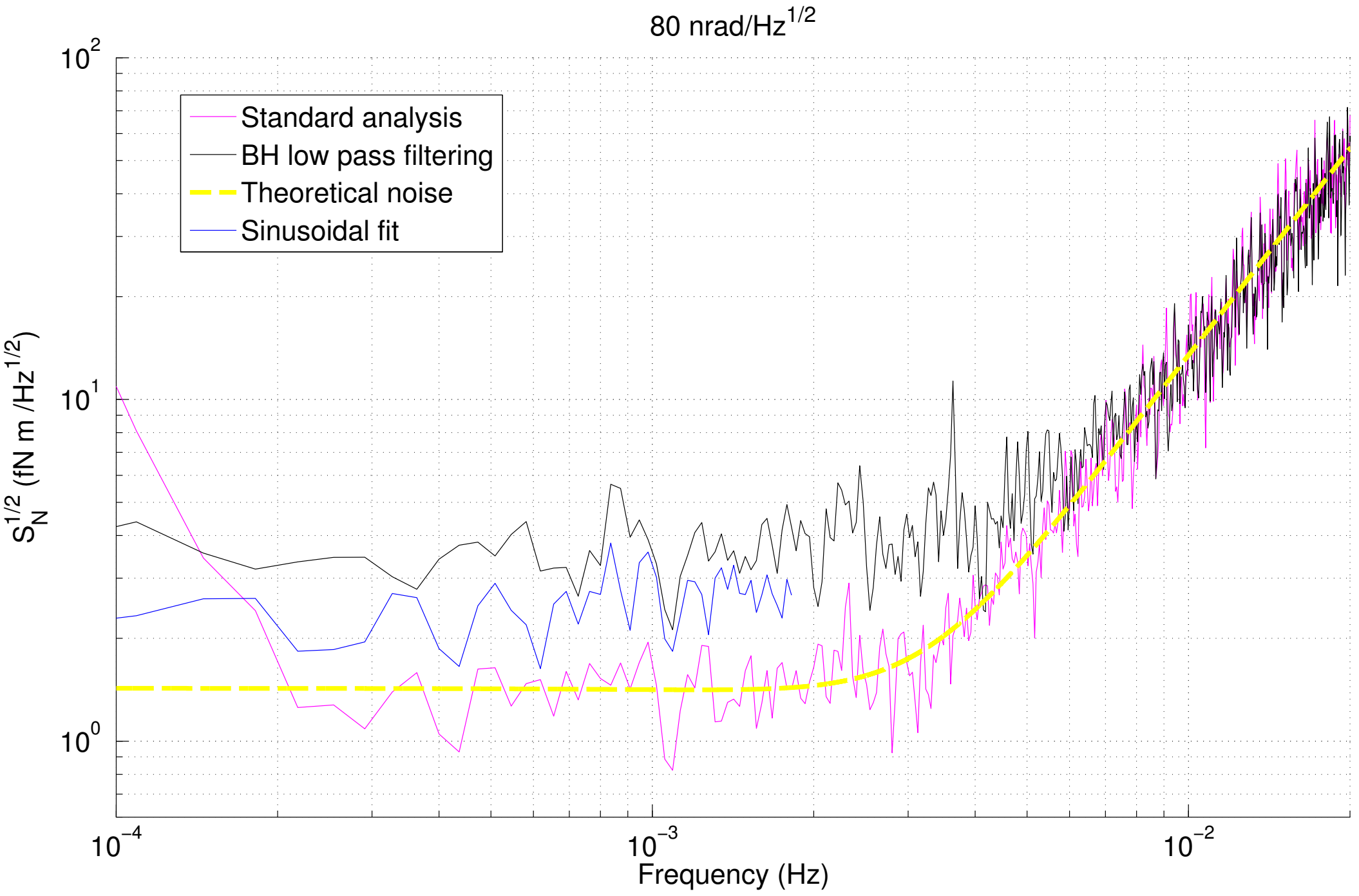

FIGURE 5.14: Torque noise spectra obtained analyzing a simulated noise run with period $T_0 = 819.69\,s$, without a DC torque and without applied impulse torques, with the two techniques described. Torque noise level injected in the data is $1.4\,fNm/\sqrt{Hz}$. Angular noise is $80\,nrad/\sqrt{Hz}$.

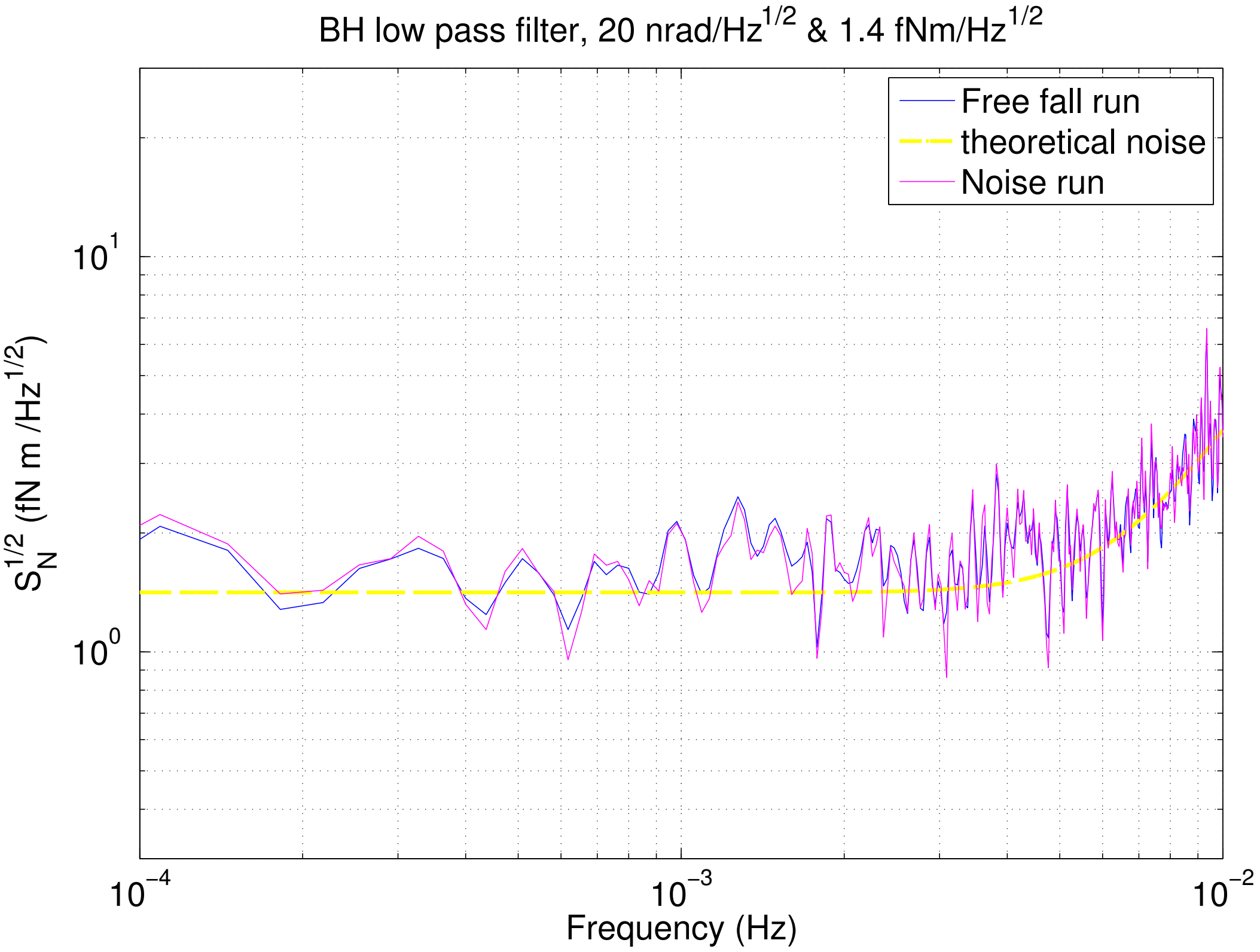

FIGURE 5.15: Torque noise spectra obtained analyzing a simulated free fall run and a simulated noise run without actuation with BH low pass filter technique. Torque noise level injected in the data is $1.4\,fNm/\sqrt{Hz}$. Angular noise is $20\,nrad/\sqrt{Hz}$.

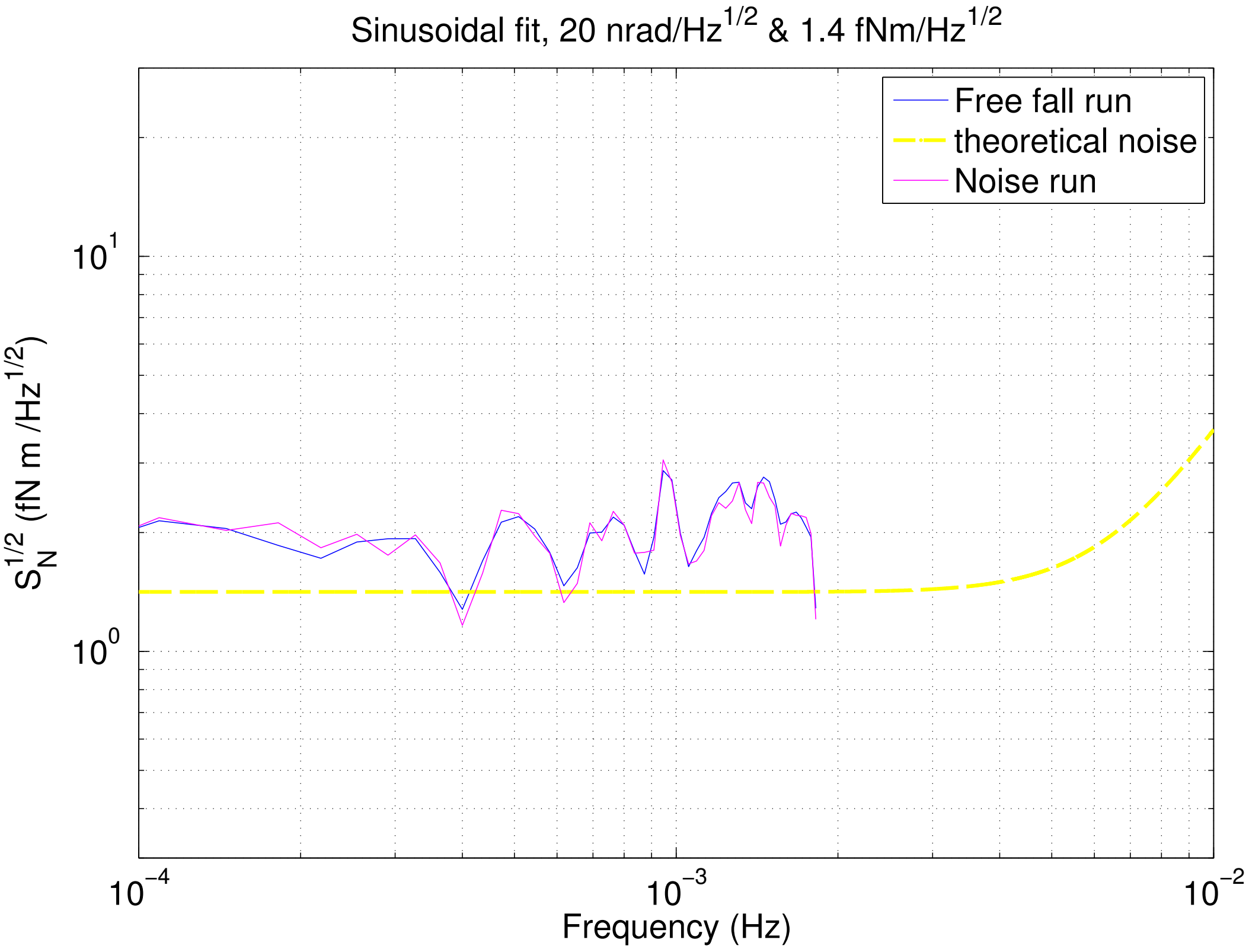

FIGURE 5.16: Torque noise spectra obtained analyzing a simulated free fall run and a simulated noise run without actuation with sinusoidal fit technique. Torque noise level injected in the data is $1.4\, fNm/\sqrt{Hz}$. Angular noise is $20\, nrad/\sqrt{Hz}$.

We simulate several runs of free fall mode, by inserting different level on noise similar to that of the interferometer, from $0.6\, pm$ to $60\, pm/\sqrt{Hz}$ and acceleration noise level of $10\, fm/s^2\sqrt{Hz}$.

Only the BH low pass filtering techniques has been used to analyze these simulations, as shown in figure 5.17. Already at $6\, pm$ of noise level, it is not possible recover the original noise injected because of the presence of impulse in the data, as already at $20\, nrad/\sqrt{Hz}$ for pendulum spectra the noise level recovered is a factor two higher then the expected. It is worth to note that the used duty cycle (1%) is lower then the others simulations. This means that the result of the analysis and the presence of the aliasing are not related with the impulse length, considering that the simulated free fall for torsion pendulum has a duty cycle of 10%.

It is possible to conclude that the amount of the effect of aliasing depends on how much high frequency sensor noise there is in the data.

We have not performed a full analysis of aliasing with the sinusoidal fit technique. However, we know that the down-sampling introduces aliasing into our measurement band by folding the spectrum at $f \pm n f_{samp}$, with $f_{samp} = 1/T_{exp}$. This means that our spectrum at, for example, 1 mHz is a sum of the true, continuous process torque noise at 1 mHz, at the output of the pendulum, plus the analysis process shown in Fig. 5.10, including both torque and readout noise, and that at 4.64 mHz and 2.64 mHz (from -2.64 mHz made positive in the single-sided spectrum), and so on.

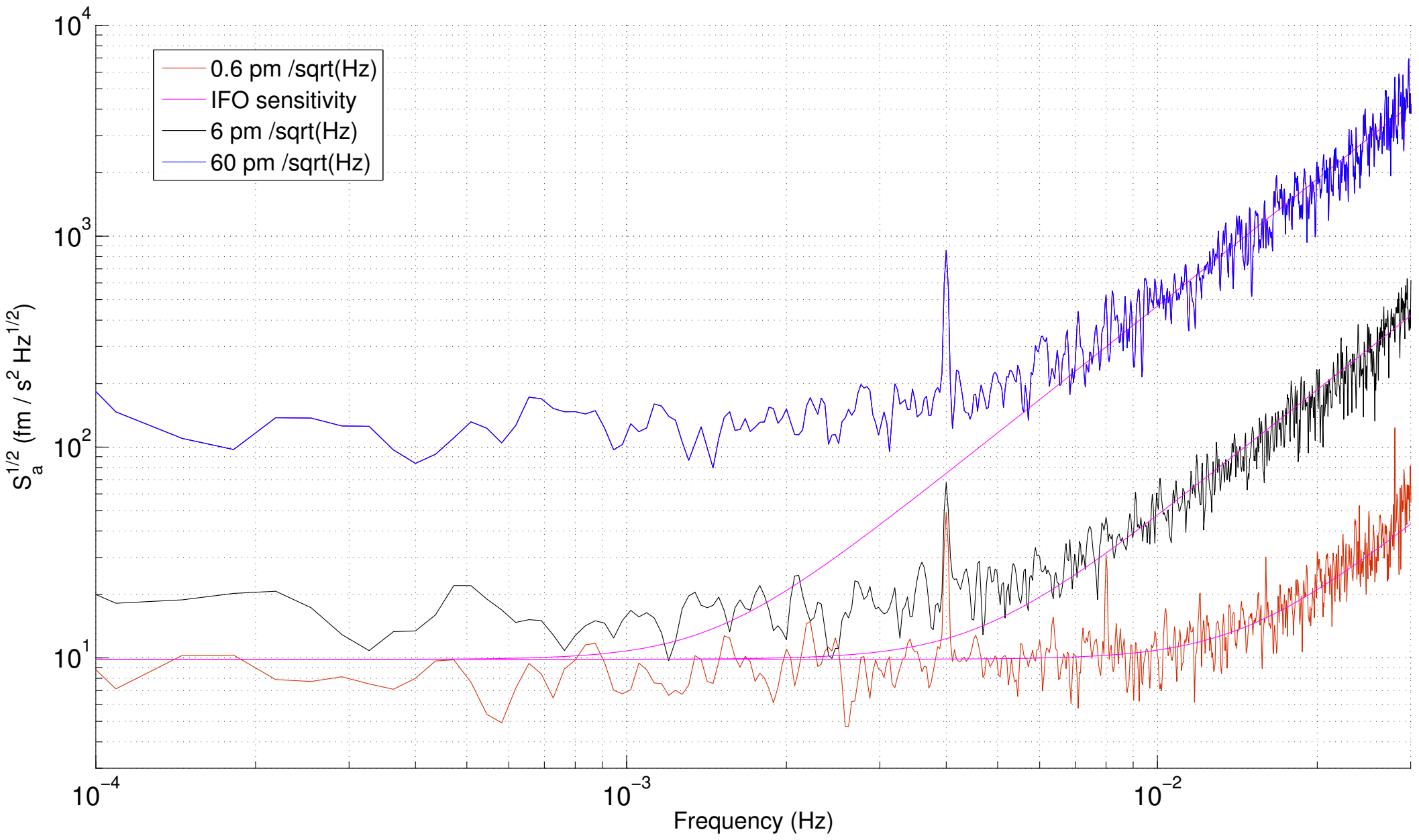

FIGURE 5.17: Differential acceleration noise spectrum obtained analyzing simulated free fall data with the low pass filtering technique. Analyzed data are produced with the harmonic oscillator simulator using noise levels similar to the LPF condition: acceleration noise of 10 $fm/s^2\sqrt{Hz}$ and from 0.6 $pm$ to 60 $pm/\sqrt{Hz}$ for IFO noise.

# Chapter 6

# The Free-fall experiment results

A long free fall testing campaign was carried out during the last year and a half with torsion pendulum facility, ranging different experimental parameters and conditions. Measurements were taken mainly during weekends in order to take advantage of the best quiet environmental conditions. These start with calibration of the apparatus and of the actuator and with pendulum rotation, to make the free fall possible. The free fall actuation mode implemented on torsion pendulum has its own non ideality. In this chapter we will present pendulum optical readout limits and non linearity and how these couple with the aliasing problem encountered in data analysis techniques developed in the previous chapter. We will explain how it is possible to correct data for non linearity and time stamp issues. We then explain how we calibrate the torque measurement, extracting the correct pendulum period and thus the stiffness, from the shift in pendulum resonance frequency due to electrostatic factors. Finally we will show results for torque time series and spectra recovered from free phases during the free fall mode, by the application of the two analysis techniques, and how we can calibrate these techniques in order to account for aliasing problems.

We are able, at present time, to recover an excess torque noise of $2\,fNm/\sqrt{Hz}$ at $1\,mHz$ from free fall measurements, a factor two above that achieved with constant DC actuation force. This corresponds to an acceleration of about $100 fm/s^2\sqrt{Hz}$, a factor 10 above the level of acceleration noise from actuation expected for LISA Pathfinder mission. The measured torque noise with constant DC actuation has a level of $0.9\,fNm/\sqrt{Hz}$ at $1\,mHz$. Attributing all of this noise to actuation would correspond to actuation fluctuations at a level of $S^{1/2}_{\delta V/V} = S^{1/2}_{\hat{N}_2}/2N_{DC} \approx 3.1 \cdot 10^{-5}\,/\sqrt{Hz}$ or roughly $30\,ppm/\sqrt{Hz}$, considering that the level of DC torque applied to hold the test mass centered is $N_{DC} = 14.62\,pNm$.

The resolution of the noise of the actuator is instead $0.2\,fNm^2/Hz$ at $1\,mHz$, that on the scale of voltage fluctuations corresponds to $S^{\frac{1}{2}}_{\delta V/V} \approx 15 ppm/\sqrt{Hz}$, a factor two above the actual measured stability with the inertial sensor Front End Electronics of LTP that gives $3 - 7ppm/\sqrt{Hz}$ at $1\,mHz$.

## 6.1 Measurement data set

In this chapter we present time series and torque noise spectra of the whole measurement data set. We briefly summarize how many run we acquired and analyzed in table 6.1. As already explained in section 5.1.1, we divide the data series in shorter stretches, that are multiplied by a Blackman Harris spectral windows and then averaged. Stretches are also overlapped by 66%. Blackman-Harris windows length is chosen to be $27500\,s$ in order to have an integer number of flights, on each stretches.
In table 6.1 we report the number of overlapped stretches for each type of measurements that are then averaged. The data set include 8 long weekend of background noise measurements performed in the summer-winter 2013 before rotating pendulum to perform the free fall measurements and 3 long weekend plus 3 night of measurements performed after rotating back the pendulum at the end of testing campaign, in March 2015. Free fall measurements presented were performed during the winter 2014-2015 and include 14 weekend of data. Each weekend was divided in two parts performing also measurements with constant DC actuation. Noise measurements weekend with DC actuation are 9. From all of these, we discarded data cuts contaminated by technical problems as time stamping that we will describe in section 6.2.2. All data are corrected for autocollimator non-linearity the we will describe in section 6.2.1.
Finally, during the week of testing campaign, we dedicate the facility also to make a calibration measurement of the free fall technique, applying a torque both during a free fall measurement than a noise run with DC actuation, as we describe in section 6.7.

## 6.2 Experimental difficulties with readout and time-stamping

At least two technical problems with the autocollimator become critical in the free-fall experiment, and must be addressed, in the data acquisition software and in the data preprocessing.
One of these problems is visible in the residuals of sinusoidal fit technique described in section 5.1. In figure 6.2 are shown residuals from sinusoidal fit applied to a free fall run with $T_{fly} = 100\,s$ and $T_{imp} = 10\,s$. The down panel shows the presence of some systematic error that must be accounted for. This problem is mainly due to the intrinsic autocollimator non linearity. Though it does not obey a simple analytic form, the autocollimator non-linear error turns out to be highly reproducible, and thus we can calibrate the effect and subtract it from our data.

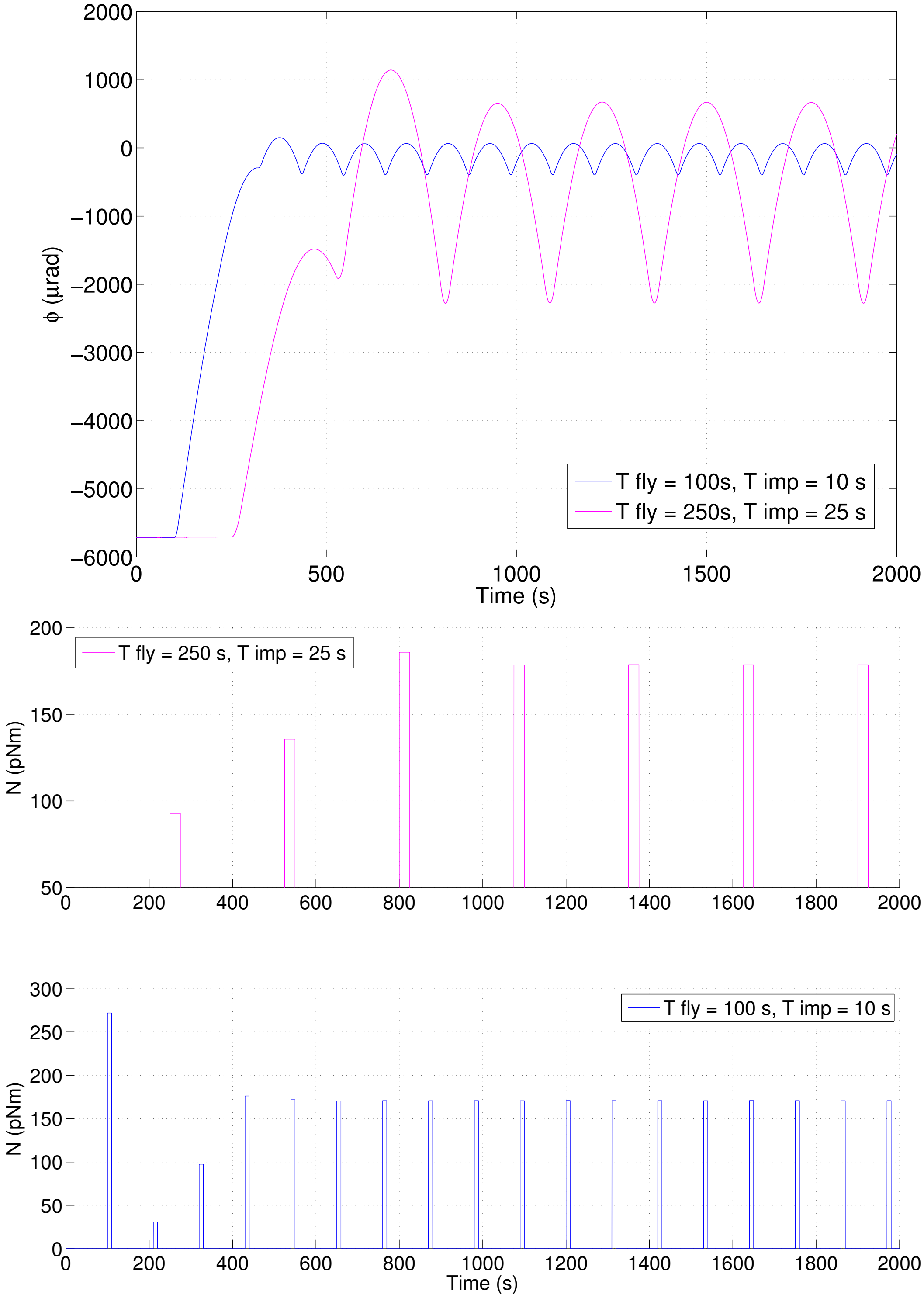

FIGURE 6.1: First panel: Angular time series from free fall measurement performed with $T_{fly} = 250\,s$, $T_{imp} = 25\,s$ and with $T_{fly} = 100\,s$, $T_{imp} = 10\,s$ (blue). Second and third panel: applied impulses.

| | Centered pendulum $\Phi_{EQ} \approx 0\,rad$ | Rotated pendulum $\Phi_{EQ} \approx -5\,mrad$ |
|---|---|---|
| Background noise | 70 of 25000 s March 2015 | |
| | 205 25000 s Winter 2013 | |
| DC actuation noise | | 40 of 27500 s Autumn 2014 |
| Free-fall | | 108 of 27500 s Winter 2014-2015 $T_{fly} = 250s$ and $T_{imp} = 25s$ set-point zero |
| | | 11 cut of 27500 s $T_{fly} = 250s$ and $T_{imp} = 25s$ set-point $-500\mu rad$ and $500\mu rad$ |
| | | 8 cut of 27500 s $T_{fly} = 100s$ and $T_{imp} = 10s$ set-point zero |
| | | 8 cut of 27500 s $T_{fly} = 140s$ and $T_{imp} = 15s$ set-point zero |
| Calibration tone free fall runs | | 60 cut of 27500 s |
| Calibration tone DC actuation runs | 4 cut of 27500 s | 44 cut of 27500 s |

Table 6.1: Measurements data set expressed in number of stretches of 27500 s or 25000 s.

Moreover, another issue can affect free fall measurements due to the high speed of pendulum at the beginning and at the end of each flight. It is related to the sampling frequency of autocollimator signal and also to the right time stamps assigned to each data points.

Both of these issues are particular to the free-fall mode, the first is due to large motions which are sensitive to the AC non-linearity, while the high velocities make time-stamping critical as well as the determination of the correct pendulum position. We address these main two problems in the next two sections.

### 6.2.1 Autocollimator non-linearity

Autocollimator response has been found non linear when pendulum has an oscillation of amplitude of few $\mu rad$. This is an effect of unknown cause, probably due to problem of CCD detector on which the light beam moves or could be due to a stray reflection issue.

To estimate this effects, it is possible to analyze measurement runs during which the pendulum has a free oscillation that covers an angular range including all the angles of the free-fall run that we want to analyze. We calculate the residual model $\Phi_{sin,mod}$ by fitting the angular time series $\Phi_{ac}$ to an ideal sine wave. This is done for a number of selected cycles of length equal to the pendulum period $T_0$, in order to analyze a number of integer cycles over the whole measurement, that are averaged at the end. Residuals $\Delta\Phi_{res}$ are calculated cycle by cycle as difference of the real autocollimator signal with the model:

$$\Delta\Phi_{res} = \Phi_{ac} - \Phi_{sin,mod}. \tag{6.1}$$

and are plotted in figure 6.3 and 6.4. Then, for each cycle, the angular range of $\Phi$ considered, is divided into a grid, whose spacing is defined to be 1 $\mu rad$ for a free fall run correction measurement, and 0.01 $\mu rad$ for a noise run correction, with DC or no actuation. Residuals are sorted in the selected bins and then summed together and averaged on the number of bins. Measured autocollimator $\Phi$ correction are shown in figures 6.3 and 6.4, respectively, for a noise run correction and for a free fall run correction, that is performed on a larger scale.

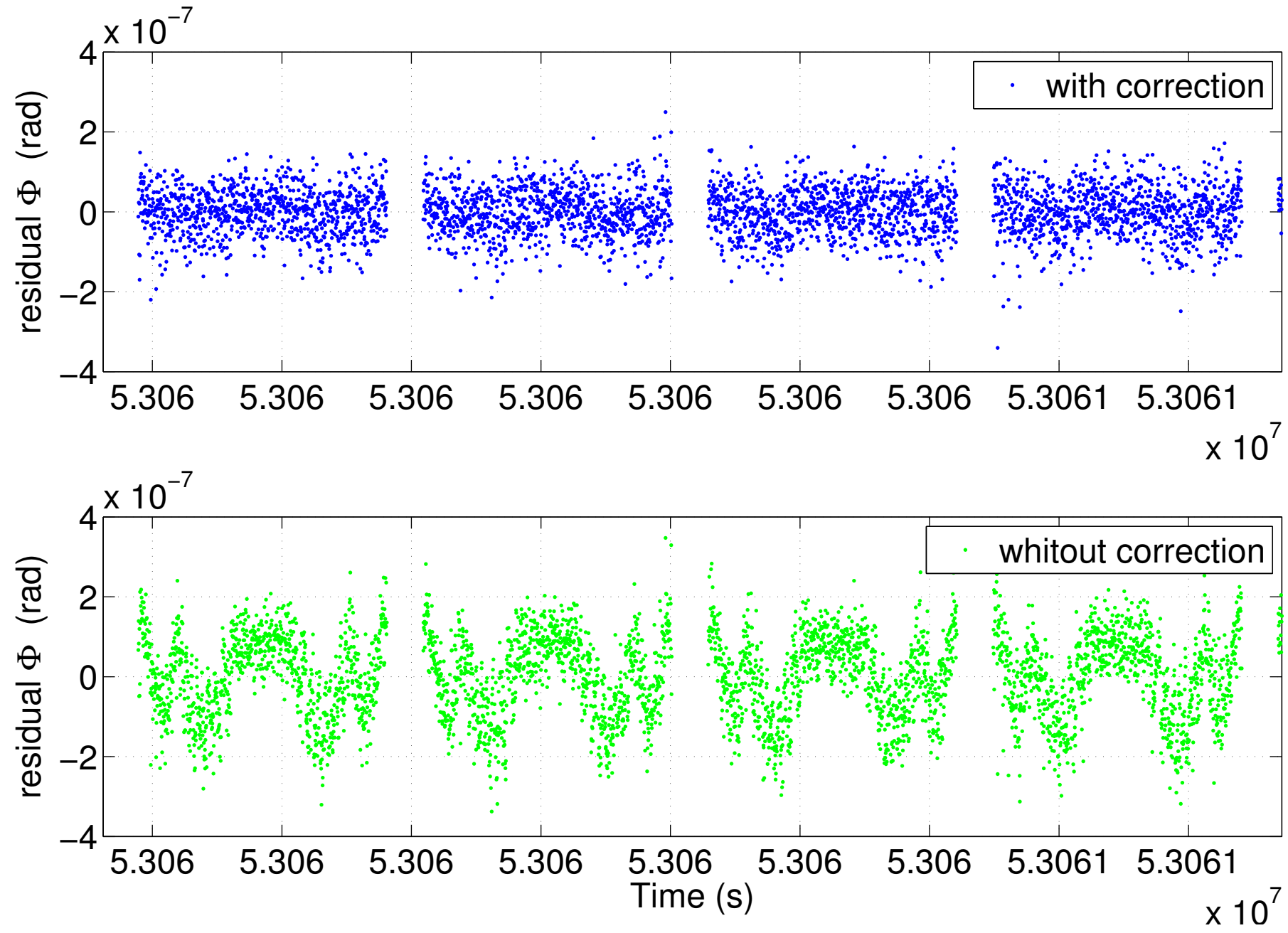

FIGURE 6.2: Residual from the sinusoidal fit for four flights. On top panel the residual are calculated after correcting data for autocollimator non linearity. On bottom panel the same data are showed without correction. A systematic error is visible.

Finally, the residual model calculated, is subtracted from pendulum angular motion $\Phi$ that needs correction, both in case of background noise measurements than for free fall runs.

In case of noise measurements, with a typical amplitude of 1 $\mu rad$, the autocollimator linear correction allows correct the final spectrum, by reducing peaks at frequency multiple of resonance, as in figure 6.4.
The correction effect is also visible on residuals from the sinusoidal fit techniques applied on free fall data, as in the top panel of figure 6.2. The systematic effect disappears in the residuals of sinusoidal fit to each flight.

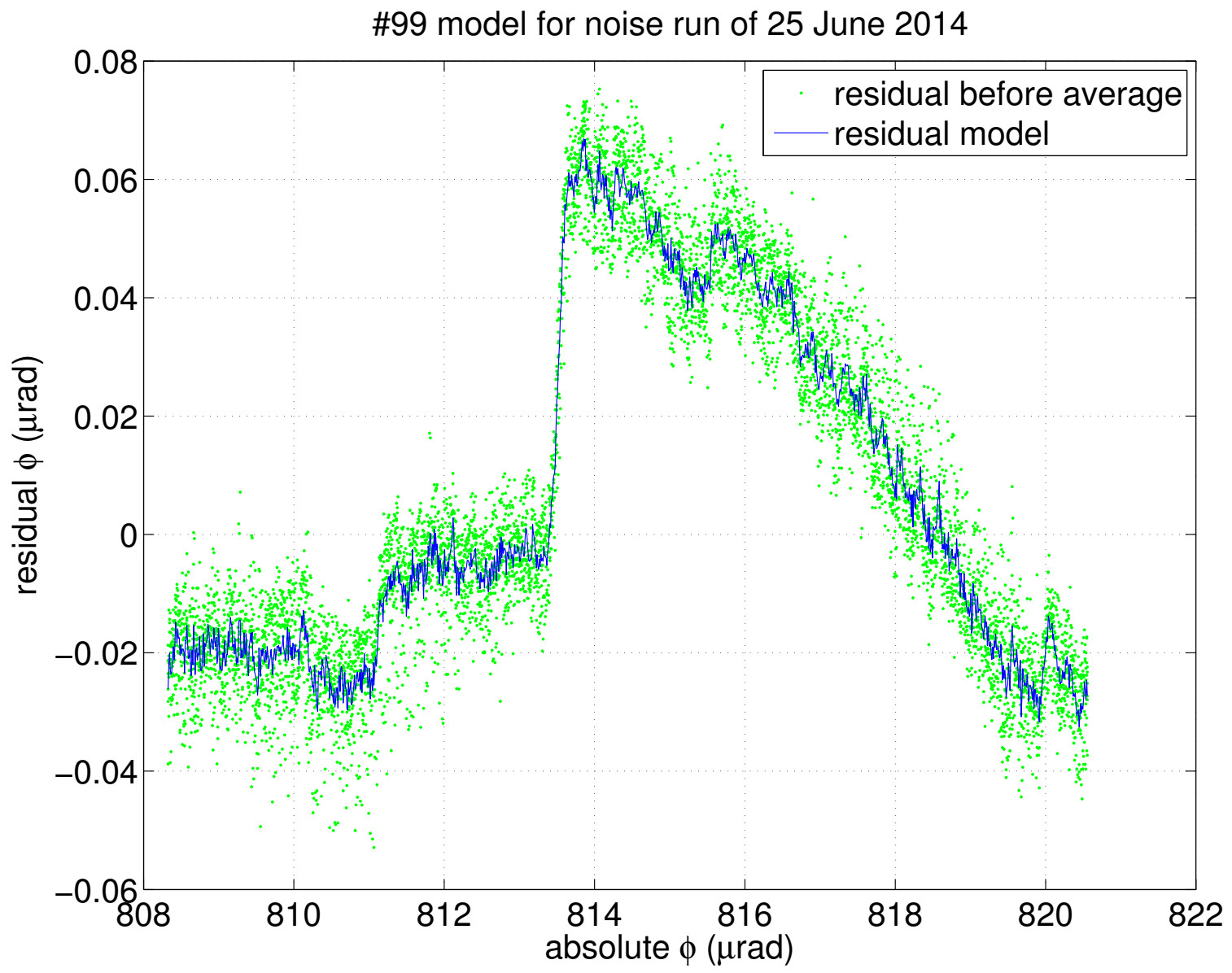

FIGURE 6.3: Residuals from the autocollimator non linearity angular correction. The analysis is performed to correct data of a background noise run.

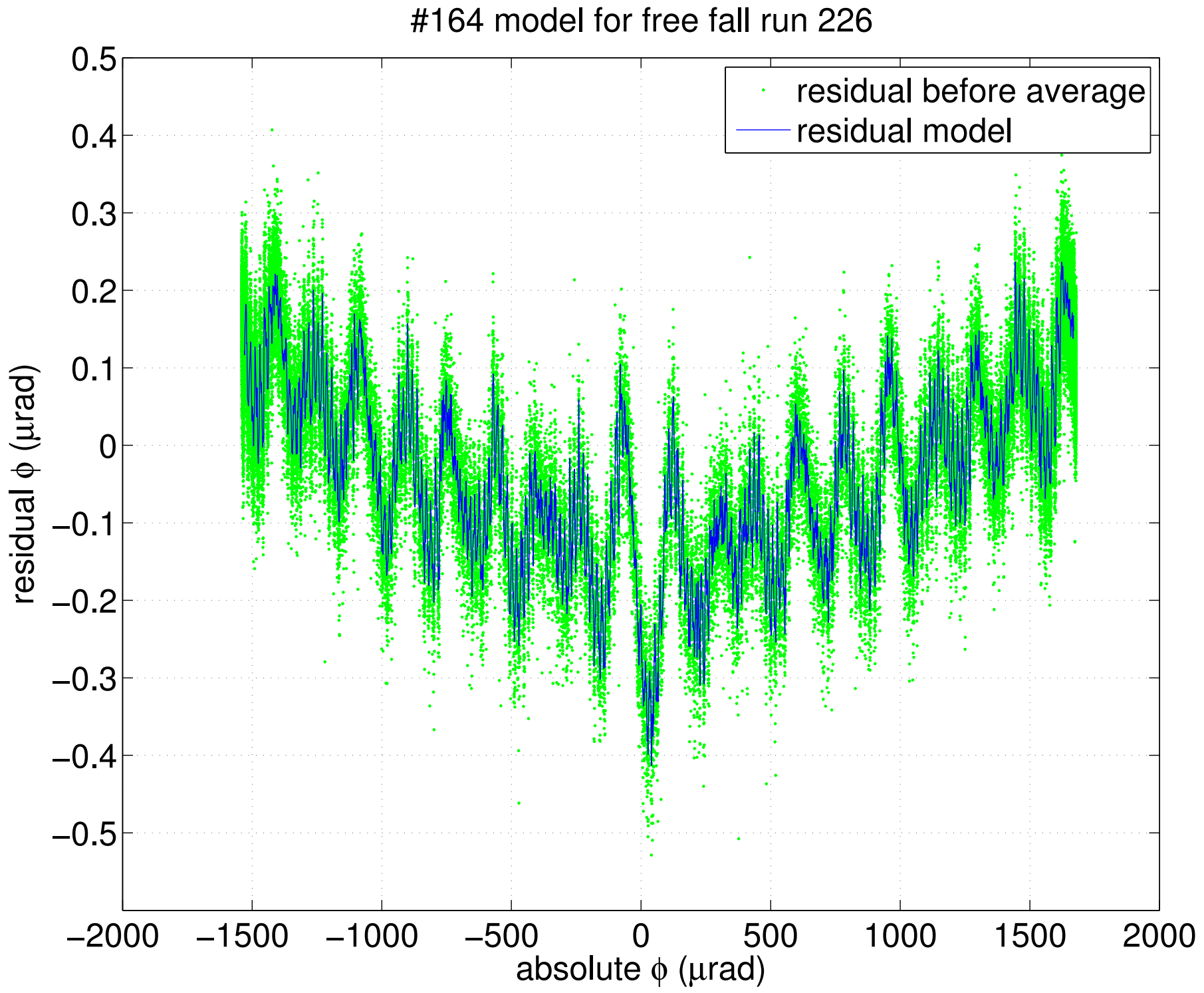

FIGURE 6.4: Residual from the autocollimator non linearity angular correction. The analysis is performed to correct data of a free fall run.

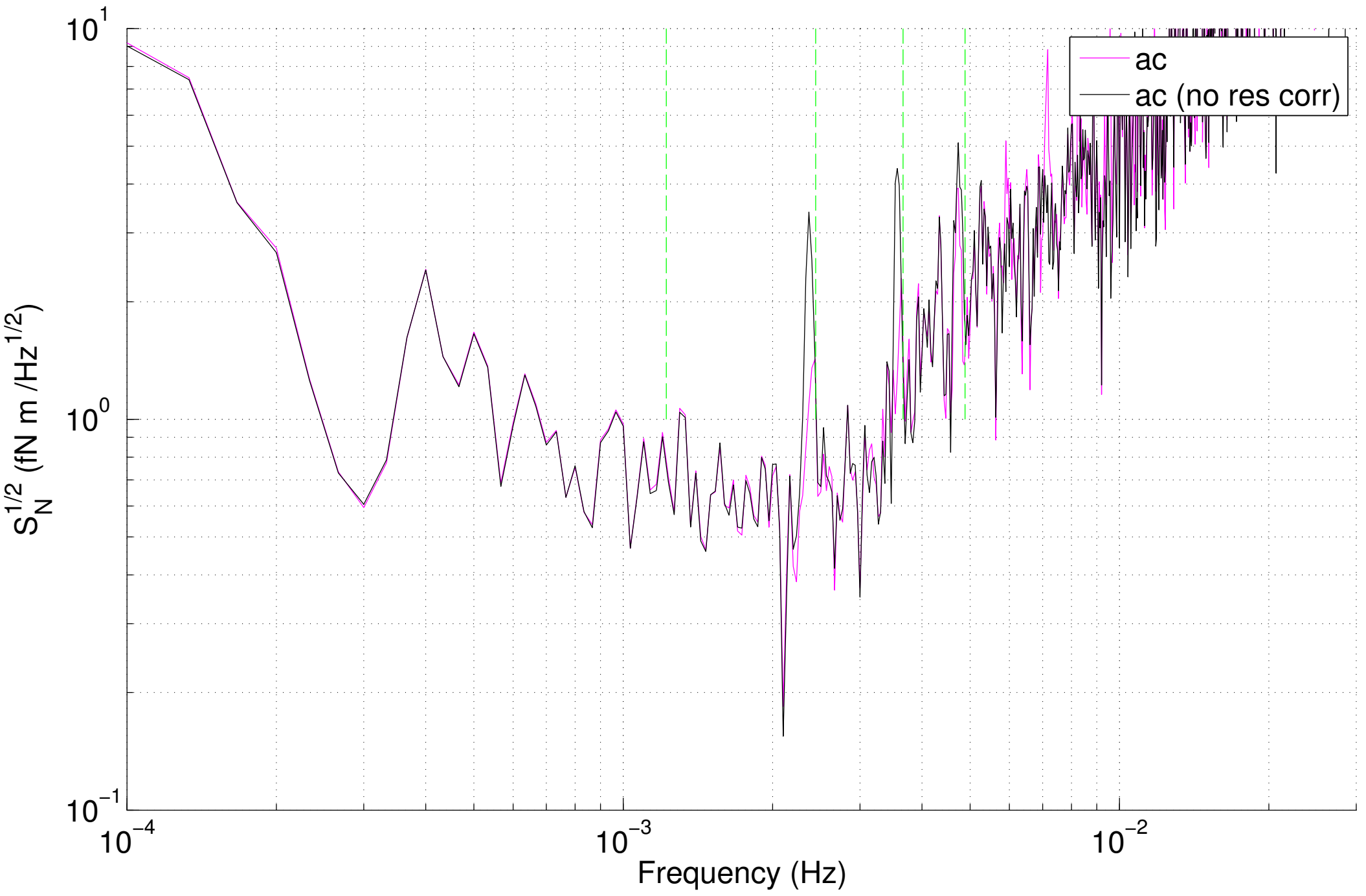

FIGURE 6.5: Torque noise spectrum of autocollimator data for a background noise measurement. Blue spectrum is corrected for autocollimator non-linearity. The green one is not corrected.

### 6.2.2 Time stamp issue

The data acquisition system that monitors the torsion pendulum experiment collects data nominally at 10 $Hz$. As we said, they are measured by a PCI-ADC NI-3032 counter. However, the autocollimator send data points to serial port at 50 Hz internal clock. The serial port is read asynchronously, based on NI clock, every 100 $ms$ with jitter of 2-3 ms due to interaction with windows, with thus 5 autocollimator data points read, in normal conditions, per sample. The autocollimator points do not have a dedicated time stamp, and our serial port sample just reads all the available data. A careful study was performed on the number of packets of bit that are sent at serial port of data acquisition computer, from autocollimator electronics, and on how many points are actually read each 100 $ms$. We found mainly two problems:

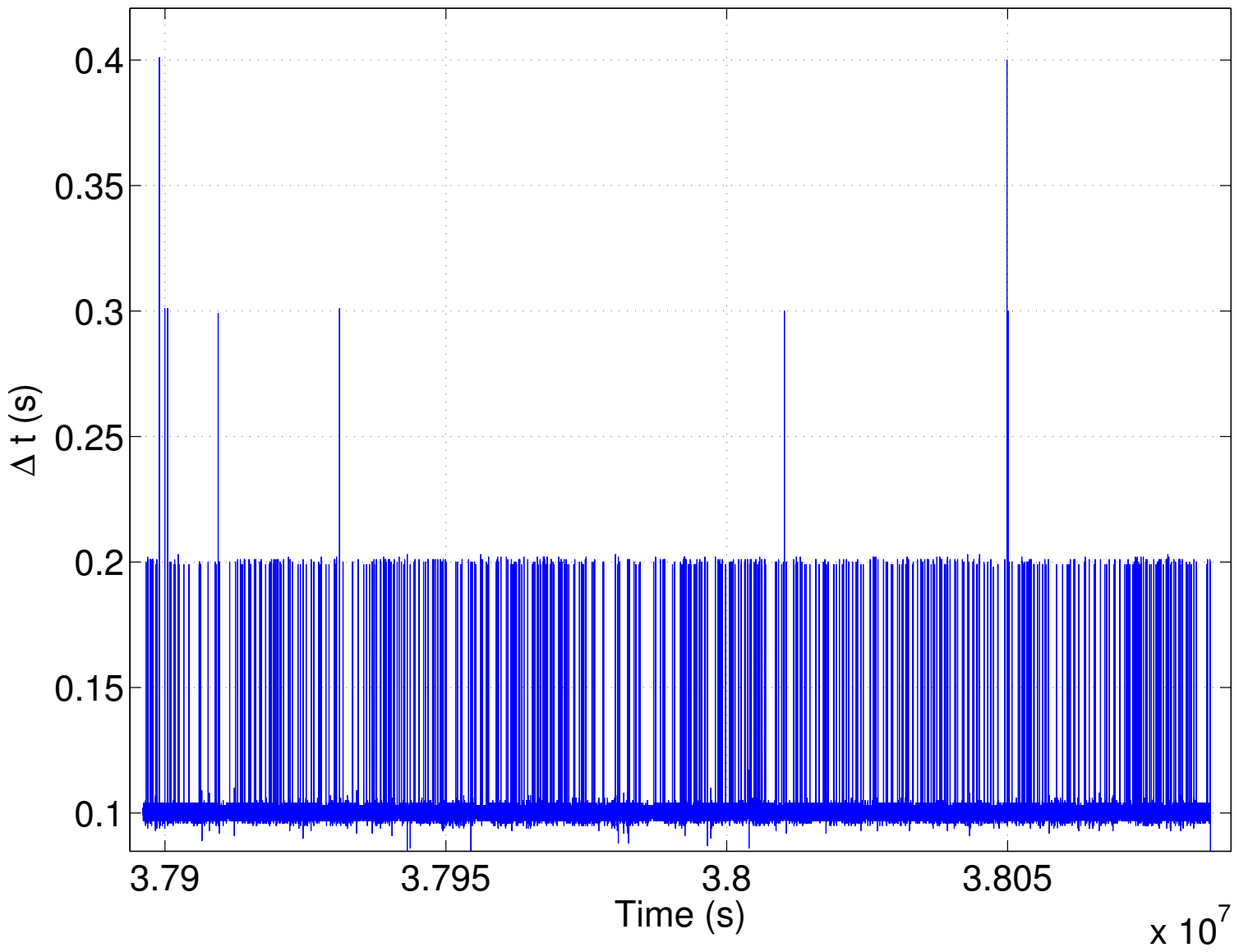

FIGURE 6.6: Interval between successive points sampled by our acquisition system, nominally 100 $ms$ but often missing one sample (0.2 $s$ interval) and rarely more missing samples.

1. missing samples from the autocollimator without any identification on which sample is missing

2. relative slippage of autocollimator internal clock compared to NI clock.

Five points are expected at serial port, 8 bytes each, if we sample each 100 $ms$, but we observe that, with our data aquisition system and NI clock programmed to sample at exactly 10 Hz, the clock slips by one sample (and thus 20 $ms$) every several hundred seconds, as if the AC internal clock were off by roughly 20 ms/300 s, or several seconds per day. This means that sometimes bouncing sequence of 4 or 6 points are available each 100 $ms$ interval if the two clock are not lined up. A faster NI clock means that it will start sampling before the 5th sample arrives, acquiring only 4 samples. The fifth sample is often available sometime in the next several, with a serial read receiving 6 samples.
This effect produce a sequence of 4/6 samples, and it will bounce forward and back to give several 4/6 sequences, (or 6/4 if the NI clock is late respect the autocollimator internal clock) before moving deterministically away from the wrong alignment, as could be seen in figure 6.7 on the top panel. The impact of this timing problems can be seen in figure 6.7, in the bottom panel. A torque time series obtained applying the BH low pass filter technique is shown and an additive noise is visible when sequence of 4/6 or 6/4 are present, that reflects as an excess noise on final data spectra at low frequency.
Making sampling frequency of the data acquisition computer variable, allows to sync up with autocollimator reading. If the sequence of 4/6 has a periodicity of $T_{4/6} = 50000\,s$ as in figure 6.7, it is possible to calculate the number of second to add or subtract from the sampling frequency in order to synchronize the two clocks as $20\,ms/T_{4/6}$. The sampling frequency we set to synchronize the clocks is around $9.9995405\,Hz$. After adjusting the sampling frequency, it is also possible to stretch the grid of sampling thresholds as desired by adding or subtract some milliseconds, depending on whether sampling is in delay or advance compared to the autocollimator clock.
This allows to have very long period without sequence of 4/6 are founded. The two clocks become synchronized, to within $20\,ms$ every $100000\,s$, would be to within a few seconds offset per year. The duration of good sampling moments depends also on environmental temperature, not always stable during weeks of testing campaign. It was necessary a continuous monitoring and adjusting of sampling frequency to obtain the very good free fall data without time stamp issues. From our final data averages on the whole set of weekends of free fall run, we will exclude periods of data in which the clock slippage problem is clear from the number of points.

These imperfections in the sampling do not affect the information content of the sampled signal in the frequency band of interest (from 1 to $3\,mHz$), in case of background noise measurements. This because of the low pendulum velocity. During free fall measurements, however, pendulum velocity is higher, around $30\,\mu rad/s$, during the upward and downward phase of each flight and the presence of a sampling delay can cause position information losses.
The missing autocollimator samples, and our lack of knowledge of when the missing samples occured, also impacts our free-fall analysis. As such, we have altered the acquisition program to record all autocollimator data arriving at the serial port, at 50 Hz, and then, in the event of missing samples, try to assign time stamps to the samples present based on a least squares fit to the preceding data. Additionally, some of the data packets were found to have incomplete autocollimator samples (see figure 6.8), that could in most cases be joined to partial information at the beginning of the next packet and we were able to reconstruct such points. This operation is done recording some data series that help us to recognize where data packets are broken and how many points are acquired. Each data point is indexed with numbers that go from 0 to 4 if we have 5 points each $0.1\,s$.
A code vector of zeros is also recorded for each point. If there are broken packets and then missing bytes, the code is sets to 1 or -1 in order to have and indication on which end the first part of packet and which packets we have to joint together.
Once the broken packets are joined, we assign time stamps to data where some samples are missing. This can be done simply by assigning a nominal time separation of 20 ms to each point, as is done for green points in figure 6.9. Otherwise, it is more precise to perform a linear fit in time considering that the number of points expected $N_{exp}$ for each sample is 5 and identifying where there are holes in the data points, if instead are found $N$ less then $N_{exp}$. A grid of times separated of 20 ms and where the position of the hole slips on the different positions in the matrix, is built. Matrix dimension is defined by the number of point founded in the sample $N$ and by the number of possible combinations that is defined as $N_{exp}!/(N!(N_{exp} - N)!)$. For instance, with 4 points received out of an expected 5, we need to decide which of the 5 times, separated by 20 ms, corresponds to the missing sample, and then place the remaining 4 samples correctly. Once identified packets of point to which reassign time stamp, a linear fit $\phi = at + b$ is performed to last 50 autocollimator points $\phi$ before the missing sample. Then is calculated the minimum $\chi^2$ as

$$\chi^2 = \sum_{i=1}^{N} [\phi_{in} - (at_{in} + b)]^2 \tag{6.2}$$

where $\phi_{in}$ and $t_{in}$ are autocollimator points and times in the interval to reposition, and $N$ the number of points in the interval. The time stamps that minimize $\chi^2$ gives the indexes of holes in the data and of the column of the time grid matrix to slip correctly points in the interval.

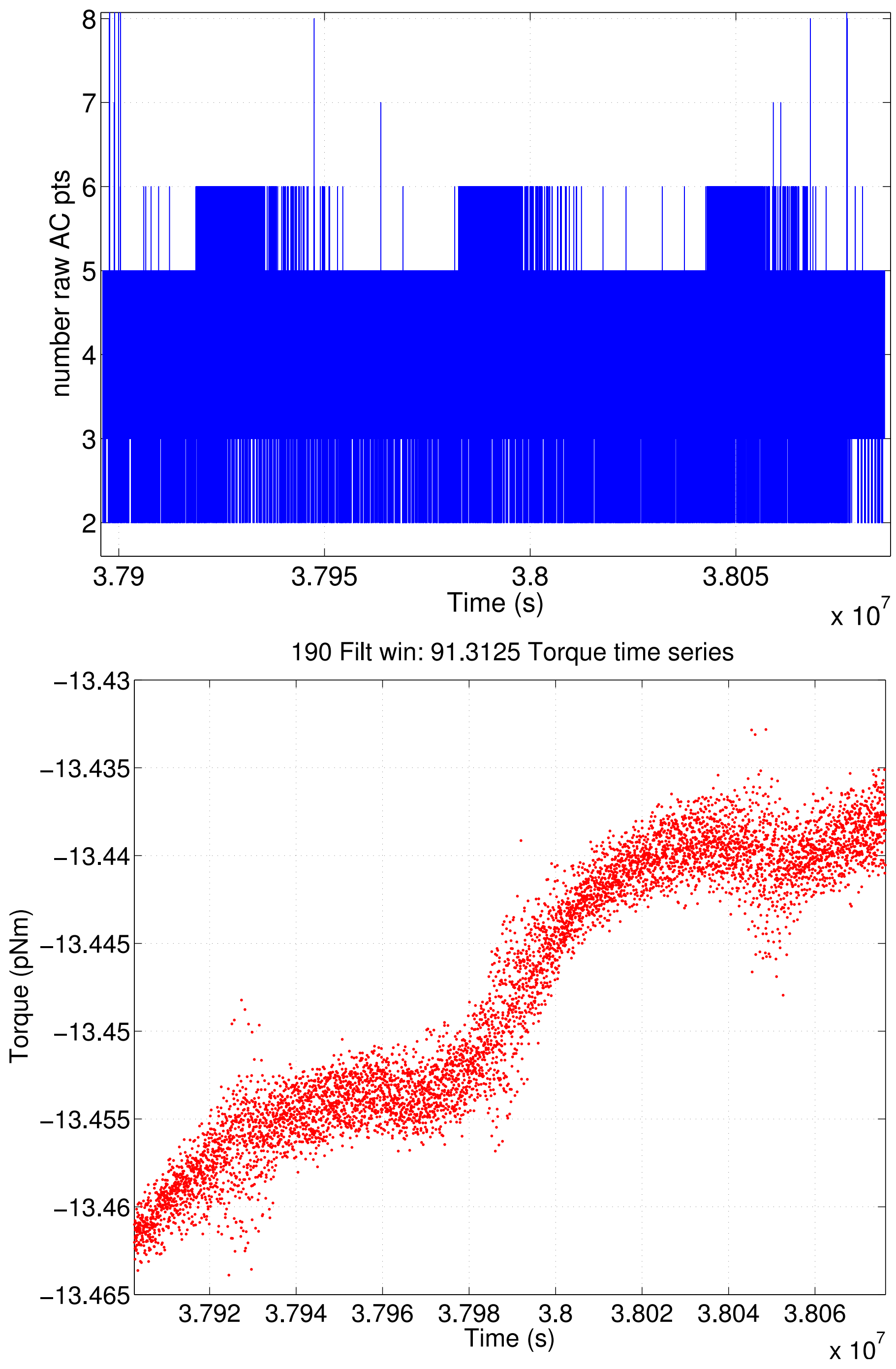

FIGURE 6.7: Top panel: sequence of points of autocollimator. Sequence of 4 or 6 points alternate with moments when 5 points arrive. Bottom panel: torque time series from a free fall measurement. Noisy points are visible during the 4/6 phases.

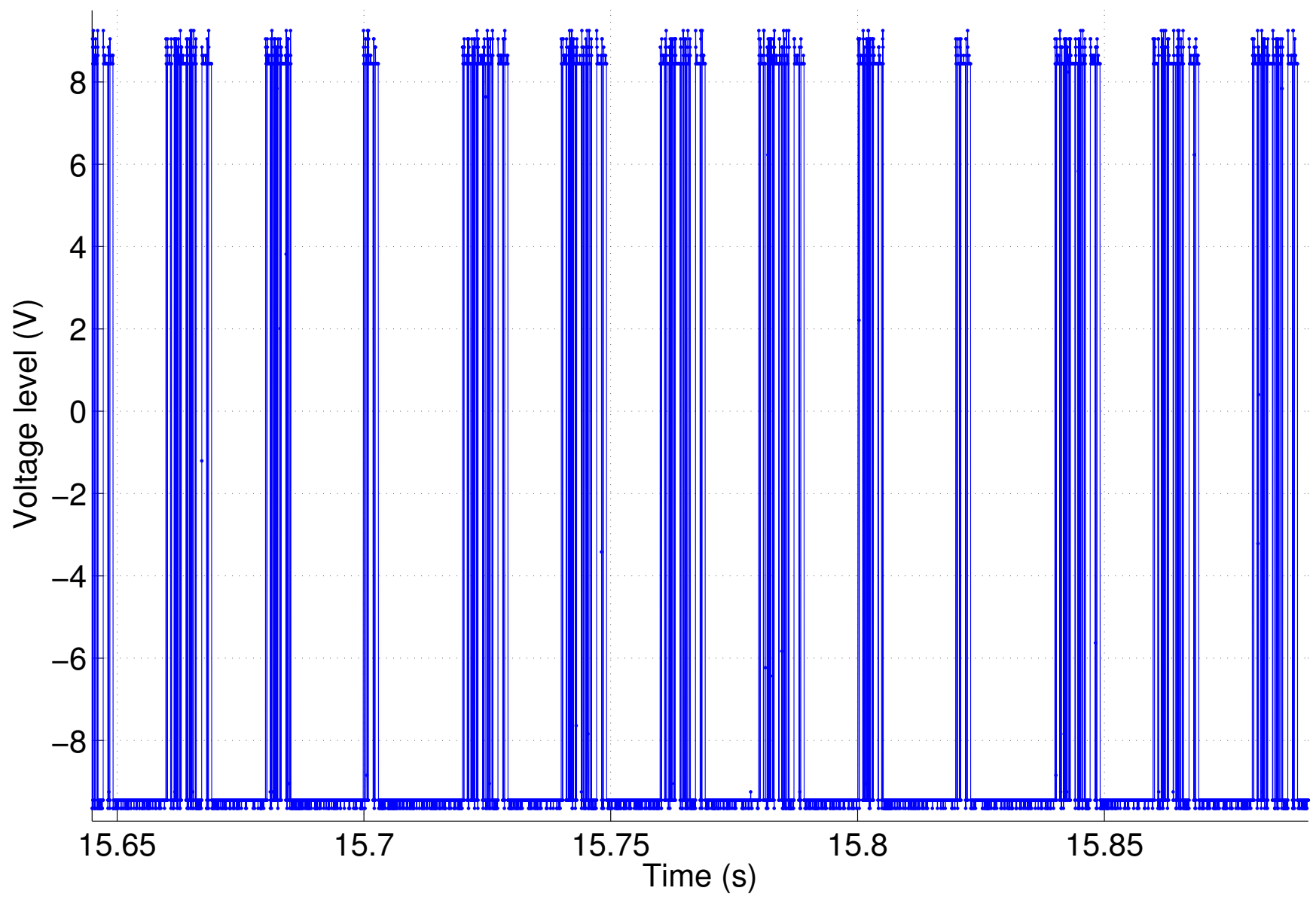

FIGURE 6.8: Autocollimator data packets as detected at serial port by using an oscilloscope. Each packet is 8 bytes. Some of these arrive corrupted, with information divided from two subsequent packets, as visible for example for packets between 15,68 and 15.7 seconds in this plot.

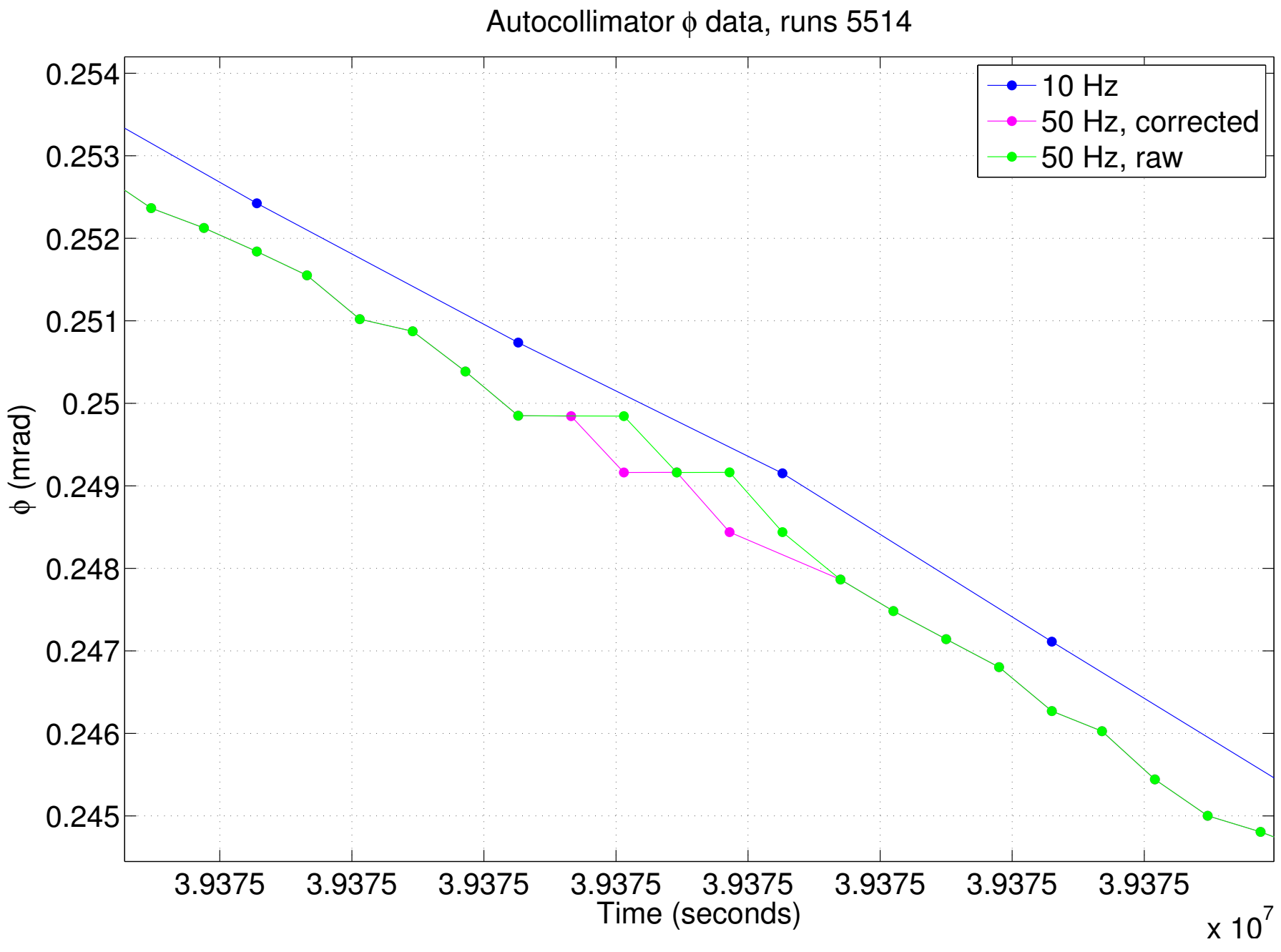

FIGURE 6.9: Autocollimator angular $\Phi$ time series. Blue data points are data sampled at $10\,Hz$. Green data points are sampled at $50\,Hz$ with no correction. Magenta points are sampled at $50\,Hz$ after performing a fit in time in order to reassign the correct time stamp.

Usually, on 10000 s of data we make 5000 corrections time stamps, including of order 1000 partial packs. Data well spaced are the magenta point in figure 6.9.

## 6.3 Calibration of the experiment: pendulum dynamical parameters

The modeled pendulum dynamic which is used to convert the angular displacement $\Phi$ in torque $N$ is linear and, as already explained, provide that:

$$N = I\ddot{\Phi} + \frac{\Gamma\delta}{\omega}\dot{\Phi} + I\left(\frac{2\pi}{T_0}\right)^2 \Phi \tag{6.3}$$

$$= I\ddot{\Phi} + \gamma\dot{\Phi} + \Gamma\Phi \tag{6.4}$$

where, for convenience, we express the dissipation factor as $\gamma = \frac{\Gamma\delta}{\omega}$ and the stiffness term $\Gamma = I\left(\frac{2\pi}{T_0}\right)^2$.

Both the analysis techniques, described in the previous chapter, employ this linear model of torque with a linear dependence of the stiffness $\Gamma$ from $\Phi$ (as seen from equation 5.2 and 5.3). The relative torque time series, as a result of the analysis techniques applied on real free fall data, are shown in figure 6.10. These are results from a free fall run with $T_{fly} = 250\,s$ and $T_{imp} = 25\,s$, considering that the pendulum is soften by applying DC bias on $Y$ electrodes of $\pm 9.5\,V$, which results in a longer pendulum period.

In case of sinusoidal fit, the only parameter employed is the stiffness, which comes from period measurement performed leaving oscillating pendulum around its equilibrium angle of $\Phi_{EQ} =\approx -5.7\,mrad$. For a measurements of free oscillation around equilibrium angle, of at least two hours, we perform a fit to stretches of data 2000 s long and typically with amplitude of $400\mu rad$ peak to peak. The performed fit estimates the period $T_0$ that minimizes the model $\chi^2$, for some range of periods $[T_{0,ran1}\ T_{0,ran2}]$ around the expected one, using quadratic fit to mean square deviation of data from fitted model. Periods calculated for each cycles are then averaged to give the final result. The measured period around the equilibrium point $-5.7\,mrad$, is $T_0 = 825.654 \pm 0.003\,s$, brings to a stiffness of $\Gamma = I(2\pi/T_0)^2 = 2.4960 \pm 0.0002\,nNm/rad$.

We note the variation of torque obtained applying the Blackman-Harris low pass filtering technique in figure 6.10. There is a difference in torque of $50\,fNm$ between the central and the side part of each flight. This residual dependence of the measured torque on $\Phi$ could correspond to an error in the value used for $\Gamma$ of roughly $20\,pNm/rad$, with $50\,fNm$ torque change over roughly $3\,mrad$ of angular range, as visible in figure 6.1. This suggests that period measurement does not accurately reflect the true stiffness relevant to the free-fall measurement. We note also that the pendulum period has been measured around the equilibrium angle of $-5.7\,mrad$, while the free fall measurement is performed between $-2.3\,mrad$ and $800\,\mu rad$. Performing the free-oscillation period measurement in this free-fall angular range would requires an applied torque, which would have a torque gradient that alters the stiffness.
This leads to think that some non linearity respect the angular position in the stiffness behavior is to take into account. A possible torque model with a quasi-elastic term could be:

$$N = I\ddot{\Phi} + \gamma\dot{\Phi} + \Gamma\left[1 + \alpha\Phi + \beta\Phi^2\right]\Phi. \tag{6.5}$$

so that the stiffness model will be

$$\Gamma_{MOD} = -\frac{\partial N}{\partial \Phi} = -\left[\Gamma + 2\alpha\Gamma\Phi + 3\beta\Gamma\Phi^2\right], \tag{6.6}$$

where the stiffness $\Gamma$ is referred to the zero angle position. Accounting for the right stiffness model parameters, $\alpha$, $\beta$, $\Gamma$, it could be possible to reduce the position dependence of the torque measured during free fall measurements, which would be accentuated here due to an higher dynamic range respect to typical background force noise measurements.
In order to calibrate the free fall experiment and estimate the dynamical parameters in the angular range relevant to the free-fall experiment and to determine if a non-linear model is necessary, we perform a least square fit to an acceleration model, calculated as the double derivative of pendulum angular deflection during the free fall measurement.

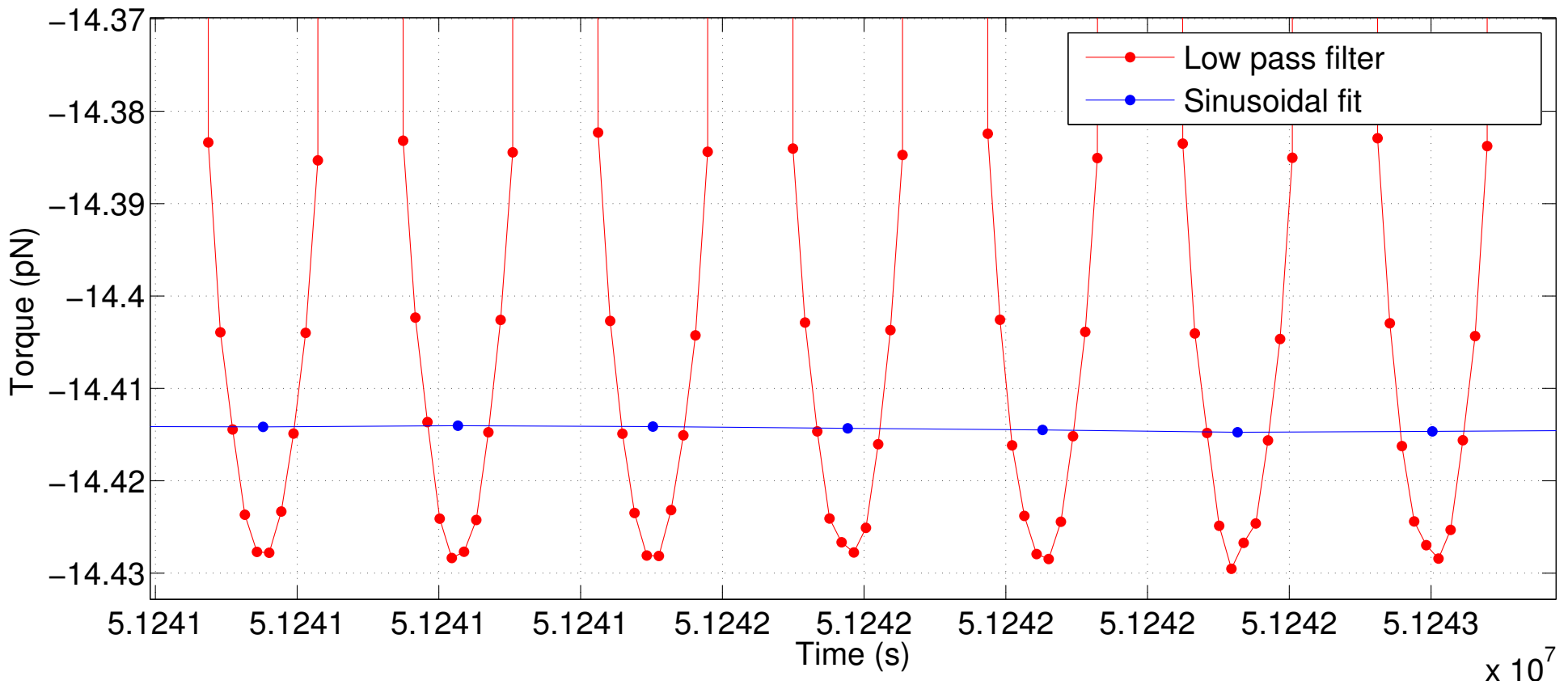

FIGURE 6.10: Torque time series for a free fall measurement as results of the two analysis technique developed. The parameter employed in the torque model come from measured period, $T_0 = 825.654 \pm 0.003s$.

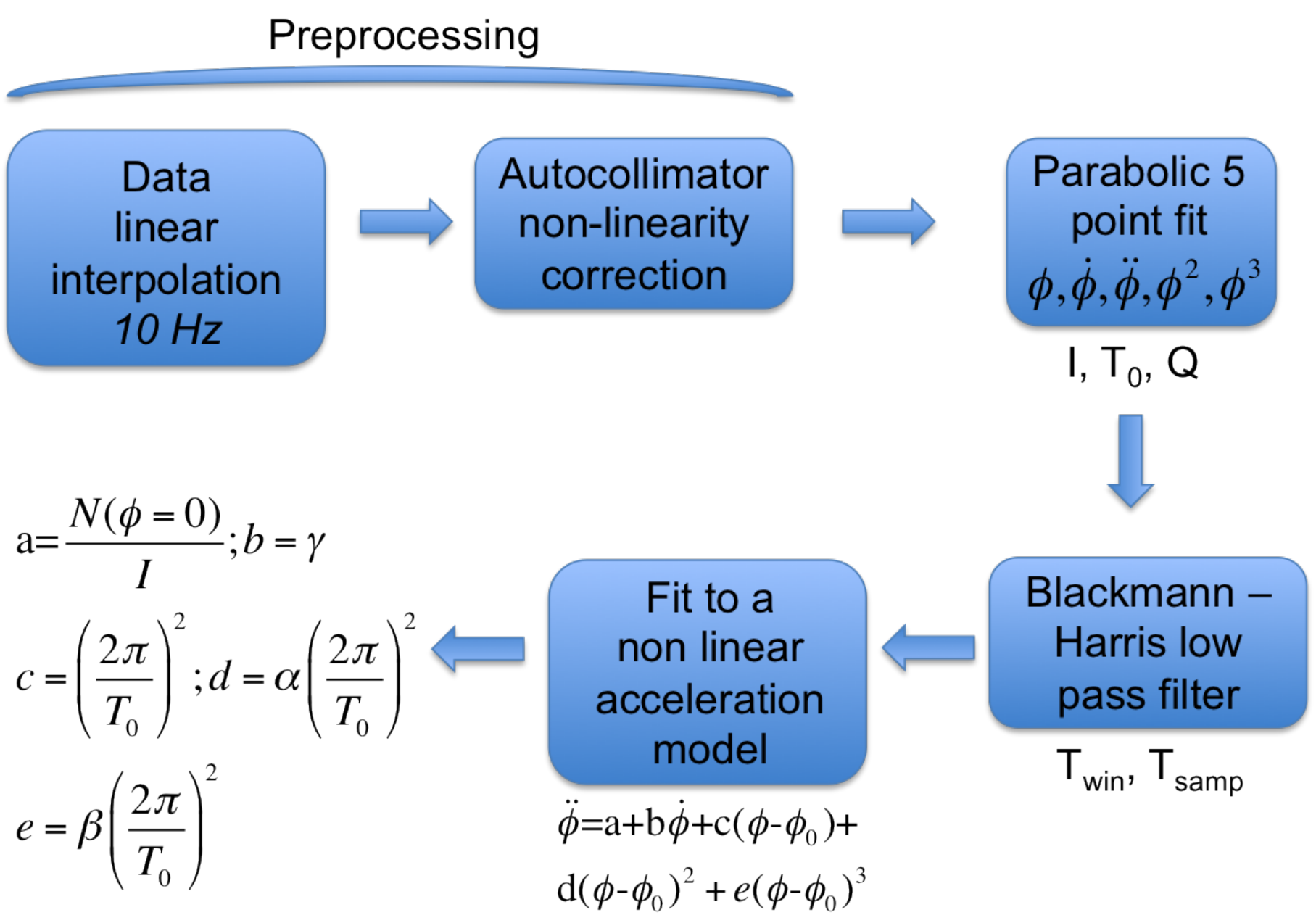

FIGURE 6.11: Flow chart of the algorithm used to estimate non linear parameters from a fit to an acceleration model.

The used model is:

$$\ddot{\Phi} = a + b\dot{\Phi} + c\Phi + d\Phi^2 + e\Phi^3 \tag{6.7}$$

where, $a = N(\Phi = 0)/I$, $b = -\gamma/I$, $c = -(2\pi/T_0)^2$ and related to the stiffness $\Gamma = I(2\pi/T_0)^2$, $d = -\alpha(2\pi/T_0)^2$ and $e = -\beta(2\pi/T_0)^2$.

The flow chart of the performed algorithm is showed in figure 6.11. First of all $\Phi$ angular data are preprocessed, by interpolating data at the nominal sampling frequency $10\,Hz$, then they are corrected for the autocollimator non-linearity. First and second derivatives of $\Phi$ are then obtained performing a parabolic fit to 5 adjoining points at close time intervals of angular time series. Before to perform the final least square fit, acceleration, velocity and angular deflection, but also $\Phi^2$ and $\Phi^3$, are filtered with the same algorithm used for the BH low pass filter technique, by choosing a Blackman-Harris window length $T_{win}$ and a sampling time $T_{samp}$. The choice of window sampling time $T_{samp}$ is done to have an integer number of finite window in one free fall cycle time $T_{exp}$ as in equation 5.5. Moreover, the BH window length $T_{win}$, has been chosen an oversampling factor equal to 5 times the sampling frequency $T_{samp}$. With the usual choice of $T_{fly} = 250\,s$ and $T_{imp} = 25\,s$, from equations 5.5 and 5.6 will be $T_{win} = 62.5\,s$ and $T_{samp} = 12.5\,s$.
The least square fit is then performed for each flight.
The same fit is performed also considering only a linear dependence from $\Phi$:

$$\ddot{\Phi} = a + c\Phi \tag{6.8}$$

with only $a = N(\Phi = 0)/I$ and $c = (2\pi/T_0)^2$. This allows to extrapolate the stiffness as the only parameters from $c$.
For a typical free fall measurement the estimated parameters, obtained performing a fit to the non linear model 6.7, averaged and scaled opportunely in order to show only quantity of interest for the final torque model 6.5, are reported in table 6.2. The analogous result for a fit to a linear acceleration model 6.8, is reported in table 6.3. Clearly, there is a discrepancy of at least 1% between the measured period $825.654 \pm 0.003\,s$ and the estimated ones from fit. This is also addressed in section 6.3.1 by comparing period, and then stiffness measurements, with the prediction of the electrostatic model, which account for an effective non linearity of the stiffness from the angular displacement.

| | Non linear parameters |
|---|---|
| $\Gamma$ | $2.5402 \pm 0.0002\, nNm/rad$ |
| $T_0$ | $818.44 \pm 0.04\, s$ |
| $2\alpha\Gamma$ | $-16.6 \pm 0.2\, nNm/rad^2$ |
| $3\beta\Gamma$ | $7.9 \pm 1.6\, \mu Nm/rad^3$ |
| $-\frac{\gamma}{I}$ | $(-1.12 \pm 0.02)10^{-7}\, 1/s$ |

Table 6.2: Non linear model parameters estimated from a typical free fall measurements performed with $T_{fly} = 250\,s$ and $T_{imp} = 25\,s$ and a corresponding range in angle displacement of around $\Delta\Phi \approx 3\,mrad$.

| | Linear parameters |
|---|---|
| $\Gamma$ | $2.5401 \pm 0.0004\, nNm/rad$ |
| $T_0$ | $818.454 \pm 0.006\, s$ |

Table 6.3: Linear model parameters estimated from a typical free fall measurements performed with $T_{fly} = 250\,s$ and $T_{imp} = 25\,s$ and a corresponding range in angle displacement of around $\Delta\Phi \approx 3\,mrad$.

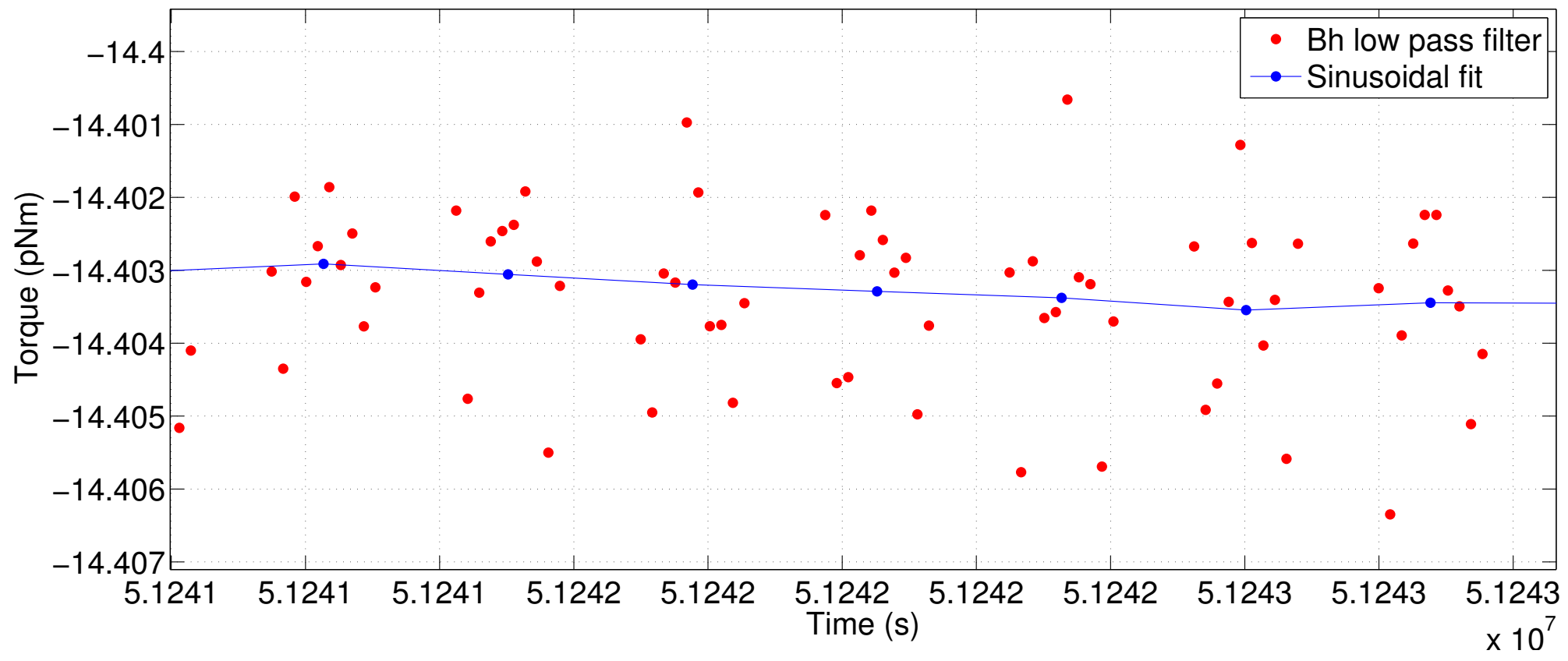

FIGURE 6.12: Torque time series for a free fall measurement as results of the two analysis technique developed. The parameter employed in the torque model 6.4, comes from linear fit at the linear acceleration model 6.8.

The determination of stiffness model parameters has an impact on the torque time series estimated during free fall measurement. Taking into account the correct stiffness obtained from the linear fit to the acceleration model 6.8, the recovered torque obtained from both analysis techniques is showed in figure 6.12. It can be noted the difference with figure 6.10 in which the parameter used in the model was the measured stiffness from period. A correct estimate of stiffness allows to reduce uncertainties in the final torque estimation.

Since the electrostatic model admits a non linear dependence of the stiffness from the angle, as we will discuss in the next section, also non linear parameters $\alpha$ and $\beta$ have been considered to calculate the stiffness and then torque model as in equation 6.5 and 6.6.

### 6.3.1 Stiffness from measurements and electrostatic model

Both to have a rough confirmation of the electrostatic stiffness and to evaluate the plausibility of a significant non-linear stiffness contribution, we performed a series of period measurements with the pendulum centered and near the equilibrium angle that we compare with measurement performed with pendulum rotated with respect to the electrode housing, near the new equilibrium angle.

The stiffness is the residual coupling of the test mass to the gravitational reference sensor and to the suspension point, defined as $\Gamma = -\partial N/\partial \Phi$. From the electrostatic point of view, this is the sum of all of components of spatial gradients due to forces acting on test mass. On torsion pendulum they sum with the stiffness due to fiber.

The linear, purely harmonic equation of motion 3.1 can be divided into different stiffness terms like $\Gamma_s$, generated by the sensor and by $\Gamma_Y$, due to the presence of DC bias applied on Y electrodes, as follows:

$$I\ddot{\Phi} + \Gamma_f(\Phi_f - \Phi_{f_0}) + \Gamma_s(\Phi_s - \Phi_{s_0}) + \Gamma_Y(\Phi_Y - \Phi_{Y_0}) = 0. \tag{6.9}$$

$\Gamma_f$ is the torsion constant of the fiber and $\Phi_f$ is the equilibrium position. Here, $\Phi_{s0}$ is the equilibrium position which nulls the torque associated with sensor induced stiffness $\Gamma_s$, while $\Phi_{Y0}$ is the equilibrium position which nulls the torque associated with Y electrodes DC bias stiffness $\Gamma_Y$, and they set the pendulum equilibrium angle to be

$$\Phi_0 = \frac{\Gamma_f\Phi_{f_0} + \Gamma_s\Phi_{s_0} + \Gamma_Y\Phi_{Y_0}}{\Gamma_f + \Gamma_s + \Gamma_Y}. \tag{6.10}$$

The overall angular spring constant $\Gamma_{tot}$ can be then considered as the sum of all single contribution, which is possible to measure by evaluating the shift each term causes in the torsion pendulum resonance frequency,

$$\Gamma_{tot} = \Gamma_f + \Gamma_s + \Gamma_Y. \tag{6.11}$$

We perform different period measurements in order to account for all of this terms, considering also the dependence of the stiffness from the angle. Period measurements are performed by letting oscillate pendulum with an amplitude around $400\mu rad$ and performing a sinusoidal fit to cycle length 2000 s, in order to calculate period on the average of different cycles.

Results are shown in table 6.4. The period $T_{0,sens}$ measured, with a centered test mass respect to the housing, and then near the equilibrium point $\Phi_{EQ} \approx 0\,rad$, is made in nominal condition with the AC sensing bias applied with an RMS amplitude of $V_{inj} = 3.6\,V$. The period decreases to $T_{0,fib} = 460.766 \pm 0.002\,s$ with the sensing bias switched off. If the contribution due to Y electrodes DC bias is taken into account, applying voltages equal to $V_Y = 9.5\,V$, the period increase, as we said, up to $T_{0,Y_{DC,bias}} = 819.73 \pm 0.01\,s$.

The same period measurements were performed in the condition in which the pendulum is out of center, with test mass rotated with respect to the electrode housing of an angle $\Phi_{EQ} \approx -2\,mrad$, as done to perform the free fall measurements. From equation 6.10, it can be noted that the equilibrium angle change when a source of stiffness is added. The new equilibrium angle with Y DC bias switched on is around $-5\,mrad$. Measurement of period around the new equilibrium angle give $T_{0,Y_{DC,bias}} = 823.43 \pm 0.05s$ when DC bias on Y electrodes are switched on.

We note again the difference of $4\,s$ in period measured around the equilibrium point zero and $-5mrad$, when Y DC bias are applied. We account for the dependence of the period, and then the stiffness, from angle, by performing a least square fit of $\Gamma$ from all the period measurements, remembering that $\Gamma = I(2\pi/T_0)^2$, to the model:

$$\Gamma_m = \Gamma_{fib} + AV^2_{inj} + BV^2_{Y_{DC,bias}} + CV^2_{Y_{DC,bias}}\Phi^2. \tag{6.12}$$

| **Period measurements** | | | | |
|---|---|---|---|---|
| | Centered pendulum $\Phi_{EQ} \approx 0\, rad$ | Rotated pendulum $\Phi_{EQ} \approx -2\, mrad$ | Rotated pendulum $\Phi_{EQ} \approx -5\, mrad$ | |
| $T_{0,fib}$ | $460.766 \pm 0.002$ | | | s |
| $T_{0,sens}$ | $468.011 \pm 0.003$ | $467.971 \pm 0.002$ | | s |
| $T_{0,Y_{DC,bias}}$ | $819.73 \pm 0.01$ | | $823.43 \pm 0.05$ | s |

| **$\Gamma$ fit results** | | **Prediction from electrostatic** | | |
|---|---|---|---|---|
| $\Gamma_{fib}$ | $8.014 \pm 0.007$ | | | nNm/rad |
| A | $-18.9 \pm 0.7$ | $\Gamma_s/V^2_{inj}$ | $-17 \pm 16$ | $pNm/V^2rad^2$ |
| B | $-58.0 \pm 0.1$ | $-2\frac{\partial^2 C_Y}{\partial \Phi^2}$ | $-45.8 \pm 1.6$ | $pNm/V^2rad^2$ |
| C | $-8.9 \pm 0.1$ | $-\frac{\partial^4 C_Y}{\partial \Phi^4}$ | $-24.0 \pm 0.3$ | $nNm/V^2rad^4$ |
| **Estimation from fit** | | | | |
| $\Gamma_{sens}(\Phi = 0) = AV^2_{inj}$ | $-245.5 \pm 9.3$ | | | pNm/rad |
| $\Gamma_{Y_{DC,bias}}(\Phi = 0) = BV^2_{Y_{DC,bias}}$ | $-5.237 \pm 0.009$ | | | nNm/rad |
| $\frac{\partial \Gamma(\Phi)}{\partial \Phi^2} = CV^2_{Y_{DC,bias}}$ | $-80 \pm 1$ | | | $nNm/rad^2$ |
| $\Gamma_{tot}(\Phi = 0) = \Gamma_f + \Gamma_{sens} + \Gamma_Y$ | $2.54 \pm 0.02$ | | | nNm/rad |

| **Capacitance derivatives from FE analysis** | | |
|---|---|---|
| $\frac{\partial^2 C_{inj}}{\partial \Phi^2}$ | $22.90 \pm 11$ | $pF/rad^2$ |
| $\frac{\partial^2 C_{tot}}{\partial \Phi^2}$ | $0.596 \pm 0.003$ | $nF/rad^2$ |
| $\frac{\partial^2 C_Y}{\partial \Phi^2}$ | $22.8 \pm 0.8$ | $pF/rad^2$ |
| $\frac{\partial^4 C_Y}{\partial \Phi^4}$ | $24.0 \pm 0.3$ | $nF/rad^4$ |

TABLE 6.4: Top panel: period measurements results. Central panel: parameters obtained from fit to measured $\Gamma$ compared with the prediction of the electrostatic model. Bottom panel: capacitance derivatives w.r.t $\Phi$ values from FE analysis.

We don't consider the dependence of the injection bias from the angle, because the Y electrodes are closer to the test mass (2.9 mm) respect the injection bias (4 mm) (see figure 1.5), and the effect can be neglected.

We obtain the stiffness contribution of the injection bias respect to $\Phi = 0rad$ as $\Gamma_{sens}(\Phi = 0) = AV^2_{inj}$. While the stiffness due to Y DC bias in zero is $\Gamma_{Y_{DC,bias}}(\Phi = 0) = BV^2_{Y_{DC,bias}}$. Results of various stiffness contribution and of coefficients of the fit, are listed in the central panel of table 6.4. All parameters errors are scaled for the $\chi^2$ of the fit that is around 100.

Each stiffness contribution, and each coefficient of stiffness fit, can be also apportioned from the electrostatic model [7], except the stiffness of fiber. Starting from equations of section 3.2.1, we can write the contribution of torque due to injection and sensing electrodes as well as that due to Y electrodes, biased with DC voltages.

- *Sensing stiffness* The total torque along $\Phi$ due to all electrodes and housing surfaces facing the test mass, can be generally written as:

$$N_\Phi = \frac{1}{2}\frac{\partial C_{inj,TM}}{\partial \Phi}(V_{inj} - V_{TM})^2 + \frac{1}{2}\sum_{i \neq inj}\frac{\partial C_{i,TM}}{\partial \Phi}(V_i - V_{TM})^2 + \frac{1}{2}\frac{\partial C_{inj,H}}{\partial \Phi}V_{inj}^2 \tag{6.13}$$

where $\frac{\partial C_{inj,TM}}{\partial \Phi}$ and $\frac{\partial C_{inj,H}}{\partial \Phi}$ are the derivative of injection electrode respect to test mass and housing surfaces. $\sum_{i \neq inj}\frac{\partial C_{i,TM}}{\partial \Phi}$ is the sum of capacitance derivative of all $x$ electrodes respect to test mass and housing, and we will named this sum of capacitance as $C_{tot}$.
With some algebra and expanding capacitance derivative with respect to a $\Phi_0$ zero position, and reducing the notation as explained in section 3.2.1

$$\frac{\partial C}{\partial \Phi} \approx \left.\frac{\partial C}{\partial \Phi}\right|_0 + \left.\frac{\partial^2 C}{\partial \Phi^2}\right|_0 (\Phi - \Phi_0), \tag{6.14}$$

the final torque expression due to electrodes sensing contribution is

$$\begin{aligned} N_\Phi = & \frac{1}{4}\alpha^2 V_{inj}^2 \left[\left.\frac{\partial C_{tot}}{\partial \Phi}\right|_0 + \left.\frac{\partial^2 C_{tot}}{\partial \Phi^2}\right|_0 (\Phi - \Phi_0)\right] \\ & + \frac{1}{2}V_{inj}^2 \left[\left.\frac{\partial C_{inj}}{\partial \Phi}\right|_0 + \left.\frac{\partial^2 C_{inj}}{\partial \Phi^2}\right|_0 (\Phi - \Phi_0)\right] \end{aligned} \tag{6.15}$$

where $\alpha$ is defined as $\alpha = C_{inj}/C_{tot}$. Now the stiffness due to the sensing bias is

$$\Gamma_s = -\frac{\partial N_\Phi}{\partial \Phi} = -\frac{1}{2}\alpha^2 V_{inj}^2 \frac{\partial^2 C_{tot}}{\partial \Phi^2} - \frac{1}{2}V_{inj}^2 \frac{\partial^2 C_{inj}}{\partial \Phi^2}. \tag{6.16}$$

This is directly comparable with the $A$ parameter coming from fit to stiffness considering

$$\Gamma_s = A V_{inj}^2. \tag{6.17}$$

If we consider an injection bias $V_{inj} = 3.6 V_{RMS}$, and we use capacitance derivatives as measured from FE analysis [7], we have model prediction of $\Gamma_s$ that is completely dominated from the error in the derivative $\frac{\partial^2 C_{inj}}{\partial \Phi^2}$ coming from FE analysis. This is still to investigate.

- *Electrostatic stiffness from Y DC bias* A similar calculation can be done for Y electrodes DC bias contribution, as already shown in equation 4.28. The torque along $\Phi$ direction is

$$N_{\Phi,Y} = \frac{1}{2}\sum_j \frac{\partial C_j}{\partial \Phi} V_{DC,Y}^2, \tag{6.18}$$

where the only contribution comes from the four Y electrodes surfaces facing the test mass and the electrode housing. This time, the capacitance derivative series expansion is performed considering also third order terms, in order to account for non linear terms in the final stiffness expression, that can explain the non linear dependence from $\Phi$ of the stiffness model 6.6. So

$$\frac{\partial C}{\partial \Phi} = \left.\frac{\partial C}{\partial \Phi}\right|_0 + \left.\frac{\partial^2 C}{\partial \Phi^2}\right|_0 (\Phi - \Phi_0) + \frac{1}{2}\left.\frac{\partial^3 C}{\partial \Phi^3}\right|_0 (\Phi - \Phi_0)^2 + \frac{1}{6}\left.\frac{\partial^4 C}{\partial \Phi^4}\right|_0 (\Phi - \Phi_0)^3. \quad (6.19)$$

The final expression for the stiffness due to Y DC bias will be

$$\Gamma_Y(\Phi) = -\frac{\partial N_{\Phi,Y}}{\partial \Phi} = -2V_{DC,Y}^2 \left[\frac{\partial^2 C_Y}{\partial \Phi^2} + \frac{1}{2}\left(\frac{\partial^4 C_Y}{\partial \Phi^4}\right)\Phi^2\right]. \quad (6.20)$$

This is comparable with parameters from fit of stiffness $B$ and $C$ so that

$$\Gamma_Y(\Phi) = BV_{DC,Y}^2 + CV_{DC,Y}^2 \Phi^2. \quad (6.21)$$

so that we can estimate B as $-2\frac{\partial^2 C_Y}{\partial \Phi^2}$ and C directly from $\partial^4 C_Y / \partial \Phi^4$. First, second and fourth order capacitance derivatives with respect to $\Phi$ of Y, X and injection electrodes, can be estimated from the electrostatic finite-element (FE) analysis, as explained in section 3.2.1. For the C parameter, we found a difference of a factor 2.7 with our model estimation. The factor B can instead be compared with the second order capacitance derivative of the Y electrodes, and a factor 1.3 of difference is found. Parameters estimated from fit are of the same order of magnitude of that expected from electrostatic. Our model is able to estimate the presence of a non-linearity at a level not comparable to the model.

It is thus possible to conclude that exist a non linearity of the stiffness respect to the angle displacement expected from the electrostatic model, as shown in equation 6.20, and this account for the discrepancy in the measured period, and consequently in stiffness, when measurements are performed out of center, as visible in table 6.4.

We can also compare these results with the estimation of non linear parameters, showed in table 6.2, obtained performing a fit of the acceleration during a free fall run at a non linear model of angle, as explained in the previous section. The estimated parameter $\beta$, that is related to the dependence on $\Phi^2$, can be compared to the estimated value of the $C$ parameter from fit to $\Gamma$, as well as the $\alpha$ parameter can be compared with $B$. There is a factor 3 of difference from $\beta$ recovered from fit to acceleration data and the value of $C$ from fit to measured period.

This can be also showed by comparing the non linear stiffness model of equation 6.6 obtained from different free fall measurements, performed in several weekend from January until March 2015, as in figure 6.13. Non linear parameters $\alpha$ and $\beta$ are well resolved inside each run (see table 6.2), but they do not agree with each other and they do not help the residuals shown in figure 6.14, as we will explain shortly. Additionally, their values differ, from run to run, from what we expect theoretically in sign, while the theoretical model for non-linear stiffness is roughly correct over larger angular scales (0 to $-5\,mrad$).

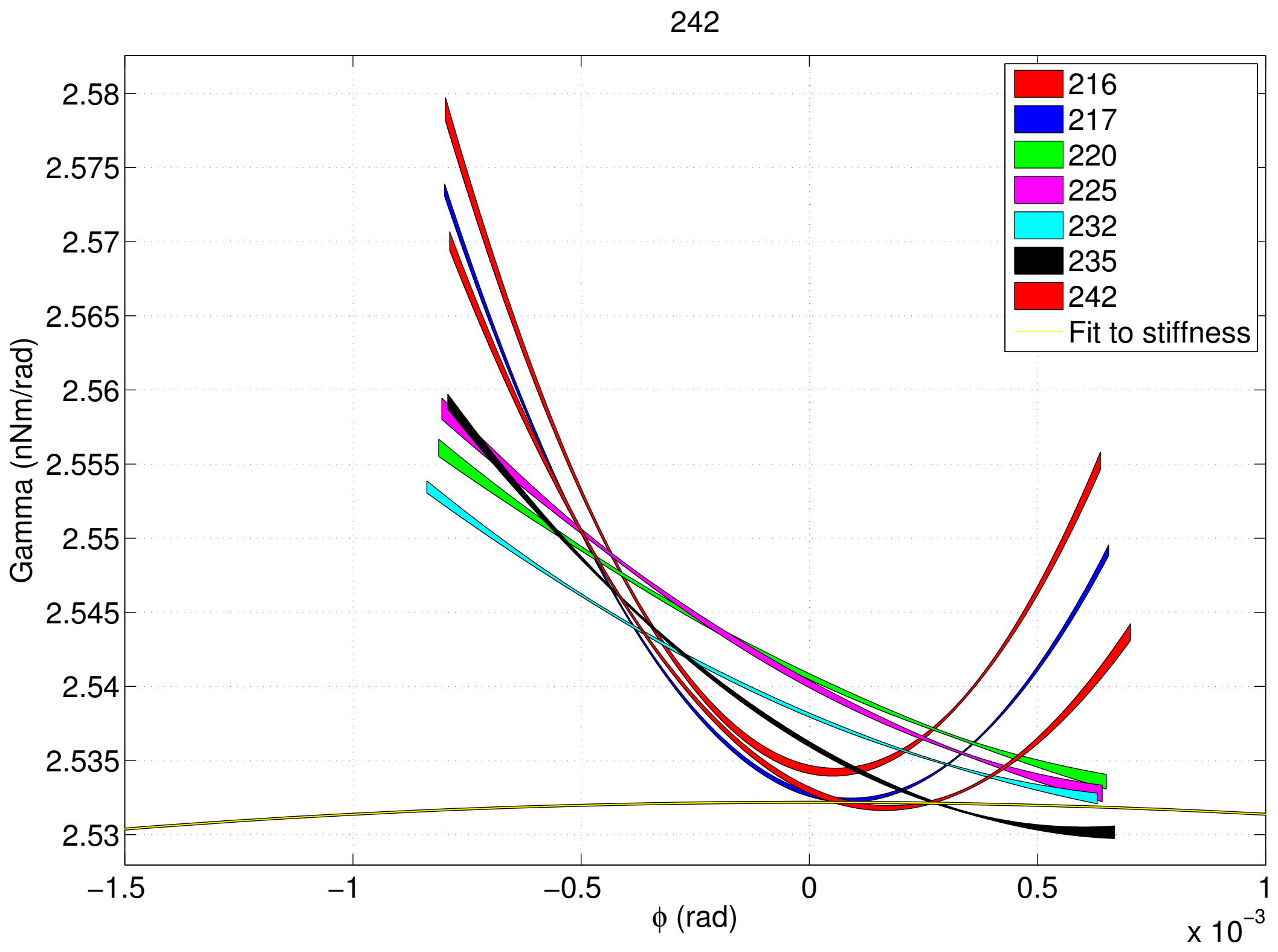

FIGURE 6.13: Plot of non linear stiffness models from equation 6.6, for different free fall measurements (numbered 216, 217, 220, ...). Non linear parameters are estimated with the calibration algorithm explained in section 6.3. We also show the model of the stiffness, for $V_{DC,Y} = 9.5V$ and $V_{inj} = 3.6V_{RMS}$, based on period measurements as in Table 6.4.

The behavior of the non linear stiffness model 6.6 is not that expected from the electrostatic (equation 6.20), because of the wrong estimated curvature. As seen, $\alpha$ and $\beta$ terms, related to the second and third power of angle, are estimated with a wrong sign from fit. The expected stiffness trend is shown as the yellow line in figure 6.13. This is estimated considering the non linear model of the measured $\Gamma$ of equation 6.12. It seems there is also a variability in time of the resulting stiffness from fit, that appears like the presence of two family of curves in figure 6.13, that can not be related to an electrostatic or a geometrical effect.

In order to show if $\alpha$ and $\beta$ parameters account for a real non linearity behavior of the stiffness measured during the free fall measurement, it is possible to show a plot of the residual non linear torque calculated for each flight as

$$N_{res,non\,lin} = I\left(\ddot{\Phi} - (a + b\dot{\Phi} + c\Phi + d\Phi^2 + e\Phi^3)\right) \tag{6.22}$$

where parameters are estimated from fit of equation 6.7. In case of linear model 6.8, this residuals reduce to $N_{res,lin} = I(\ddot{\Phi} - (a + c\Phi))$. Residuals from non linear fit are plotted with respect to $\Phi$ displacement in figure 6.14 on bottom panel. On top panel are shown the same residuals obtained by calculating parameters $a$ and $c$ as in $N_{res,lin}$. The angle time series $\Phi$ used to calculate residuals, is obtained after the Blackman-Harris filtering as in the algorithm explained in section 6.3 (shown schematically in figure 6.11). All displacement data are corrected for the autocollimator non linearity.

From these figures it is clear that the non linear model 6.7 doesn't help to account for all the non linearity effect at angle near $600\mu rad$, where there is a systematic variation in the torque residual in a small range of angle.

The same plot has been done without considering the autocollimator correction, in order to see what is the impact of this correction on the final data, as it is shown in figure 6.15. The use of the autocollimator correction gives an improvement in the residuals especially in the range of $\Phi$ on the top part of the flight, near $600\,\mu rad$, where the pendulum is slower and remains there for more time, scanning slowly over the periodic errors of the autocollimator.

As the residual torque in figure 6.14 varies systematically in angle on small scales, as does that measured without the AC correction (6.15), it is likely that the systematic short scale non-linearity is due to some imperfection in our autocollimator correction. Residuals indicate an error that varies on much shorter angular scales than could be explained by an electrostatic non-linearity.

In conclusion, an application of the non linear torque model 6.5, with non linear parameter $\alpha$ and $\beta$ in the final estimation of torque in free fall measurements, has been found not a real advantage in the torque spectral estimation. For all our torque estimation, we use the linear model 6.4.

## 6.4 Actuator calibration

The calibration of the electrostatic actuator, with a scale factor $\alpha$ defined in equation 4.7, is important for the accuracy of the torque estimation with DC actuation (as already described in section 4.1 in equations 4.7-4.9). We note that the actuator in the DC torque experiment is used with $\Phi = 0$, so that is the angle at which we must calibrate the actuator. Though not relevant for the free-fall mode analysis, the calibration does also affect the accuracy of the free-fall mode torque impulse controller.

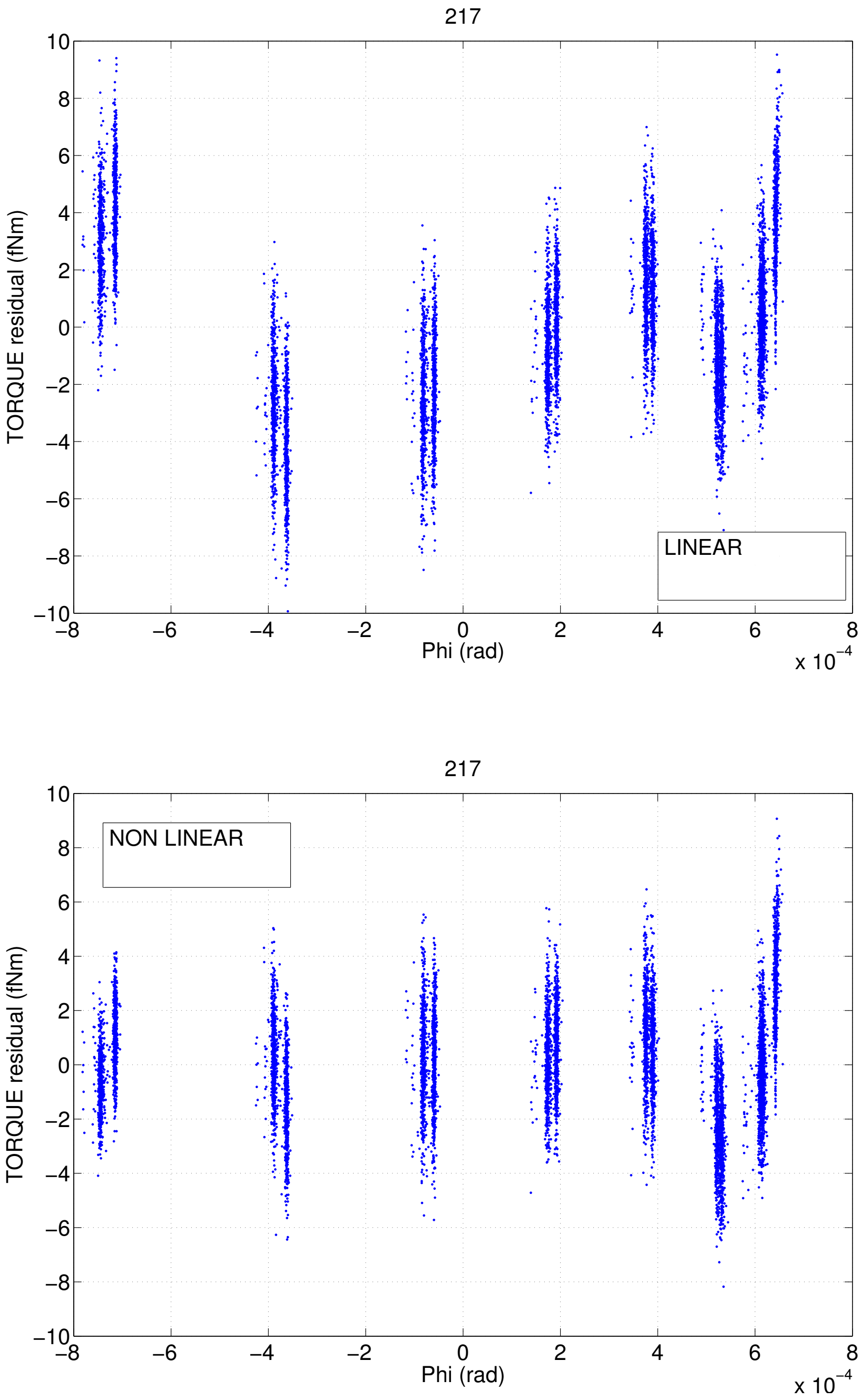

FIGURE 6.14: Residual torque from equation 6.22. Fit parameters are calculated with the calibration method explained in section 6.3 in the linear and not linear case. Data are corrected for autocollimator non linearity.

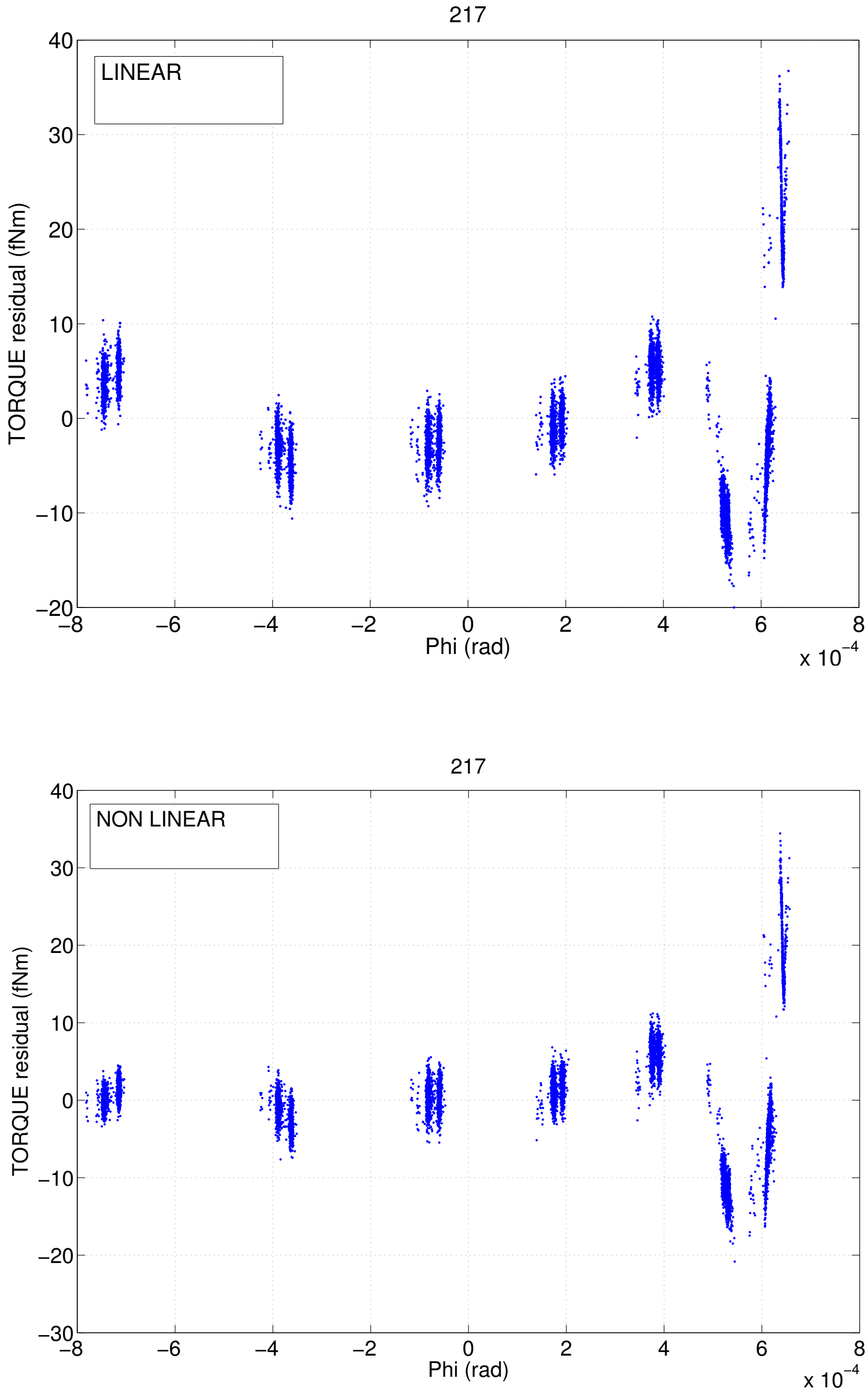

FIGURE 6.15: Residual torque from equation 6.22 for data not corrected for the autocollimator non linearity. Fit parameters are calculated with the calibration method explained in section 6.3 in the linear and not linear case.

As already seen in section 4.2.3, pendulum actuation scheme allows a range of torque $N_{ACT}$ while holding the stiffness $\Gamma_{ACT}$ constant, applying voltages $\pm V_{i,\Phi}^2$ on diagonal couples of $x$ electrodes so that:

$$N_{ACT} = \left| \frac{\partial C}{\partial \Phi} \right| \left( < V_{1\Phi}^2 > - < V_{2\Phi}^2 > \right). \tag{6.23}$$

We remember here that the square wave voltages applied are always attenuated of a low pass filter factor measured to be $f_{att} = 0.85$, so that the real applied voltage is :

$$N_{ACT} = \left| \frac{\partial C}{\partial \Phi} \right| f_{att}^2 \left( V_{com,1\Phi}^2 - V_{com,2\Phi}^2 \right) \equiv \left( V_{com,1\Phi}^2 - V_{com,2\Phi}^2 \right) \frac{\partial N}{\partial (V_{com}^2)}. \tag{6.24}$$

The actuator calibration consists to measure partial derivatives with respect to $\Phi$ of the capacitance of $x$ electrode in case of AC actuation, in order to obtain the final factor $\frac{\partial N}{\partial (V_{com}^2)}$. This was done before, during and after pendulum rotation with respect to the electrode housing.

The technique usually employed consist to produce a square wave modulation on the diagonal couple of electrodes $EL1$ and $EL3$, as in figure 4.5, polarizing them by switching on and off an AC signal of commanded amplitude $V_{MOD}$ chosen to be an exact integer multiple of the DAC LSB ($\delta V_{bit} = 312.5\,\mu V$). The chosen frequency of modulation is $f_{MOD} = 3\,mHz$.

In case of measurement with pendulum rotated with respect to the electrode housing, there is also an offset $V_{OFF}$ that has to be considered, applied on the used electrodes, in order to keep the test mass centered while the modulation is on. This offset corresponds to $V_{OFF} = \pm 2.65857\,V$ in case of a DC torque applied during a noise measurements with actuation, that is $N_{DC} = \left| \frac{\partial C}{\partial \Phi} \right| f_{att}^2 V_{OFF}^2 = 14.52\,pNm$.

This operation induces a torque signal on test mass with squarewave amplitude

$$N_{MOD} = 2 \frac{\partial C}{\partial \Phi} f_{att}^2 V_{OFF} V_{MOD}. \tag{6.25}$$

The first harmonic is $4/\pi$ times this value and we applied a $V_{MOD}$ corresponding to roughly one bit of the DAC, $\delta V_{bit} = 3.125\mu V$, for a modulated torque of order of $4\,fNm$, as also shown in figure 6.16. This allows the extraction of the calibration factor between the DAC commanded squarewave actuation amplitude and torque as $\partial C / \partial \Phi = 2.04 \pm 0.02\,pNm/V^2$.

At the end of the experiment, with the pendulum rotated near to $\Phi = 0$ (and thus with no torque needed to hold the pendulum centered), we perform the same calibration by applying $V_{MOD} = 200 \delta V_{bit} = 0.0625\,V$ (again modulating at frequency $f_{MOD} = 3\,mHz$) and an offset of $V_{OFF} = \pm 0.0625\,V$, obtaining a modulated torque of order of $15\,fNm$. The calibration factor measured with pendulum centered is $\frac{\partial N}{\partial (V_{com}^2)} = 2.021 \pm$

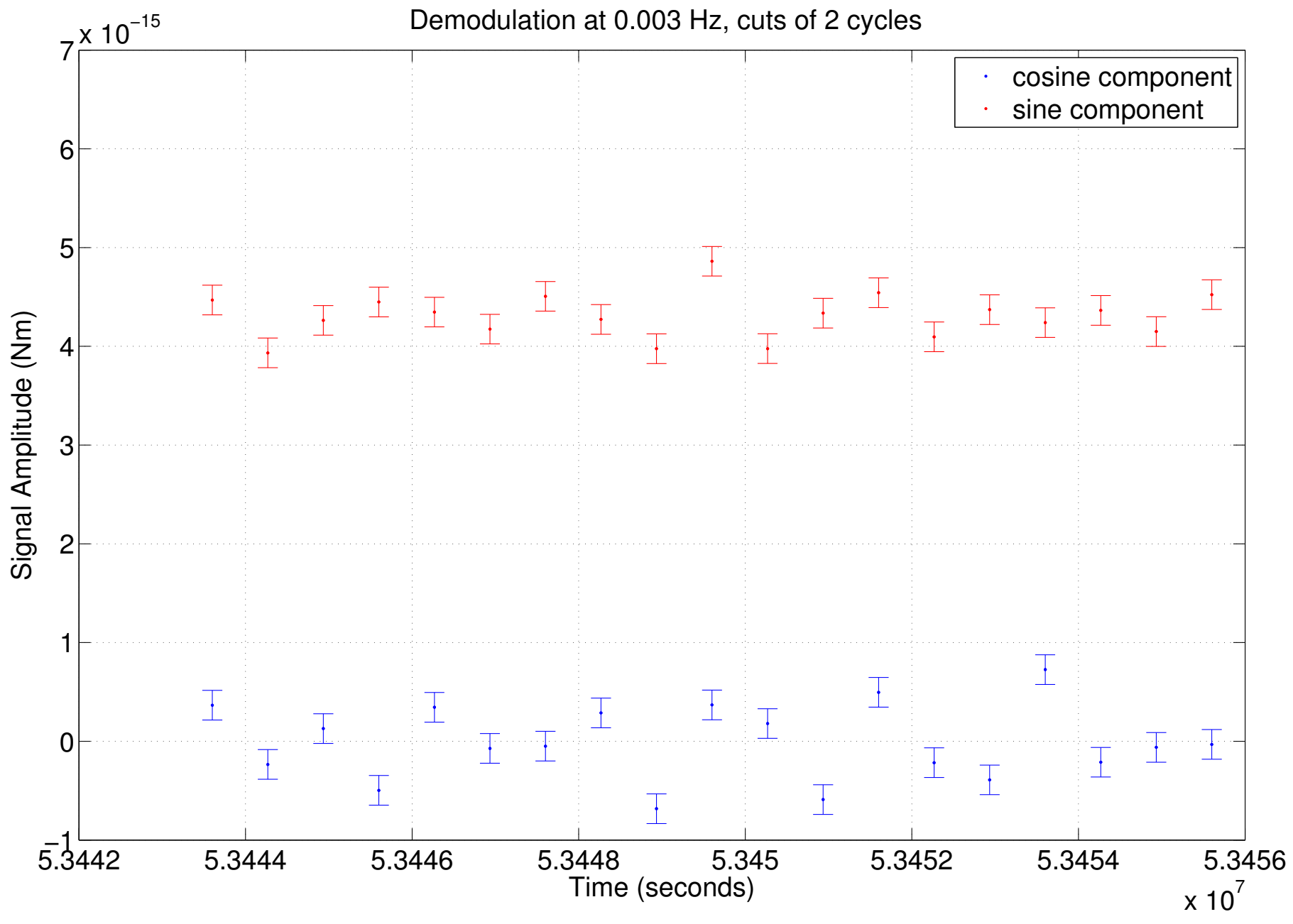

FIGURE 6.16: Torque amplitude signal of order $4\, fNm$, obtained after demodulation of a measurement of the actuator calibration factor with pendulum out of center of $-5\, mrad$.

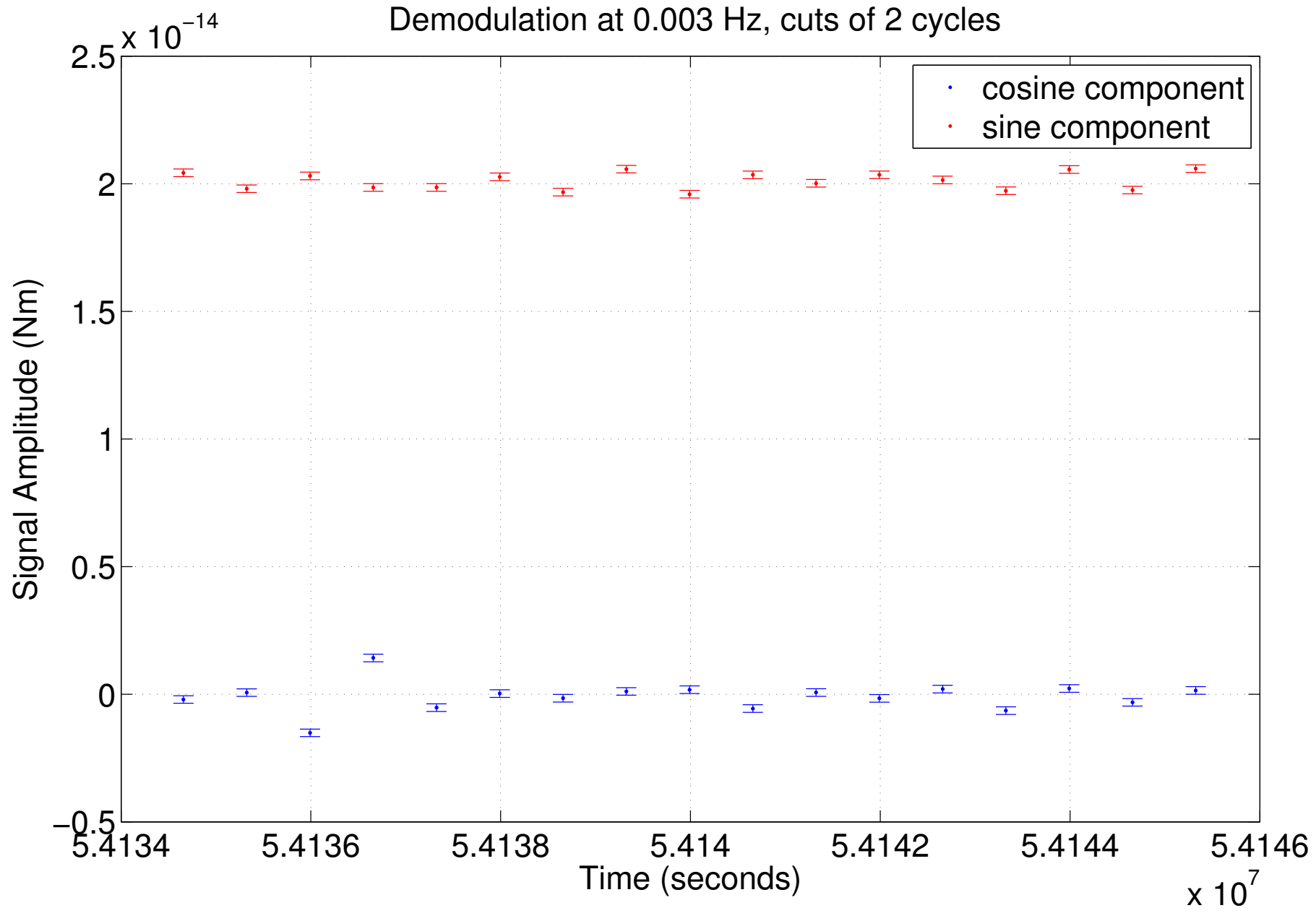

FIGURE 6.17: Torque amplitude signal of order $20\, fNm$, obtained after demodulation of a measurement of the actuator calibration factor, with pendulum centered around $0\, rad$.

$0.008\,pNm/V^2$.

Results and measured conditions are summarized in table 6.5

Both measurements are sensitive to possible electronics non-linearities in a different way with respect to the noise measurements with DC torque applied. The centered pendulum measurements use very small offset voltage, while in the DC actuation noise measurements there is a voltage of 2.6 V.

| | Centered pendulum $\Phi_{EQ} \approx 0\,rad$ | Rotated pendulum $\Phi_{EQ} \approx -5\,mrad$ | |
|---|---|---|---|
| $V_{MOD}$ | $200\delta V_{bit} = 0.0625$ | $\delta V_{bit} = 0.0003125$ | V |
| $V_{OFF}$ | $\pm 0.0625$ | $\pm 2.65857$ | V |
| $f_{MOD}$ | 3 | 3 | mHz |
| $N_{MOD}$ | 20 | 4 | fNm |
| $\frac{\partial C}{\partial \Phi} f_{att}^2$ | $2.021 \pm 0.008$ | $2.04 \pm 0.02$ | $pNm/V^2$ |

TABLE 6.5: Actuator calibration measurement parameters and results.

Finally, the used actuator calibration factor is obtained with a weighted average from the two values measured:

$$\left|\frac{\partial C}{\partial \Phi}\right| f_{att}^2 = 2.024 \pm 0.007\,\frac{pNm}{V^2}. \tag{6.26}$$

## 6.5 Measured time series

Finally, in figures 6.18, we show angular time series of different free fall measurements performed in sequence, at different combinations of $T_{fly}$ and $T_{imp}$. In order to show the accuracy and the repeatability of the free fall measurement on time, we show also the corresponding torque time series as coming from the two analysis techniques, as in figure 6.20 and 6.21.

As discussed above, we are using a harmonic model with linear stiffness term. This means that the torque has a linear model like in equation 6.4, and the parameter $\Gamma$ is obtained from the linear fit in equation 6.8. The parameter is calculated separately for each run.

In order from left to right, showed measurements have been performed with $T_{fly} =$ 100, 140, 100, 250, $s$ and $T_{imp} = 10,\ 15,\ 10,\ 25\,s$.

The last time series, in red on each plot, is a noise measurement performed with constant DC actuation applied to hold the test mass centered around zero (as zoomed in the bottom panel of figure 6.19 that shown also the difference in the dynamic range inspected by pendulum during the two type of measurements). As already said, the torque measured with DC actuation depends on the subtracted value of the commanded torque as in equation 4.6, which we calculate to be $N_{DC} = 14.28\, pNm$, considering that the applied AC voltage on each x electrodes is $2.659\, V$ and the calibration factor used above $2.024 \pm 0.007\, \frac{pNm}{V^2}$. To compare the torque measured with the results coming from free fall analysis, we subtract this DC value from the estimated torque time series. There is a discrepancy of 0.5% in the torque between free-fall and the continuous actuation case, which could be compatible with the uncertainties in our actuator calibrator mentioned above.

The torque estimated from free fall measurements operated in different conditions, is in line and the visible drift in time is due to an angular drift related to laboratory temperature variation.

Data in figure 6.21, analyzed with BH low pass filter techniques for $T_{fly} = 100s$ (magenta curve), appears divided into two bands. The BH windows length in this case is chosen to be $T_{win} = 50.1667s$ (with a $T_{samp} = 9.1667s$ and an oversampling factor of 5.4727). There is a still a residual systematic position dependence of the measured torque still to correct. It is consistent with an effective miscalibration of $\Gamma$ by of order $0.1\, nN/rad$, that could be related to the local errors in the autocollimator residual, locally, at the angles relevant to the 100 s free-fall dynamics or related to the different window sizes used on the shorter data. This needs to be still investigated.

Another test of the free-fall mode accuracy, particularly regarding the ability to accurately estimate the external torque at $\Phi = 0$, independently of the pendulum angle during the measurement, is shown in figure 6.22 where the torque time series are shown for different set points angles. The corresponding torque recovered with the BH low pass filter technique is shown in figure 6.23. The used set points are 0, 500 and -500 $\mu rad$.

Points of the time series with noisier periods around $50720000s$ for cyan curve or $50900000s$ for red curve are due to the moments of clock slippage mentioned in section 6.2.2 and that can't be reduced. This is the reason why we discard this data stretches from the grand averages performed in the final analysis.

## 6.6 Torque noise spectra

In this section we present all torque noise power spectral densities obtained for our free fall, DC actuation and background noise measurements data set. All the spectra showed in this section, are obtained considering a linear model for the stiffness with all data corrected for the autocollimator non-linearity. The use of 50 Hz sampling, with the correction scheme discussed in section 6.2.2, has only been implemented with the sine-fit analysis, and is still being investigated for the BH analysis.

Moreover, all torque time series have been divided in stretches of $27500\, s$ with 66% overlapping each multiplied by a Blackman Harris window, as explained in section 5.1.1,

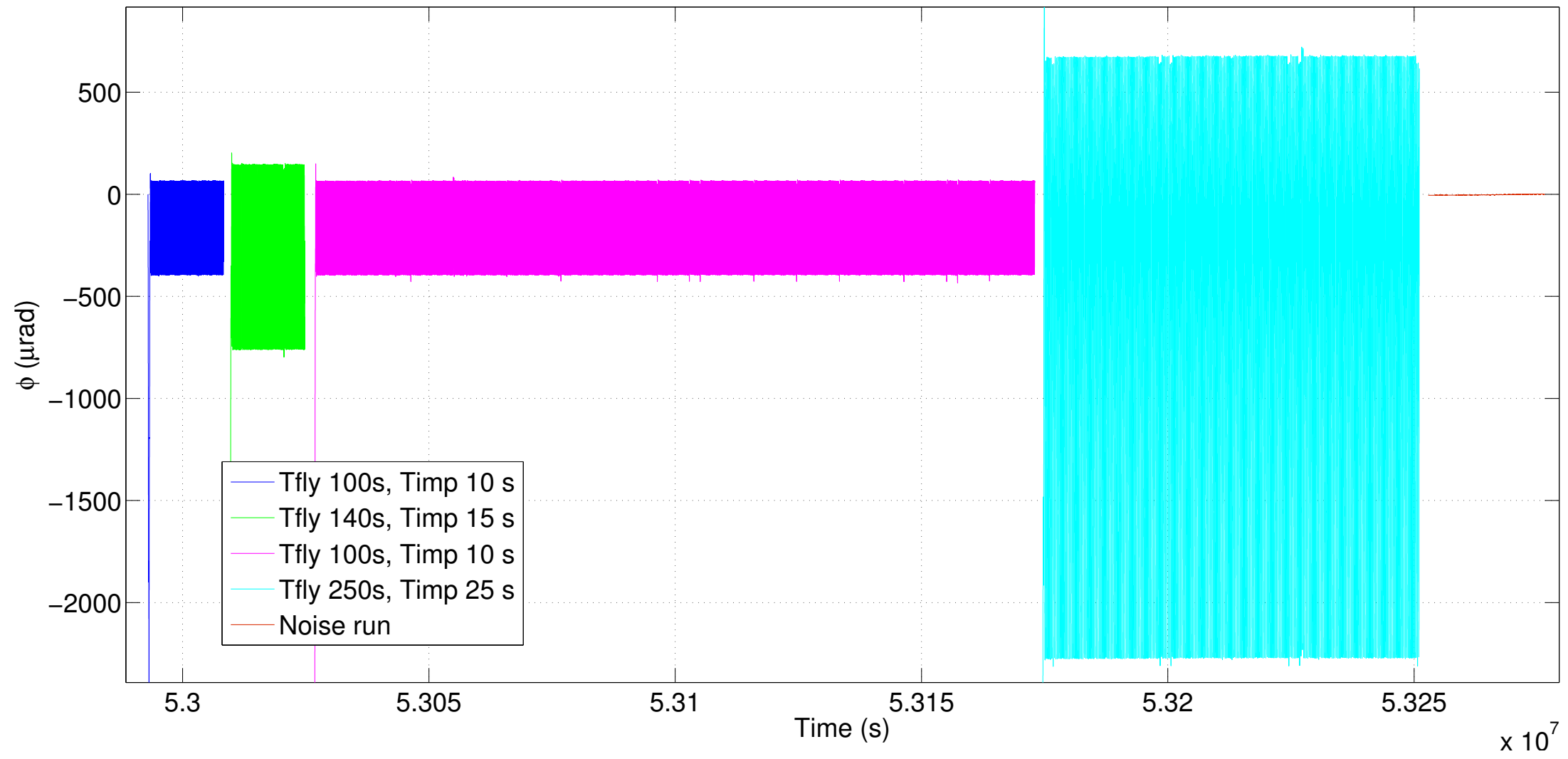

FIGURE 6.18: Angular time series of different free fall measurements performed with, in order from left to right, $T_{fly} = 100,\ 140,\ 100,\ 250,\ s$ and $T_{imp} = 10,\ 15,\ 10,\ 25\,s$. The last red time series on the right is a noise measurement with constant DC actuation with $N_{DC} = 14.56\,pNm$.

when we performed spectra on simulated data.

The cross correlation technique between the sensor and autocollimator data was not used to analyze free fall data because of the high noise level in sensor data. As we said, we didn't analyze in detail non linearity and sensor issues. We still use this noise estimator for noise run with constant DC actuation and without actuation.

Moreover, we analyze the DC actuation and no actuation data, which are intrinsically continuous and without gaps, both with a "standard" analysis (following equation 4.6) and with the sine and BH analysis techniques used for the free-fall data.

In figure 6.24 we report torque noise power spectral densities of a background noise run, with electrostatic constant DC actuation applied to hold the test mass centered, compared with a free fall measurement analyzed with both the techniques developed. As we said in section 4.1, we want to verify that the free fall can allow a torque noise measurement at the background levels measured in the absence of actuation, lower then that possible with the commanded DC actuation force. At the current state, this is not totally true at frequency of interest of the experiment. This appears to be due to the effect of the aliasing of high frequency noise components, and to the down conversion at low frequency, as this was already observed by analyzing simulated torsion pendulum data in section 5.1.1.

It can be useful to compare the same free fall measurement of figure 6.24 with the showed noise run with DC actuation, analyzed with the sine-fit and BH free-fall analysis techniques, essentially analyzing the continuous data as if it had gaps for the impulses.

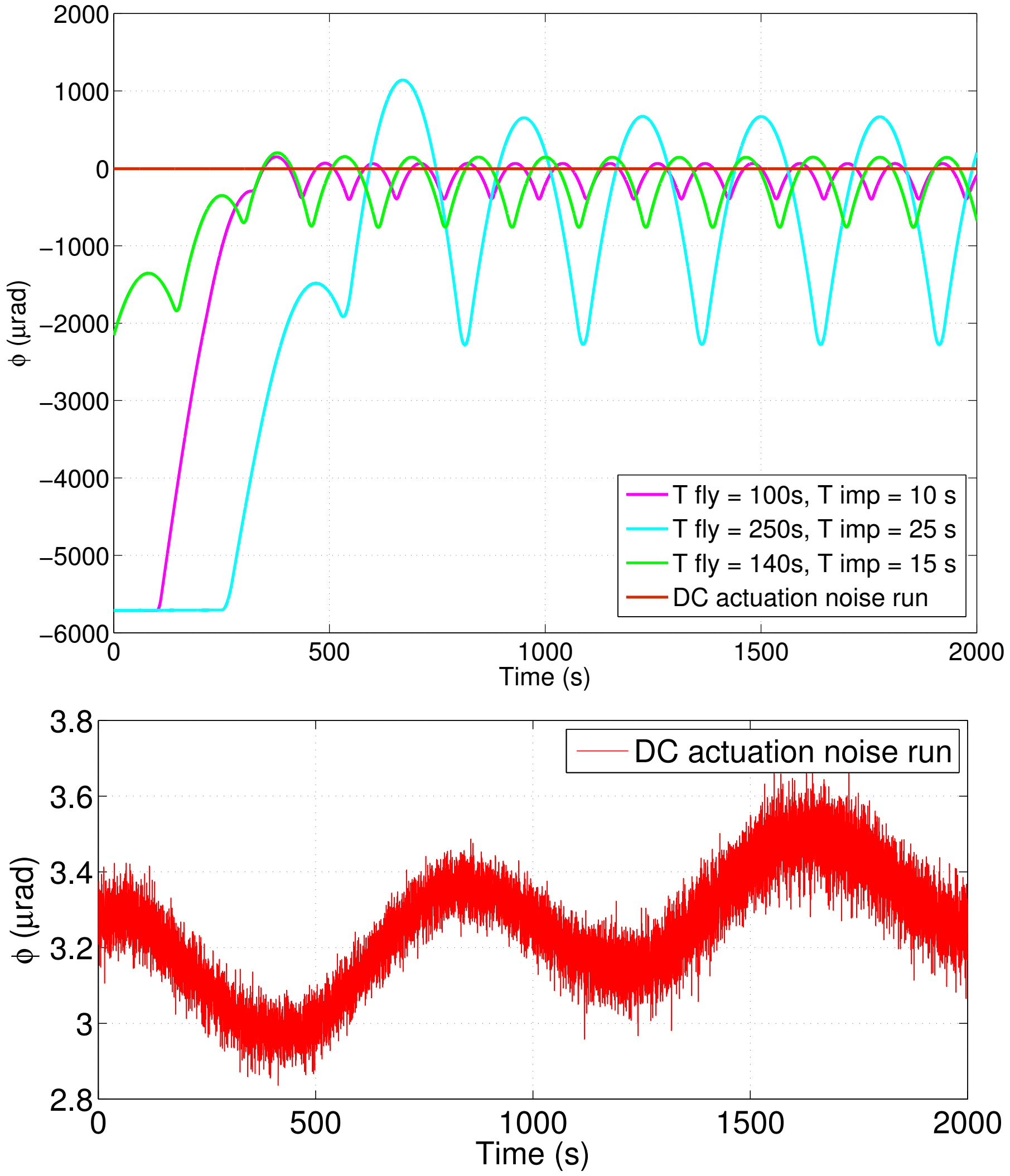

FIGURE 6.19: Top panel: Angular time series of different free fall measurements performed with different $T_{fly} = 100, 140, 100, 250, s$ and $T_{imp} = 10, 15, 10, 25\,s$. They are overlying respect to the plot in figure 6.18. The red time series is a noise measurement with constant DC actuation with $N_{DC} = 14.28\,pNm$. Bottom panel: zoom on DC actuation noise run time series.

In figure 6.25 we analyzed the free fall run and the noise run with the BH low pass filtering technique. This technique works better with continuous data, but there is still an effect of residual aliasing. Even though the detrending operated on filtered data, peaks at frequency of the experiment $T_{exp} = 275\,s$ are still present.

A similar comparison is done with data analyzed with the sinusoidal fit technique in figure 6.26. At frequency below $1\,mHz$ this technique works well with noise data with DC actuation, recovering the same torque noise level obtained with the standard analysis. Moreover, gaps in analyzed data produce excess noise, but not enough to explain the noise floor in our free-fall experiment.

To illustrate the variation in time of the noise level achieved, at least in one relevant frequency window, we can integrate the spectrum in the range from 1 to 3 mHz to obtain a mean square torque. Uncertainty has been estimated from the standard deviation of the samples in the frequency bin. We associate a single data point with each 27500 s cut and plot this as a time series, depending on the day of the week in which it was taken, as in figure 6.27.

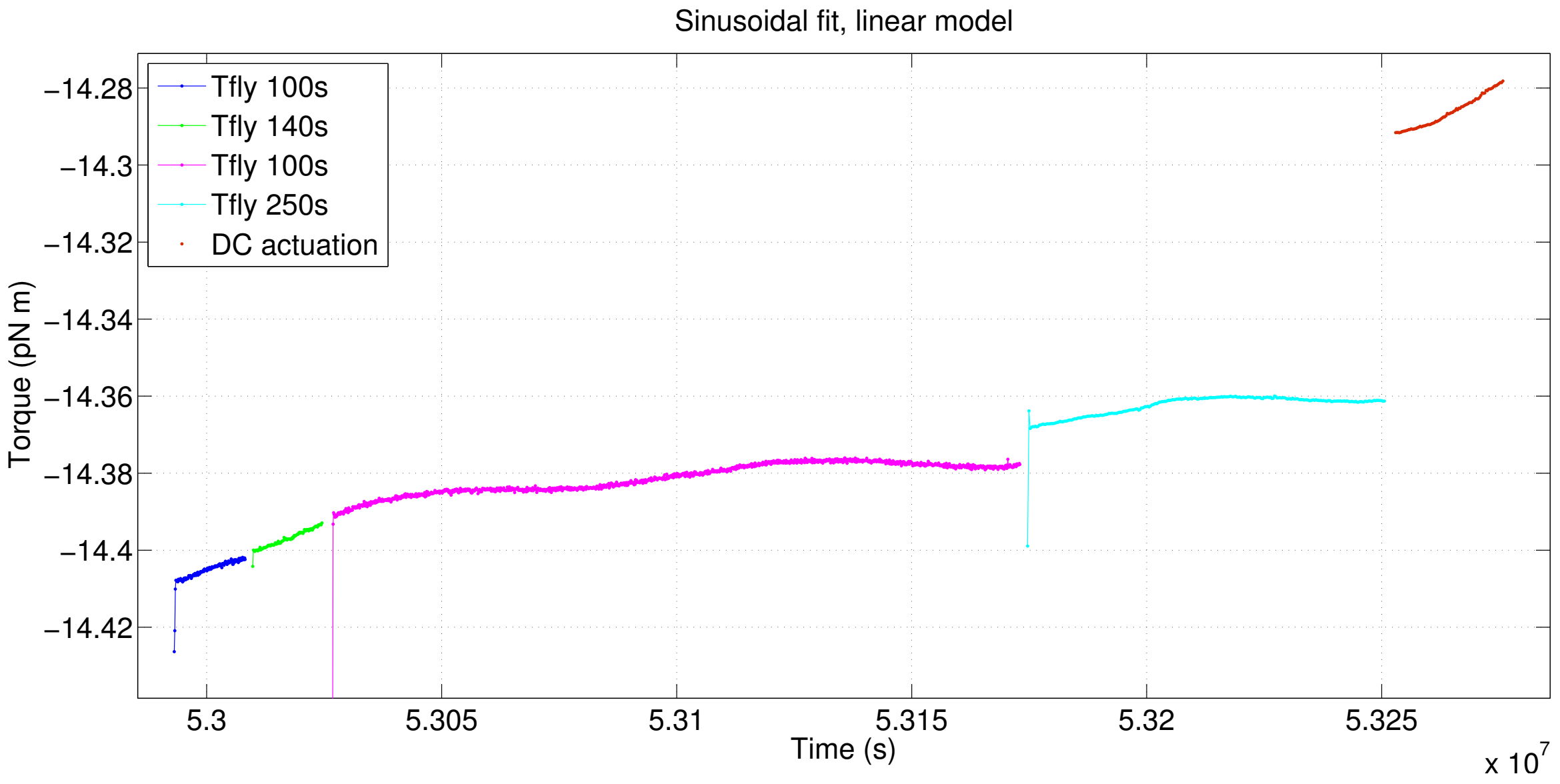

FIGURE 6.20: Torque time series as results of sinusoidal fit techniques, for different free fall measurements performed with, in order from left to right, $T_{fly} = 100, 140, 100, 250, s$ and $T_{imp} = 10, 15, 10, 25\,s$. The last red series on the right is torque difference between a noise measurement with constant actuation and the calibrated DC offset in torque $N_{DC} = 14.28\,pNm$.

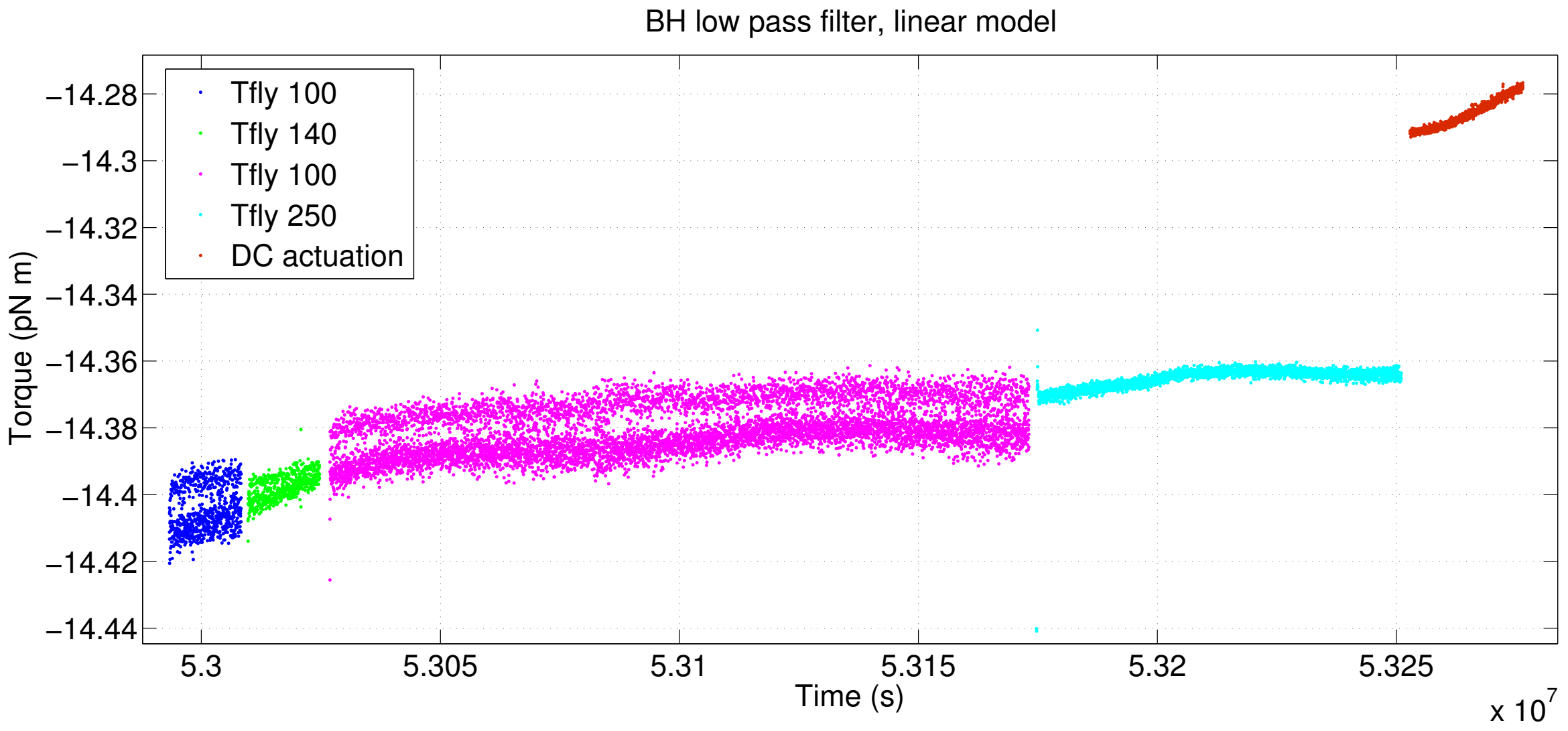

FIGURE 6.21: Torque time series as results of Blackman-Harris low pass filtering techniques, for different free fall measurements performed with, in order from left to right, $T_{fly} = 100, 140, 100, 250, s$ and $T_{imp} = 10, 15, 10, 25\,s$. The last red series on the right is torque difference calculated using Eqn 4.6 with applied DC torque $N_{DC} = 14.28\,pNm$.

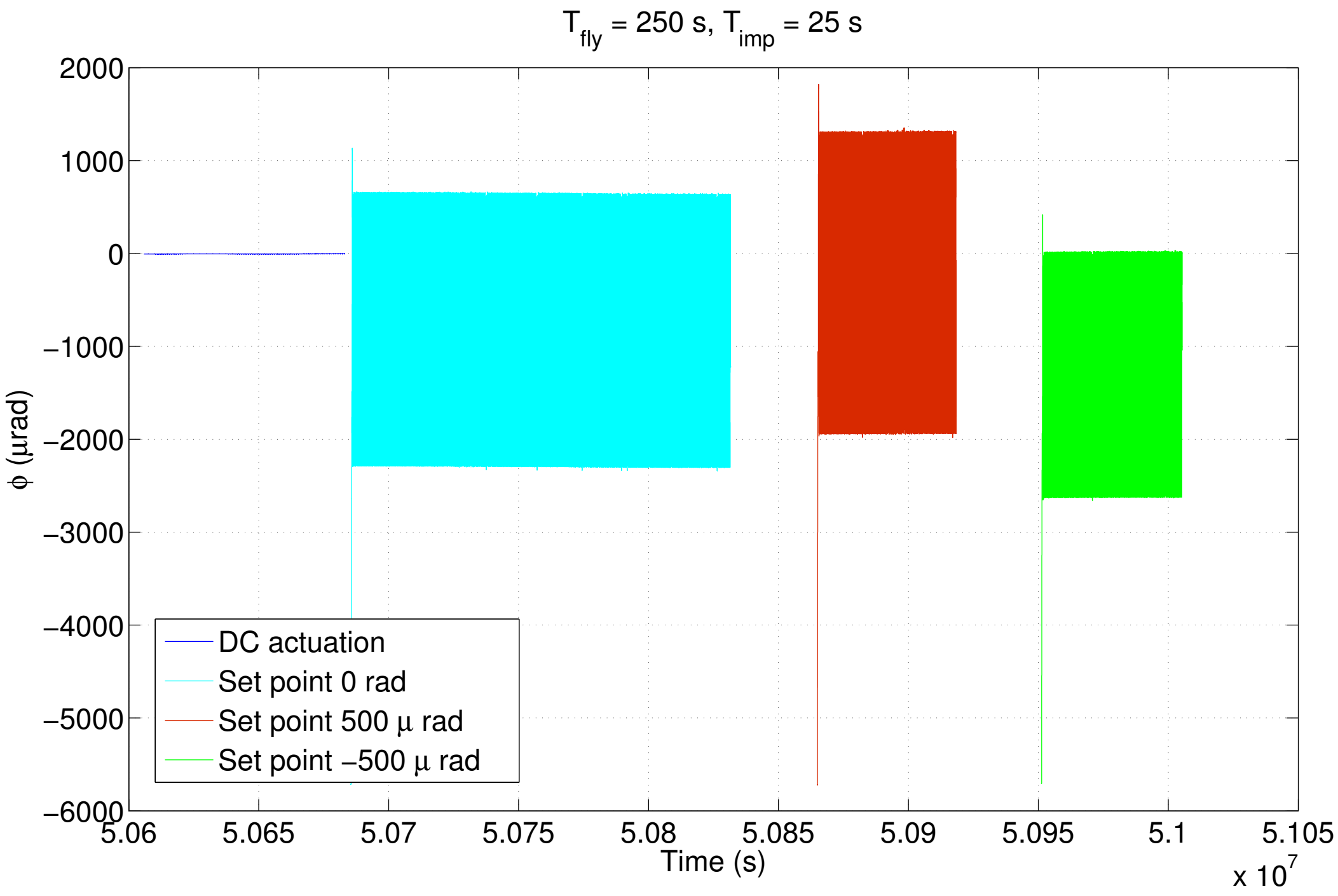

FIGURE 6.22: Angular time series of different free fall measurements performed with different set points, in order from left to right, $0\,rad$, $-500\,\mu rad$ and $500\,\mu rad$. Experimental parameters are $T_{fly} = 250$, $s$ and $T_{imp} = 25\,s$. The first blue series on the left is from a noise measurement with constant actuation on with $N_{DC} = 14.28\,pNm$.

In this figure, all free fall run stretches are shown, for measurements performed between October 2014 and March 2015, mainly during weekends, as function of experimental time scale. The spectra can be averaged to obtain a final spectrum. Before average, some cuts is rejected from final average according to a certain criterion (gray points in figure 6.27). Criteria developed to discriminate data are based on the implementation problem of data acquisition explained in section 6.2. Cuts in which data presents a time interval between consecutive points longer then $dt > 1\,s$ are rejected. The same is done for data that have the problem of clock slippage explained in section 6.2. This is done by considering the number of 4/6 points that are collected for each stretch. If this number exceeds 100 the cut is rejected.

All rejected stretches based on the explained criteria are shown as faded points in figure 6.28. A zoom on the same data is shown in figure 6.29.

Finally, in figure 6.30, free fall integrated torque spectra are compared to those of noise measurements performed with constant DC actuation. Time variations in the torque

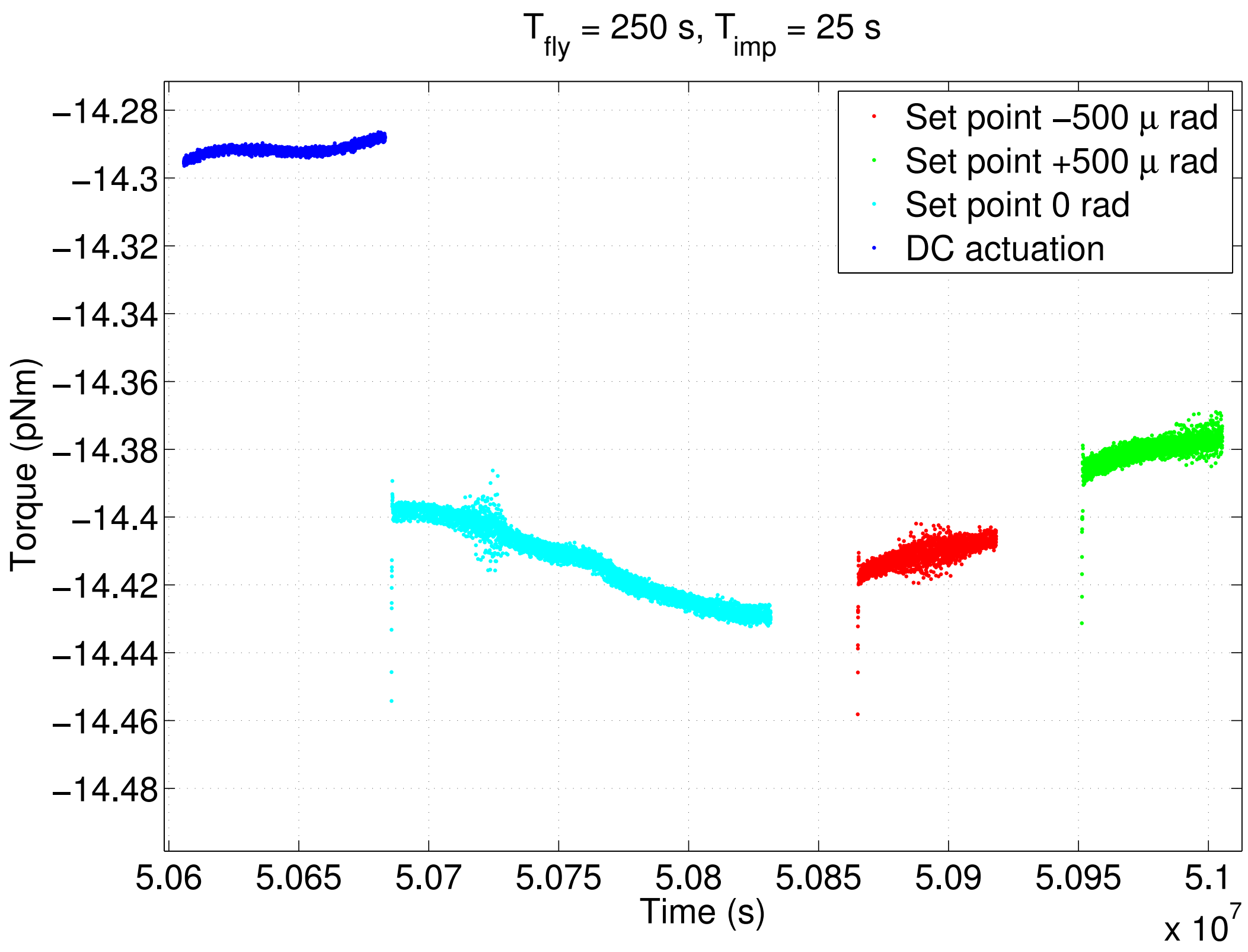

FIGURE 6.23: Torque time series as results of Blackman-Harris low pass filtering techniques, for different free fall measurements performed with different set points, in order from left to right,$0\,rad$, $-500\,\mu rad$ and $500\,\mu rad$. Experimental parameters are $T_{fly} = 250$, $s$ and $T_{imp} = 25\,s$. The first blue series on the left, is torque difference between a noise measurement with constant actuation and the applied DC offset in torque $N_{DC} = 14.28\,pNm$. Points of the time series with noisier periods around $50720000s$ for cyan curve or $50900000s$ for red curve are due to the moments of clock slippage mentioned in section 6.2.2

noise power integrated in the frequency band from 0.45 to 1 $mHz$, for both cases, are shown.

The full data set of free fall runs have been then averaged and is showed in figure 6.31, analyzed with both analysis techniques and compared with the noise with DC actuation measurements. The average is performed by pre-averaging spectra from groups of 10 cuts of each run of measurements and then taking the final average. The error is the standard deviation of the real part of spectra of all stretches, divided by the number of spectra. Noise run with DC actuation are averaged on BH windows length of $25000s$ and overlapped of 66%.
We are able, at present time, to recover an excess torque noise of $2\,fNm/\sqrt{Hz}$ at $1\,mHz$ for data analyzed with both the free fall analysis techniques, a factor two larger then that achieved with constant DC actuation force. This represents a level of acceleration noise of $100\, fm/s^2\sqrt{Hz}$ that we can compared to LPF differential acceleration requirement. This is factor 10 above the level of actuation noise expected for LISA Pathfinder mission showed in figure 2.3 in section 2.2, that is $7.5\, fm/s^2\sqrt{Hz}$ at $1\, mHz$.
We note here the presence of a peak near the pendulum resonance frequency for both free fall data, that is not presents in DC actuation data. This is not only due to aliasing effect, but it is also related to the dynamics during the free fall not completely accounted for. The BH low pass filter data shown also a peak around $1.5\, mHz$, related again to aliasing problem.

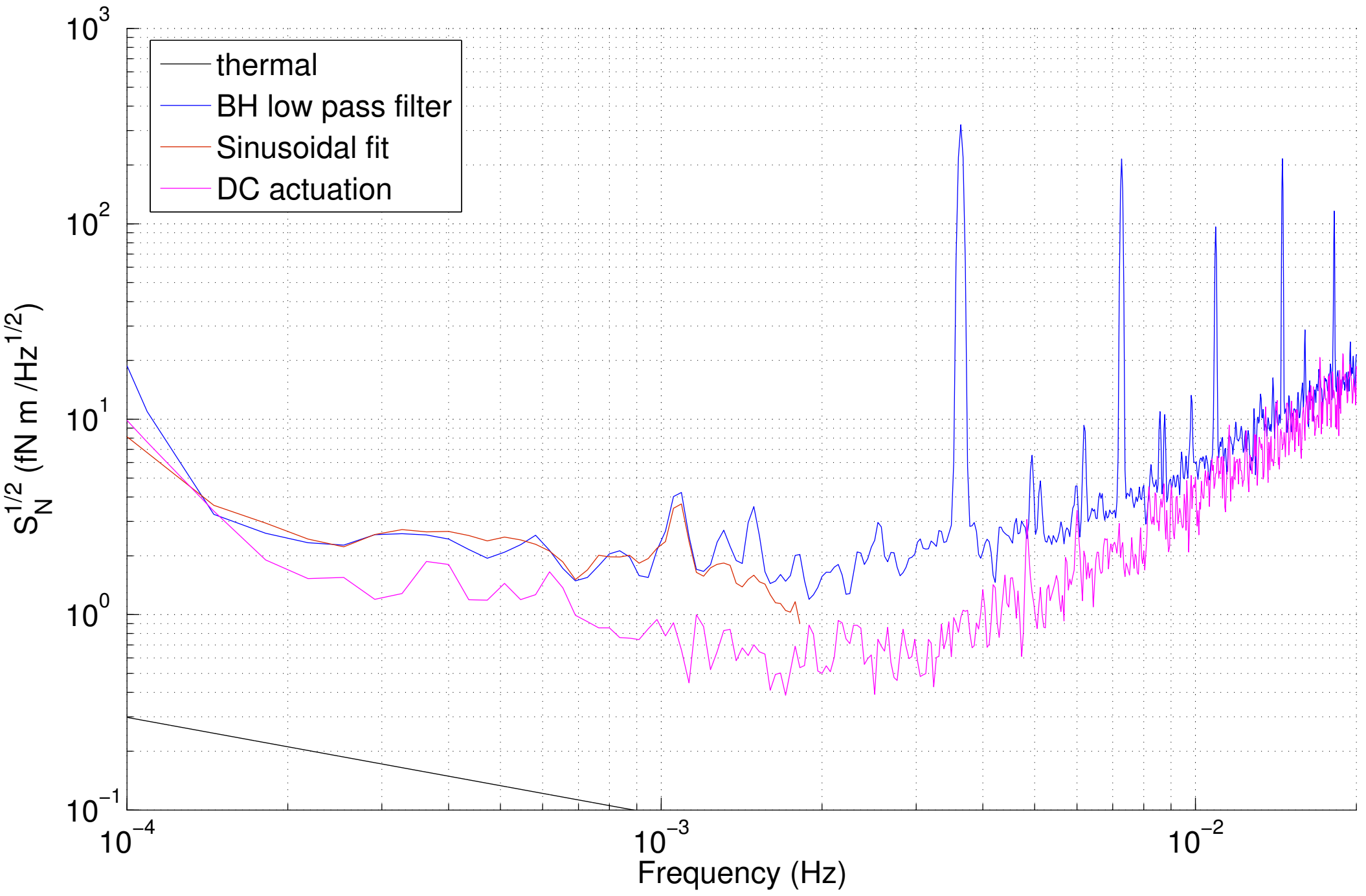

FIGURE 6.24: Torque power spectral densities of a free fall measurements analyzed with sinusoidal fit techniques (red) and BH low pass filter technique (blue). The magenta spectrum is a noise measurement performed applying a constant DC electrostatic actuation. BH low pass filter windows length is $T_{win} = 91.3125s$.

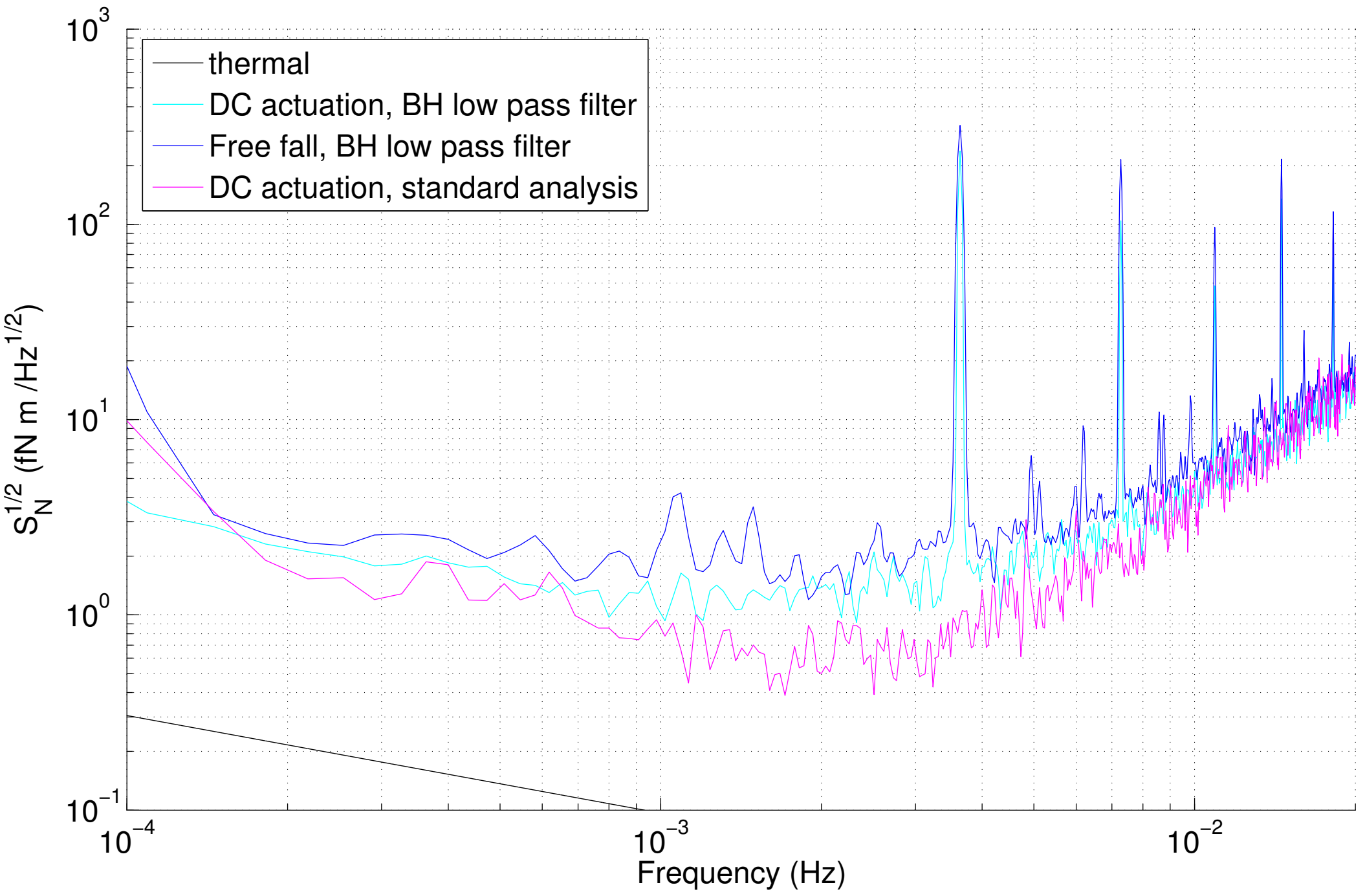

FIGURE 6.25: Torque noise for DC actuation run measured with standard analysis technique (magenta) and with the free-fall BH analysis technique, with artificial gaps (cyan). Free-fall run from fig 6.24 with BH analysis shown for comparison (blue).

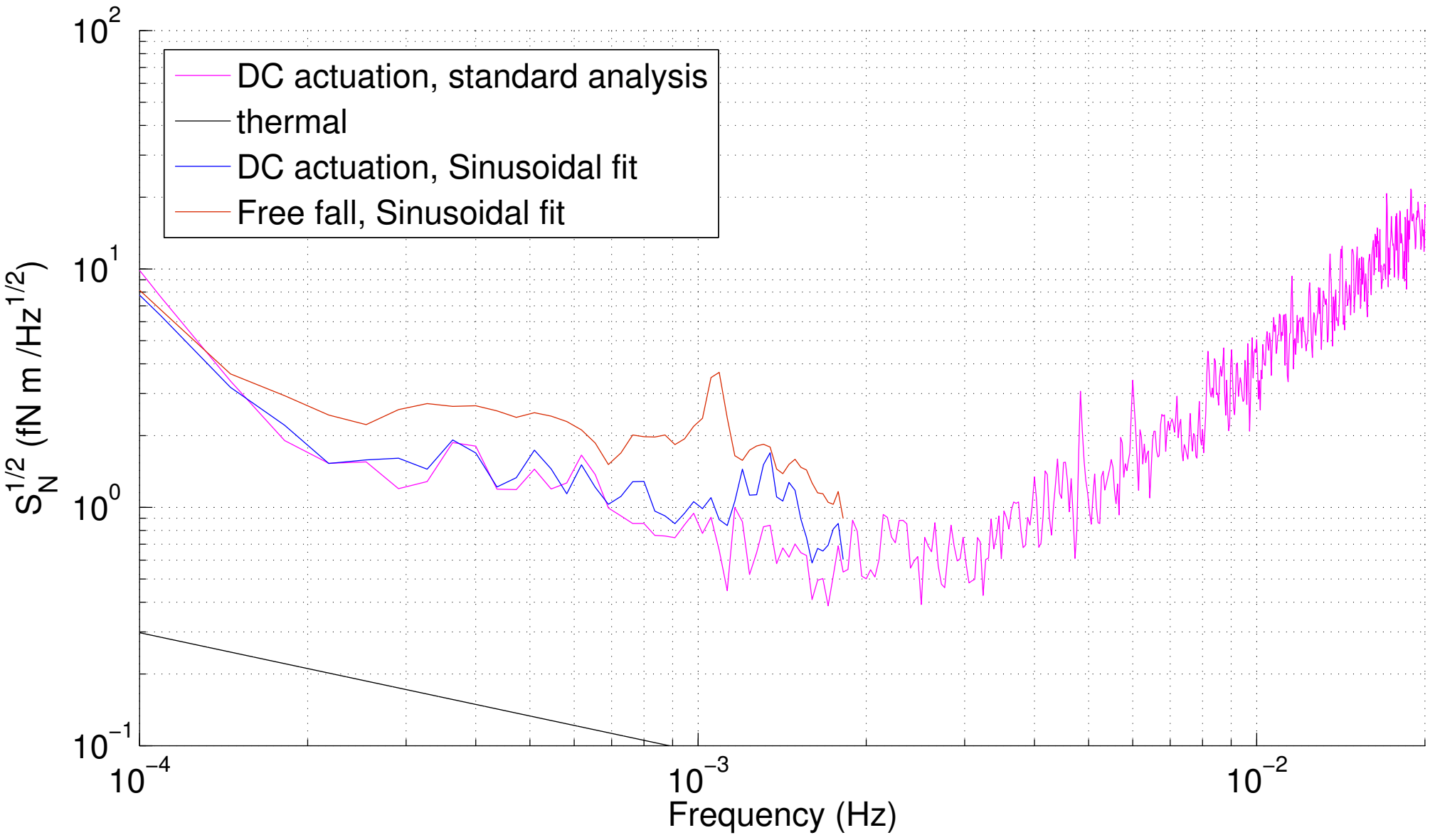

FIGURE 6.26: Torque noise for DC actuation run measured with standard analysis technique (magenta) and with the free-fall sinusoidal fit analysis technique, with artificial gaps (blue). Free-fall run from fig 6.24 with sinusoidal fit analysis shown for comparison (red).

### 6.6.1 Noise with constant DC actuation

Measurements performed with constant DC actuation were alternated with free fall measurements during weekends of test campaign. The measured torque noise in this phases includes the contribution from the actuation fluctuations, as explained in section 4.1

$$S_{N_{ACT}}^{1/2} = 2N_{DC}S_{\delta V/V}^{1/2}. \tag{6.27}$$

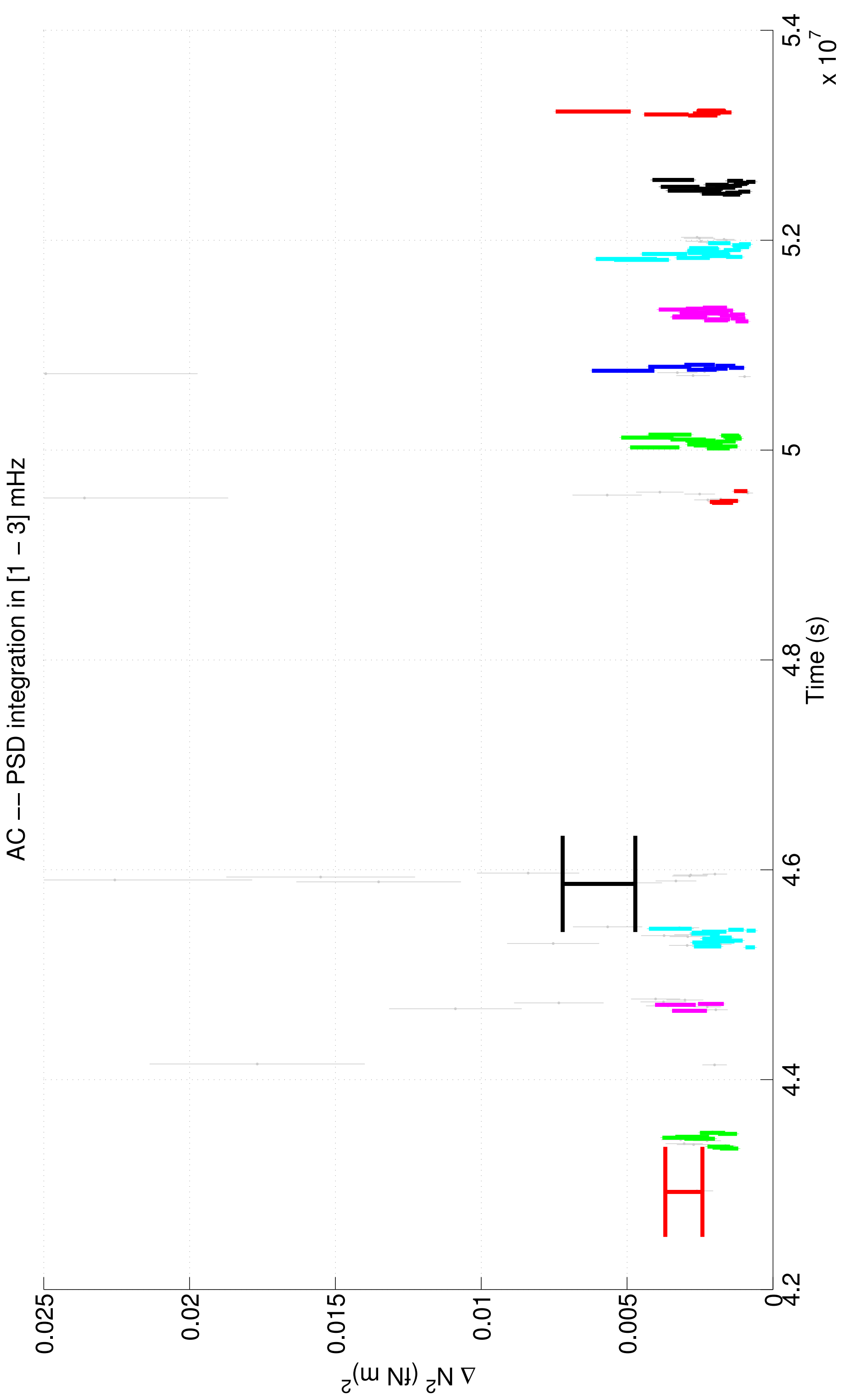

FIGURE 6.27: Integrated torque noise power spectral density in the frequency band from $1\,mHz$ to $3\,mHz$ obtained from all free fall measurements, mainly performed during weekends. Faded points represents rejected cuts based on time stamp issues. Data points that have big error bars are due to an effect of the rendering of the screen.

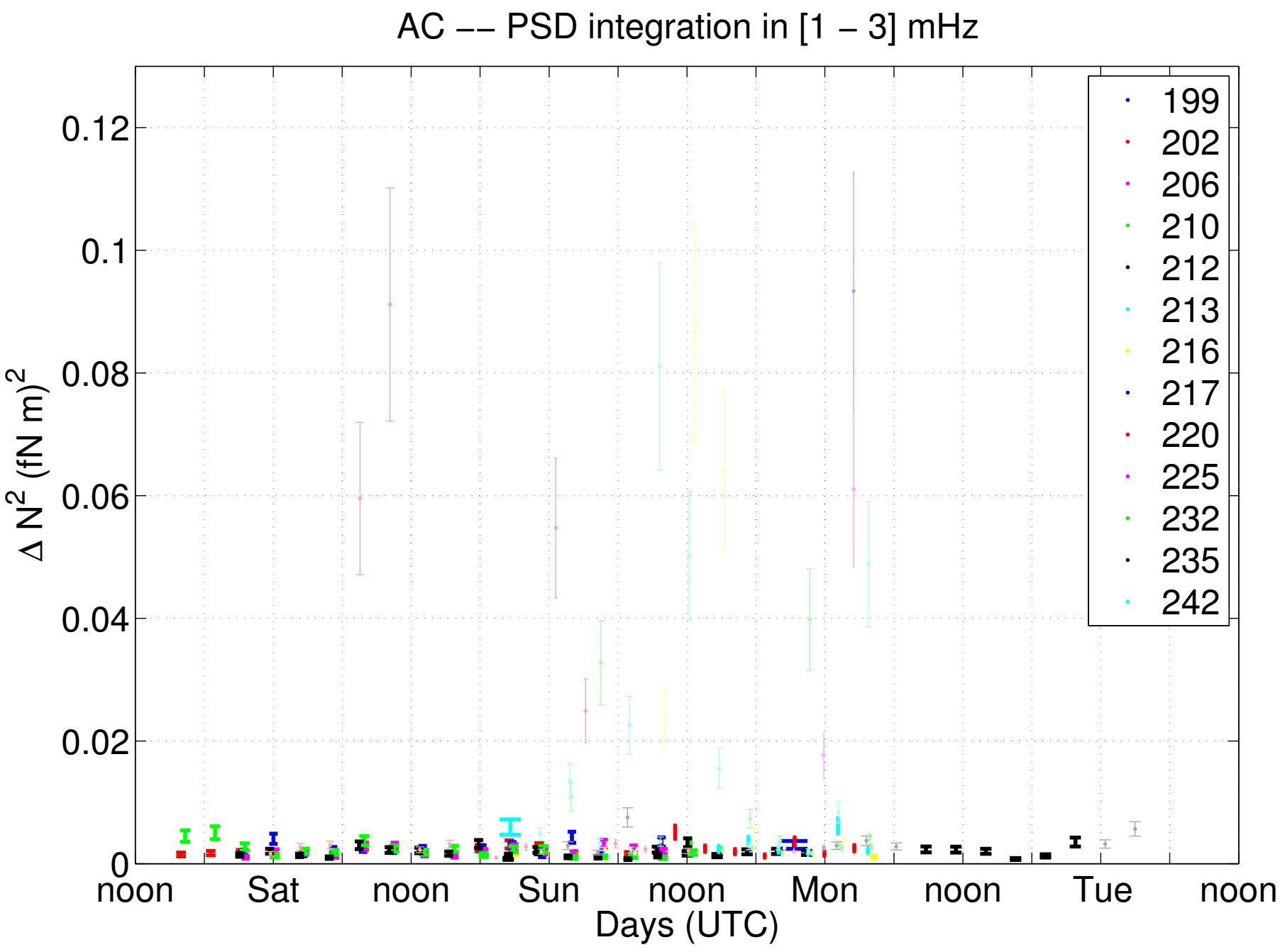

FIGURE 6.28: Integrated torque noise power spectral density in the frequency band from $1\,mHz$ to $3\,mHz$ obtained from all free fall measurements (numbered 199, 202, ...), mainly performed during weekends. Faded points represents rejected cuts based on the experimental implementation problem criteria.

The measured total torque noise is showed in figure 6.33, and has a level of $S^{1/2}_{\hat{N}_2} \approx 0.9\,fNm/\sqrt{Hz}$ at $1\,mHz$. This means that, attributing all of this noise to actuation would correspond to actuation fluctuations at a level of $S^{1/2}_{\delta V/V} = S^{1/2}_{\hat{N}_2}/2N_{DC} \approx 3 \cdot 10^{-5}\,/\sqrt{Hz}$, considering that the level of DC torque applied to hold the test mass centered is $N_{DC} = 14.28\,pNm$. This is roughly equivalent to $30\,ppm/\sqrt{Hz}$ of voltage stability at mHz, a factor 15 above that expected for LPF that is $S^{\frac{1}{2}}_{\delta V/V} < 2 \cdot 10^{-6}/\sqrt{Hz}$. We also show in figure 6.34 background torque noise measured in March 2015, after the free fall test campaign, and after rotate again the suspended test mass respect to the electrode housing to center it.

Make the difference between torque noise without actuation and that measured with constant DC actuation allow to estimate the excess actuation noise as defined in equation 4.6. The difference from power spectral densities is shown in figure 6.35. We plot also the difference in the cross spectral densities between torque measured by the two readout, compared to the LPF requirements converted into an equivalent torque noise. At frequency near resonance (1-1.5 mHz) we are able to resolve an excess actuation noise of $0.2 fN^2m^2/Hz$ that correspond to a voltage noise at level of $15ppm/\sqrt{Hz}$. At higher frequency, as well as frequency below 1 mHz, we don't resolve this effect. There is not a noise stationarity, at low frequency this difference is negative, highlighting an excess in noise without actuation.

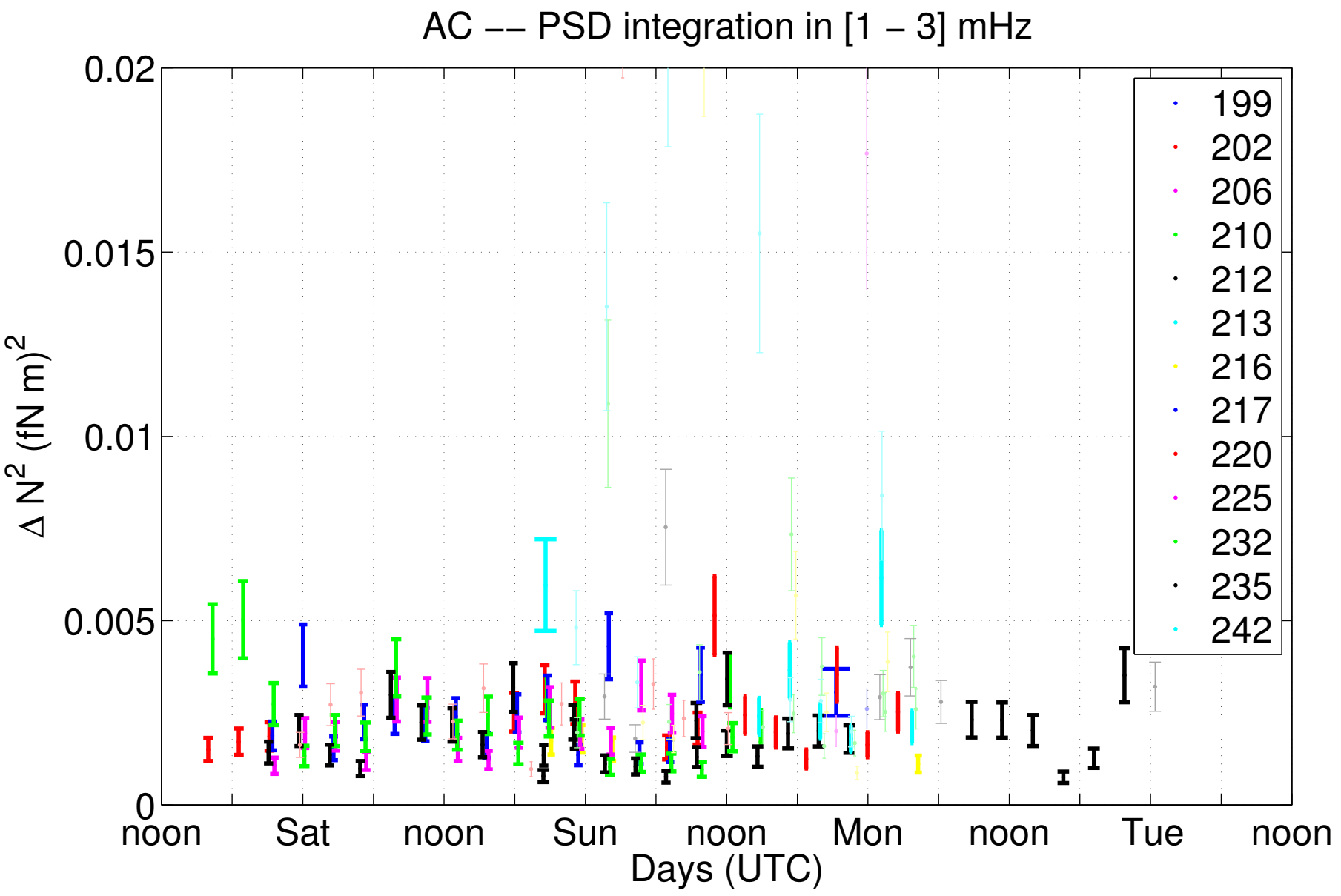

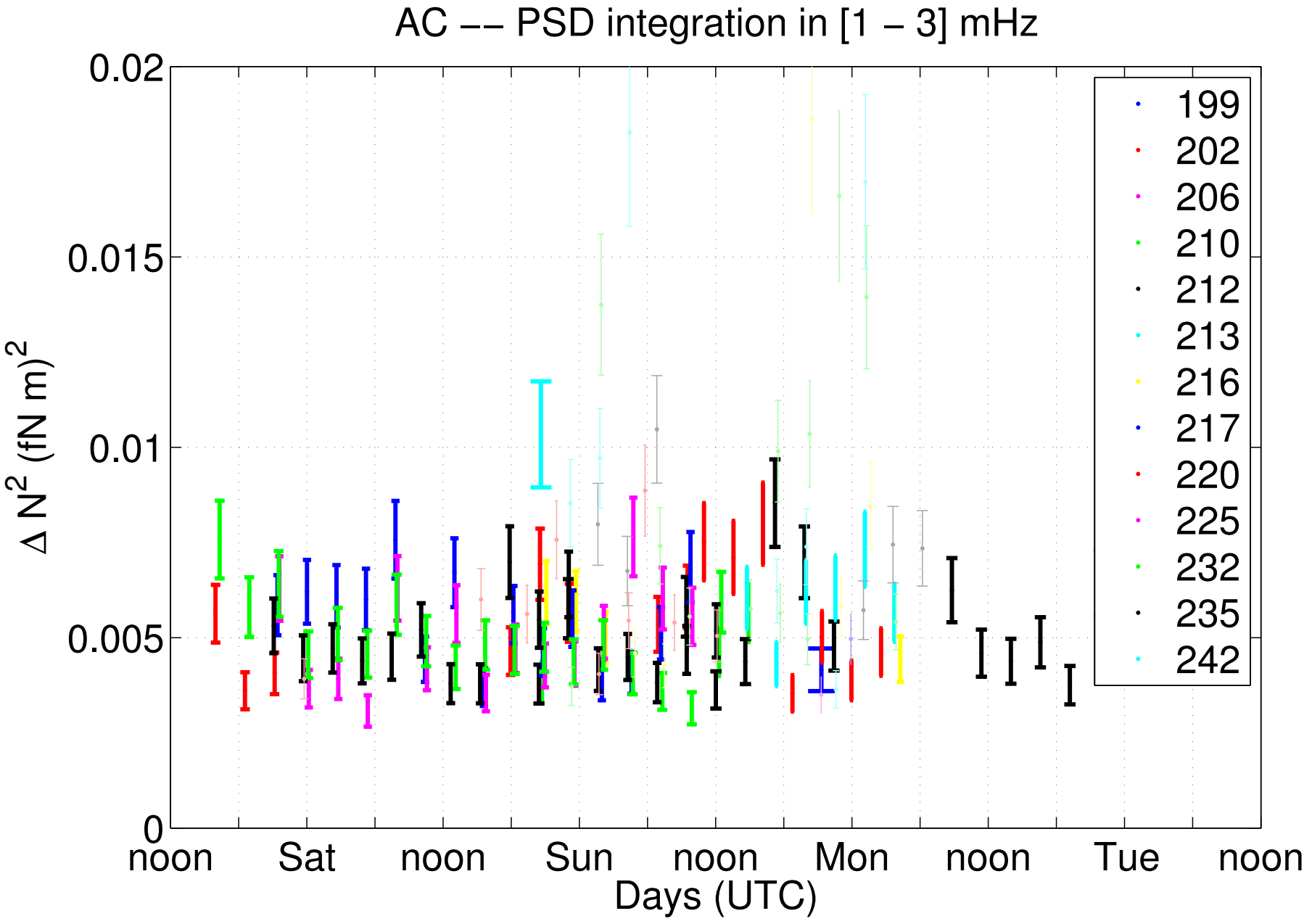

FIGURE 6.29: Integrated torque noise power spectral density in the frequency band from 1 $mHz$ to 3 $mHz$ obtained from all free fall measurements (numbered 199, 202, ...), analyzed with sinusoidal fit techniques(first panel) and BH low pass filter techniques (second panel).

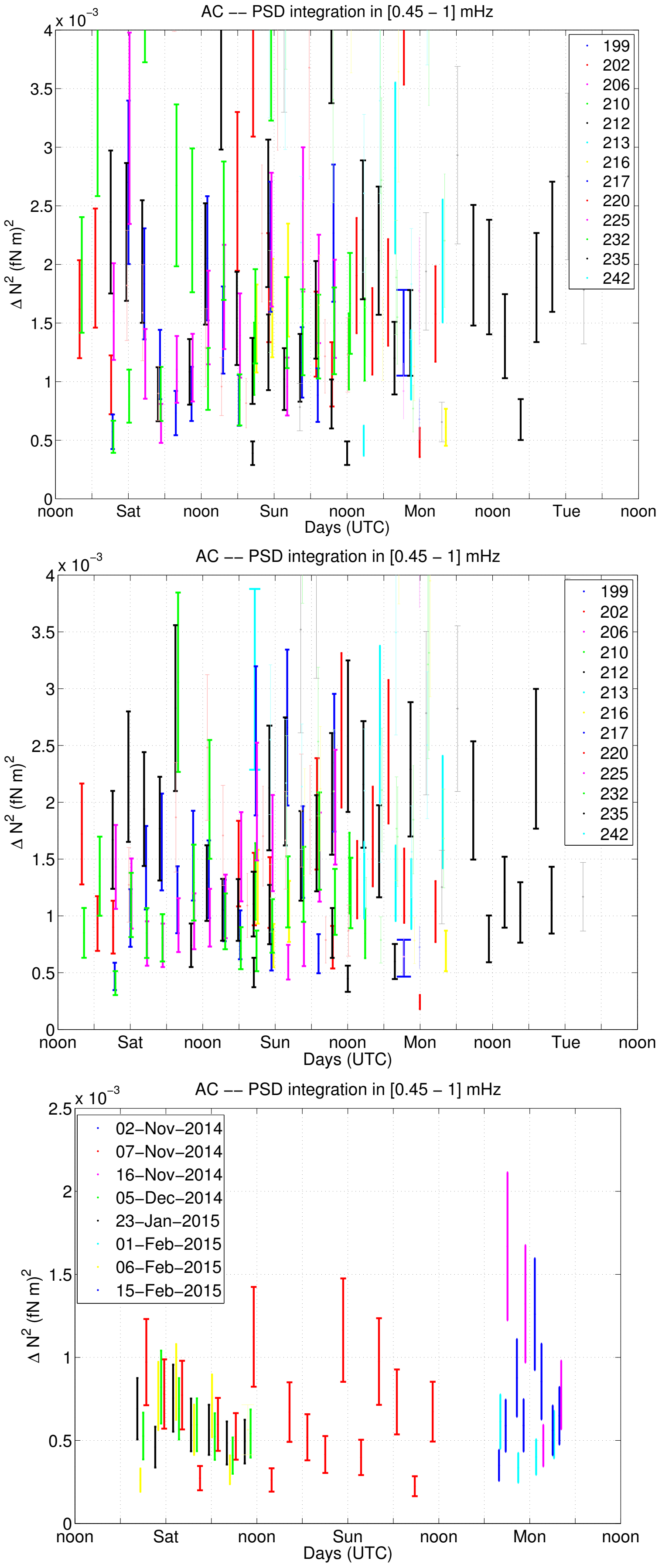

FIGURE 6.30: First and second panel: Integrated torque noise power spectral density in the frequency band from $0.45\,mHz$ to $1\,mHz$ obtained from all free fall measurements, analyzed with sinusoidal fit techniques(first panel) and BH low pass filter techniques (second panel). Bottom panel: Integrated torque noise power spectral density in the same frequency band obtained from noise measurements with constant DC actuation with standard analysis.

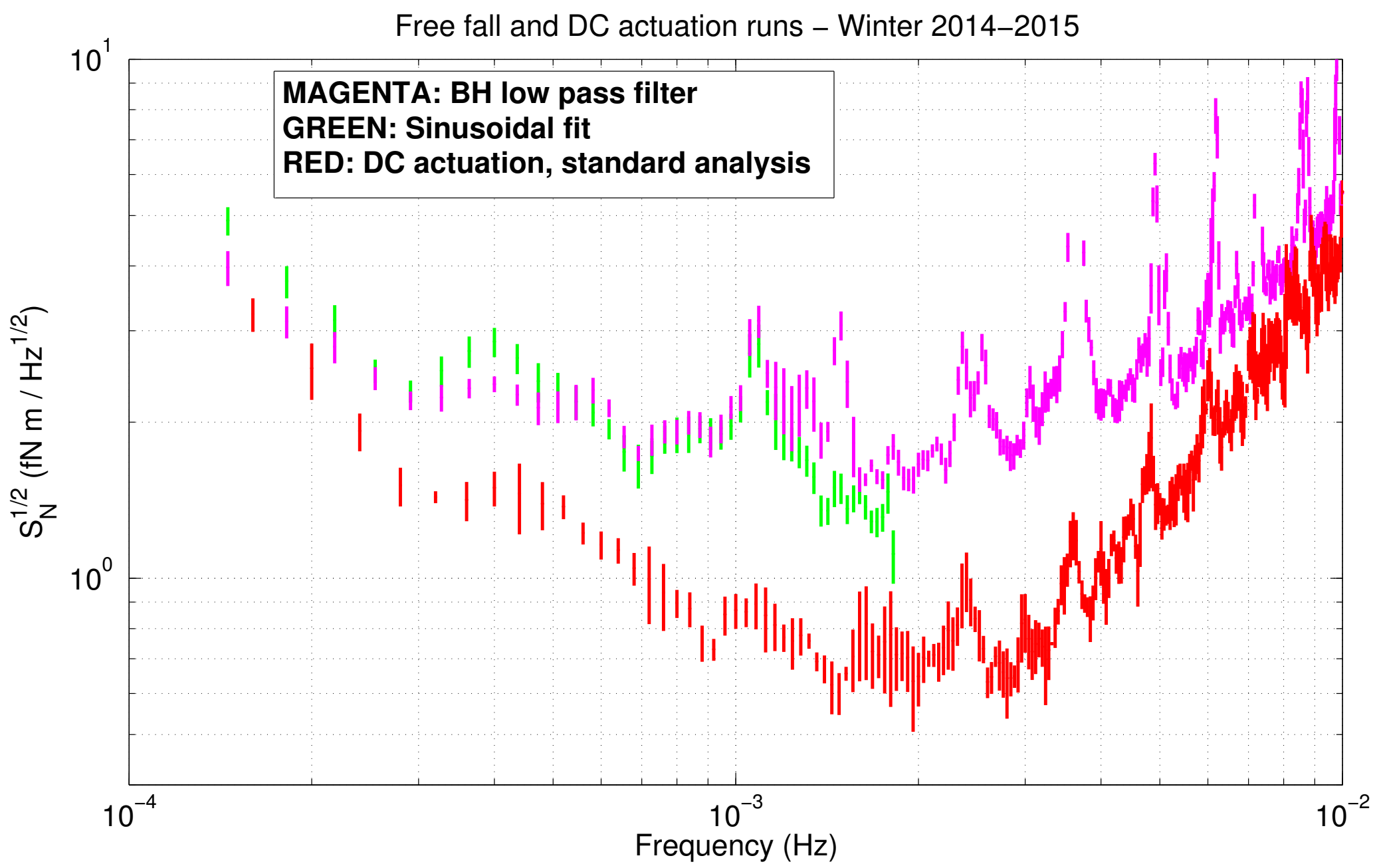

FIGURE 6.31: Torque power spectral densities of all free fall measurements with $T_{fly} = 250\,s$ and $T_{imp} = 25\,s$, averaged on 10 groups of 10 cuts of 27500 s. Magenta points are analyzed with BH low pass filtering technique. Green points are analyzed with the sinusoidal fit techniques. Red points are average of 4 groups of 10 cuts of noise run with constant DC actuation measurements analyzed with the standard technique.

## 6.7 Calibration tone

In order to both test the accuracy of the torque measurement in free-fall mode and to test our ability to resolve a coherent torque signal, we have implemented a calibration tone experiment that was applied to the data in both free-fall and continuous actuation modes.

We do this by performing a measurement of the test mass potential, which is very nearly constant during our measurements, at a level of $V_{TM} \approx -37.5mV$. We can generate a torque of nearly constant amplitude with a standard charge measurement. This means to modulate a sinusoidal bias potential $\pm V_{MOD} \sin(2\pi f_{MOD} t)$, with amplitude $V_{MOD} = 10\,mV$ at frequency $f_{MOD} = 0.5\,mHz$, applied to $x$ sensing electrodes of figure 4.5 with polarization positive on EL2, EL4 electrodes and negative on EL1, EL3.

The torque exerted on the test mass with this bias configuration is, from electrostatic considerations of section 3.2.1, equal to:

$$N_\Phi = \frac{1}{2} \sum_i \left[ \frac{\partial C_i}{\partial \Phi} + \frac{\partial^2 C_i}{\partial \Phi^2} \Phi \right] (V_i - V_{TM})^2, \tag{6.28}$$

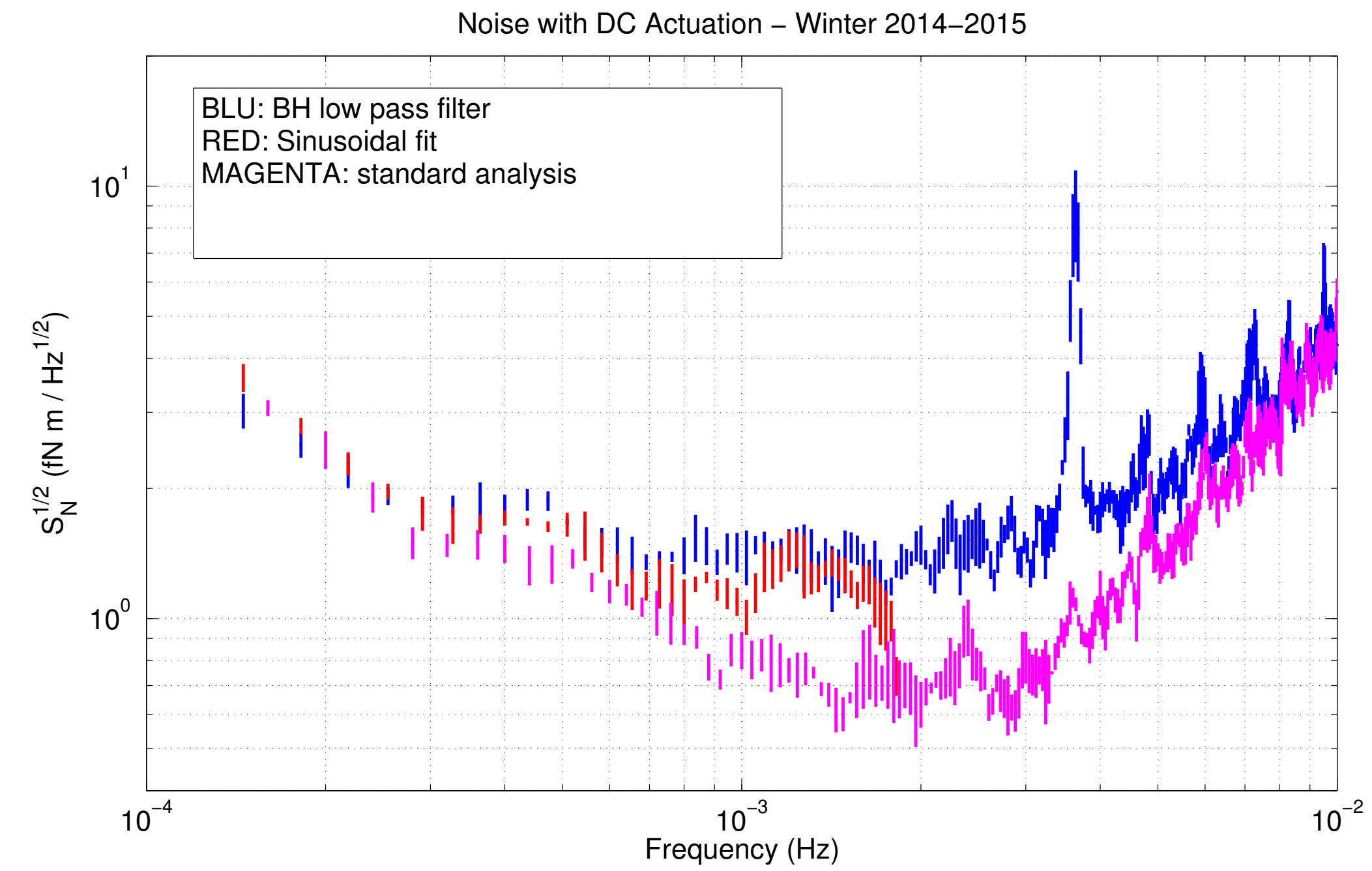

FIGURE 6.32: Torque power spectral densities of DC actuation measurements averaged on 4 groups of 10 cuts, analyzed with the two free fall analysis techniques. Blue points are analyzed with BH low pass filtering technique. Red points are analyzed with the sinusoidal fit techniques.

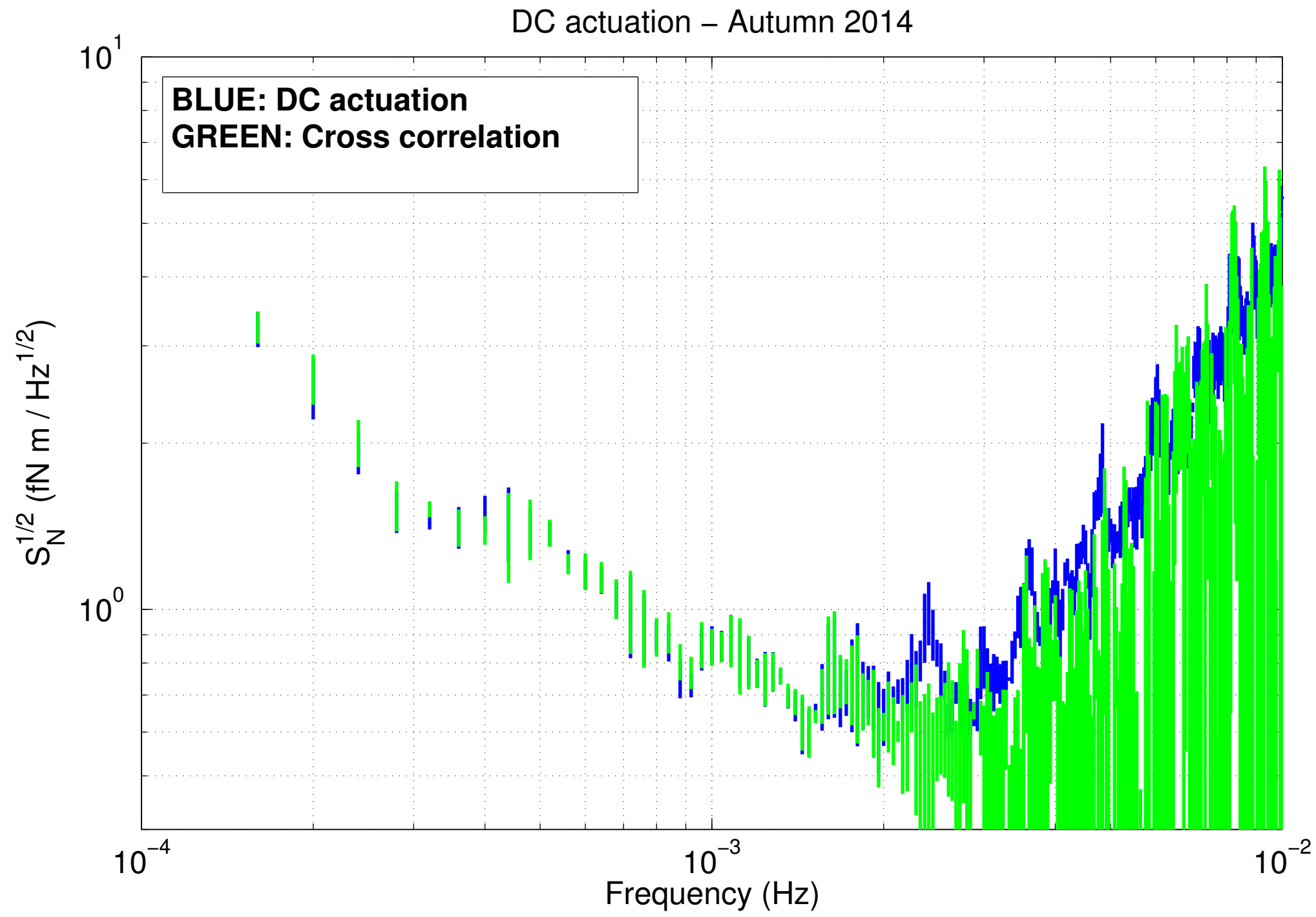

FIGURE 6.33: Torque power spectral densities of all noise measurements performed with constant DC actuation, averaged on 4 groups of 10 cuts of 25000 s. Blue data are autocollimator averaged data, green points are cross spectra between sensor and autocollimator signal.

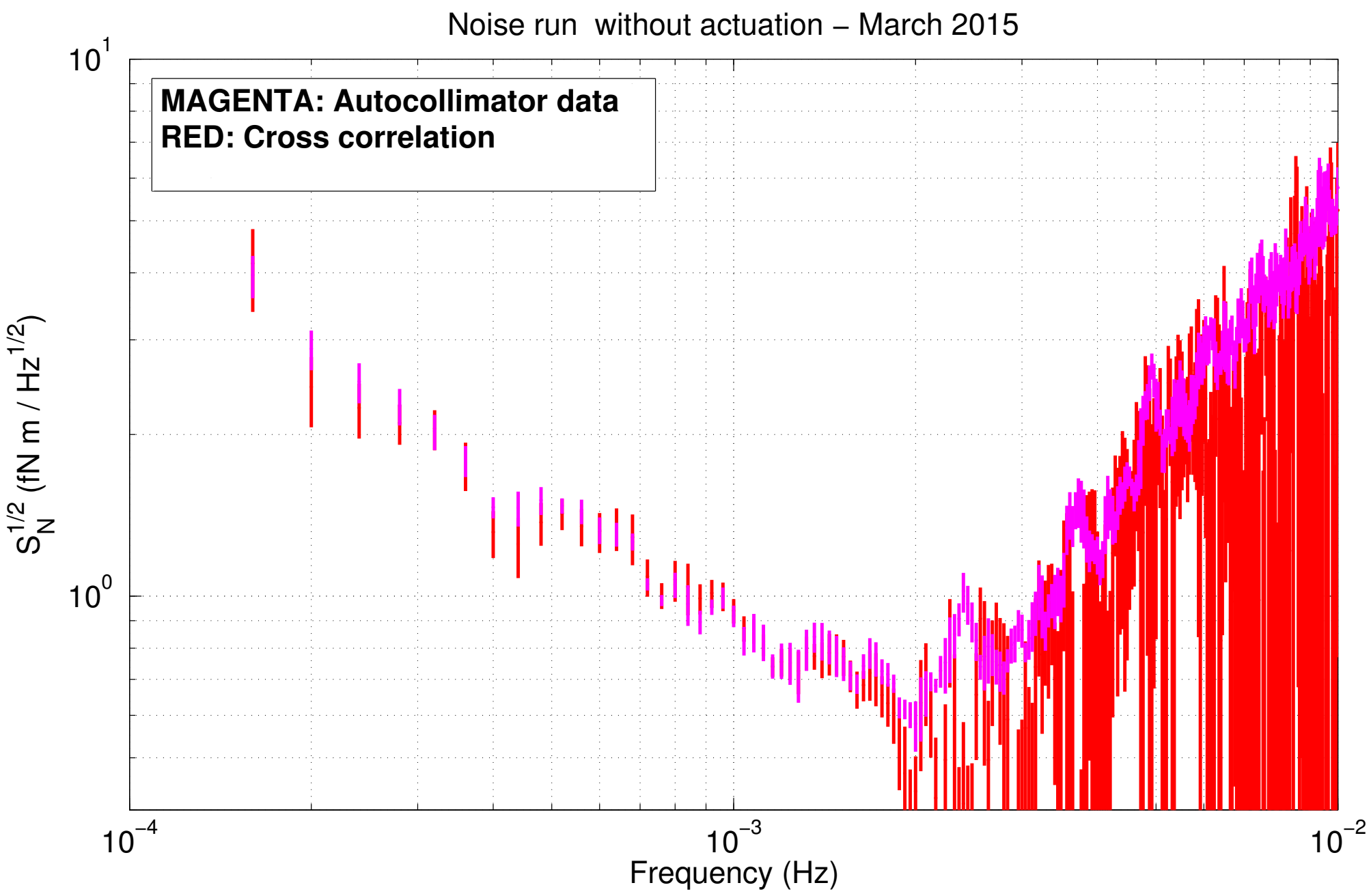

FIGURE 6.34: Torque power spectral densities of all background noise measurements averaged on 15 groups of 10 cuts of 25000 s. Magenta data are autocollimator averaged data, red points are cross spectra between sensor and autocollimator signal.

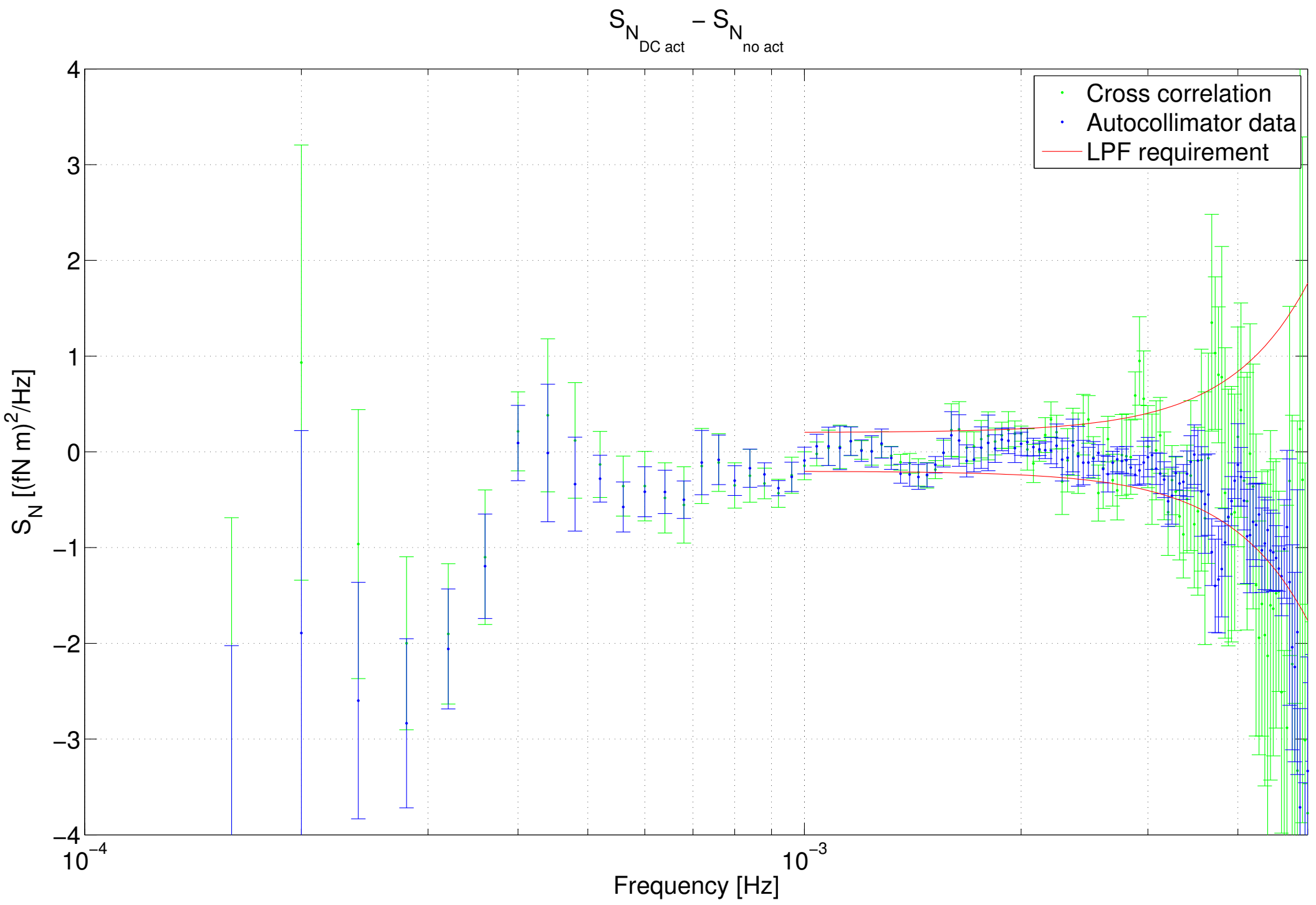

FIGURE 6.35: Actuation torque noise power as calculated by the difference in noise with and without DC actuation. The LPF requirements converted into an equivalent torque noise is also shown.

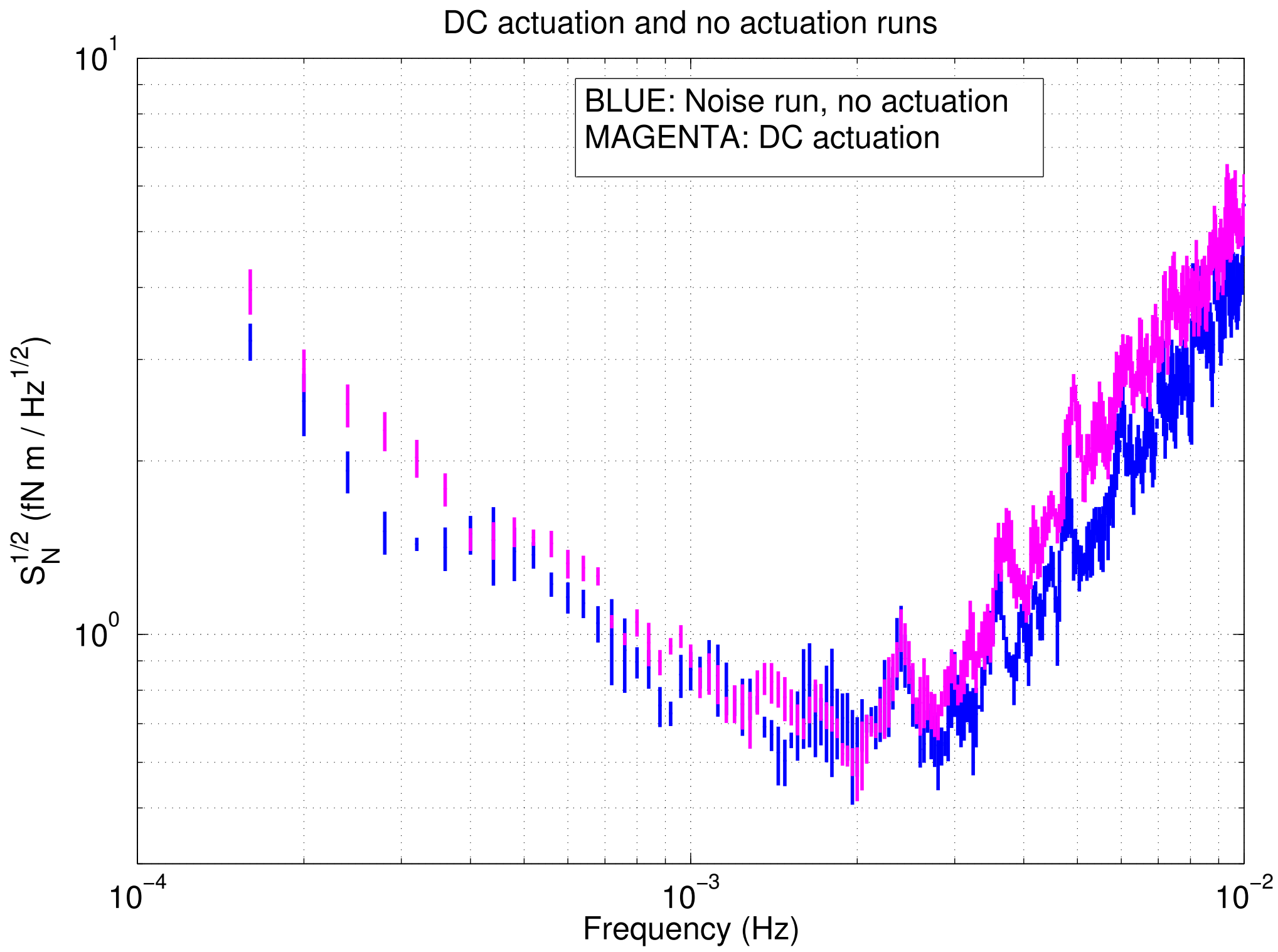

FIGURE 6.36: Torque noise of background noise measurements without actuation (blue) and with constant DC actuation (magenta).

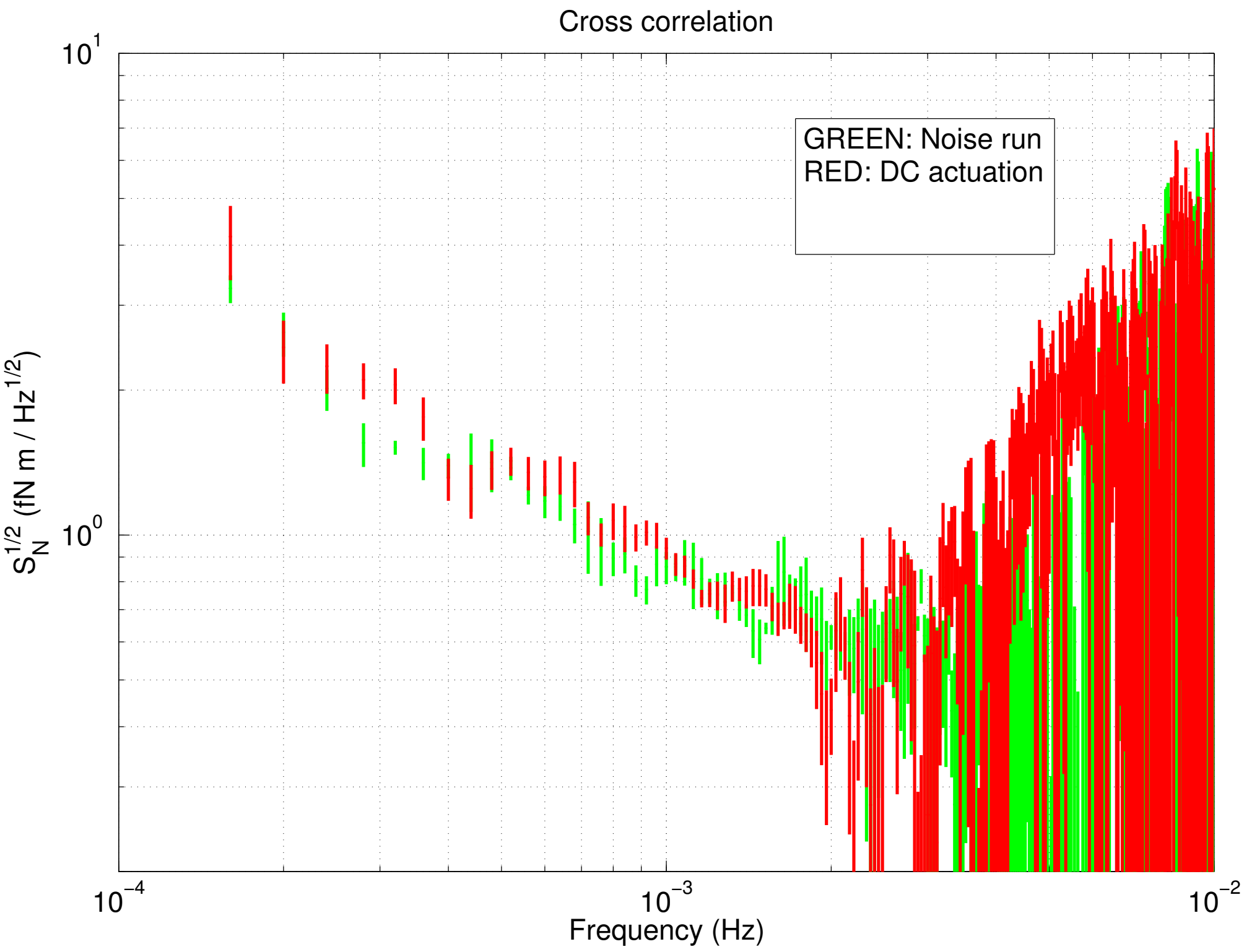

FIGURE 6.37: Cross correlation.

where $V_i$ are voltages applied on each electrodes, that in this case are sum of a DC voltage $\delta V_i$ plus the modulation $V_{MOD}\sin(2\pi f_{MOD}t)$.
Additionally, to have a torque signal that is independent of angle, we try to null the rotational electrostatic DC bias imbalance to remove the angular dependence of the coherent torque [32].
Because pendulum was rotated with respect to the electrode housing, in this phase, the test mass voltage $V_{TM}$ has an additional component depending on the angle of rotation $\Phi$:

$$V_{TM} = V_{TM,0} + 4\frac{\mid \frac{\partial C}{\partial \Phi} \mid \Phi V_{MOD}}{C_{TOT}} \tag{6.29}$$

By substituting in the equation 6.28, the component of torque at the first harmonic will be

$$N_{\Phi,1f} = V_{MOD}\sin(2\pi f_{MOD}t)\left[-4\Delta V_{TM}\frac{\partial C}{\partial \Phi} + \Phi\frac{\partial^2 C_i}{\partial \Phi^2}\left(1 - 4\frac{\mid \frac{\partial C}{\partial \Phi}\mid^2}{C_{TOT}\frac{\partial^2 C_i}{\partial \Phi^2}}\right)\Delta_{\Phi_{(x)}}\right], \tag{6.30}$$

where we define

$$\Delta V_{TM} = V_{TM,0} - \sum_i \delta V_i \tag{6.31}$$

$$\Delta_{\Phi_{(x)}} = \delta V_{EL1} - \delta V_{EL2} + \delta V_{EL3} - \delta V_{EL4}. \tag{6.32}$$

where $\Delta_{\Phi_{(x)}}$ is the rotational DC bias imbalance. To measure and null $\Delta_{\Phi_{(x)}}$, we performed measurement of test mass potential with the same sinusoidal charge modulation applied on x electrodes, at different $\Phi$ angles $0, -1, -2, -3, mrad$, as we show in figure 6.38. In this case $V_{MOD} = 200\,mV$ and $f_{MOD} = 3\,mHz$. The corresponding angular deflection is converted in torque and is then demodulated, and the resulting torque components at modulation frequency, are averaged for each steps. Finally, a linear fit of torque as a function of $\Phi$ is implemented, as visible in figure 6.39. The slope is proportional to $\Delta\Phi_{(x)}$

$$\frac{\partial N_{\Phi,1f}}{\partial \Phi} = \left(V_{MOD}\frac{\partial^2 C_i}{\partial \Phi^2} - 4\frac{V_{MOD}(\frac{\partial C}{\partial \Phi})^2}{C_{TOT}}\right)\Delta_{\Phi_{(x)}}. \tag{6.33}$$

In order to compensate the $\Phi$ dependence of the applied force 6.28 during measurements of free fall with calibration tone switched on, it is possible to apply to each electrode $x$ a compensation voltage proportional to $\Delta_{\Phi_{(x)}}$

$$V_{comp} = -\frac{\Delta_{\Phi_{(x)}}}{4}. \tag{6.34}$$

We obtain a $\Delta_{\Phi_{(x)}} \approx 0.275\,V$, considering that $\partial C_x/\partial\Phi = 3.3\,pNm/rad$ and $\partial^2 C_x/\partial\Phi^2 = 28\,pNm/rad^2$ from FE analysis [7]. We apply a $V_{comp} \approx 69mV$ on $x$ electrodes with sign plus or minus depending on the electrodes sign polarization, in order to null the potential applied.

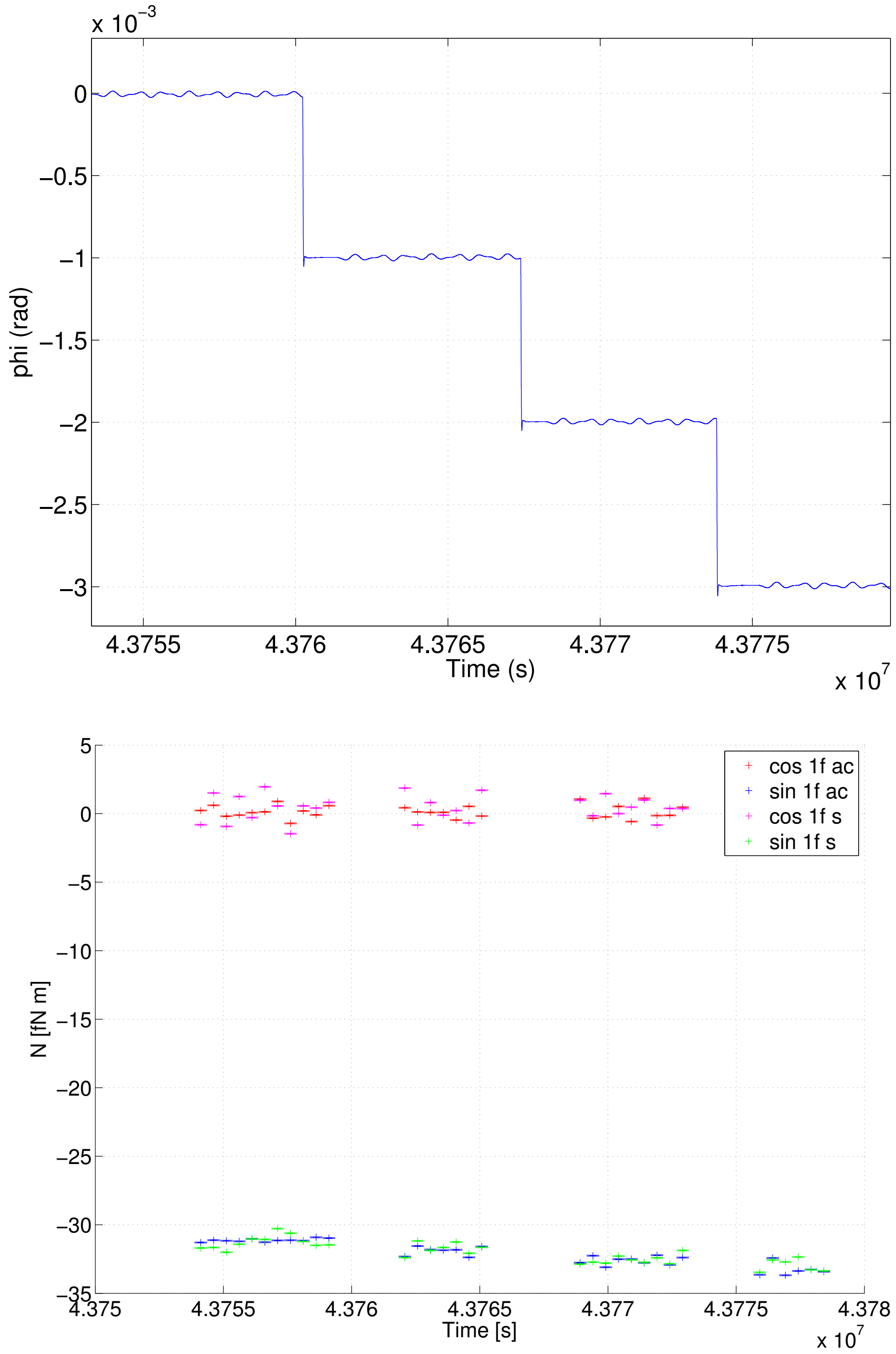

FIGURE 6.38: Top panel: Time series of modulated signal at different angle 0, −1, −2, −3, *mrad*. Bottom panel: The resulting sine and cosine component of torque after demodulation at 1 $f_{MOD}$.

With this compensation applied, we perform measurements of free fall and background noise with DC actuation force, both with the calibration tone switched on. It is possible to analyze relative runs either with sinusoidal fit techniques then with Blackman-Harris low pass filter, to recover the corresponding torque time series. In figure 6.40 we show the recovered torque time series of a free fall run, where the modulation applied is clearly visible. We want to recover the resolution with which we measure the torque signal applied during the free fall mode and during the DC actuation noise measurement. To do this we perform a sinusoidal fit to torque time series obtained applying free fall analysis techniques on both type of measurements, on cycles length $1/f_{MOD}$, using the model

$$N_{cal} = N_0 + A\cos(2\pi f_{MOD} t) + B\sin(2\pi f_{MOD} t). \tag{6.35}$$

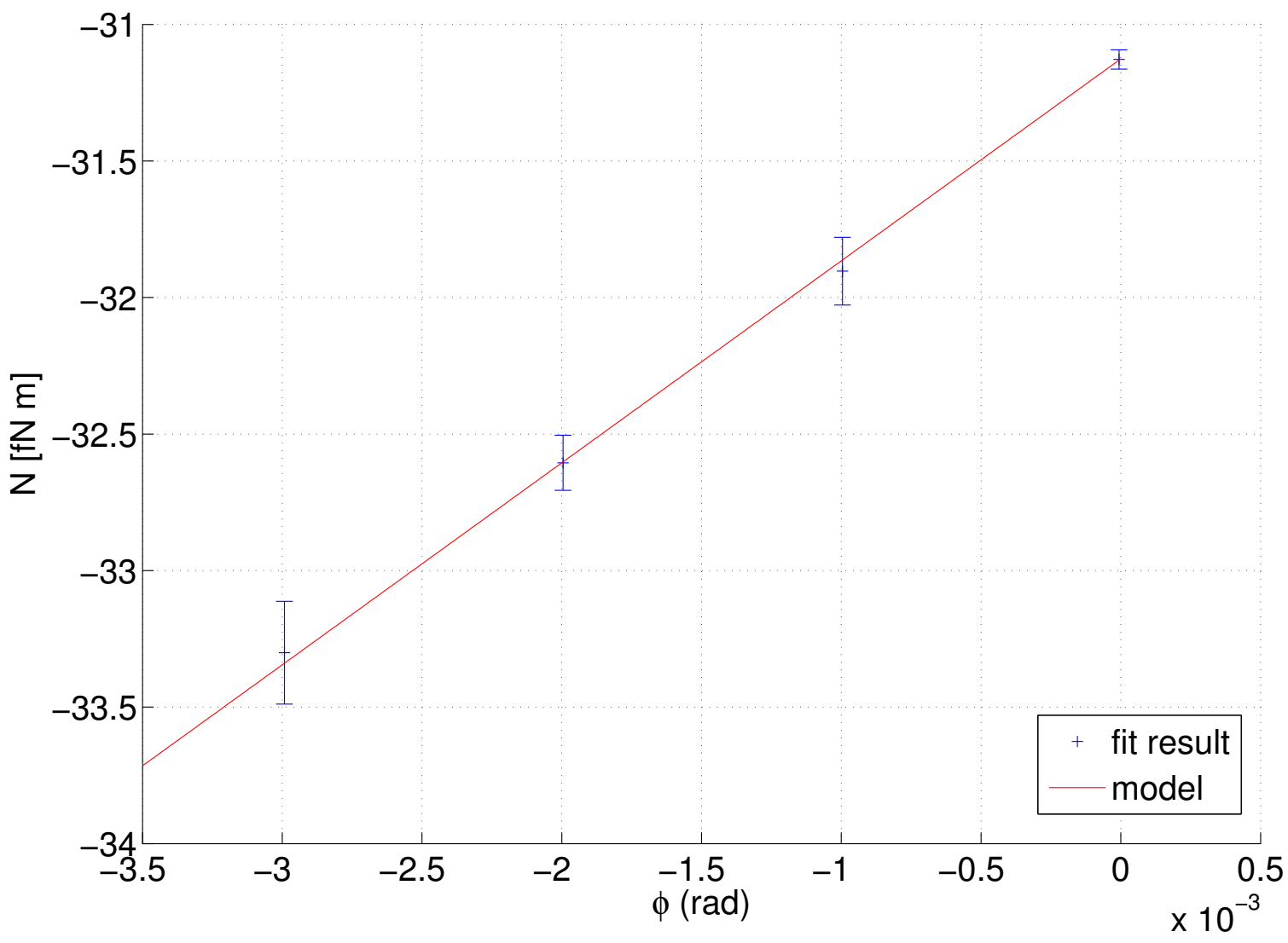

FIGURE 6.39: Linear fit of torque component measured at different $\Phi$ position.

Results of fit components are listed in table 6.6 with their standard deviation, as is also shown in figure 6.41. We are able to extract with this technique, the coherent torque applied, from all analysis techniques within 1%. The RMS deviation obtained cycle by cycle is related to the torque noise at frequency of modulation $f_{MOD}$ as $\sigma = \sqrt{S_N(f_{MOD}) f_{MOD}}$. From figure 6.31 the torque noise expected at $f_{MOD} = 0.5\, mHz$ for the measurement analyzed with the BH low pass technique, or with the sinusoidal fit, is around $2\, fNm/\sqrt{Hz}$. This means that we have a resolution of $0.05\, fNm$. This is a factor two below the resolution with which we resolve the coherent torque signal applied on our pendulum with the same techniques of analysis, shown in table 6.6, that are $\sigma_{sin} = 0.11$ for BH low pass and $\sigma_{sin} = 0.09$. For the torque noise with DC actuation showed in figure 6.32, analyzed with standard analysis, the resolution at $f_{MOD} = 0.5\, mHz$ is $\approx 0.03\, fNm$, compatible with $\sigma_{sin} = 0.04$ obtained with standard analysis during the DC actuation noise measurement with the calibration tone switched on.

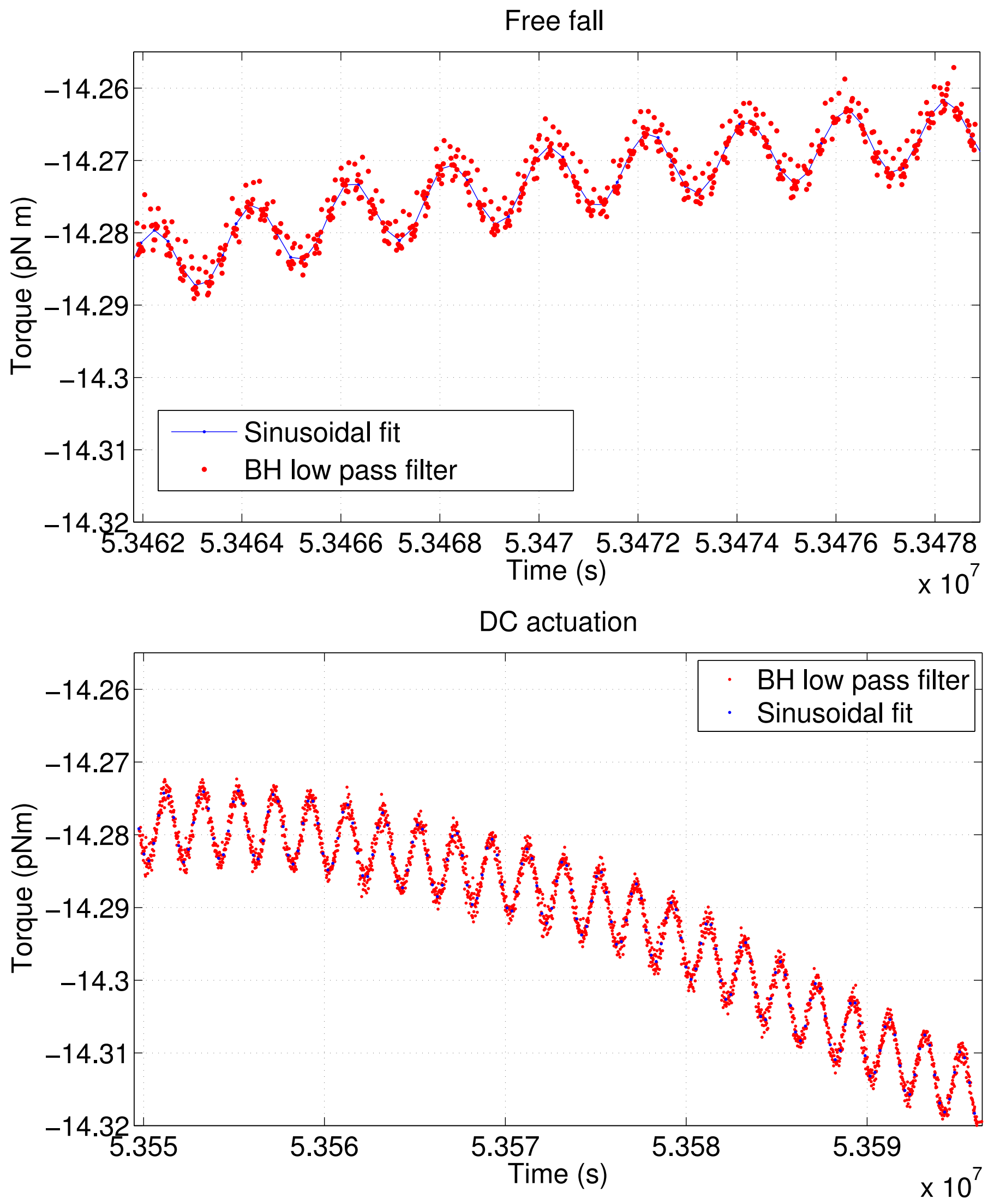

FIGURE 6.40: Torque time series recovered with free fall analysis techniques, during the application of a calibration tone at frequency $f_{MOD} = 0.5\,mHz$ during a free fall measurement (top panel) and a noise run with DC actuation on (bottom panel).

| | | | | | |
|---|---|---|---|---|---|
| **Free fall** | | | | | |
| Sinusoidal fit | Cos | $0.015 \pm 0.007$ | $\sigma_{cos}$ | 0.04 | fNm |
| | Sin | $-4.84 \pm 0.02$ | $\sigma_{sin}$ | 0.09 | fNm |
| BH low pass | Cos | $0.021 \pm 0.016$ | $\sigma_{cos}$ | 0.08 | fNm |
| | Sin | $-4.86 \pm 0.02$ | $\sigma_{sin}$ | 0.11 | fNm |
| **DC actuation** | | | | | |
| Sinusoidal fit | Cos | $-0.014 \pm 0.009$ | $\sigma_{cos}$ | 0.05 | fNm |
| | Sin | $-4.79 \pm 0.01$ | $\sigma_{sin}$ | 0.07 | fNm |
| BH low pass | Cos | $-0.03 \pm 0.02$ | $\sigma_{cos}$ | 0.08 | fNm |
| | Sin | $-4.79 \pm 0.02$ | $\sigma_{sin}$ | 0.12 | fNm |
| Standard analysis | Cos | $-0.014 \pm 0.007$ | $\sigma_{cos}$ | 0.03 | fNm |
| | Sin | $-4.795 \pm 0.008$ | $\sigma_{sin}$ | 0.04 | fNm |

TABLE 6.6: Charge measurement parameters and their resolution for free fall and DC actuation runs. For both measurements were used 23 cycles.

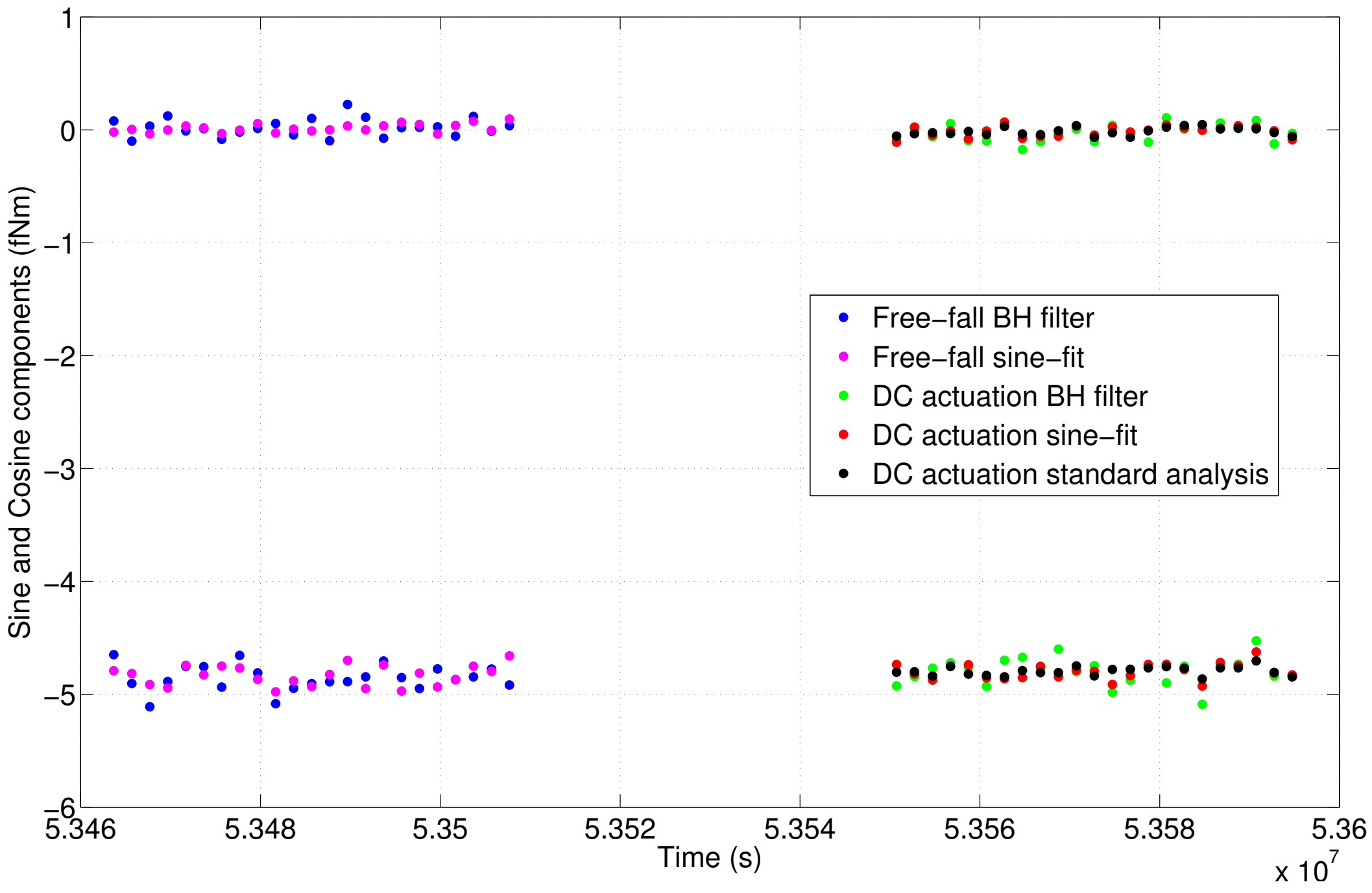

FIGURE 6.41: Sine and cosine components.

# Chapter 7

# Conclusions and future steps

The LISA Pathfinder mission has a very challenging task in the measurement of residual differential acceleration between two free falling test masses at level of $30 fm/s^2\sqrt{Hz}$ at $1\,mHz$. The actuation forces necessary in flight introduce a source of disturbance internal to the orbiting apparatus, that affects the acceleration noise budget of the mission. The current best estimate of the LPF differential acceleration noise considers a stray acceleration noise level of $7.5\,fm/s^2\sqrt{Hz}$ at $1\,mHz$ considering a gravitational balancing tolerances along the sensitive axis of $\Delta g_{DC} \approx 0.65\,nm/s^2$, as shown in figure 2.3.
This is the motivation that led to the search for a way of reduce the actuation noise at least along the sensitive axis of the interferometer on board of LTP. This is the Free-fall mode.
The possibility of test the free fall mode with a quantitatively interesting accuracy and sensitivity on ground has been provided by means of torsion pendulum test bench. It is possible to mimic the LPF gravity gradients that must be compensated in orbit as external torques acting on the suspended test mass.
Key device that allows to implement the free fall mode on pendulum, is the home made front end electronics. The actuation circuit of the capacitive position sensor gives the authority and the sensitivity with which the free fall mode can be implemented on torsion pendulum facility. The maximum voltage that can be applied by DAC to each electrode is $\pm 10\,V$ that means the possibility to apply a maximum torque on test mass of around $200\,pNm$. This translates into an equivalent differential applied force corresponding to $20\,nN$ forces applied by each electrodes. This means that with 10% of duty cycle, we can't have more than $20\,pNm$ of DC torque, equivalent to $2\,nN$ of differential force. This is similar to LPF level ($\approx 1nN$ for 1% of duty cycle), but applied with a higher duty cycle. Torque authority is reduced respect to that in flight, where it is possible to apply roughly $11\,V$ in science mode and $135\,V$ AC, applied half the time, in wide range. Despite the difference in the authority, the study of DC actuation noise on ground has allowed to put an upper limit on actuation noise up to $1fNm/\sqrt{Hz}$ at $1mHz$. The electrostatic actuator calibration has allowed to set the accuracy in the determination of the torque with DC actuation with respect to the torque measured with the free fall technique. We found a discrepancy of the 0.5% as visible in figure 6.20, compatible with the uncertainties in our actuator calibration.

We also put an upper limit on the actuator noise, performing the difference in noise with and without DC torque actuation for our system based on a commercial DAC card and homemade switching circuitry, and as shown in figure 6.35. The resolution of the actuation noise is $0.2\,fN^2m^2/Hz$ at $1\,mHz$, that on the scale of voltage fluctuations corresponds to $S^{\frac{1}{2}}_{\delta V/V} \approx 15ppm/\sqrt{Hz}$. This is a factor two above the actual measured stability with the inertial sensor Front End Electronics of LTP that gives $3-7ppm/\sqrt{Hz}$ at $1\,mHz$.

The torsion pendulum facility is a high sensitivity apparatus to measure small and parasitic forces acting on the suspended test mass at low frequencies. The measured torque noise is $0.8\,fNm/\sqrt{Hz}$ at $1\,mHz$ that corresponds to an effective LPF TM acceleration noise level of $40\,fm/s^2\sqrt{Hz}$. We are able to resolve an extra acceleration noise acting on our instrument within LISA Pathfinder specifications in the frequency region of interest near $1\,mHz$, either from actuation noise or by the free fall experiment.

One advantage of the ground testing facility is the possibility to tune the effective DC gravitational imbalance, to allow more flexibility to explore different control strategies, by varying flight and impulse time or control points, and different dynamic configurations made possible by having a variable stiffness. This means to introduce high angular and dynamics ranges never inspected before. The typical amplitude of a noise floor measurement is around the microradiant, while the range inspected during the free flight can be from $300\,\mu rad$ up to $3\,mrad$, as also visible in figure 6.18.

Thanks to the study of the pendulum dynamics and the authority of its front-end electronics, it has been possible to choose the optimal experimental configuration to implement the free fall mode in a way most similar to that in flight. To achieve similar conditions, we softened electrostatically the pendulum in order to reduce the stiffness applying DC constant voltages on $Y$ electrodes in order to lengthen the pendulum period, to allow more longer flights without to be near the condition of pendulum amplitude divergence. Pendulum softening had an impact on the total stiffness of the system, with the addition of an electrostatic stiffness that involves a non linearity and a quadratic dependence of the stiffness on the angular displacement. This is also confirmed by the electrostatic model of the gravitational reference sensor.

An important part of this work was dedicated to the calibration of the experiment in order to estimate pendulum dynamical parameters which are used in the dynamical model that convert pendulum angular displacement in torque and to the measurement of the stiffness from pendulum period in order to estimate the dependence from the angular displacement. We account for a non linear dependence of the stiffness from the angle that allow to explain 1% difference in the period measured with the pendulum centered with respect to the electrode housing and rotated by 5 $mrad$ and that has an impact on the torque estimation during free fall as visible by comparing figure 6.10 and 6.12. Non linear parameters, however, are not used in the torque model (eqn 6.4) used to convert the pendulum angular motion in the free fall data analysis, because they do not agree from one run of measurement to another and are inconsistent with stiffness estimated from period measurements over a larger angular range.

Moreover, the free fall mode implementation has allowed to reach a deep knowledge of pendulum dynamics and data acquisition system, bringing to light problems that otherwise would not have been investigated, and that was founded to be critical for the free fall experiment on ground. These are related to an optical readout non linearity and to a problem in the time stamping of the data acquisition system, that must be corrected in order to improve the torque noise estimation during free fall measurement. Similar issues could be potentially relevant also for the implementation in flight and could be depend on the interferometer linearity over the 5 micron length scales explored in the orbit free-fall test. A testing campaign about inertial sensor non linearity could be implemented in order to reduce possible in-flight issues.

Another important challenge of this experiment is the data analysis of free fall data because of the introduction of gaps. The two developed analysis techniques have been tested on free fall data as well as on continuous data, with and without DC actuation, whit no gaps and impulses, as finally compared in figure 6.31 and 6.33. The main idea of the testing campaign was to verify that the free fall can allow a torque noise measurement at the background levels measured in the absence of actuation, lower then that possible with the commanded DC actuation force. At the current state, the free-fall mode has been tested at the level of $2fNm/\sqrt{Hz}$, at the frequencies of $1mHz$, a factor 2 larger than from the measured torque noise with DC actuation at the same frequency, and corresponding to an acceleration of about $100fm/s^2\sqrt{Hz}$. This is factor 10 above the level of actuation noise expected for LISA Pathfinder mission, that is $7.5\,fm/s^2\sqrt{Hz}$ at $1\,mHz$.

The discrepancy observed is still under investigation and will be the subject of future developments in particular on the aspects of the aliasing from the high frequency components. There is a noise component, at low frequency, that scales with level of high frequency readout noise. This is due to the presence of an aliasing effect. This is related to the presence of gaps in the data, because the same effect was found also on continuous data analyzed with the same techniques (figure 6.32). This was also confirmed by the test on simulated data, as shown by comparing different free fall simulations with growing readout noise in figure 5.12, and need still investigation.

However, we are able to test the accuracy of the torque measurement in free-fall mode and to test our ability to resolve a coherent torque signal, thanks to the implementation of a calibration tone experiment that has been applied with a charge modulation to the data in both free-fall and continuous actuation modes. We are able to resolve a torque signal applied at frequency $0.5\,mHz$ in the free-fall and DC actuation conditions, by performing a fit to torque estimated with the free fall analysis techniques, within 1%, as reported in table 6.6. We can compare this resolution with the estimate of torque noise from the free fall and DC actuation spectrum at the same frequency of the modulation, and analyzed with the same techniques, as in figure 6.31 and 6.33, we found comparable resolutions.

In conclusion, we are able to implement a ground testing of a noise reduction technique like the free fall mode, never implemented on a torsion pendulum facility, usually used to measure small forces and not large DC torque as during the free fall mode. We produce a big data set, 554 stretches of $25000s$ overlapped of 66%, around $9^6\,s$, that produce a statistics on the free fall mode experiment but also on DC actuation measurements as well as the pendulum noise floor, and that are ready for further analysis. Results from the testing campaign will be of support to the implementation of the free fall mode in flight, eventually testing the data analysis techniques developed for the mission on real data.

*A* ***Bill****, per la dedizione, per avermi donato il suo punto di vista, per la pazienza e gli infiniti schizzi ed appunti su carta. Grazie!*
*A* ***Rita****, perchè fare fisica sorridendo (e ridendo) è un dono speciale.*
*Ad* ***Antonella****, per avermi 'adottato', per i giovedì al mercato, per la compagnia e perchè è un'amica speciale.*
*A* ***Stefano****, il nostro capitano!*
*Alla mia* ***famiglia****, perchè ci sono sempre, anche ad 800 km di distanza.*
*A* ***Paola*** *per il sostegno, le cenette e per essere sempre nel mio taschino.*
*A* ***Manu e Dani****, la mia famiglia trentina.*
*Ad* ***Andrea, Daniel, Paolo e Valerio****, ottimi compagni di viaggio e non solo.*
*Al* ***Lab BT****, perchè è una casa accogliente.*
*Al* ***Pendolo 1 TM****, perchè ha un 'modo' davvero swing!*